\documentclass[%
 aip,
 amsmath,amssymb,
 reprint,%
]{revtex4-1}

\usepackage{graphicx}
\usepackage{dcolumn}
\usepackage{bm}
\usepackage{xcolor}
\usepackage[utf8]{inputenc}
\usepackage[T1]{fontenc}
\usepackage{mathptmx}
\usepackage{etoolbox}
\usepackage{amsmath}
\usepackage{lipsum}
\usepackage{graphics}
\makeatletter
\def\@email#1#2{%
 \endgroup
 \patchcmd{\titleblock@produce}
  {\frontmatter@RRAPformat}
  {\frontmatter@RRAPformat{\produce@RRAP{*#1\href{mailto:#2}{#2}}}\frontmatter@RRAPformat}
  {}{}
}%
\makeatother
\begin{document}

\preprint{AIP/123-QED}

\title{Rijke tube: A nonlinear oscillator}


\author{Krishna Manoj}
 \affiliation{
 Department of Mechanical Engineering, Massachusetts Institute of Technology, Cambridge, MA 02139, USA
 }
 \author{Samadhan A. Pawar}%
 \affiliation{Department of Aerospace Engineering, Indian Institute of Technology Madras, Chennai 600036, India}
  \author{ J\"urgen Kurths}%
 \affiliation{
 Potsdam Institute for Climate Impact Research, Potsdam, Germany
 }
 \affiliation{
 Department of Physics, Humboldt University of Berlin, Berlin, Germany
 }
\author{R. I. Sujith}%
  \affiliation{Department of Aerospace Engineering, Indian Institute of Technology Madras, Chennai 600036, India}
  \email{sujith@iitm.ac.in}

\date{\today}

\begin{abstract}
Dynamical systems theory has emerged as an interdisciplinary area of research to characterize the complex dynamical transitions in real-world systems. Various nonlinear dynamical phenomena and bifurcations have been discovered over the decades using different reduced-order models of oscillators. Different measures and methodologies have been developed theoretically to detect, control, or suppress the nonlinear oscillations. However, obtaining such phenomena experimentally is often challenging, time-consuming, and risky, mainly due to the limited control of certain parameters during experiments. With this review, we aim to introduce a paradigmatic and easily configurable Rijke tube oscillator to the dynamical systems community. The Rijke tube is commonly used by the combustion community as a prototype to investigate the detrimental phenomena of thermoacoustic instability. Recent investigations in such Rijke tubes have utilized various methodologies from dynamical systems theory to better understand the occurrence of thermoacoustic oscillations, their prediction and mitigation, both experimentally and theoretically. The existence of various dynamical behaviors has been reported in single as well as coupled Rijke tube oscillators. These behaviors include bifurcations, routes to chaos, noise-induced transitions, synchronization, and suppression of oscillations. Various early warning measures have been established to predict thermoacoustic instabilities. Therefore, this review paper consolidates the usefulness of a Rijke tube oscillator in terms of experimentally discovering and modeling different nonlinear phenomena observed in physics; thus, transcending the boundaries between the physics and the engineering communities.
\end{abstract}

\maketitle

\begin{quotation}
The occurrence of various nonlinear self-sustained oscillations in different systems observed in our day-to-day life has been studied from a dynamical systems perspective. Many such systems that mesmerize the human mind have been modeled as an oscillator. Theoretical reduced-order models have been developed for oscillators, e.g., Stuart-Landau, Van der Pol, Rossler, Lorenz, etc., to study and predict a plethora of dynamical behaviors observed in natural systems. The experimental validations of these theoretically discovered dynamical phenomena however are limited to oscillators involving electronic circuits including Chua’s circuit, lasers, pendulums, chemical oscillators, etc. In the present study, we introduce the Rijke tube as a paradigmatic member to the family of nonlinear oscillators. Rijke tube systems are prototypical thermoacoustic oscillators and have been extensively studied to understand the occurrence of complex thermoacoustic instabilities observed in gas turbines and rocket engines used for propulsion and power generation applications. Recent studies on the Rijke tube have shown the existence of numerous dynamical states, bifurcations, and nonlinear behaviors such as synchronization and oscillation quenching in coupled systems that are often observed in nonlinear oscillators. Different nonlinear measures have been used to predict critical transitions in a Rijke tube system. Therefore, through this review paper, we introduce the dynamical systems community to the Rijke tube oscillator to experimentally validate their novel theoretical findings, and thus bridge the gap between the physics and the engineering communities.
\end{quotation}

\section{\label{sec1}Introduction}

Most observations in our daily life can in one way or the other be studied from a dynamical systems perspective. Any system whose behavior evolves with time, such as a moving bicycle \cite{aastrom2005bicycle}, the flowing riverbed \cite{robert2014river,kundu2002fluid}, flocks of birds flying in the sky \cite{o2017oscillators, nagy2010hierarchical, hemelrijk2012schools}, the changing climate \cite{zhisheng2015global}, the beating heart \cite{ivanov1999multifractality}, varying population densities of animals \cite{royama2012analytical, turchin2013complex}, and many other systems that we come across in our day-to-day life can be considered as dynamical systems. These systems can be mathematically modeled through differential equations by applying various physical laws \cite{lakshmanan2012nonlinear}. For example, the motion of an object can be described using Newton’s laws of motion, planetary dynamics using gravitational laws, and the power output from electronic circuits using electrostatic and electrodynamical system equations \cite{griffiths2005introduction}. Essentially, any system that evolves with time can be investigated from a dynamical systems perspective, using a governing equation of the form:
\begin{equation}
     \dot{x} = f(x,t),
\end{equation}
where $x$ refers to the state variable (or a vector of state variables) of the system and $f$ indicates a function that governs the evolution of the variable in time. The dynamical behavior of a system can manifest as various dynamical states. For instance, in the trivial case when $f(x,t)=0$, the system is always considered to be at a steady state where the dynamics of the state variable $x$ saturates to a fixed value. On the contrary, when $f(x,t)$  constitutes linear and nonlinear terms, the behavior of the system becomes complicated and exhibits a wide variety of dynamical states. One commonly observed dynamical state is the self-sustained oscillatory state wherein the dynamical behavior of a state variable shows fluctuations about a mean value. The occurrence of various self-sustained nonlinear behaviors has been studied from a dynamical systems perspective by modeling the system as a network of oscillators \cite{friedland2012control,strogatz2004sync}.

Oscillations fascinate the human mind from a very young age, starting from the joyful oscillations in a swing to the monotonous motion of a pendulum bob. Our knowledge on such oscillations grows as we learn about the spring-mass systems from physics textbooks \cite{kovacic2020nonlinear,hagedorn1981non}. The simple back and forth repeated fluctuations turn into intricate linear and nonlinear differential equations. 
Oscillations can vary from being a mind-soothing tone from musical instruments \cite{mcintyre1983oscillations} such as a flute or the vibrations in the string of a guitar, to the loud destructive sounds from the roaring of gas turbines or rocket engines \cite{culick2006unsteady}. In biological systems, oscillations can be associated with the sustenance of life in the form of respiratory cycles, neural networks in the brain, rhythmic beating of the heart, etc. \cite{ouguztoreli1976effects,cardon1970oscillations}. Furthermore, hazardous disease spread models \cite{duncan1997dynamics} and structural oscillations in bridges \cite{green2006failure,strogatz2005crowd} and skyscrapers \cite{singhose1997method,ucke2008oscillating} are also represented by oscillators. Oscillations, therefore, are ubiquitous in nature and engineering, and their characteristics and desirability vary from system to system. 

\begin{figure}
\centering
\includegraphics[width=0.49\textwidth]{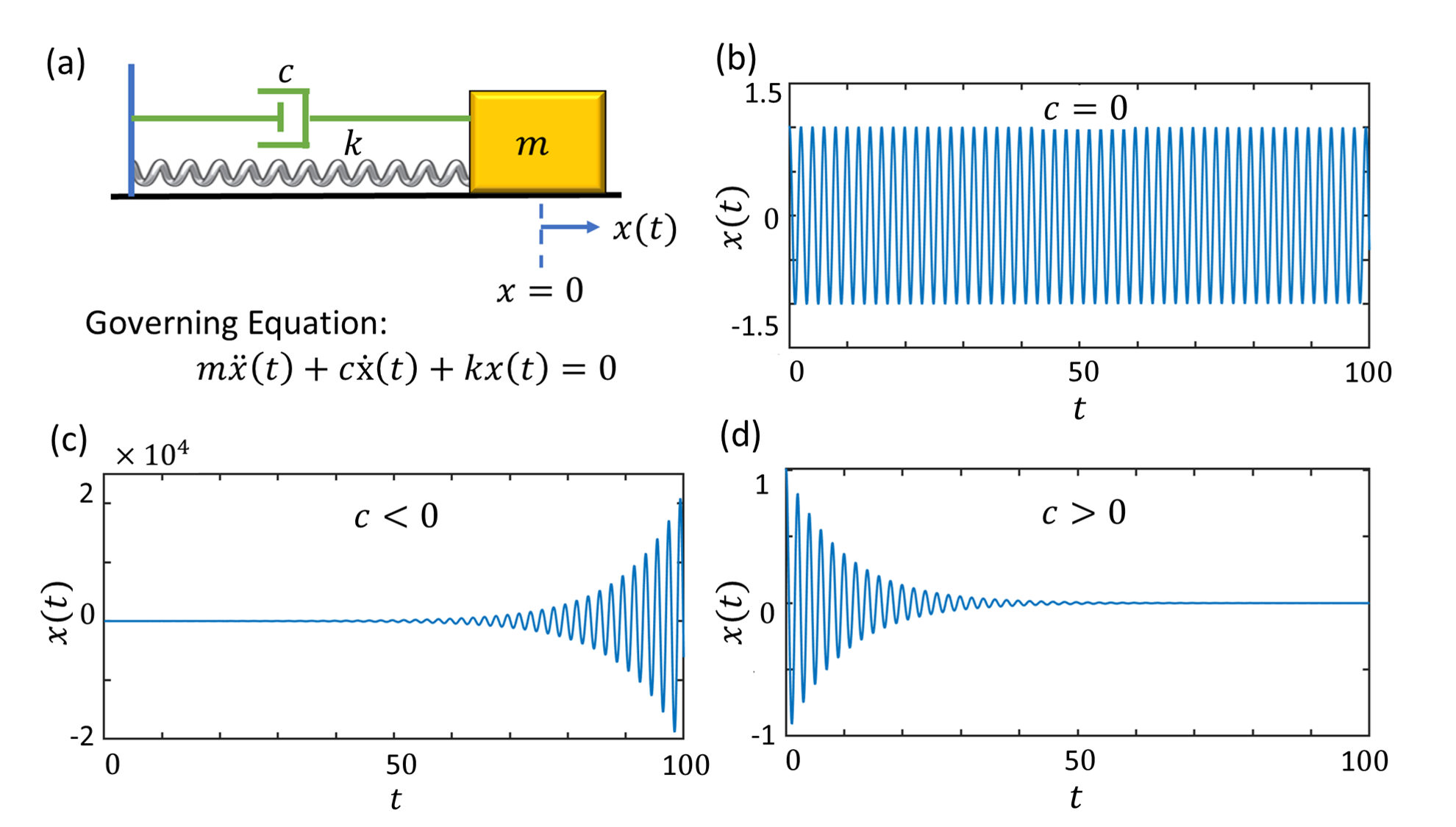}
\caption{\label{fig:1}(a) Schematic of a spring-mass system with state variable, $x(t)$, defined as the distance from mean position $x=0$, mass $m$, damping coefficient $c$, and spring constant $k$. Dynamics of the state variable shows (b) an oscillatory behavior for $c=0$, (c) a linearly unstable behavior for $c=-0.2<0$ and (d) a stable steady state for $c=0.2>0$. The other parameters are kept at $m=1$, $k=10$ and $x(0)=1$.}
\end{figure}

Although the nature of these aforementioned systems may seem very different, the inherent equation behind the oscillations remains the same. For example, let us consider a spring-mass-damper system, governed by the following equations:
\begin{equation}
    m \ddot{x} + c \dot{x} + k x =0
\end{equation}
where $m$ is the mass of the system, $c$ and $k$ are the damping coefficient and the spring constant, respectively (Fig. \ref{fig:1}a). The oscillations in the system are driven due to the restoring force ($kx$) of the spring, while the damper ($c\dot{x}$) damps the oscillations. 
 Simple harmonic oscillations are observed in the undamped case for $c=0$ (Fig. \ref{fig:1}b), while negatively damped oscillations are observed for $c<0$ (Fig. \ref{fig:1}c). 
For $c>0$, the system ultimately attains a steady state in time (Fig. \ref{fig:1}d), where $x=0$ can be referred to as an equilibrium state (or fixed point). To analytically obtain the equilibrium points, we need to set all the equations of linearized time derivatives of state variables to zero, where the roots of these equations indicate fixed points. The stability of these fixed points can be obtained by computing the first derivatives of these linearized equations (i.e., $f’(x)$) about the fixed points. Depending on the value of $f’(x)$, i.e., $f’(x)<0$ or $f’(x)>0$, the fixed point is classified as stable or unstable, respectively. A stable fixed point tends to attract all the neighbouring trajectories towards it - similar to a sink. In contrast, an unstable fixed point tends to repel all the trajectories nearby - similar to a source. 

Apart from fixed points, there exists another set of attractors and repellers for the trajectories in the phase space for systems that exhibit oscillatory behavior. These attractors are often classified as regular or strange \cite{lakshmanan2012nonlinear,landa2001regular,puu2013attractors}. Regular attractors possess a distinct closed-looped shape for a particular dynamical state, whose examples include limit cycle and frequency-locked oscillations. A regular attractor is also observed for quasiperiodic oscillations, where the trajectory is bounded by a torus in the phase space. In contrast, strange attractors are observed for chaotic oscillations  \cite{strogatz1994nonlinear,landa2001regular,gleick1987chaos}. Such oscillations are deterministic and exhibit sensitive dependence on the change in initial conditions. The dimension of regular attractors is an integer number, while that of strange attractors is a non-integer number \cite{landa2001regular}. Regular oscillations are often modelled using Vand der Pol or Stuart-Landau oscillators, while chaotic oscillations are modeled using Lorenz or Rössler oscillators \cite{strogatz1994nonlinear,lakshmanan2012nonlinear,hilborn2000chaos,nayfeh2008applied}.  

Extensive research in dynamical systems theory has been carried out to characterize the nonlinear behavior of oscillators. Bifurcation analysis is one commonly used approach developed to study the occurrence of qualitative changes in the behavior of a system of oscillators on the variation of a control parameter \cite{strogatz1994nonlinear,hale2012dynamics}. These qualitative changes include emergence or change in the stability of fixed points \cite{wiggins2013global}, presence of tipping \cite{scheffer2020critical}, bistability and hysteresis \cite{berglund1998adiabatic}, etc. Other approaches that have been developed to detect the dynamical properties of a system include Poincar\'{e} map, recurrence plots, calculating measures such as Lyapunov exponents, correlation dimension, etc. \cite{marwan2007recurrence,parker1987chaos}. In addition to studies on characterizing the dynamical behavior of individual oscillators, many studies have been devoted towards understanding the coupled dynamics arising due to the interaction of two or more oscillators. Furthermore, several studies have focused on developing various control strategies based on self-coupling, mutual coupling, and external forcing to control or quench self-sustained oscillations in coupled systems \cite{balanov2009synchronization,pikovsky2003synchronization,strogatz1993coupled,awrejcewicz1991bifurcation}.

Over the last three decades, various researchers have used different reduced-order nonlinear models and coupling schemes to analyze the behavior of coupled oscillators. Towards this purpose, commonly used oscillator models include the Van der Pol, Lorenz, Stuart-Landau, Duffing, Chua, relay oscillators, etc. \cite{lakshmanan2011dynamics,balanov2009synchronization,madan1993chua,kovacic2011duffing,thompson2002nonlinear}. Various coupling schemes \cite{zou2021quenching} that have been invented, including time-delay coupling, dissipative coupling, relay coupling, conjugate coupling, environment coupling, etc. A coupled system of such oscillators exhibits a plethora of dynamics depending on the number of oscillators and their coupling scheme \cite{manoj2021experimental,wickramasinghe2013spatially,wickramasinghe2013synchronization}. These dynamical states include homogenous states such as synchronization \cite{pikovsky2003synchronization,lakshmanan2011dynamics}, amplitude or oscillation death \cite{saxena2012amplitude,koseska2013oscillation,zou2021quenching}, symmetry-breaking states such as chimera \cite{abrams2004chimera,mondal2017onset,hart2016experimental}, weak chimera \cite{manoj2019synchronization,ashwin2015weak,wojewoda2016smallest} and clustering \cite{pecora2014cluster}, etc. However, the experimental evidence of coupled dynamical behaviors of these oscillators are limited to a few systems including electronic circuits \cite{crowley1986electrically,carroll1995nonlinear}, lasers \cite{wunsche2005synchronization,erzgraber2009locking}, chemical oscillators \cite{nkomo2013chimera,tinsley2012chimera}, and thermo-fluid systems \cite{manoj2018experimental,manoj2019synchronization}. Although these experimental systems provide limited controllability and a reduced number of control parameters, they are extensively used due to the demand for experimental verification.

In the present review, we introduce the Rijke tube, a prototypical thermoacoustic oscillator, as a paradigmatic oscillator in the family of the aforementioned nonlinear oscillators. A typical thermoacoustic system consists of a heat source placed at a particular location inside a duct. The heat source comprises a single flame, multiple flamelets, or an electrically heated wire mesh.  In such systems, positive feedback between the acoustic field in the duct and the heat release rate fluctuations across the heat source often lead to the occurrence of large amplitude self-sustained acoustic oscillations, known as thermoacoustic instability.  Earlier review papers on Rijke tubes in the engineering literature \cite{raun1993review,feldman1968reviewa,feldman1968reviewb,sarpotdar2003rijke,bisio1999sondhauss,mcmanus1993review,oran1985chemical} highlight the application and relevance of such systems in the aerospace and rocket industry from the perspective of investigating mechanisms and control of thermoacoustic instability. Here, we will cover numerous recent experimental and theoretical studies performed on Rijke tube systems in the last decade from a dynamical systems perspective. These studies have investigated various dynamical transitions (bifurcations) leading to the occurrence of thermoacoustic instability, different nonlinear states observed during such instabilities, and a variety of methodologies based on coupling and external forcing used to mitigate these instabilities, and measures to predict the occurrence of thermoacoustic instability in the system. Similar studies on characterizing and controlling the dynamical behavior of oscillators are usually performed with phenomenological models in the dynamical systems literature. Here, we aim at attracting the attention of the dynamical systems community to the Rijke tube oscillator, which is known only in the thermoacoustic community, with its potential applications in advancing experimental research on nonlinear oscillators. 

Rijke tube systems are rather simple in design, easy to fabricate and operate, and also allow us to perform strictly controlled experiments. Furthermore, the presence of numerous control parameters in such systems and their individual control facilitate the investigation of various phenomena observed in general dynamical systems theory. The effect of external fluctuations (both harmonic and stochastic) on the nonlinear behavior of a bistable oscillator can be easily demonstrated through experiments by installing various additional external subsystems such as actuators. Coupled phenomena such as synchronization and amplitude death observed due to the interaction of oscillators can be easily studied and verified by connecting two or more Rijke tubes using simple tubes. Thermoacoustic instability in Rijke tubes often portrays itself as dancing flames along with a rhythmic sound production during the states of limit cycle, quasiperiodicity, frequency-locked and chaotic oscillations \cite{kabiraj2012bifurcations,kashinath2014nonlinear,vishnu2015role}.  

The outline of the paper is as follows. In Sec. \ref{sec2}, we describe the discovery of thermoacoustic oscillations in the original Rijke tube system and various advances that have been made in the study of Rijke tubes over the years in brief. Subsequently, we explain various dynamical states exhibited by the Rijke tubes and, thereby, justify the claim of it being an excellent example for an oscillator. We also present different types of Rijke tube systems and briefly describe each of their experimental setups. In Sec. \ref{sec3}, we present the various bifurcations exhibited in a Rijke tube oscillator by varying the control parameter along with a description of the dynamical states exhibited by the system. This is followed by a discussion on various routes to chaos observed in Rijke tube systems. Section \ref{sec4} describes bistability along with different noise-induced dynamical behaviors, such as coherence resonance, stochastic bifurcations, and pulsed instabilities. The interaction between coupled Rijke tube oscillators leading to synchronization and phase-flip bifurcation, and different states of forced synchronization of the Rijke tube oscillator are presented in Sec. \ref{sec5}, followed by a discussion on control strategies implemented to mitigate thermoacoustic instability in Sec. \ref{sec6}. Finally, in Sec. \ref{sec7}, we conclude the study and provide insights on possible future advancements and developments in the field along with its applications to other streams of science and technology. Hence, we summarize relevant works considering the oscillatory behavior of the Rijke tube and the various dynamical behaviors exhibited by the oscillator. Before we dive into delineating the simple experimental Rijke tube as an oscillator and explaining its distinguished characteristics, let us explore the various types of Rijke tube oscillators.

\section{\label{sec2} A brief history on Rijke systems}

\subsection{Thermoacoustic instability and its challenges}
The occurrence of thermoacoustic instability in rocket and gas turbine engines has hindered the development of the energy and aviation industry as well as the space and defense programs for decades \cite{culick2006unsteady,juniper2018sensitivity,lieuwen2012unsteady}. The issue of thermoacoustic instability emerged with deadly consequences in the rocket industry especially in the 1960’s during the testing phase of the Apollo launch \cite{fisher2009remembering,sujith2016non}. When testing the F1 engine for powering the Saturn V rocket, the Apollo team at NASA found that the gases in the engine developed violent pressure oscillations (known as “combustion instability” or “thermoacoustic instability” in the parlance of propulsion engineers), which causes significant harm to the engine. Although combustion instability refers to stable limit cycle oscillations, engineers refer to it as ‘instability’ or the ‘unstable state of operation' due to its disastrous consequences. The consequences of thermoacoustic instability include loss of structural integrity resulting from the increased vibrations, overwhelming the thermal protection systems, damage to electronic systems including guidance and navigation systems, performance losses due to thrust oscillations, loss of controllability of the vehicle, and sometimes even failure of the mission, causing an impediment in the engine development amounting to billions of dollars of losses annually to engine manufacturers \cite{lieuwen2005combustion,culick2006unsteady,sujith2016non,sujith2020dynamical}. 

Scientists from all around the world have invested considerable time and effort to suppress thermoacoustic instability and thereby reduce the financial losses associated with it. Various theoretical and experimental studies on thermoacoustic instability have been performed over the years to understand the thermoacoustic phenomena, characterize the various dynamical behaviors, and develop methodologies to suppress these large amplitude thermoacoustic oscillations. To understand the complex interactions between subsystems that lead to the occurrence of thermoacoustic instability, it is essential to begin the process from a simple prototypical system and gradually work our way towards more complex systems by adding individual complexities. Hence, fundamental research on thermoacoustic instability began on prototypical thermoacoustic systems known as Rijke tubes. Next, we present a historical perspective on the development of Rijke tube systems.

\subsection{History of Rijke tube systems}
\begin{figure}
\centering
\includegraphics[width=0.5\textwidth]{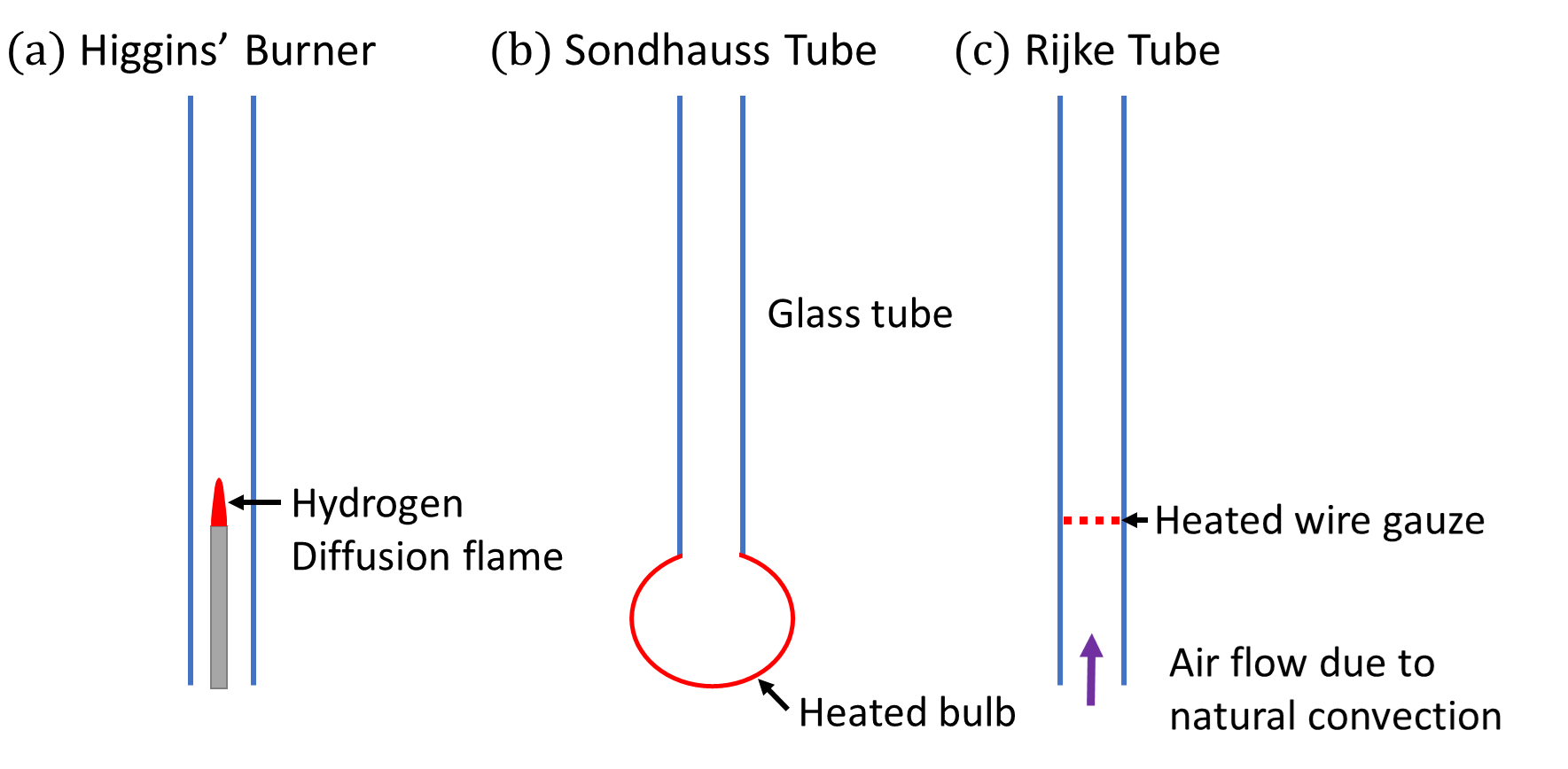}
\caption{\label{fig:2}Schematic of the experimental setups used for investigating thermoacoustic instability in the pioneering studies by (a) Higgins \cite{higgins1802sound}, (b) Sondhauss \cite{sondhauss1850ueber}, and (c) Rijke \cite{rijke1859lxxi}.}
\end{figure}

Higgins \cite{higgins1802sound} was the first to report the generation of combustion-driven acoustic oscillations by a hydrogen diffusion flame enclosed in a tube (Fig. \ref{fig:2}a). He referred to this phenomenon as the ‘singing flame’. However, recent reports \cite{noda2013thermoacoustic,ueda1974ugetsu,lawn2018common} points to the existence of such oscillations prior to Higgins in a devise called ‘Kibitsunokama' (or the iron bowl of Kibitsu) which was mentioned by a Buddhist monk in his diary in 1568. Subsequently, Sondhauss \cite{sondhauss1850ueber} observed the occurrence of acoustic oscillations in a glass tube with a heated closed bulb at one end and the other end open to the atmosphere (Fig. \ref{fig:2}b). Later, in 1859, Rijke \cite{rijke1859lxxi} discovered the production of a tonal sound from a metal gauze, heated using a burner in a vertical duct. Such a setup using the vertical duct with the concentrated heat source located in the lower half was thereafter referred commonly as the ‘\textit{Rijke tube}’ (Fig. \ref{fig:2}c). He observed the production of a loud sound soon after the removal of the flame from the duct, which gradually decayed as the gauze cooled. He inferred that the production of sound was due to the direct conduction of heat from the metal gauze to the surrounding air in the tube. Rijke further observed that the sound was absent when the tube is placed horizontally or when the gauze is located in the upper half of the tube. He reasoned that the upward flow of air in the vertical tube, due to the natural convection of air, is necessary for the production of sound. The rapid expansion of the air as it passes through the hot gauze and the gradual contraction after engenders the sound in the tube \cite{rijke1859lxxi,sarpotdar2003rijke}. However, his conclusion was incomplete and was unable to explain the relevance of locating the wire gauze in the lower half of the Rijke tube in the production of sound.

Subsequent analysis of heat-driven oscillations by Rayleigh filled this void \cite{rayleigh1878explanation,rayleigh1878instability,rayleigh1896theoretical,rayleigh1896theory}. He proposed that the addition of heat at the point of highest compression or the extraction of heat at the point of highest expansion in an acoustic cycle promoted the generation of tonal sound waves in the system. On the other hand, the heat addition during the maximum expansion and the heat extraction during the maximum compression resulted in the damping of acoustic oscillations in the system. Thus, to generate thermoacoustic instability, both acoustic pressure and heat release rate fluctuations should be in-phase with each other. This description of the condition for acoustic driving by a heat source is now popularly referred to as the \textit{Rayleigh criterion} \cite{poinsot2005theoretical}, as it explains the promotion of thermo-acoustic oscillations in a Rijke tube\cite{sarpotdar2003rijke}. Subsequently, the Rayleigh criterion was generalized to account for the acoustic losses in the system, whose expression can be given as follows\cite{putnam1971combustion,chu1965energy}, 
\begin{equation}
\label{eqtn:RC}
  \frac{1}{T} \int_0^T \int_0^V p'(t) \dot{q}'(t) dVdt > \text{Acoustic} ~ \text{damping},   
\end{equation}
where $p'(t)$  and  $\dot{q}'(t)$ correspond to the acoustic pressure and the global heat release rate fluctuations in the flame, $t$, $V$ and $T$ correspond to the time variable, combustor volume, and the time period of oscillations, respectively. Thus, thermoacoustic instability is established in a system only if the acoustic driving caused by the unsteady heat release rate fluctuations overbalances the acoustic damping in the system. A detailed description of the history and the development of the Rijke tube can be found in refs. \cite{putnam1971combustion,sujiththermoacoustic,rayleigh1896theory,reynst1961pulsating, feldman1968reviewa,feldman1968reviewb, raun1993review,sarpotdar2003rijke}. Hereon, we will discuss various modern variants of Rijke tube configurations developed recently for studying the nonlinear behavior of a Rijke tube oscillator.

\subsection{Types of Rijke tube systems}
\subsubsection{Horizontal Rijke tube}

\begin{figure}
\centering
\includegraphics[width=0.5\textwidth]{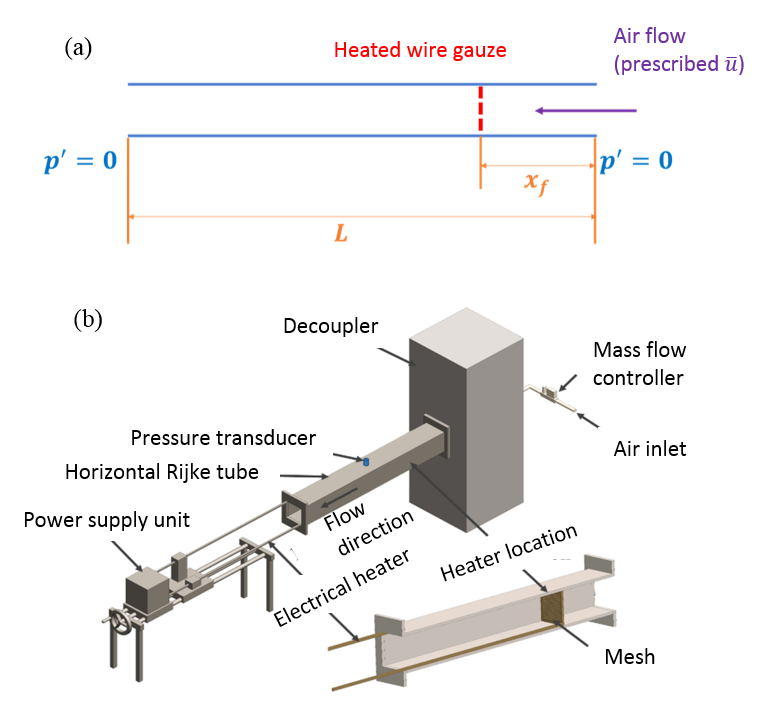}
\caption{\label{fig:3}Schematics illustrating (a) the boundary conditions of a duct that is open at both the ends and (b) the experimental setup of a horizontal Rijke tube. In (a), $x_f$ indicates the location of the heater wire gauze from the inlet, $L$ is the duct length, $p'$ is the acoustic pressure fluctuations, and $\Bar{u}$ is the flow velocity. (b) Adapted with permission from Tandon \it et al.\rm \cite{tandon2020bursting}. }
\end{figure}

The horizontal Rijke tube is a recent variant of the original Rijke tube that we discussed before. It consists of a horizontal duct with an electrically heated wire mesh as a compact heat source, located at the quarter location from the inlet of the duct (Fig. \ref{fig:3}). An external power supply is used to control the heat input to the wire mesh; thus, it controls the heat release rate fluctuations in the system. As mentioned above, the natural convection of the air flow is necessary for the generation of acoustic oscillations in the vertical configuration of a traditional Rijke tube. It, therefore, causes an intrinsic dependency between the heat release rate fluctuations in the flame and the upward air flow. As a result, it is difficult to obtain an independent control over the supply of air and the generation of heat release rate fluctuations in the traditional vertical Rijke tube. The ingenious invention of the horizontal Rijke tube by Matveev \cite{matveev2003energy} in 2003 brought a radical change, making the research performed on Rijke tubes far less complicated. In this system, a continuous mean air flow is established using external devices such as a blower \cite{matveev2003energy,mariappan2015experimental} or a compressor \cite{etikyala2017change}. This decouples the mean flow and the heat release rate fluctuations in the system, that in turn, helps in independently studying the effect of an increase in the mean air flow rate in the Rijke tube system. This simplification in the setup further enables us to evade the need to model natural convention (seen in traditional Rijke tubes), facilitating much easier modeling of Rijke tube systems.

The duct used in the horizontal Rijke tube is long and maintains an open-open boundary condition for the acoustic standing wave established inside the duct. Mathematically speaking, the total pressure $p(x,t)$ in a Rijke tube can be described as $p(x,t)=\Bar{p} +p' (x,t)$, where $\Bar{p}$ is the atmospheric pressure, $p'(x,t)$ are the acoustic pressure oscillations, and $x$ and $t$ are the space and time variables. At both the ends of the Rijke tube, we have $p(x=0,t)=p(x=L,t)=\Bar{p}$, where $L$ is the length of the duct. Therefore, at the boundary, we observe $p'(x=0,t)=p'(x=L,t)=0$. This boundary condition where the acoustic pressure is zero is referred to as an acoustically open boundary condition. On the other hand, in the case of acoustically closed boundary conditions, the acoustic velocity ($u'$) is zero at the boundary \cite{munjal1987acoustics,kinsler2000fundamentals,rienstra2004introduction}. 

Furthermore, air is passed through a decoupler prior to entering the system. The decoupler is a large chamber used to dampen the fluctuations in the air flow and supply a steady flow into the system. The acoustic pressure oscillations established in the Rijke tube can be measured using microphones/piezoelectric transducers mounted on the duct. In a horizontal Rijke tube, we can vary different control parameters, such as the heater power supplied to the mesh, the heater location in the duct, and the mass flow rate of air, to study the occurrence of limit cycle oscillations (i.e., thermoacoustic instability) in the system. In addition, we can study the effect of external perturbations (e.g., noise or harmonic forcing), facilitated through loudspeakers, on the transition of the system behavior from a steady state to limit cycle oscillations. Electrical heaters are also used in the vertical configurations of Rijke tubes in recent theoretical studies by Andrade \it et al. \rm \cite{de2018backstepping,de2017boundary} and Wilhelmsen and Meglio \cite{wilhelmsen2020observer}.  

\subsubsection{Vertical Rijke tube burners}

In addition to the previously discussed Rijke tube configuration consisting of a heated wire mesh as a compact heat source, another widely used configuration uses the flame as a compact heat source. Here, the flame indicates the region in the space where chemical reactions take place that converts cold unburnt reactants (i.e., fuel and air) into hot burnt products. By saying compact, we mean that the length of the heat source (i.e., the flame or the mesh) is much smaller than the wavelength of the acoustic standing wave established in the duct (i.e., $l_{flame}<<\lambda$). We refer to such systems as Rijke tube burners in this paper. Depending on how the fuel and air enter into the combustion chamber, the type of flame in vertical Rijke tube burners is usually classified as a diffusion flame or a premixed flame. 

\begin{figure}
\centering
\includegraphics[width=0.45\textwidth]{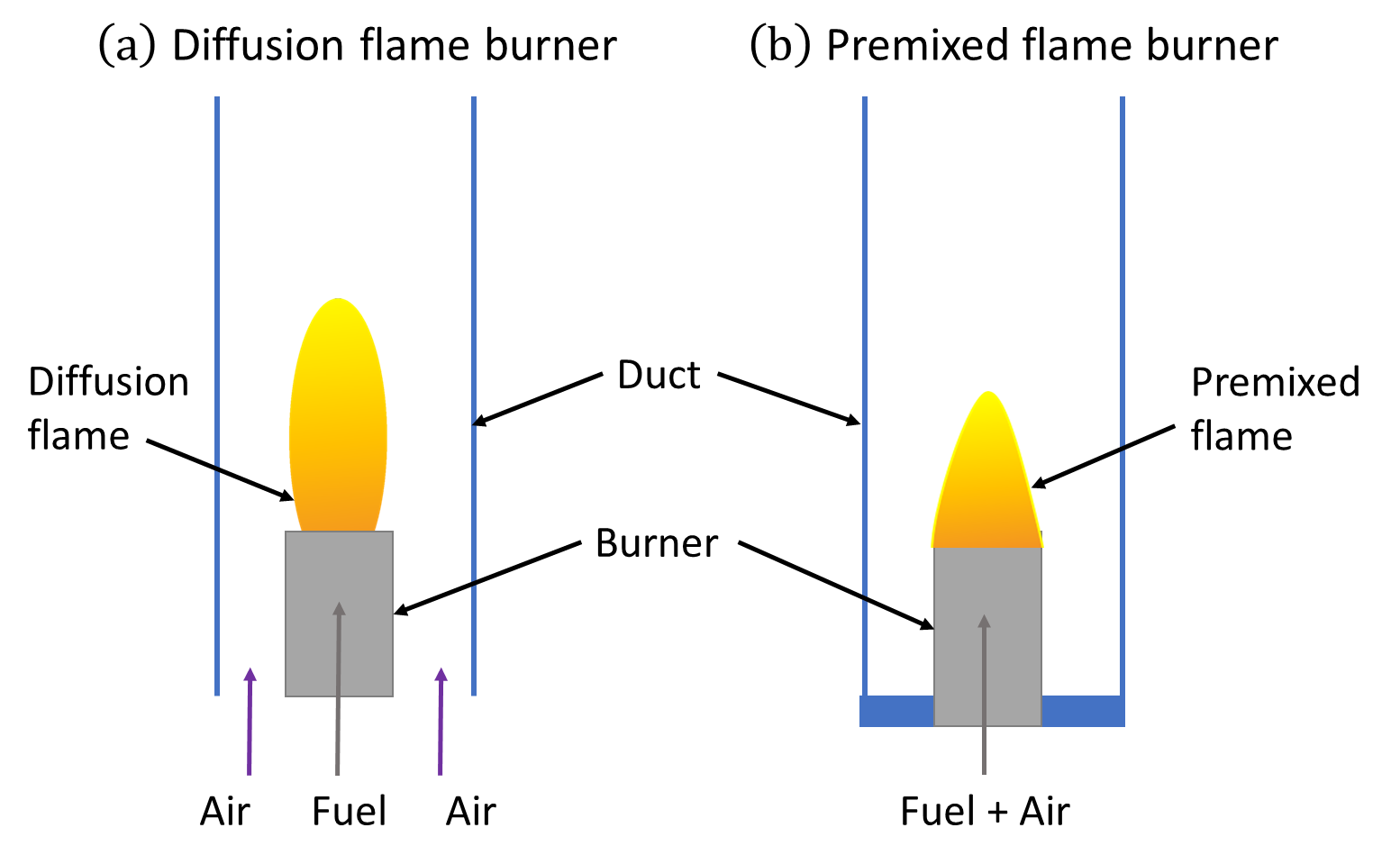}
\caption{\label{fig:4}Schematic representation of (a) a diffusion and (b) a premixed flame vertical Rijke tube burner.}
\end{figure}

In a diffusion flame Rijke tube burner (Fig. \ref{fig:4}a), the fuel and the oxidizer (air) are supplied through separate feed lines in the Rijke tube. The fuel is supplied through the burner tube, whereas the oxidizer is supplied through the annular space between the burner tube and the Rijke tube. Both the fuel and the oxidizer enter the system via separate decouplers that suppress the fluctuations, providing a quiet flow. The diffusion flame is established at the interface where the fuel (in gaseous form) meets the air. Previous experimental studies showed that a conical laminar flame \cite{jegadeesan2013experimental} or a turbulent flame \cite{murugesan2018physical} can be established in this type of burner. 

On the other hand, in a premixed flame Rijke tube burner (Fig. \ref{fig:4}b), the fuel and air are injected into a common mixing chamber and this well-mixed fuel-air mixture is then fed to the burner tube through a decoupler. The fuel-air mixture is ignited in the system using a spark plug or a small pilot flame. In this setup, we can study the interaction of the acoustic field with different configurations of laminar flames including conical flame \cite{kabiraj2012route,guan2020intermittency}, V-flame \cite{vishnu2015role,mukherjee2015nonlinear, durox2009experimental}, and also multiple conical flames \cite{durox2009experimental, kabiraj2012route, kasthuri2019bursting}. 

In experiments with such Rijke tube burners, we can vary different control parameters, such as the equivalence ratio (i.e., the ratio of the actual fuel/air ratio to the ideal/stoichiometric fuel/air ratio for combustion) and the location of the flame in the tube, to study the occurrence of thermoacoustic instability in the system. The acoustic pressure in the duct can be measured using condenser microphones or piezoelectric transducers. The heat release rate fluctuations in the flame can be measured in terms of temporal or spatiotemporal fluctuations in CH* or OH* radicals \cite{sardeshmukh2017use,kiefer2008investigation,hardalupas2004local} emitted by the flame using a photomultiplier tube or a high-speed camera. 

\subsubsection{Other Rijke-type combustors}

Other than the aforementioned two basic types of Rijke-type combustors, there are a few more novel Rijke-type combustors utilized to investigate thermoacoustic instability. These include the spray combustor \cite{pawar2016intermittency,pawar2019phase}, two-heater Rijke tubes \cite{mondal2019forced,bhattacharya2020data}, loop tubes \cite{sakamoto2007improvement,sakamoto2007improvement,yu2010fishbone}, segmented Rijke tube \cite{hernandez2010transition}, and Rijke-Zhao tubes \cite{zhao2009tuned,zhao2012energy,zhao2013waste}. The spray combustor used by Pawar \it et al. \rm \cite{pawar2016intermittency,pawar2019phase} consists of needle spray injectors producing tiny droplets of fuel into the resonator tube. The droplets are further passed through a mesh unit, where secondary atomization takes place. The mesh unit also serves as a flame holder, facilitating the variation of the location of the flame in the duct. The two-heater Rijke tube \cite{mondal2019forced} consists of a horizontal aluminum duct with a square cross-section having two heating elements: a stationary primary heater and a movable secondary heater. A segmented tube \cite{hernandez2010transition} is a Rijke tube consisting of two segments having different cross-sectional areas for the upstream and the downstream of the tube. A Rijke-Zhao tube \cite{zhao2009tuned} consists of a mother tube having a Bunsen burner that splits into two daughter tubes having different lengths. We will discuss various dynamical behaviors and bifurcations observed experimentally in the aforementioned configurations of Rijke tubes in detail in Secs. \ref{sec3} to \ref{sec5}. Having discussed various experimental configurations of a Rijke tube oscillator, we next move our attention towards their mathematical modeling.

\subsection{Theoretical studies}

Ever since the discovery of thermoacoustic oscillations in the Rijke tube system, experimental studies on such systems were reported investigating various characteristics of the system. This was followed by theoretical studies to enhance the understanding of the dynamics exhibited by the system. The model based on the friction interaction between the heated gauze and the convective updraft proposed by Pflaum \cite{pflaum1909versuche} in 1909 set the beginning of such an analysis. The first attempt at quantitative modeling was put forth by Lehmann \cite{lehmann1937theorie} in 1937 based on flawed assumptions. This, in turn, led to the inaccurate conclusion that an increase in the convective velocity of the system would indefinitely increase the intensity of sound. Later, the study by Neuringer and Hudson \cite{neuringer1952investigation} adopted a different approach by starting from the equations of pressure and velocity followed by a linear perturbation analysis. In this manner, they derived the equation for the complex frequencies in a Rijke tube. Their analysis verified the experimental observation of growth of oscillations when the heater is located in the lower half of the tube and its dampening when the heater is placed in the upper half of the tube.

Subsequent studies focused on the development of flame transfer functions \cite{merk1957analysis}, deriving equations for the growth rate of the oscillations \cite{mugridge1980combustion}, and the robustness of such oscillations to changes in parameters, such as flow velocities and heater temperature \cite{madarame1981thermally}. Successful predictions of stability limits were obtained using the analysis with flame transfer functions and growth rates \cite{merk1957analysis,madarame1981thermally}. A similar theoretical analysis was performed on premixed flames by obtaining the transfer functions for a conical flame and thereby obtaining the stability limits \cite{merk1958analysis,merk1957analysisp}. A series of investigations by McIntosh and his colleagues \cite{clarke1980influence,mcintosh1985cellular,mcintosh1991pressure} investigated premixed flames using the large activation energy theory to simplify the differential equations in the flame zone. They obtained the flame response for various parameter combinations of flame location, mean flow rate, temperature, and finite tube lengths.

Another approach extensively used to model Rijke tube systems was through investigations on the Rayleigh criterion (Eq. \ref{eqtn:RC}), which requires the addition of heat during maximum compression or minimum expansion to promote oscillations, and vice versa to dampen oscillations. Putnam and Dennis \cite{putnam1953organ,putnam1954burner} theoretically verified this criterion, starting from the linearized gas equations to investigate the phasing between the heat addition and the pressure fluctuations. Clarke \textit{et al.} \cite{clarke1984shocks} obtained an analogy of the phasing relation using a piston configuration, where they concluded that driving of oscillations can be obtained when the phase difference between the heat and the pressure fluctuations remains bounded between $\pm 90^{\circ}$. They inferred that damping of oscillations occurs when the phase difference between the heat and the pressure fluctuations are beyond these set limits. The study by Culick \cite{culick1987note} produced a general proof for the Rayleigh criterion applicable to both linear and nonlinear thermoacoustic oscillations. These studies, therefore, marked the beginning of investigations on utilizing the phase relations and the coupling between the pressure and the heat release rate fluctuations to understand thermoacoustic instability deeper.

A particular form of investigation of this coupling was developed by Crocco and Cheng \cite{crocco1956theory}, commonly referred to as the \textit{$n-\tau$ model}, to investigate the linear stability of combustion systems. Nicoli and Pelce \cite{nicoli1989one} derived a relation for the heat transfer between the heater and the surroundings in a low Mach number flow by taking the instantaneous mass flow rate perturbations into account. Using the modified King’s Law \cite{king1914xii}, Heckl \cite{heckl1990non} developed a correlation between the unsteady heat release rate at time $t$ to the acoustic velocity fluctuations at the time, $t-\tau$. Zinn and co-workers \cite{zinn1971application,zinn1970application,zinn1971nonlinear} and Culick and co-workers \cite{culick2006unsteady} introduced a Galerkin approach and its extension to solve the nonlinear models of the thermoacoustic system. Balasubramanian and Sujith \cite{balasubramanian2008thermoacoustic} constructed a reduced-order model for a horizontal Rijke tube exhibiting a subcritical Hopf bifurcation. This model utilizes the modified King’s law and the Galerkin technique to get the temporal evolution of the acoustic perturbations in a Rijke tube. 

Next, we will describe the derivation of a mathematical model in the time domain for a horizontal Rijke tube system from momentum and energy conservation laws proposed by Balasubramanian and Sujith \cite{balasubramanian2008thermoacoustic}. The conservation laws for a one-dimensional acoustic field are: 	
 \begin{align}
\begin{split}
    \Bar{\rho} \frac{\partial \tilde{u}'}{\partial \tilde{t}} + \frac{\partial \tilde{p}'}{\partial \tilde{x}} & =0, \\ 
    \frac{\partial \tilde{p}'}{\partial \tilde{t}}  + \gamma \Bar{p} \frac{\partial \tilde{u}'}{\partial \tilde{x}} & = (\gamma -1)\dot{\tilde{Q}}',
\end{split}
 \end{align}
where $\tilde{p}'$ and $\tilde{u}'$ are the dimensional acoustic pressure and velocity fluctuations, and $\gamma$ is the heat capacity ratio. Here, $\dot{\tilde{Q'}}$ is the heat release rate modeled using the modified King’s law and follows the empirical model suggested by Heckl \cite{heckl1990non}:  
\begin{align}
\begin{split}
    \dot{\tilde{Q}}' & = \frac{2 L_w (T_w - \Bar{T})}{S\sqrt{3}} \sqrt{\frac{\pi \lambda C_v \Bar{\rho} d_w}{2}} \\
    & \hspace{4ex} \times \left[\sqrt{\left|\frac{u_0}{3} + u'_f (t-\tau)\right|} - \sqrt{\frac{u_0}{3}}\right] \delta (\tilde{x}- \tilde{x_f}).
\end{split}
\end{align}
Here, $L_w$ refers to the equivalent length of the wire, $(T_w - \Bar{T})$ is the temperature difference between the wire and the ambient temperature, $S$ is the cross-sectional area of the duct, $\lambda, C_v,\tau$ and $\Bar{\rho}$ are the heat conductivity, the specific heat of air at constant volume, time lag accounting for the thermal inertia of the medium and mean density of air, respectively. The above sets of equations are normalized as follows:
\begin{align}
\begin{split}
    x = \frac{\tilde{x}}{L};\hspace{5ex} t = \frac{c_o}{L\tilde{t}};\hspace{5ex} u' = \frac{\tilde{u}'}{\tilde{u}}; \\ p' = \frac{\tilde{p}'}{\tilde{p}};\hspace{3ex} \dot{Q}' = \frac{\dot{\tilde{Q}}'}{c_o \tilde{p}};\hspace{3ex} M = \frac{\tilde{u}}{c_o},
\end{split}
\end{align}
using the length of the duct, $L$, speed of sound, $c_o$ to infer the non-dimensional set of equations:
\begin{align}
\begin{split}\centering
    \gamma M \frac{\partial u'}{\partial t} + \frac{\partial p'}{\partial x} &  = 0,\\
    \frac{\partial p'}{\partial t} + \gamma M \frac{\partial u'}{\partial x} & = (\gamma-1) \frac{2 L_w (T_w - \Bar{T})}{S c_o \Bar{p} \sqrt{3}} \sqrt{\frac{\pi \lambda C_v \Bar{\rho} d_w u_0}{2}}\\ 
    & \hspace{1ex} \times \left[\sqrt{\left|\frac{1}{3} + u'(t-\tau)\right|} - \sqrt{\frac{1}{3}}\right] \delta (x- x_f).
\end{split}
\end{align}
On reducing the above set of partial differential equations to ordinary differential equations using the Galerkin technique \cite{zinn1971nonlinear}, the velocity and the pressure field can be written as 
\begin{equation}
    u'= \sum_{j=1}^\infty \eta_j \cos{(j\pi x)} \hspace{2ex}
\rm and \it \hspace{2ex}
    p'=- \sum_{j=1}^\infty \frac{\gamma M}{j \pi} \dot{\eta_j} \sin{(j\pi x)}.
\end{equation}
Therefore, we obtain the following set of equations after accounting for the damping in the system:
\begin{align}
    \label{eqtn:BS}
    \begin{split}
     & \hspace{3ex} \frac{d \eta_j}{d t}  = \dot{\eta}_j, \\
    \frac{d \dot{\eta}_j}{d t} + 2 \zeta_j \omega_j \dot{\eta}_j & + k_j^2 \eta_j   = - \frac{2 K}{\gamma M} j \pi \sin{(j \pi x_f)} \\
    & \hspace{8ex} \times \left[ \sqrt{\left| \frac{1}{3} + u_f^{'} (t-\tau)\right|} - \sqrt{\frac{1}{3}}\right] ,
    \end{split}
\end{align}
where $\zeta_j = \frac{1}{2 \pi} \left[c_1 \frac{\omega_j}{\omega_1} + c_2 \sqrt{\frac{\omega_1}{\omega_j}}\right]$. Here, $c_1$ and $c_2$ are the damping coefficients and the expression of non-dimensional heater power is given by:
\begin{equation}
    K = \frac{4 (\gamma - 1) L_w}{\gamma M c_o \Bar{p} S \sqrt{3}} (T_w - \Bar{T}) \sqrt{\pi \lambda C_v u_0 \Bar{\rho} l_c}.
\end{equation}
The above set of equations (Eq. \ref{eqtn:BS}) indicate the final second-order equation and is hereafter referred to as the \textit{Balasubramanian-Sujith oscillator}.

Numerical integration of Eq. (\ref{eqtn:BS}) generates the acoustic pressure and velocity time series from the model. The variation of parameters, such as the heater power ($K$), the time lag ($\tau$), the heater location ($x_f$), and the damping coefficient ($c_1$), are utilized to study the onset of thermoacoustic oscillations. Subramanian \textit{et al.} \cite{subramanian2010bifurcation} used the method of numerical continuation to conduct a thorough bifurcation analysis, and obtained regions of global stability, instability, and bistability. Subsequently, Subramanian \textit{et al.} \cite{subramanian2013subcritical} employed the method of multiple scales to get the slow flow equations from Eq. (\ref{eqtn:BS}) and recast it into the Stuart-Landau equation \cite{garcia2012complex}. Furthermore, linear and nonlinear stability analyses were performed using the method of harmonic balance and numerical continuation \cite{orchini2016weakly}.

Magri and Juniper \cite{magri2013sensitivity,magri2014adjoint} proposed a mathematical framework of an adjoint sensitivity analysis to detect the most influential components of the system that is responsible for the occurrence of thermoacoustic instability and quantified their influence on the frequency and growth rate of oscillations. This method, in turn, helps in creating changes in a thermoacoustic system or developing passive controls that can extend its linearly stable region. They performed two types of analysis, i.e., structural sensitivity analysis and a base-state sensitivity analysis, on the adjoint equations obtained from the linear stability analysis of the Balasubramanian and Sujith \cite{balasubramanian2008thermoacoustic} model. Through a structural sensitivity analysis, they quantified the effect of feedback mechanisms possessed by any component of the system on the frequency and growth rate of oscillations. On the other hand, through a base state sensitivity analysis, they examined the effect of a change in different parameters in Eq. (\ref{eqtn:BS}) on the stability of the Rijke tube system. 

\section{\label{sec3}Nonlinear behavior of a Rijke tube oscillator}

Having discussed the various types of Rijke tube systems and models for the Rijke tube oscillators, in the present section, we characterize various bifurcations, nonlinear phenomena, and dynamical states exhibited by such oscillators due to a change in the system parameter. We start our discussion with primary bifurcations observed during the transition from the steady state to thermoacoustic instability in Rijke tube systems.   

\subsection{Hopf bifurcations}



\begin{figure}[t!]
\centering
\includegraphics[width=0.5\textwidth]{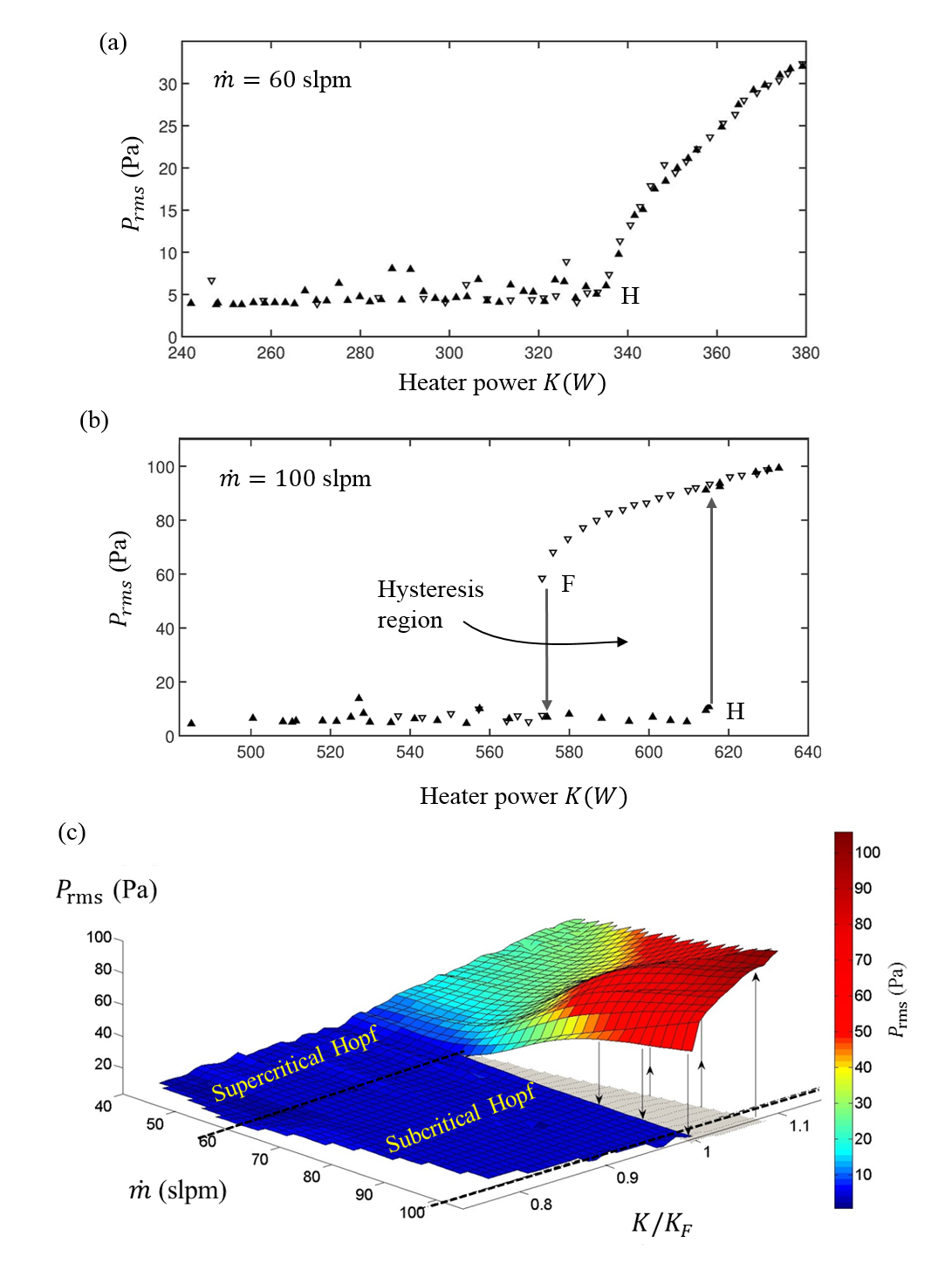}
\caption{\label{fig:5}One-parameter bifurcation diagram showing the variation of root-mean-square of acoustic pressure fluctuations ($P_{rms}$) with the heater power ($
K$) for (a) supercritical and (b) subcritical bifurcation observed in a horizontal Rijke tube. (c) Two-parameter bifurcation diagram between the mass flow rate of air ($\dot{m}$) and the normalized heater power ($K/K_F$) showing a change in the criticality of the system as the mass flow rate of air is varied in the same system, where $K_F$ indicates the heater power at the fold point. Reproduced with permission from  Etikyala \textit{et al.} \cite{etikyala2017change}. }
\end{figure}

As we know from the fundamentals of dynamical systems theory, variation in the control parameter can induce a change in the stability of fixed points (or closed orbits) that, in turn, leads to the creation of new fixed points (or closed orbits) or the destruction of the existing ones in a phase space \cite{hilborn2000chaos,strogatz1994nonlinear}. Such a qualitative change in the dynamics of the system due to a small change in the control parameter is referred to as bifurcation. There are four types of local bifurcations, i.e., saddle-node, transcritical, pitchfork, and Hopf bifurcations, which are commonly studied using reduced-order models in dynamical systems theory  \cite{strogatz1994nonlinear,guckenheimer2013nonlinear,hale2012dynamics}. Similar to the bifurcations observed in paradigmatic models \cite{marsden2012hopf,hilborn2000chaos,strogatz1994nonlinear}, most of the Rijke tube systems undergo a Hopf bifurcation due to the variation of different control parameters, such as the heater power, the heater location, and the damping coefficient \cite{kabiraj2012route,subramanian2010bifurcation,matveev2003thermoacoustic,balasubramanian2008thermoacoustic,mariappan2012theoretical,juniper2011triggering,gopalakrishnan2014influence}. During this bifurcation, a change in the control parameter leads to the transition of the system behavior from a fixed point to an oscillatory state (often limit cycle oscillations). Hopf bifurcations are primarily classified into two types: supercritical Hopf and subcritical Hopf bifurcation. 

In Fig. \ref{fig:5}, we show the Hopf bifurcation characteristics of a horizontal Rijke tube system during the transition from steady state to limit cycle oscillations (thermoacoustic instability). The bifurcation diagram is obtained by plotting the variation of the root-mean-square value of acoustic pressure fluctuations ($P_{rms}$) against the electric power supplied to the heater ($K$) in a quasi-static manner \cite{etikyala2017change}. We notice that the nature of Hopf bifurcation observed in the horizontal Rijke tube depends on the value of the mass flow rate of air supplied to the system. For low or high values of the mass flow rate of air, the system exhibits a supercritical Hopf bifurcation or a subcritical Hopf bifurcation, respectively, for the variation of heater power ($K$) as the control parameter.

For the supercritical Hopf bifurcation (Fig. \ref{fig:5}a), we observe a continuous (i.e., a second-order) transition in the pressure amplitude as the system behavior changes from steady state to limit cycle oscillations. Furthermore, the variation in the pressure amplitude is nearly the same in both the forward (increasing $K$) and the reverse (decreasing $K$) variation of the heater power. On the other hand, during the subcritical Hopf bifurcation (Fig. \ref{fig:5}b), for the forward path (increasing $K$), we observe an abrupt jump (i.e., explosive, first-order transition) in the amplitude of acoustic pressure fluctuations during the transition from steady state to limit cycle oscillations. While for the reverse path (decreasing $K$), the system remains in the limit cycle state even after the Hopf point and transitions abruptly to the steady state at a lower value of the heater power compared to the Hopf point. This bifurcation from limit cycle oscillations to steady state is called fold bifurcation \cite{hilborn2000chaos,strogatz1994nonlinear}. Thus, we notice the existence of hysteresis in the parameter space of the heater power for subcritical Hopf bifurcation (Fig. \ref{fig:5}b). Furthermore, we observe that the variation of the mass flow rate of air causes a change in criticality \cite{etikyala2017change} of the horizontal Rijke tube (Fig. \ref{fig:5}c). Here, a change of criticality refers to the switching from supercritical to subcritical Hopf bifurcation or vice versa with varying mass flow rates of air in the same system. Note that the transition between these bifurcations is gradual.

\begin{figure}
\centering
\includegraphics[width=0.5\textwidth]{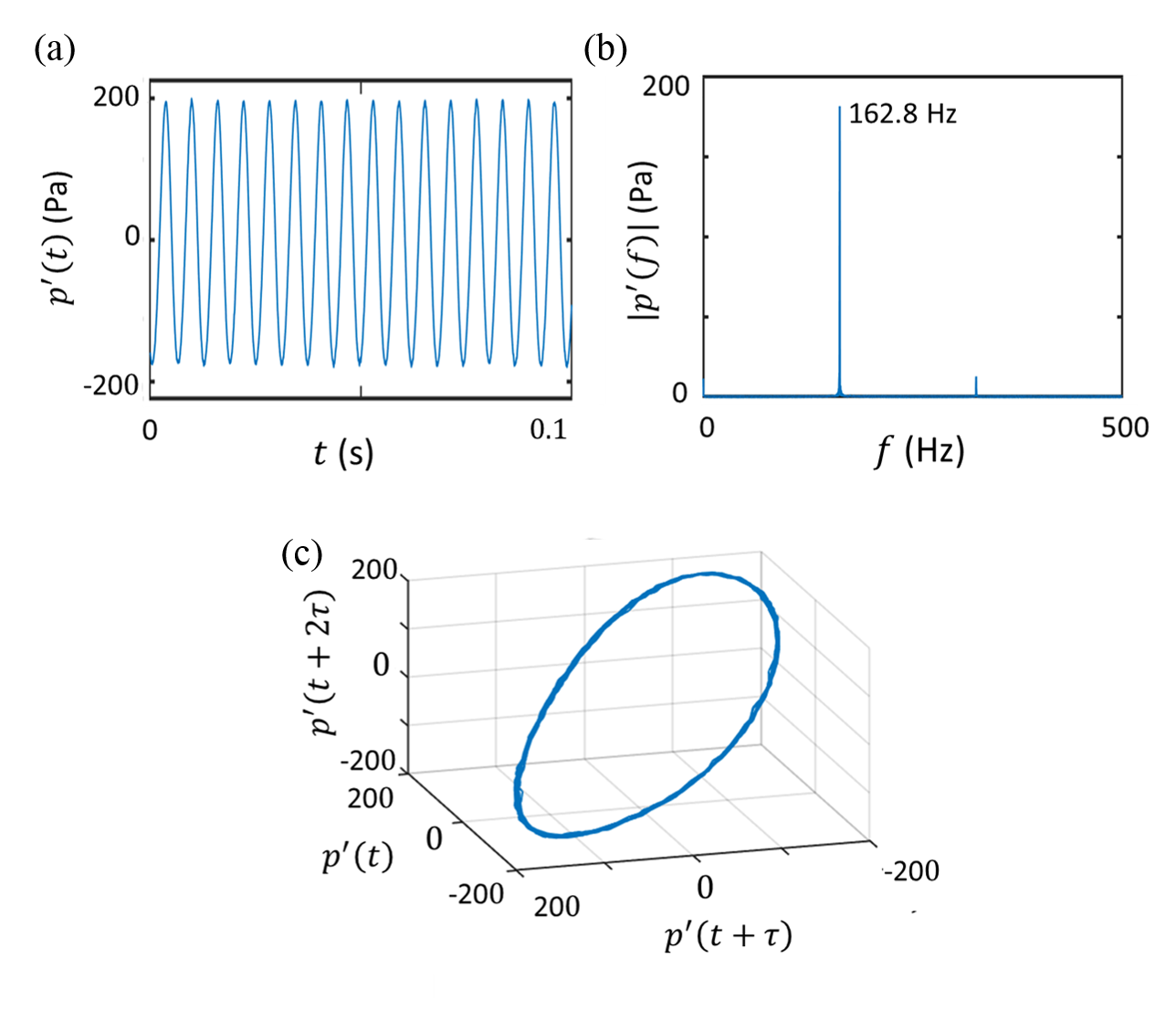}
\caption{\label{fig:6}(a) Time series, (b) amplitude spectrum, and (c) phase portrait corresponding to the state of limit cycle oscillations in a horizontal Rijke tube, highlighting the presence of high amplitude thermoacoustic oscillations with a dominant frequency of 162.8 Hz and a single closed loop in the phase portrait.}
\end{figure}

Figure \ref{fig:6} shows the properties of limit cycle oscillations observed in a horizontal Rijke tube system during the state of thermoacoustic instability. For limit cycle oscillations, we observe constant amplitude periodic oscillations (Fig. \ref{fig:6}a). During this state, the system emits a very loud tonal sound having a specific frequency corresponding to the unstable acoustic mode of the duct (Fig. \ref{fig:6}b). Such oscillations manifest as a single closed loop attractor in the embedded phase space (Fig. \ref{fig:6}c). 

\subsection{Tipping}

In the previous subsection, we discussed the bifurcation induced transition from steady state to limit cycle oscillations, or more specifically bifurcation induced tipping \cite{ashwin2012tipping}. Tipping (alternatively known as critical transition) is a general classification of a phenomenon where a small change in the control parameter across a critical value leads to a qualitative change in the state of the system. The value of the parameter at which such a transition happens is referred to as the critical point or the tipping point \cite{scheffer2009early}.

\begin{figure}[t!]
\centering
\includegraphics[width=0.45\textwidth]{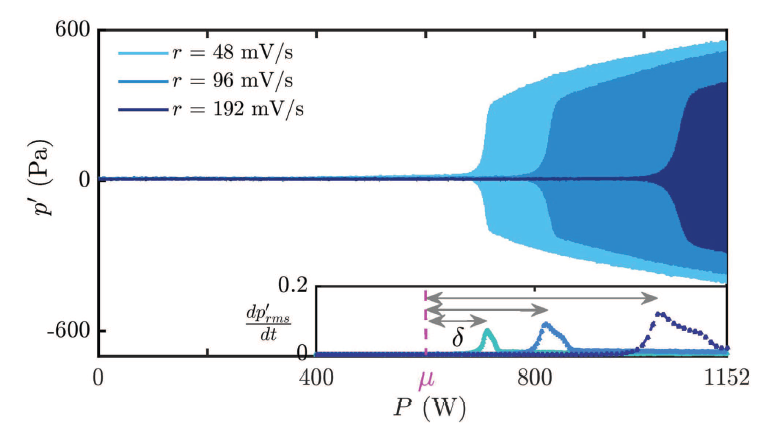}
\caption{\label{fig:7}Time series of acoustic pressure fluctuations ($p^\prime$) as a function of the time-varying control parameter (i.e., heater power $P$) during the occurrence of thermoacoustic instability in a horizontal Rijke tube, for three different rates ($r$) of change of voltage inputted to the heater. The inset indicates the rate of change of root-mean-square of pressure fluctuations ($p^\prime_0$), whose value is maximum at the onset of thermoacoustic instability. We can notice an increase in the delay ($\delta$) in the transition to thermoacoustic instability with an increase in $r$. Here, $\mu$ indicates the Hopf point of the system from quasi-static experiments. Reproduced with permission from Pavitran and Sujith \cite{pavithran2021effect}.}
\end{figure}

Ashwin \textit{et al.} \cite{ashwin2012tipping} classified critical transitions in a dynamical system into three types, where the classification is based on the mechanism of tipping. Bifurcation-induced tipping (B-tipping) occurs when the system parameter gradually crosses the critical point (the bifurcation point) resulting in a bifurcation, as discussed in the previous section. On the other hand, noise-induced tipping (N-tipping) refers to the switching of the state of a system due to the presence of stochastic perturbations. Rate-induced tipping (R-tipping) occurs when the system parameter is considered to be a time-dependent variable. Tipping occurs when the rate exceeds the critical value leading to a qualitative change in the system dynamics. The study by Thompson and Sieber \cite{thompson2011predicting,thompson2011climate} classified tipping based on the different levels of consequences as safe, explosive, and dangerous. These classifications of tipping, i.e., based on either the mechanism or consequences of tipping, were derived from investigations on climate change models.

In addition to B-tipping discussed in the previous subsection, there are a few studies in the thermoacoustic literature that focus on investigating R-tipping and N-tipping in Rijke tube systems \cite{tony2017experimental,unni2019interplay}. Tony \textit{et al.} \cite{tony2017experimental} were the first to study rate induced tipping in horizontal Rijke tubes where they demonstrated preconditioned R-tipping both experimentally and theoretically. They observed that the critical rate of change of the control parameter is a function of the initial condition. Later, Unni \textit{et al.} \cite{unni2019interplay} examined the effect of noise on rate-dependent transitions and observed high variability in the critical transitions due to the presence of noise. They observed the transition from R-tipping to N-tipping as the amplitude of the pressure oscillations approached the noise floor and delayed transition due to varying rates. A subsequent study by Zhang \textit{et al}. \cite{zhang2020rate} investigated the R-tipping delay phenomenon in a thermoacoustic model, where the rate of parameter variation is observed to delay the tipping point (see Fig. \ref{fig:7}). They noticed that the characteristics of additive and multiplicative exponential colored noise, such as initial values, ramp rate, etc., have considerable influence on the R-tipping delay phenomenon \cite{bury2021deep,pavithran2021effect}. We will discuss the studies on tipping that serve as early warning signals to thermoacoustic systems in detail in Sec. \ref{sec6}.4.

\subsection{Transition from steady state to limit cycle via intermittent oscillations }

Unlike direct transitions observed from steady state to thermoacoustic instability through Hopf bifurcation in the previous subsections, we come across a few studies on Rijke tube systems that report the transition to occur via an intermediate dynamical state. Various types of oscillatory dynamics such as intermittency \cite{pawar2016intermittency}, bursting \cite{tandon2020bursting,kasthuri2019bursting}, beating \cite{weng2016investigation}, and mixed-mode oscillations \cite{kasthuri2019bursting} have been observed as the intermediate state in different Rijke tube systems. Intermittency is characterized by the occurrence of bursts of high amplitude periodic oscillations amidst epochs of low amplitude aperiodic ones. Similarly, bursting oscillations refer to the alternating occurrence of large amplitude periodic oscillations and a quiescent state. Mixed-mode oscillations refer to the switching of the system behavior between two or more distinct amplitudes of periodic oscillations and timescales, whereas beating refers to the occurrence of amplitude-modulated periodic oscillations in the system. These oscillations are conjectured to arise due to the coexistence of subsystems with multiple time scales of oscillations and such systems are usually referred to as slow-fast systems \cite{omelchenko2010synchronization, bertram2017multi, kasthuri2020recurrence}. 

\begin{figure}
\centering
\includegraphics[width=0.5\textwidth]{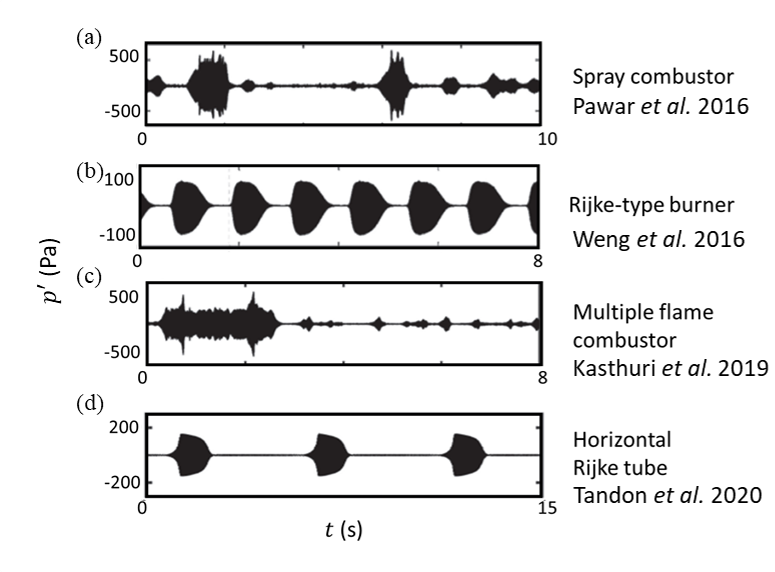}
\caption{\label{fig:8} Different types of intermediate states observed in acoustic pressure fluctuations ($p'$) during the transition from steady state to limit cycle oscillations such as (a) intermittency in the spray combustor\cite{pawar2016intermittency}, (b) beating dynamics in the Rijke-type burner\cite{weng2016investigation}, (c) bursting in the multiple flames matrix burner\cite{kasthuri2019bursting}, and (d) bursting in the horizontal Rijke tube\cite{tandon2020bursting}. Reproduced with permission from Tandon \textit{et al}. \cite{tandon2020bursting}.}
\end{figure}

Pawar \textit{et al.} \cite{pawar2016intermittency} reported the existence of intermittency\footnote{Kabiraj and Sujith \cite{kabiraj2012nonlinearJFM} were the first to report the term intermittency in the context of thermoacoustics in a ducted laminar premixed flame Rijke tube burner. They observed the occurrence of intermittency prior to flame blowout in the system and not prior to the onset of limit cycle oscillations.} in a Rijke-type laboratory spray burner during the transition from stable operation to thermoacoustic instability when the flame location is varied (Fig. \ref{fig:8}a). Using various measures from dynamical systems theory, they confirmed the presence of type-II intermittency. Furthermore, their study suggests that intermittency could be more dangerous as compared to limit cycle oscillations, as the maximum amplitude of bursts during intermittency is nearly thrice the amplitude of limit cycle oscillations. Weng \textit{et al.} \cite{weng2014beat,weng2016investigation} reported the presence of beating dynamics between the steady state and limit cycle oscillations in a porous plug stabilized laminar premixed flame Rijke tube burner (Fig. \ref{fig:8}b). The amplitude-modulated oscillations were accompanied by low frequency flame pulsations having a frequency lower than 1 Hz; thereby, creating a time scale difference of $10^2 - 10^3$ between the pulsations in the flame and the thermoacoustic oscillations. Subsequently, Kasthuri \textit{et al.} \cite{kasthuri2019bursting} observed the presence of bursting and mixed-mode oscillations in a premixed matrix burner with several interacting laminar flames (Fig. \ref{fig:8}c). They found that these oscillations occur due to the interaction of a slow timescale associated with temperature fluctuations and a fast timescale with acoustic pressure fluctuations.

Tandon \textit{et al}. \cite{tandon2020bursting} systematically investigated the role of slow and fast timescales on the occurrence of intermittent oscillations prior to thermoacoustic instability in a horizontal Rijke tube system. Towards this purpose, they modeled slow oscillations in the control parameter and studied the interaction of these oscillations with a fast oscillating acoustic pressure field as the system dynamics transitions from steady state to limit cycle oscillations. When slow and fast subsystems are uncoupled, they observed regular occurrence of bursting in the pressure signal prior to thermoacoustic instability (Fig. \ref{fig:8}d). On the other hand, when slow and fast subsystems are coupled with each other, they noticed the creation of amplitude-modulated bursting in the pressure oscillations. 

So far, we discussed the transition of a Rijke tube system from steady state to limit cycle oscillations and their corresponding bifurcations. In the following subsection, we will describe the dynamics of such systems beyond the state of limit cycle oscillations and associated bifurcations leading to the occurrence of different dynamical regimes.


\subsection{Secondary Hopf bifurcations in thermoacoustic systems}

In many dynamical systems, increasing the control parameter further in the regime of limit cycle oscillations engenders the possibility of secondary Hopf bifurcations, leading to the emergence of new frequencies in the system \cite{nayfeh2008applied,hilborn2000chaos}. The interaction between the former and the newly generated frequencies post bifurcation gives rise to various complex dynamical states that are different from period-1 limit cycle oscillations. These states include period-2, period-3, period-$k$, frequency-locked, quasiperiodic, strange nonchaotic, intermittent, and chaotic oscillations. There are many experimental as well as theoretical studies in the thermoacoustic literature that report the existence of these dynamical behaviors in Rijke tube systems\cite{lei2009nonlinear,subramanian2010bifurcation,kabiraj2012route,premraj2020strange,guan2020intermittency,kashinath2014nonlinear,kabiraj2012nonlinearJFM,vishnu2015role,premraj2021dragon}. Sometimes, a secondary Hopf bifurcation observed due to a change in the control parameter leads to the transition from low amplitude limit cycle oscillations to high amplitude limit cycle oscillations, where both the limit cycle oscillations exhibit the same frequency\cite{strogatz1994nonlinear}. Mukherjee \textit{et al}. \cite{mukherjee2015nonlinear} reported the presence of such secondary bifurcation of limit cycle oscillations in a laminar Rijke type burner.
Furthermore, as the dynamical behavior of many systems ultimately tends to reach a state of chaotic oscillations with a change in the control parameter, the dynamical transitions associated with the occurrence of chaos are often referred to as \textit{routes to chaos} \cite{hilborn2000chaos,lakshmanan2012nonlinear,lakshmanan2012nonlinear,schuster2006deterministic}. The system finally reaches the state of chaotic oscillations either through period-doubling route to chaos, via Ruelle-Takens-Newhouse route to chaos or through intermittency route to chaos \cite{parker1987chaos}. A plethora of nonlinear dynamical states observed during each of these routes to chaos have been reported in Rijke tube systems as well \cite{kabiraj2012nonlinearJFM,kashinath2014nonlinear,guan2020intermittency,lei2009nonlinear,subramanian2010bifurcation}. 

\subsubsection{Rich nonlinear behavior of thermoacoustic systems}

\begin{figure}
\centering
\includegraphics[width=0.5\textwidth]{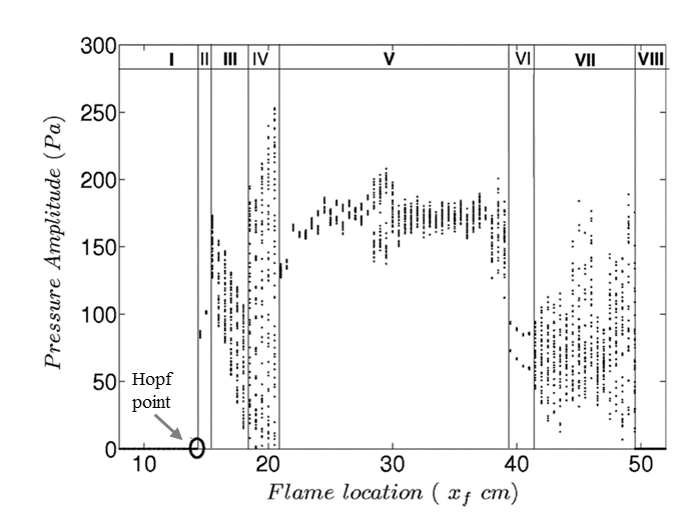}
\caption{\label{fig:9}Secondary bifurcations of acoustic pressure fluctuations observed from experiments in a
laminar premixed flame Rijke tube burner, as the flame location ($x_f$) is varied as the control parameter. Regions (I)–(VIII) indicate steady state, limit cycle, quasiperiodicity,  frequency-locked, quasiperiodicity, period-2, chaos, and steady state, respectively. Reproduced with permission from Kabiraj \textit{et al}. \cite{kabiraj2012bifurcations}.}
\end{figure}

In laminar premixed flame Rijke tube burners (Fig. \ref{fig:4}), we witness rich dynamical behavior resulting from a secondary Hopf bifurcation of limit cycle oscillations (see Fig. \ref{fig:9}) due to the variation of different control parameters \cite{kabiraj2012bifurcations,kabiraj2012investigating,kabiraj2012nonlinearJFM,kabiraj2012route,kashinath2014nonlinear,vishnu2015role,guan2019chaos,premraj2020strange,premraj2021effect,mukherjee2015nonlinear}. In this section, we will discuss the characteristics of these dynamical states and then elaborate different routes to chaos observed in Rijke tube systems. 

\begin{enumerate}
\item \textbf{Period-1 limit cycle oscillations:} Limit cycle oscillations are characterized by constant-amplitude periodic oscillations (Fig. \ref{fig:10}a). Such oscillations have a single dominant frequency in the power spectrum; hence, often referred to as period-1 limit cycle oscillations. As a result, these signals possess a distinct single closed-loop attractor in the phase space, where the phase space trajectory repeats its behavior after each time period of the oscillation. The Poincaré section of limit cycle oscillations shows a single point. 
    
\item \textbf{Frequency-locked or period-$k$ oscillations:} Unlike period-1 limit cycle oscillations, frequency-locked oscillations possess more than one narrow band peaks (say, $f_1$ and $f_2$), which are rationally related to each other (i.e., $f_1/f_2=p/q$, where $p$ and $q$ are integer numbers) in the power spectrum (Fig. \ref{fig:10}b). These signals are periodic and repeat their behavior in the phase space, depending on the ratio of frequencies ($f_1/f_2$). When this ratio is an integer number (say, $k$), we observe period-$k$ oscillations in the signal with $k$ orbits in the phase space. For example, during period-2 oscillations, we observe two dominant frequencies in the spectrum, where the low amplitude frequency (say $f_2/2$) is observed at the subharmonic of the dominant frequency (say $f_2$). We notice the presence of two loops for the phase space trajectory (Fig. \ref{fig:10}c); hence, two distinct points in the Poincaré section. In Rijke tube systems, many theoretical \cite{subramanian2010bifurcation,kashinath2014nonlinear} and experimental \cite{kabiraj2012intermittencyThesis,gopalakrishnan2014influence} studies have reported the presence of period-2 oscillations. The experimental evidence of frequency-locked oscillations has been reported by Kabiraj \textit{et al}. \cite{kabiraj2012bifurcations, kabiraj2012route} and Vishnu \textit{et al}. \cite{vishnu2015role}.

\begin{figure}[t!]
\centering
\includegraphics[width=0.5\textwidth]{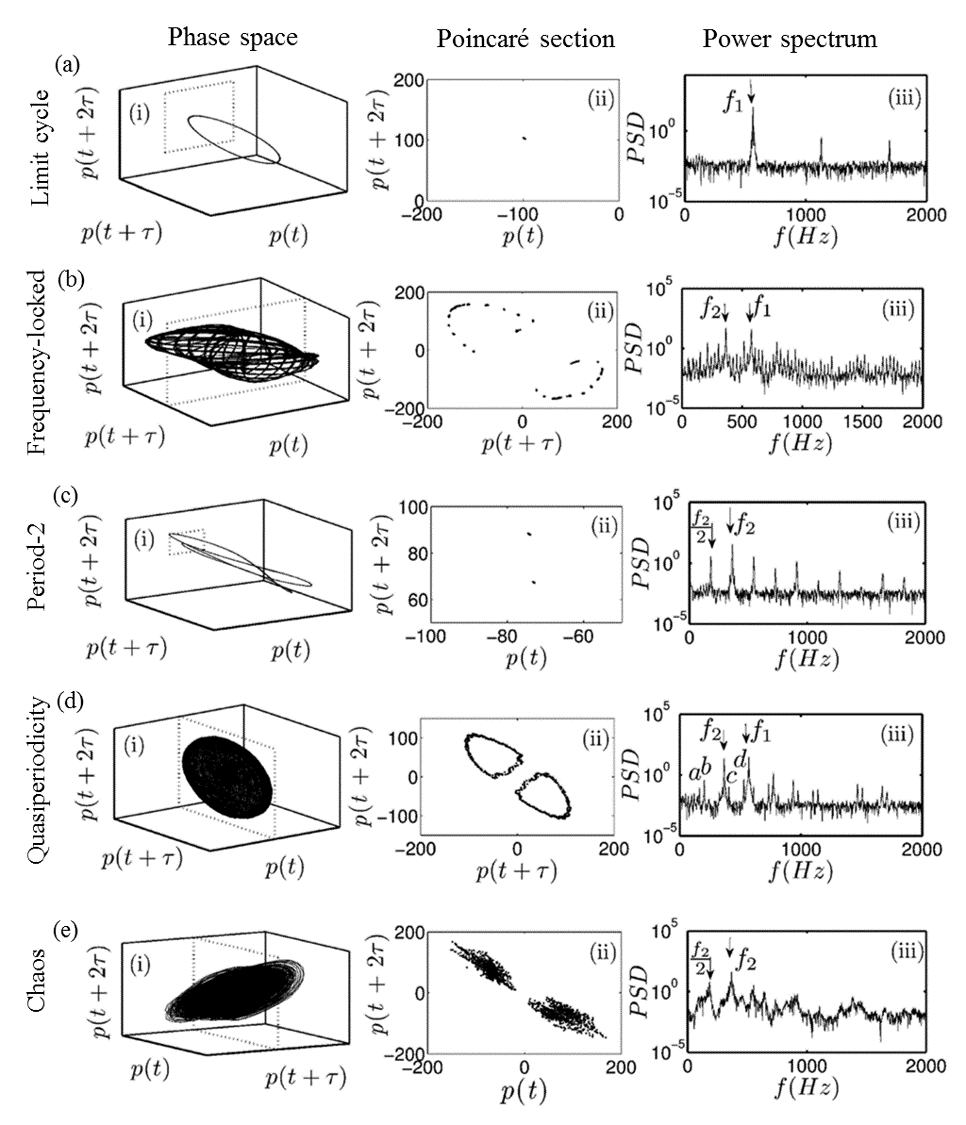}
\caption{\label{fig:10} Three-dimensional phase portrait, Poincaré map, and the power spectrum corresponding to various dynamical states observed after secondary bifurcation in Fig. \ref{fig:9} including (a) limit-cycle, (b) frequency-locked, (c) period-2, (d) quasiperiodicity, and (e) chaotic oscillations. Adapted with permission from Kabiraj \textit{et al.} \cite{kabiraj2012bifurcations}. }
\end{figure}
    
    \item \textbf{Quasiperiodic Oscillations:} For quasiperiodic oscillations, we observe two dominant frequencies (say, $f_1$ and $f_2$) and frequencies corresponding to their linear combinations (say, $nf_1+mf_2$, where $n$ and $m$ are integer numbers) in the spectrum (Fig. \ref{fig:10}d). These two dominant frequencies are irrationally related to each other ($f_1/f_2 \neq p/q$). As a result, quasiperiodic oscillations are aperiodic oscillations, and their properties never repeat after a finite duration of time. The phase space trajectory of quasiperiodic oscillations lies on a torus structure in the phase space (Fig. \ref{fig:10}d) and its Poincaré section shows a closed structure  \cite{hilborn2000chaos}. Quasiperiodic oscillations have been reported in a theoretical study on a two-dimensional ducted premixed flame by Kashinath \textit{et al}. \cite{kashinath2014nonlinear} and in experimental studies on a laminar premixed flame Rijke tube burner by \cite{kabiraj2012bifurcations, kabiraj2012route, vishnu2015role,guan2020intermittency}.
    
    \item \textbf{Chaotic Oscillations:} Chaotic oscillations are characterized by an exponential divergence of nearby trajectories in the phase space. A power spectrum of these oscillations possesses more than two irrationally related frequencies and their linear combinations, which eventually manifests as a broadband spectrum. As a consequence, chaotic oscillations are aperiodic in time. The phase space of such oscillations shows the existence of a strange attractor, where the behavior of the phase trajectory is highly unstable, while their Poincaré section exhibits a scatter of points (Fig. \ref{fig:10}e). The maximum Lyapunov exponent of chaotic oscillations is always positive. These oscillations have been reported in theoretical studies on a two-dimensional ducted premixed flame by Kashinath \textit{et al.} \cite{kashinath2014nonlinear} and on a horizontal Rijke tube by Subramanian \textit{et al.} \cite{subramanian2010bifurcation}, and experimental studies on laminar Rijke tube burner by \cite{kabiraj2012bifurcations, kabiraj2012route, vishnu2015role,guan2020intermittency}.
    
    \begin{figure}[t!]
    \centering
    \includegraphics[width=0.47\textwidth]{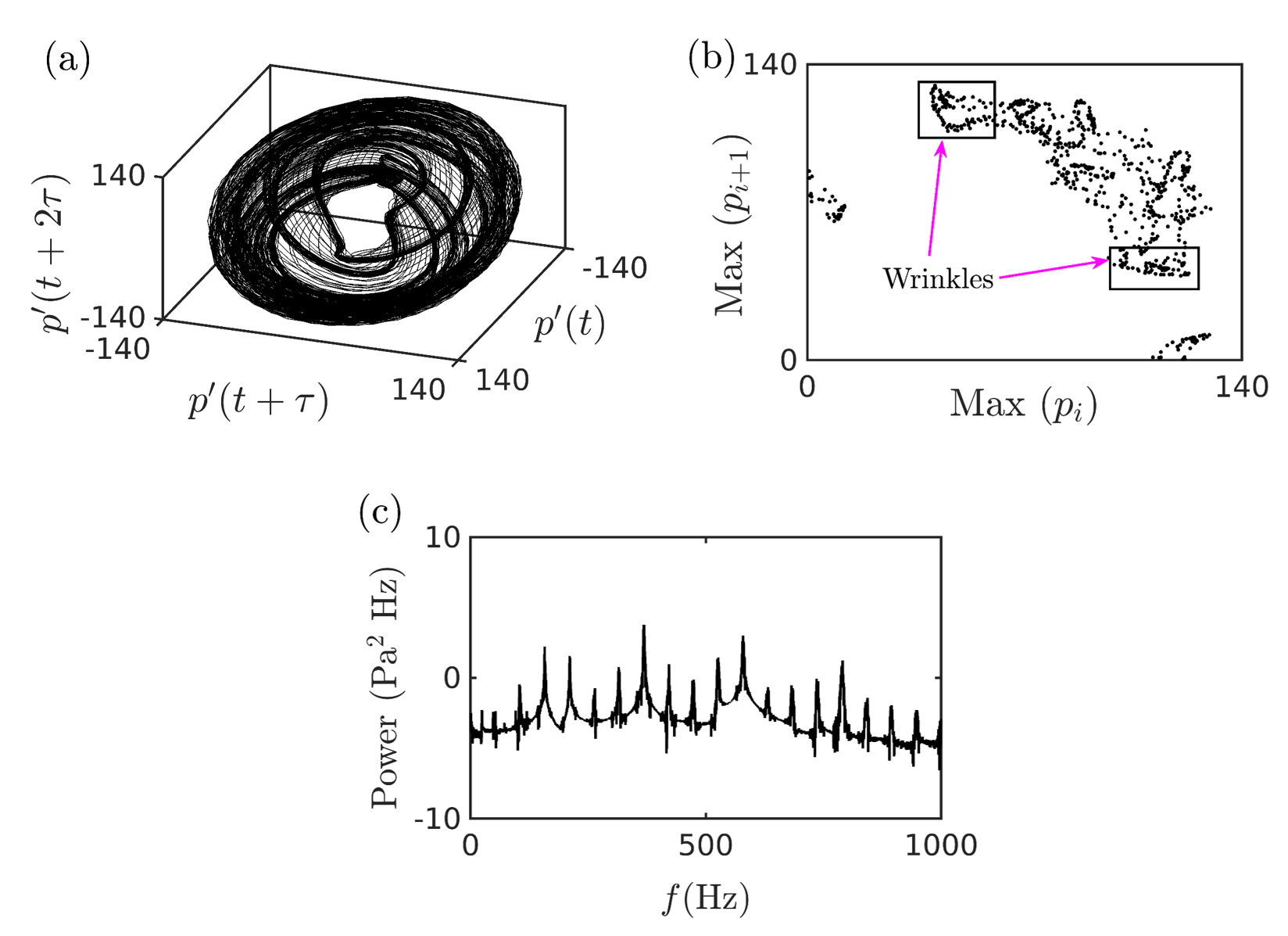}
    \caption{\label{fig:11}(a)–(c) Phase portrait, Poincaré section, and power spectrum, respectively, of strange nonchaotic oscillations observed from experiments in a laminar premixed  Rijke tube burner. Adapted with permission from Premraj \textit{et al}. \cite{premraj2020strange}.}
    \end{figure}
    
    \item \textbf{Strange nonchaotic oscillations:} Strange nonchaotic oscillations point towards the existence of a fractal attractor, similar to that observed for chaotic oscillations; however, unlike chaos, they do not possess sensitivity to initial conditions. Hence, the maximum Lyapunov exponent of strange nonchaotic oscillations is always negative. The Poincaré section of strange nonchaotic oscillations presents a wrinkled torus (Fig. \ref{fig:11}). The power spectrum of strange nonchaotic oscillations is broadband. These oscillations are often observed in systems with quasiperiodically forced oscillations \cite{pikovsky1995characterizing,heagy1994birth,ditto1990experimental}. Although the evidence of such oscillations in self-excited dynamics is rare, they have been observed in a pulsating star network by Lindner \textit{et al}. \cite{lindner2015strange} and recently in experiments on laminar premixed flame Rijke tube burner by Premraj \textit{et al.} \cite{premraj2020strange}. Guan \textit{et al.} \cite{guan2018strange} reported the existence of strange nonchaos in forced limit cycle oscillations of acoustic pressure in a premixed flame Rijke tube burner. On the other hand, Weng \textit{et al}. \cite{weng2020synchronization} provided the theoretical evidence of strange nonchaos in a model of nonlinearly coupled damped oscillators of a laminar Rijke tube burner. 
    
\end{enumerate}

\begin{figure}[t!]
\centering
\includegraphics[width=0.5\textwidth]{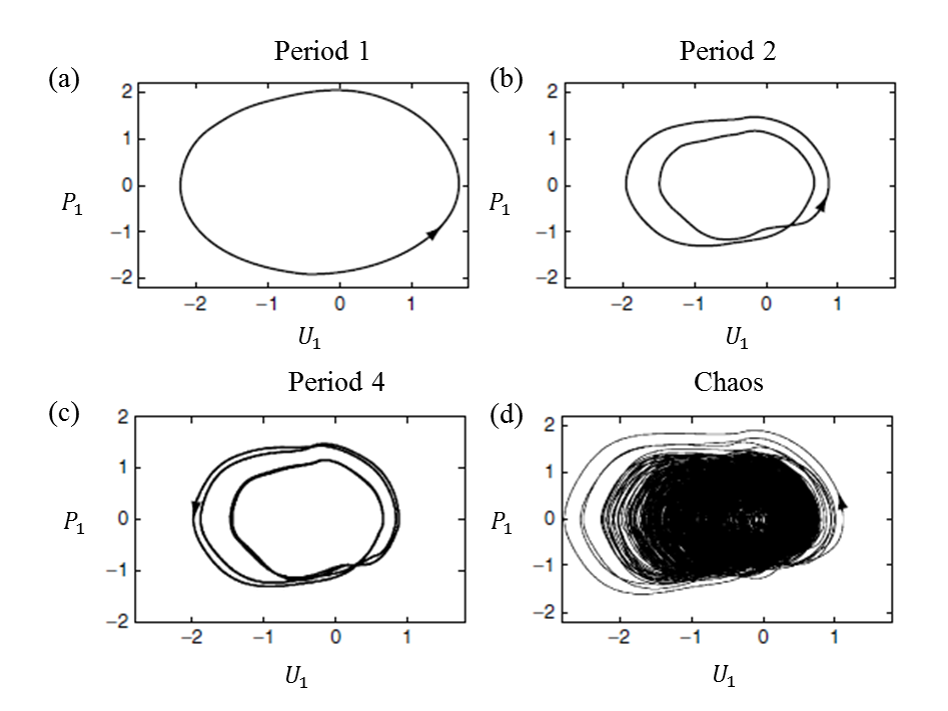}
\caption{\label{fig:12}Period- doubling route to chaos reported in a mathematical model of the horizontal Rijke tube \cite{balasubramanian2008thermoacoustic}, showing the presence of (a) period-1 limit cycle oscillations, followed by (b) period-2 and (c) period-4 oscillations, ultimately reaching the state of (d) chaotic oscillations. Reproduced with permission from Subramanian \textit{et al}. \cite{subramanian2010bifurcation}.}
\end{figure}

\subsubsection{Various routes to chaos in thermoacoustic systems}

As mentioned before, route to chaos refers to the fundamental mechanism by which a regular attractor becomes a chaotic attractor as the control parameter is varied \cite{sastry1981bifurcation,hilborn2000chaos,lakshmanan2012nonlinear}. Various numerical studies have focused on studying different routes to chaos in order to clearly understand the enigma of chaotic oscillations itself. In laminar Rijke-type thermoacoustic systems, three routes to chaos have been reported, which we describe as follows.

\begin{enumerate}
\item \textbf{Period-doubling route to chaos:} This route to chaos is the most commonly studied scenario in the dynamical systems literature \cite{sleeman1988period, geest1992period, cheung1987chaotic, ye1993period, stone1993period, simpson1994period,hilborn2000chaos}. It was first discovered by Feigenbaum \cite{feigenbaum1979universal}, and hence referred to as Feigenbaum scenario. 
    
Lei and Turan \cite{lei2009nonlinear} reported the presence of a period-doubling route to chaos in a time-lag model of a combustion system. Subramanian \textit{et al}. \cite{subramanian2010bifurcation} showed the existence of this route to chaos for the variation of heater power in the Balasubramanian-Sujith oscillator model of the Rijke tube oscillator (Fig. \ref{fig:12}). During period-doubling bifurcations, the system behavior initially transitions from a steady state to limit cycle oscillations via Hopf bifurcation (Fig. \ref{fig:12}a). Such limit cycle oscillations undergo a sequence of secondary Hopf bifurcations, causing their transition to period-2 (Fig. \ref{fig:12}b), period-4 (Fig. \ref{fig:12}c), period-8 oscillations, etc. until chaotic oscillations are observed (Fig. \ref{fig:12}d). Similar results were observed in a numerical study on slot stabilized two-dimensional premixed flame by Kashinath \textit{et al.} \cite{kashinath2014nonlinear} for the variation of flame location as the control parameter. As per our knowledge, experimental evidence on the period-doubling route to chaos is still unreported in Rijke tube systems. An experimental study on a horizontal Rijke tube with an electrically heated wire mesh as the heat source by Gopalakrishnan and Sujith \cite{gopalakrishnan2014influence} reported the presence of period-2 oscillations. However, further period-doubling bifurcations were not observed in the system due to limitations in the experimental configuration. The usage of the wire mesh as the heat source restricted the increase in the heater power above a limit, above which the mesh melts. Therefore, future investigations on a horizontal Rijke tube consisting of a plate-type heat source may provide the possibility of observing multiple period-doubling bifurcations leading to chaos experimentally. 

\begin{figure}[t!]
\centering
\includegraphics[width=0.5\textwidth]{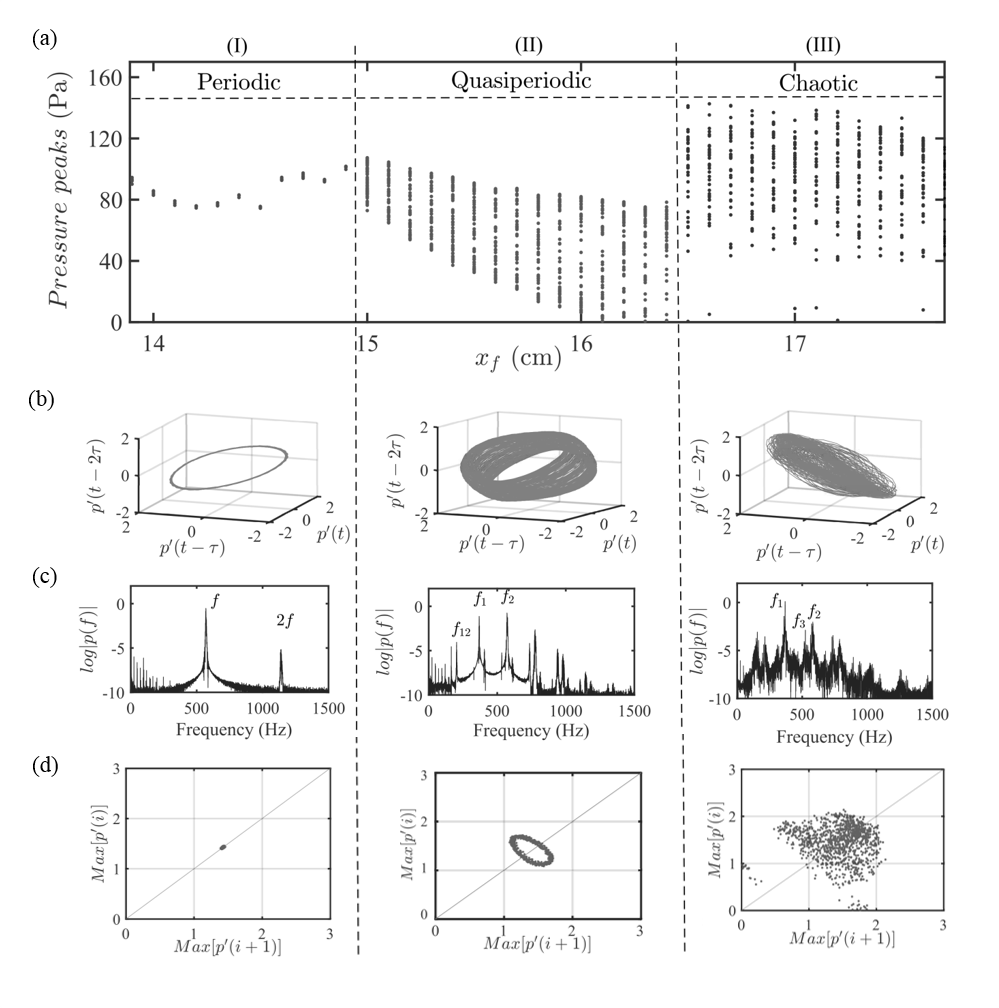}
\caption{\label{fig:13}(a) Quasiperiodicity route to chaos highlighting the transition from (I) limit cycle oscillations to (II) quasiperiodic oscillations, ultimately leading to (III) chaotic oscillations. (b) Phase portraits, (c) amplitude spectra and (d) Poincaré sections corresponding to the three dynamical states. Reproduced with permission from Mondal \textit{et al}. \cite{mondal2017synchronous}.}
\end{figure}

\item \textbf{Ruelle-Takens-Newhouse route to chaos:} In the Ruelle-Takens scenario \cite{hilborn2000chaos}, the system exhibiting limit cycle oscillations undergoes another Hopf bifurcation leading to the appearance of a second frequency in the signal. Contrary to the period-doubling route, where the second frequency is rationally related to the first frequency, here the system acquires a second frequency that is irrationally related to the first, and hence exhibiting quasiperiodic oscillations. Further increase in the control parameter leads to the occurrence of another frequency that is incommensurate with the other two frequencies. The presence of three frequencies leads to the transition from quasiperiodic oscillations to chaotic oscillations. This route was first discovered by Ruelle and Takens \cite{ruelle1971nature} and  Newhouse \textit{et al.} \cite{newhouse1978occurrence} independently, and is also referred to as a quasiperiodic route to chaos.

Kabiraj \textit{et al}. \cite{kabiraj2012route,kabiraj2012bifurcations} observed a quasiperiodic route to chaos in an experimental study on a laminar premixed flame Rijke tube burner as the location of the flame in the duct is varied as the control parameter (Fig. \ref{fig:13}). They observed the transition from limit-cycle oscillations (Fig. \ref{fig:13}-I) to chaotic oscillations (Fig. \ref{fig:13}-III) via the intermediate states of quasiperiodic oscillations (Fig. \ref{fig:13}-II). Furthermore, Kashinath \textit{et al}. \cite{kashinath2014nonlinear} reported the presence of Ruelle-Takens-Newhouse route to chaos on the variation of the flame position in a numerical study on slot stabilized two-dimensional premixed flame. 
    
\begin{figure}[t!]
\centering
\includegraphics[width=0.45\textwidth]{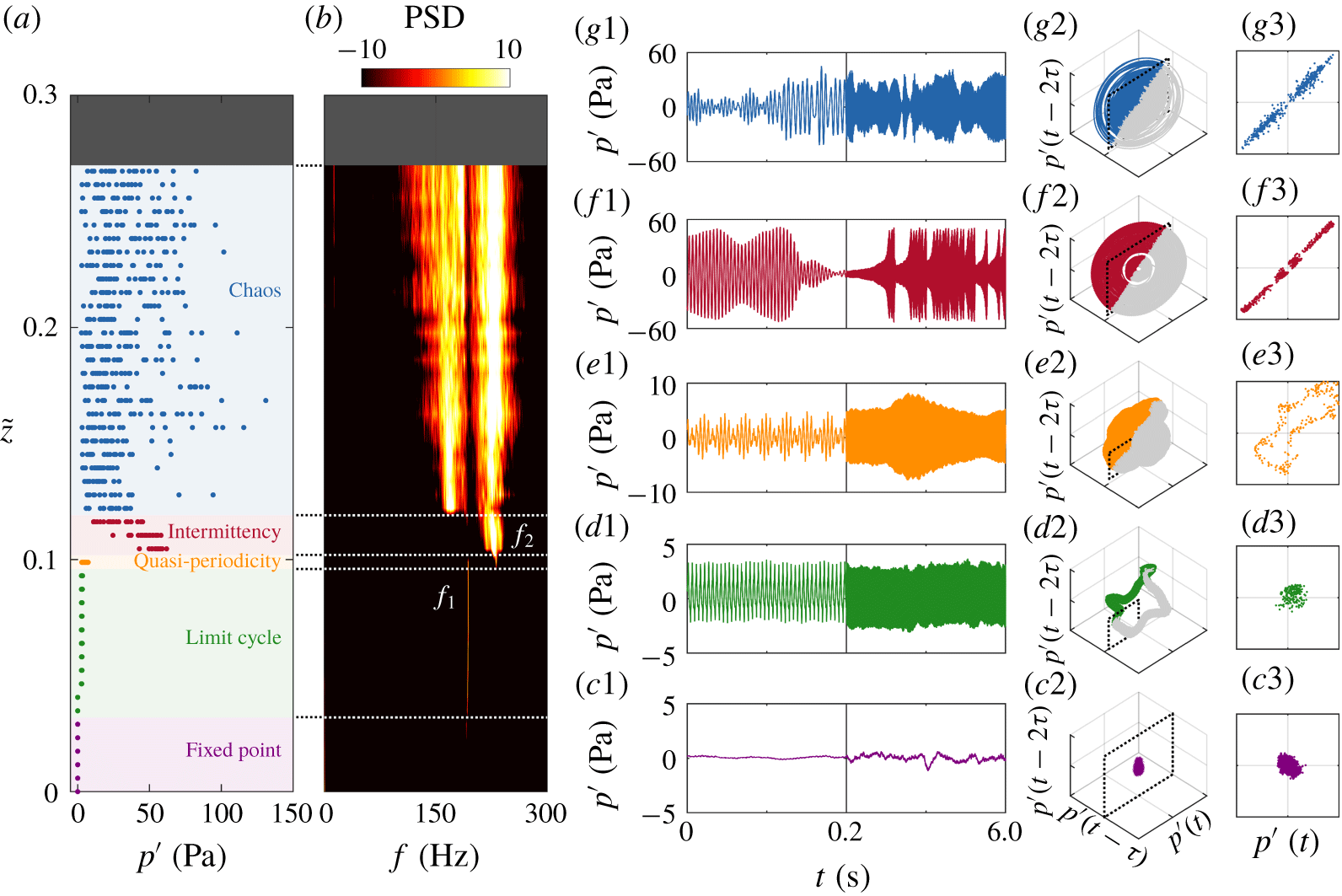}
\caption{\label{fig:14}(a) Bifurcation diagram and (b) power spectral density variation during the intermittency route to chaos when the flame location ($\overline{z}$ measured from the bottom of the combustor) is varied as the control parameter in a laminar premixed Rijke tube burner. During this route to chaos, the system behavior transitions from (c) fixed point, (d) limit cycle oscillations, (e) quasiperiodicity, (f) intermittency to (g) chaotic oscillations. Here, subplots 1 to 3 correspond to the time series, phase portraits, and Poincaré section, respectively, for the corresponding dynamical states shown in c to g. Reproduced with permission from Guan \textit{et al}. \cite{guan2020intermittency}.}
\end{figure}
    
    \item \textbf{Intermittency route to chaos:} During the intermittency route to chaos, as we change the control parameter, the limit cycle oscillations transition to chaotic oscillations via intermittency \cite{hilborn2000chaos,nayfeh2008applied,pomeau1980intermittent}. During the state of intermittency, the system dynamics alternates between irregularly occurring bursts of chaotic oscillations and epochs of periodic oscillations. As the system approaches the onset of chaotic oscillations, the number of occurrences of such bursts in the signal is observed to be increasing; ultimately leading to chaotic oscillations in the system. This route was first discovered by Pomeau and Manneville \cite{pomeau1980intermittent} in dissipative dynamical systems and is therefore also called the Pomeau-Manneville scenario. 

    In thermoacoustic systems, Guan \textit{et al}. \cite{guan2020intermittency} reported the presence of an intermittency route to chaos in an experimental study on a premixed flame Rijke tube burner as the location of the flame is varied as the control parameter. They observed the transition from steady state (Fig. \ref{fig:14}c) to limit-cycle oscillations (Fig. \ref{fig:14}d), followed by quasiperiodicity (Fig. \ref{fig:14}e), to intermittency (Fig. \ref{fig:14}f) and then to chaos (Fig. \ref{fig:14}g). The intermittency observed in the system consists of epochs of high amplitude chaos amidst bursts of medium-amplitude quasiperiodicity. 
\end{enumerate}

To summarize, Rijke-type thermoacoustic systems, similar to other phenomenological oscillators in dynamical systems theory, exhibit complex nonlinear behaviors and bifurcations. Hence, we confirm the nonlinear nature of Rijke tube oscillators and encourage the application of the Rijke tube oscillator as a general nonlinear oscillator. Next, we discuss the bistable nature of the Rijke tube oscillator and present different nonlinear behaviors that can arise in such systems due to the influence of external stochastic perturbations in the system.


\section{\label{sec4}Noise-induced dynamics in the subthreshold and bistable regions of thermoacoustic systems}

\begin{figure}
\centering
\includegraphics[width=0.5\textwidth]{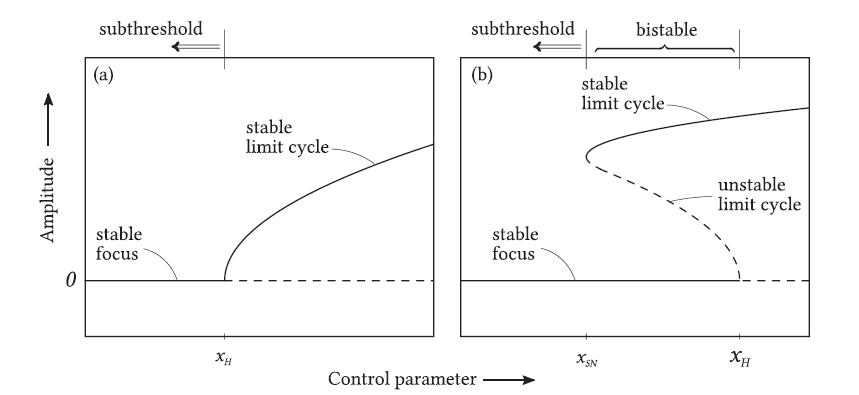}
\caption{\label{fig:15}Bifurcation diagrams of (a) supercritical and (b) subcritical Hopf highlighting the subthreshold and bistable regions. Reproduced with permission from Gupta \textit{et al}. \cite{gupta2017numerical}.}
\end{figure}

Most systems in nature are inherently noisy and, therefore, exhibit many noise-induced phenomena and bifurcations \cite{horsthemke1984noise,moss1989noise,van1994noise,gammaitoni1998stochastic,garcia2012noise,arnold1995random}. These dynamical changes include modification in the stability margins, occurrence of coherence and stochastic resonance, and excitation of new dynamical states \cite{arnold1995random,kabiraj2020review,lindner2000coherence,gang1993stochastic}. In this section, we discuss various noise-induced dynamics in the sub-threshold and bistable regimes of Rijke tube oscillators \cite{waugh2011triggeringa,waugh2011triggeringb,kabiraj2015coherence,saurabh2017noise,gopalakrishnan2014influence,gupta2017numerical,jegadeesan2013experimental,lee2020input,gopalakrishnan2016stochastic,li2019coherence}. The regime corresponding to a single stable fixed-point solution (stable focus), observed prior to the Hopf point in supercritical Hopf bifurcation (Fig. \ref{fig:15}a) and the fold point (saddle-node) in subcritical Hopf bifurcation (Fig. \ref{fig:15}b), is referred as the subthreshold regime \cite{kabiraj2020review}. A bistable region is observed for subcritical Hopf bifurcation and lies between the Hopf and fold points of the system parameter space (Fig. \ref{fig:15}b), wherein a stable fixed point coexists with stable and unstable solutions of limit cycle oscillations. In the upcoming section, we discuss some kinds of noise-induced dynamics namely coherence resonance and stochastic bifurcations observed in the subthreshold regime of the Rijke tube oscillator. Subsequently, we present the discussion on noise-induced dynamics in the bistable region of such oscillators.

\subsection{Coherence resonance}
The addition of noise in the subthreshold regime of an excitable system (or an oscillator) has a counter-intuitive effect of increasing the coherent nature of its oscillatory response rather than deteriorating it \cite{balanov2009synchronization,zakharova2013coherence}. Coherence resonance refers to a noise-induced coherence characterized with a resonance-like dependence on the strength of noise as the system approaches the bistable region. It was first described and analyzed by Pikovsky and Kurths \cite{pikovsky1997coherence} in a noise-driven excitable FitzHugh–Nagumo system. During coherence resonance, the degree of regularity in the dynamics of the system is observed to be maximum at intermediate values of external noise intensity. This phenomenon has been studied in many oscillators including Stuart-Landau \cite{ushakov2005coherence} and Van der Pol oscillators \cite{zakharova2010stochastic,zakharova2013coherence}, under the influence of additive white noise, indicated by $\sqrt{2D\zeta(t)}$ where $D$ represents the noise intensity and $\zeta(t)$ highlights the noise characteristics. 

\begin{figure}
\centering
\includegraphics[width=0.5\textwidth]{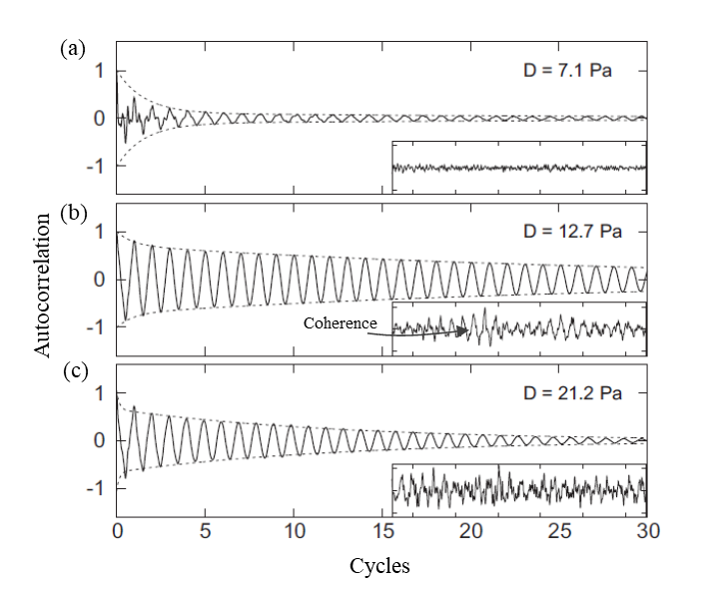}
\caption{\label{fig:16}(a)-(c) Variation of the autocorrelation function and the acoustic pressure signal (see inset) depicting the characteristics of the coherence resonance phenomenon observed in a laminar premixed flame Rijke tube burner for low, medium, and high levels of noise intensity $D$, respectively. The maximum coherence is observed at intermediate noise levels ($D=12.7$ Pa). Reproduced with permission from Kabiraj \textit{et al}. \cite{kabiraj2015coherence}.}
\end{figure}

Coherence resonance has been studied in various Rijke tube systems both experimentally \cite{kabiraj2015coherence} and theoretically \cite{gupta2017numerical,li2019coherence}. Figure \ref{fig:16} shows the occurrence of coherence resonance in a laminar premixed flame Rijke tube burner for increasing values of the noise intensity $D$ \cite{kabiraj2015coherence}. For low and high values of $D$, the noisy fluctuations in the system induce transient coherence which dies down as time progresses (insets of Fig. \ref{fig:16}a,c). On the contrary, for intermediate values of $D$, we observe the noise-induced emergence of coherent (periodic) oscillations in the system (inset of Fig. \ref{fig:16}b). 

The existence of coherence resonance has an important application in thermoacoustic systems. We can use the increase in the coherent nature of pressure signals in the steady state regime of the system prior to the Hopf point as a precursor to an impending thermoacoustic instability \cite{kabiraj2015coherence,gupta2017numerical,li2019coherence}. Furthermore, the existence of coherence resonance has been examined for subcritical Hopf bifurcation \cite{kabiraj2015coherence,saurabh2017noise} and supercritical Hopf bifurcation \cite{lee2020input} individually as well as collectively \cite{gupta2017numerical}. The comparative analysis between these bifurcations showed the existence of a qualitative difference in the variation of different measures such as autocorrelation factor and spectral width and height of the coherence resonance phenomenon. Such a qualitative difference exists due to the inherent difference in the type of nonlinearity in the system \cite{gupta2017numerical}.

Similar to coherence resonance, where maximum coherence is observed in the signal at intermediate levels of noise, we can also observe the maximum amplification in the signal for intermediate levels of noise due to stochastic resonance \cite{gammaitoni1998stochastic,wellens2003stochastic}. Stochastic resonance is one of the well-known noise-induced phenomena in bistable systems that correspond to the enhancement of amplitude response of the system due to the addition of external periodic forcing in the presence of noise. This phenomenon has potential applications in various fields including physics, engineering, sensory systems, biology, and medicine \cite{moss2004stochastic,gammaitoni1998stochastic,benzi1982stochastic,hanggi2002stochastic}.  The existence of stochastic resonance has however not yet been discovered in the thermoacoustic system as per the authors’ knowledge.


\subsection{Stochastic bifurcations and hysteresis}
\label{Stochastic}
The presence of high intensity additive noise in a system could lead to the disappearance of sharp transitions to limit cycle oscillations during Hopf bifurcations, which are otherwise observed in deterministic systems \cite{juel1997effect,namachchivaya1990stochastic}. Hence, it is indeed very challenging to obtain the Hopf points (or transition boundaries) in the system, as the system transitions from being a deterministic system to a stochastic system \cite{gopalakrishnan2016stochastic}. Therefore, we resort to tracking the probability distribution of the variables rather than calculating their absolute values \cite{arnold1995random}.

Furthermore, systems with noise may undergo stochastic bifurcations, while transitioning from one dynamical state to another. Stochastic bifurcations are classified into two types: phenomenological bifurcation and dynamic bifurcation, commonly referred to as P-bifurcation and D-bifurcation, respectively \cite{crauel1999stochastic,arnold1995random}. P-bifurcation describes qualitative changes observed in the probability density function (PDF) of the variable, whereas D-bifurcation is associated with the change in the measure of a system variable (as discussed in Sec. \ref{sec3}) or with the sign change of the Lyapunov exponent, due to a change in the control parameter. 

Stochastic bifurcations are observed in various nonlinear systems  \cite{crauel1999stochastic,arnold1995random} such as Van der Pol oscillators \cite{zakharova2010stochastic}, biological systems \cite{song2010estimating}, and laser systems \cite{billings2004stochastic}. Rijke tube systems also tend to exhibit stochastic behavior and stochastic bifurcations in the presence of noise. As a result, such systems are modeled using stochastic differential equations \cite{jin2022stochastic}.  The probability density function (PDF) is calculated by solving the Fokker-Planck equation of stochastic systems, which was first introduced to thermoacoustic systems by Clavin \textit{et al.} \cite{clavin1994turbulence}. Noiray and Schuermans \cite{noiray2013deterministic} introduced the Fokker-Planck equation to identify the deterministic characteristics of noise perturbed limit cycle oscillations in a turbulent thermoacoustic system undergoing a supercritical Hopf bifurcation. Gopalakrishnan \textit{et al.} \cite{gopalakrishnan2016stochastic} derived the stationary amplitude distribution from the Fokker-Planck equation of a stochastic Balasubramanian-Sujith oscillator model \cite{balasubramanian2008thermoacoustic} for the horizontal Rijke tube undergoing a subcritical Hopf bifurcation. They observed the presence of stochastic P-bifurcations at low levels of noise as well as their absence at high levels. At a low noise level, the transition of the system behavior from the subthreshold to the bistable (or hysteresis) region is associated with the occurrence of a P-bifurcation, where the PDF changes from being unimodal to a bimodal form. While the transition from the bistable to limit cycle region is associated with the occurrence of a second P-bifurcation, where the PDF changes from being bimodal to a unimodal form. With an increase in the noise intensity, the width of the hysteresis region correspondingly decreases following a power-law behavior, while the transition from steady to limit cycle oscillations becomes continuous \cite{gopalakrishnan2015effect}. As a result, at a very high noise level, we do not observe any hysteresis region; hence, the PDF always remains unimodal, leading to the absence of P-bifurcation in the system. 

Saurab \textit{et al.} \cite{saurabh2017noise} experimentally investigated the effects of noise in a Rijke tube with laminar premixed flame and observed the presence of a P-bifurcation along with coherence resonance. Li \textit{et al}. \cite{li2019coherence} analytically studied the stability of the stochastic one-dimensional self-excited nonlinear standing wave thermoacoustic system \cite{li2017experimental}. Moreover, Li \textit{et al}. \cite{li2018effects} identified the presence of two different types of P-bifurcations in this system, one with a crater-like PDF and the other had two peaks and one trough. 


\subsection{Noise-induced limit cycle oscillations: effect of bistability}

\begin{figure}
\centering
\includegraphics[width=0.48\textwidth]{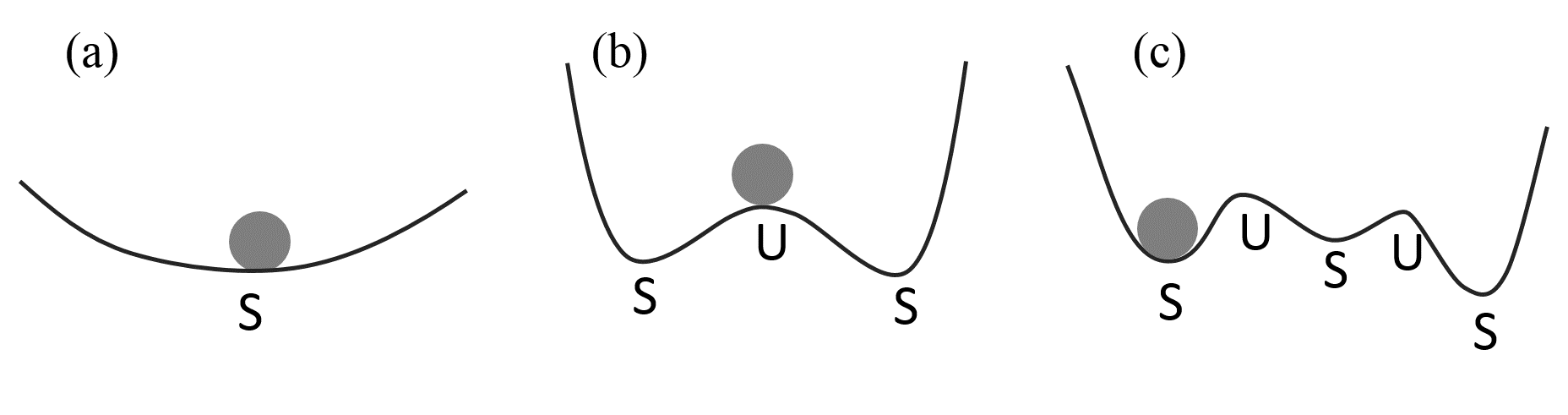}
\caption{\label{fig:17}Schematic diagrams representing a ball resting on a surface that is (a) globally stable, and (b), (c) having multiple stable states. The introduction of finite amplitude perturbations can change the stability of the system in (b) and (c). The stable and unstable positions are marked by S and U, respectively.}
\end{figure}

As discussed in Sec. \ref{sec3}.1, Rijke tubes undergo a subcritical Hopf bifurcation for certain parameter ranges. This bifurcation is accompanied by the formation of a hysteresis loop and a bistable region (Fig. \ref{fig:15}b). In the bistable region, the system dynamics can exist as two possible stable states, and the choice of the state acquired by the system depends on the initial amplitude or the energy possessed by it \cite{subramanian2010bifurcation,ananthkrishnan2005reduced}. For example, let us consider a simple nonlinearly unstable system of a ball resting in the depression. In this scenario, small perturbations to the ball's displacement would die down making the system linearly stable. On the other hand, large perturbations would cause the ball to become unstable, causing it to fall to another stable state (Fig. \ref{fig:17}). Hence, depending on the amplitude of initial perturbations, the system would either remain in the same stable state or transition to another stable state.

For subcritical Hopf bifurcation, if the system is operating at the stable state in the bistable zone and the perturbations induced are below a certain threshold, the system approaches the same stable steady state after the transients subside. On the other hand, when the amplitude of perturbations or the corresponding energy is higher than the threshold, the system switches its dynamical behavior and transitions to stable limit cycle oscillations. Such a phenomenon is commonly known as subcritical transition  \cite{chapman2002subcritical,baggett1997low}.

For example, in a fluid flow through a pipe, the transition from laminar to turbulent flow occurs when the value of Reynolds number is greater than 5000 ($Re_{cr}>5000$), and the unstable eigenvalues emerge in the system \cite{ryan1959transistion,rott1990note}. Therefore, for $Re>5000$, any external perturbations introduced in the system grows in time. However, nonlinear perturbation analysis  \cite{schmid2007nonmodal} shows that the highest Reynolds number at which external perturbations decay is between $100 < Re^* <1000$, which is significantly less than $Re_{cr} = 5000$. Such behavior in turbulent flow systems is referred to as a \textit{bypass transition} and such transitions are different for normal and non-normal systems. In the case of normal systems, the two Reynolds numbers coincide ($Re^*=Re_{cr}$), whereas the difference between the two critical Reynolds numbers arises leading to a bistable zone. For a normal system, when the individual eigenvectors decay, the resultant also decays. In contrast, for a non-normal system, the individual eigenvectors may decay, but the resultant can grow.
Such a non-normal system may show a transient growth in the amplitude of perturbations in a linearly stable regime and the amplitude of the fluctuations may grow due to linear mechanisms to a level where the nonlinearities are important. When the system is operating in a bistable state, it may switch to other dynamical states as a result of nonlinear driving. Therefore, if a system operates in the linearly stable steady state of the bistable regime of a subcritical Hopf bifurcation, the presence of both nonlinearity and non-normality plays a role in exciting the system to limit cycle oscillations from small but finite amplitude disturbances \cite{juniper2011triggering,sujith2016non}.

The progress in non-modal stability analysis and non-normal behavior have recently enabled us to study the impact of short-term behavior on the occurrence of thermoacoustic instabilities from a new perspective \cite{sujith2016non}. The influence of non-normality has been investigated theoretically in a horizontal Rijke tube \cite{balasubramanian2008thermoacoustic,kedia2008impact,juniper2011triggering, zhao2012transient,selimefendigil2011identification}, premixed flame Rijke tube burners \cite{subramanian2010bifurcation,balasubramanian2008non,zhang2015transient}, resonator tubes \cite{mangesius2011discrete}, one-dimensional thermoacoustic systems \cite{wieczorek2011assessing,li2015mean}, and entropy waves \cite{li2016effect}. Experimental verification of non-normality in thermoacoustic systems was performed by Mariappan and Sujith \cite{mariappan2015experimental} in a horizontal Rijke tube. Further investigations on the non-normal behavior in thermoacoustic systems highlighted its effects on various control strategies. For example, nonlinear driving that drives the system to an unstable behavior can be prevented by controlling the transient growth through active control \cite{kulkarni2011non} or feedback control \cite{zhang2015feedback}. For a comprehensive discussion on the non-normal and nonlinear nature of the thermoacoustic system, the readers may refer to Juniper \cite{juniper2011triggering} and Sujith \textit{et al.} \cite{sujith2016non}.

Furthermore, the phenomenon of subcritical transition to limit cycle oscillations in the bistable region due to external perturbations has been examined in different genres of thermoacoustic systems including a horizontal Rijke tube \cite{gopalakrishnan2015effect,mariappan2015experimental}, premixed flame Rijke tube burner \cite{kabiraj2011investigation,zhao2012transient}, and ducted non-premixed flame burner \cite{jegadeesan2013experimental}, and Rijke-Zhao tubes \cite{zhao2012energy}. The external perturbations (both harmonic and noise) required for a subcritical transition in thermoacoustic systems are often generated through loudspeakers. We note that such a subcritical transition of the system behavior from stable steady state to stable limit cycle oscillations (i.e., thermoacoustic instability) due to external perturbations has been traditionally referred to as ‘triggering’ in the parlance of aerospace and rocket propulsion systems \cite{culick2006unsteady,juniper2011triggering}. 

In addition to excitation through periodic perturbations, noise-induced excitation of limit cycle oscillations in the bistable region has also received immense interest, due to its practical applicability in excitation of thermoacoustic instability in solid rocket combustors \cite{culick2006unsteady,lieuwen2002experimental,lieuwen2005background,clavin1994turbulence}. This phenomenon of noise-induced limit cycle oscillations has been studied theoretically \cite{waugh2011triggeringa,waugh2011triggeringb} as well as experimentally \cite{jegadeesan2013experimental} in different Rijke tube oscillators. An excitation to limit cycle oscillations in such systems is strongly dependent on the strength of the noise. Hence, when the strength of noise added to the system exceeds a certain threshold value, the system undergoes nonlinear driving to a limit cycle state  \cite{waugh2011triggeringa,waugh2011triggeringb}. Further studies examined the influence of the frequency and the type of noise on the subcritical transition of a thermoacoustic system and concluded that low frequency noise is more effective in facilitating the transition in the system than high frequency noise \cite{juniper2011triggering}. Furthermore, pink noise is found to be more effective than white noise or blue noise in exciting the system to thermoacoustic instability  \cite{waugh2011triggeringa}. Jegadeesan and Sujith \cite{jegadeesan2013experimental} found that the noise strength required for exciting limit cycle oscillations in a diffusion flame Rijke tube burner is significantly lower than that required for harmonic perturbations in a deterministic system.

So far, we have discussed the effect of perturbations on the stability and dynamical characteristics of the acoustic pressure field in the subthreshold and bistable regimes of different Rijke tube oscillators. In the upcoming section, we move our attention towards synchronization in coupled thermoacoustic oscillators. 

\section{\label{sec5} Synchronization in thermoacoustic oscillators}

In this section, we will present the synchronization characteristics of coupled or forced Rijke tube oscillators. As mentioned above, during the state of thermoacoustic instability, such as limit cycle, quasiperiodic or chaotic oscillations, a Rijke tube system behaves as a nonlinear oscillator. Coupling or forcing of such Rijke tube oscillators can cause the system to exhibit a wide variety of synchronization phenomena. Before going into the details of synchronization of Rijke tube oscillators, we first provide a brief discussion on synchronization of general oscillators.

Synchronization is a ubiquitous phenomenon observed due to the interaction between two or more oscillators in many natural and engineering systems \cite{pikovsky2003synchronization,balanov2009synchronization,pikovsky2012synchronization}. It refers to the adjustment of motions of the constituent oscillators to a common phase and frequency upon coupling \cite{pikovsky2003synchronization}. Interaction between oscillators has been studied primarily through two mechanisms, i.e., mutual coupling and forcing \cite{balanov2009synchronization}. The corresponding types of synchronization are classified as mutual and forced synchronization, respectively. During mutual coupling, the constituent oscillators change their behavior due to the presence of bidirectional coupling between them. Various types of local, non-local, and global coupling schemes have been employed to study mutual synchronization of oscillators \cite{lakshmanan2011dynamics,boccaletti2018synchronization}. These couplings include time-delay, dissipative, conjugate, diffusive, environment, on-off coupling, etc. \cite{saxena2012amplitude,boccaletti2018synchronization,zou2021quenching}. On the other hand, in the forced coupling, a unidirectional coupling exists between the forcing and the oscillator \cite{pikovsky2003synchronization}. The forcing parameters such as the amplitude and frequency of the forcing signal are varied as control parameters in studies on forced synchronization, as elaborated in Sec. \ref{FS}. 

\begin{figure}[t!]
\centering
\includegraphics[width=0.45\textwidth]{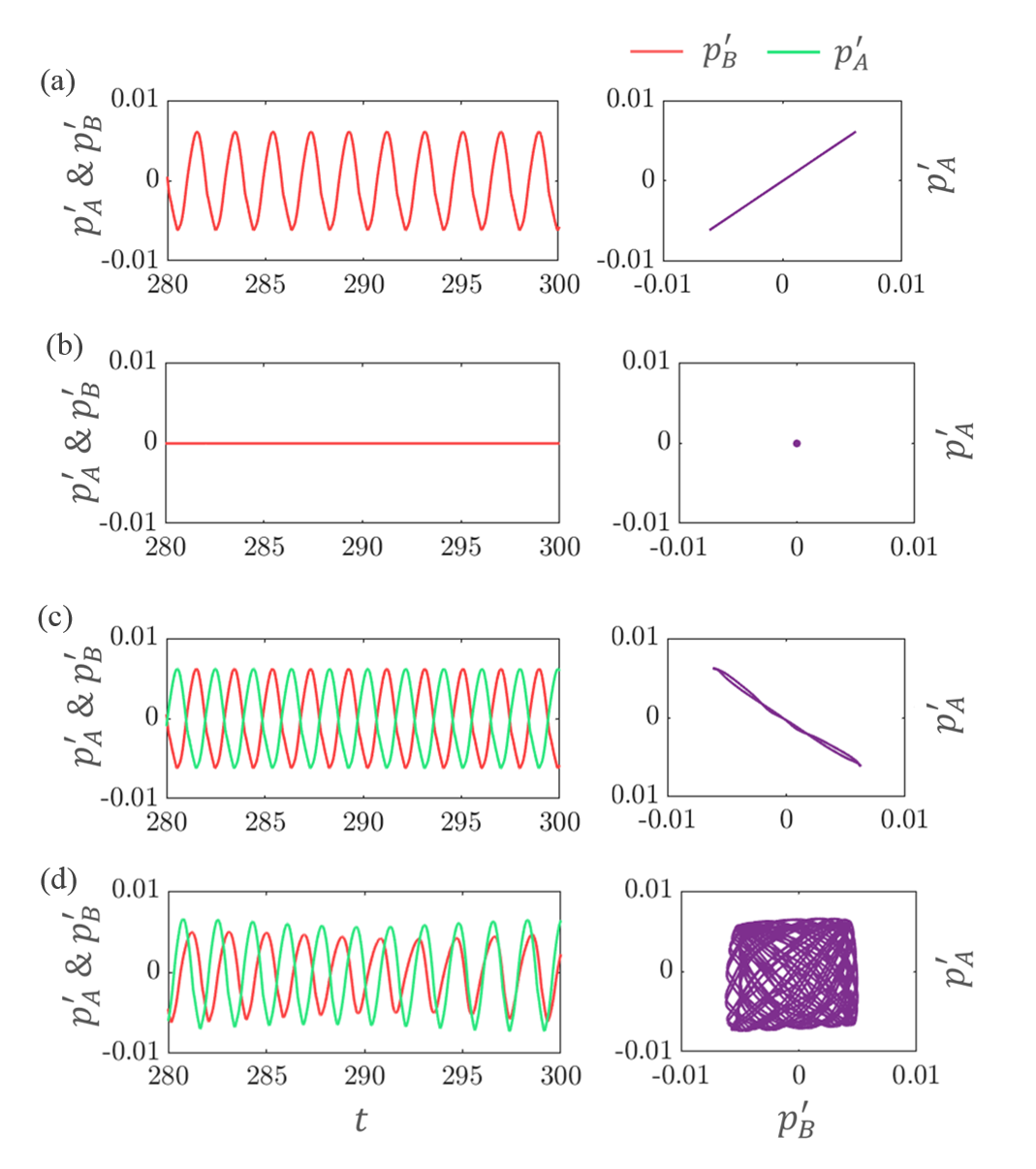}
\caption{\label{fig:18}Time series and the amplitude correlation plot corresponding to (a) in-phase synchronization, (b) amplitude death, (c) anti-phase synchronization, and (d) desynchronization in a model of two coupled Rijke tube oscillators \cite{srikanth2021dynamical}.}
\end{figure}

The mutual interaction of two oscillators prominently gives rise to three distinct states of coupled behavior: synchronized oscillations, desynchronized oscillations, and quenching of oscillations. Depending on the value of phase difference between the synchronized oscillations, the coupled behavior can be classified as in-phase (0 deg phase difference) and anti-phase (180 deg phase difference) synchronization, as shown in Fig. \ref{fig:18}a and \ref{fig:18}c, respectively. If both the oscillators possess different (non-identical) frequencies, the phase difference between them drifts in time, and the corresponding oscillators are characterized as desynchronized (Fig. \ref{fig:18}d). 
The interaction between the coupled oscillators sometimes leads to complete suppression of their oscillations, known as ‘oscillation quenching'. Oscillation quenching has been mainly classified into two types: amplitude death and oscillation death \cite{koseska2013oscillation,zou2021quenching}. During amplitude death, all the oscillators reach a homogenous steady state (Fig. \ref{fig:18}b), while during oscillation death, both the oscillators stabilize to different steady states (non-homogenous steady states).  In the case of forced interaction, quenching of oscillations in the forced system occurs through a phenomenon of asynchronous quenching \cite{dewan1972harmonic,keen1970suppression,mondal2019forced}, wherein the amplitude of the oscillator drops to a minimum value equal to that of external forcing. Thus, the methodologies based on the mechanisms of coupling or forcing an oscillator can help in mitigating thermoacoustic instabilities \cite{sujiththermoacoustic}. We have provided an elaborate discussion on amplitude death and asynchronous quenching in Sec. \ref{sec6}.

There have been extensive studies performed on mutual or forced synchronization of phenomenological oscillators, such as Stuart-Landau, Van der Pol, R\"{o}ssler, and Lorenz oscillators, neural networks, ecological models, population models, disease spread models, etc. \cite{balanov2009synchronization,lakshmanan2012nonlinear,pikovsky2003synchronization,quiroga2002performance,granovetter1978threshold,pikovsky2012synchronization,mosekilde2002chaotic,blasius1999complex}. These studies have shed light on many hidden features of interacting systems. In the upcoming discussion, we show that the experimental and theoretical investigations on Rijke tube oscillators also demonstrate the characteristics of mutual and forced synchronization as observed for paradigmatic oscillators. 

\subsection{Mutual synchronization of coupled Rijke tube oscillators}

In Sec. \ref{sec3}, we discussed the bifurcation characteristics of a single Rijke tube oscillator during the transition from steady state to limit cycle oscillations. The introduction of coupling between two such oscillators significantly changes their dynamical properties. In this case, Srikanth \textit{et al}. \cite{srikanth2021dynamical,srikanth2021self} found a forward shift in the occurrence of Hopf bifurcation along with a reduction in the amplitude of limit cycle oscillations when compared to these properties for an isolated Rijke tube oscillator. Recently, there has been an increased interest in studying the behavior of two coupled Rijke tube oscillators \cite{thomas2018effecta,thomas2018effectb,dange2019oscillation,hyodo2020suppression,sahay2020dynamics,srikanth2021dynamical}. These studies have potential applications in understanding the interaction between multiple combustion systems of can-annular type combustors used in gas turbine engines \cite{jegal2019mutual,ghirardo2019thermoacoustics,farisco2017thermo,moon2020mutual,moon2019combustion}. 

The coupled behavior of two thermoacoustic systems has been studied under two coupling schemes: time-delay and dissipative \cite{thomas2018effecta}. Time-delay coupling accounts for the finite time required for the propagation of information (or acoustic oscillations) from one oscillator to another. This type of coupling is introduced in an experimental system by connecting the two oscillators using a coupling tube, whose diameter is smaller than the diameter of the Rijke tube \cite{biwa2015amplitude,dange2019oscillation}. The increase in the length ($l_c$) and the diameter ($d$) of the coupling tube in experiments have direct correspondence with an increase in the time delay ($\tau$) and the strength of coupling ($\mathcal{K}_\tau$) between the oscillators in the model \cite{dange2019oscillation,srikanth2021dynamical}. On the other hand, dissipative coupling accounts for the dissipation of energy during the transfer of information from one oscillator to another as a consequence of mutual interaction. The sources of dissipation of energy could arise due to the direct flow transfer from one system to another through the coupling tube or from the loss of acoustic energy due to the introduction of the coupling tube \cite{biwa2015amplitude,dange2019oscillation}. In addition to the variation of coupling parameters, the effect of change in system parameters, such as the amplitude and natural frequency of each oscillator in the uncoupled state, has been shown to play an important role in the coupled dynamics of Rijke tube oscillators \cite{dange2019oscillation,premraj2021effect,srikanth2021dynamical}.

\begin{figure}
\centering
\includegraphics[width=0.45\textwidth]{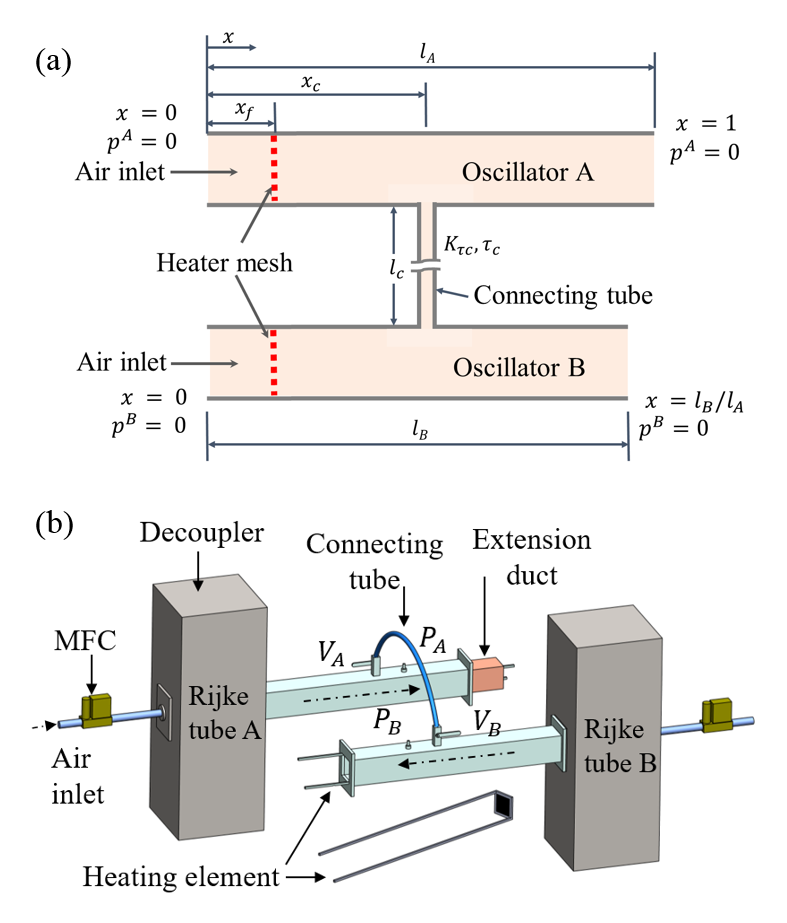}
\caption{\label{fig:19} Schematics of the model and the experimental setup of two coupled Rijke tube oscillators. Coupling parameters in the model are coupling strength ($\mathcal{K}$) and coupling delay ($\tau$), while in experiments these parameters are the length and diameter of the connecting tube. (b) Reproduced with permission from  Dange \textit{et al}. \cite{dange2019oscillation}.}
\end{figure}

Thomas \textit{et al.} \cite{thomas2018effecta} used the mathematical model of the thermoacoustic oscillator \cite{balasubramanian2008thermoacoustic} explained in Sec. \ref{sec2}.4, to study the coupled interaction of two thermoacoustic oscillators (see Fig. \ref{fig:19}a). The equations  for a pair of Balasubramanian-Sujith oscillators coupled through both time-delay and dissipative coupling are as follows \cite{srikanth2021dynamical}:
\begin{equation}
    \label{eqtn:BSC}
    \frac{d \eta_j^a}{dt} = \dot{\eta}_j^a 
\end{equation}
\begin{align*}
       \frac{d \eta_j^a}{d t}  + 2 \zeta_j \omega_j \dot{\eta_j^a} & + \omega_j^2 \eta_j^a = -j\pi K \sin(j \pi x_f) \\
       & \times \left[  \sqrt{\left| \frac{1}{3} + \cos(j \pi x_f) \eta_j^a(t - \tau_1) \right|} - \sqrt{\frac{1}{3}}\right] 
\end{align*}
\begin{equation*}
     \hspace{16ex} + \underbrace{\mathcal{K}_d (\dot{\eta_j^b} - \dot{\eta_j^a})}_\text{Dissipative coupling} + \underbrace{\mathcal{K}_\tau (\dot{\eta_j^b}(t - \tau) - \dot{\eta_j^a}(t))}_\text{Time-delay coupling}
\end{equation*}
where $a$ and $b$ indicate the oscillators in the coupled system, $\mathcal{K}_d$ and $\mathcal{K}_\tau$ denote the dissipative coupling strength and time-delay coupling strength, respectively, and $\tau$ denotes the time delay between the oscillators. Keeping either of the coupling strengths ($\mathcal{K}_d$ or $\mathcal{K}_\tau$) as zero makes the system purely time-delay coupled or dissipative coupled, respectively. The remaining terms in the equation are similar to those discussed in Sec. \ref{sec2}.4.


In an experimental study, Dange \textit{et al.} \cite{dange2019oscillation} observed the synchronization of large amplitude limit cycle oscillations in two identical Rijke tube oscillators coupled using a single coupling tube (see Fig. \ref{fig:19}b). With an increase in the length of the coupling tube ($l_c$), they found that the oscillators suddenly change their synchronized dynamics from in-phase to anti-phase synchronization or vice-versa, at a critical value of the coupling tube length. This abrupt change in the phase of these oscillators from one form of synchronization to another is commonly referred to as phase-flip bifurcation \cite{prasad2006phase,prasad2008universal}. 

Srikanth \textit{et al}. \cite{srikanth2021dynamical} studied the characteristics of coupled Rijke-tube oscillators during phase-flip bifurcation for a wider range of coupling parameters numerically and analytically. They found a recurring occurrence of phase-flip bifurcation in a system at odd multiples of half the time period of limit cycle oscillations, i.e., for $\tau_c=\tau/T=n/2$, where $\tau$ is the coupling delay, $T$ is the time period of oscillations in the uncoupled state, and $n=1,3,5,...$. In addition to the change in the relative phase of oscillators, they detected an abrupt jump in the frequency of the oscillators during the phase-flip bifurcation. The amplitude of limit cycle oscillations also shows an oscillatory pattern, where the maximum amplitude suppression is observed during the occurrence of phase-flip bifurcation in the system. 
Having discussed mutual synchronization in coupled thermoacoustic oscillators, we present the forced synchronization characteristics of such oscillators in the following subsection.



\subsection{Forced synchronization of a Rijke tube oscillator \label{FS}}

Studies on forced synchronization highlight the behavior of a Rijke tube oscillator in response to external harmonic forcing, as modeled by the following equation \cite{mondal2019forced}:
\begin{equation}
    \label{eqtn:BSF}
    \frac{d \eta_j}{dt} = \dot{\eta_j} 
\end{equation}
\begin{align*}
       \frac{d \eta_j}{d t}  + 2 \zeta_j \omega_j \dot{\eta_j} + &  \omega_j^2 \eta_j  = -j\pi K \sin(j \pi x_f) \\
       & \hspace{2ex} \times \left[  \sqrt{\left| \frac{1}{3} + \cos(j \pi x_f) \eta_j(t - \tau_1) \right|} - \sqrt{\frac{1}{3}}\right]  \\
       & \hspace{2ex} + \hspace{1ex} \underbrace{A_f\sin(2 \pi f_f t)}_\text{Forcing term} 
\end{align*}
where $A_f$ is the forcing amplitude and $f_f$ is the forcing frequency.

As mentioned previously, the occurrence of forced synchronization is characterized by the state where the forced oscillator exhibits the same frequency as the forcing system and the relative phase between these systems remains constant in time. The response of a forced oscillator to forcing depends on $A_f$ and $f_f$ of the forcing. Generally, such responses of a forced oscillator are effectively represented using an Arnold tongue, which is the synchronization boundary in the parameter space of amplitude and frequency of forcing. Nearby the onset of forced synchronization, a plethora of dynamical states are observed in the relative phase dynamics of the forced system, which include phase drifting, intermittent phase locking, phase trapping, and phase locking \cite{balanov2009synchronization}. Furthermore, a forced oscillator shows resonance amplification and synchronous quenching when forcing is applied near the natural frequency of the oscillator \cite{odajima1974synchronous}.  

Recently, forced synchronization of limit cycle, quasiperiodic, and chaotic oscillations in a Rijke-tube oscillator have been studied in great detail. All the aforementioned phenomena, which have been previously observed in forced paradigmatic oscillators such as Van der pol and Stuart-Landau oscillators, are witnessed in the dynamics of a forced Rijke tube oscillator \cite{kashinath2018forced,mondal2019forced,guan2019control,guan2019forced,guan2020intermittency,premraj2021effect}. Kashinath \textit{et al.} \cite{kashinath2018forced} studied forced synchronization of limit cycle, quasiperiodic and chaotic oscillations in a model for a laminar premixed flame Rijke tube burner. In an experimental study, Mondal \textit{et al}. \cite{mondal2019forced} examined forced synchronization of limit cycle oscillations in the acoustic pressure of a horizontal Rijke tube system, while Guan \textit{et al.} \cite{guan2019forced} and Roy \textit{et al.} \cite{roy2020mechanism} investigated forced synchronization of limit cycle oscillations in both the acoustic pressure and heat release rate fluctuations of a laminar premixed flame Rijke tube burner. Furthermore, Guan \textit{et al.} \cite{guan2019forced} and Sato \textit{et al.} \cite{sato2020synchronization} extended the experimental investigation to study forced synchronization of quasiperiodic oscillations in a laminar premixed flame Rijke tube burner and a gas filled resonance tube, respectively. In a different study, Sahay \textit{et al}. \cite{sahay2020dynamics} examined forced synchronization of two coupled identical and non-identical horizontal Rijke tube oscillators with forcing being applied to one of the oscillators, both experimentally and numerically.

\begin{figure}[t!]
\centering
\includegraphics[width=0.48\textwidth]{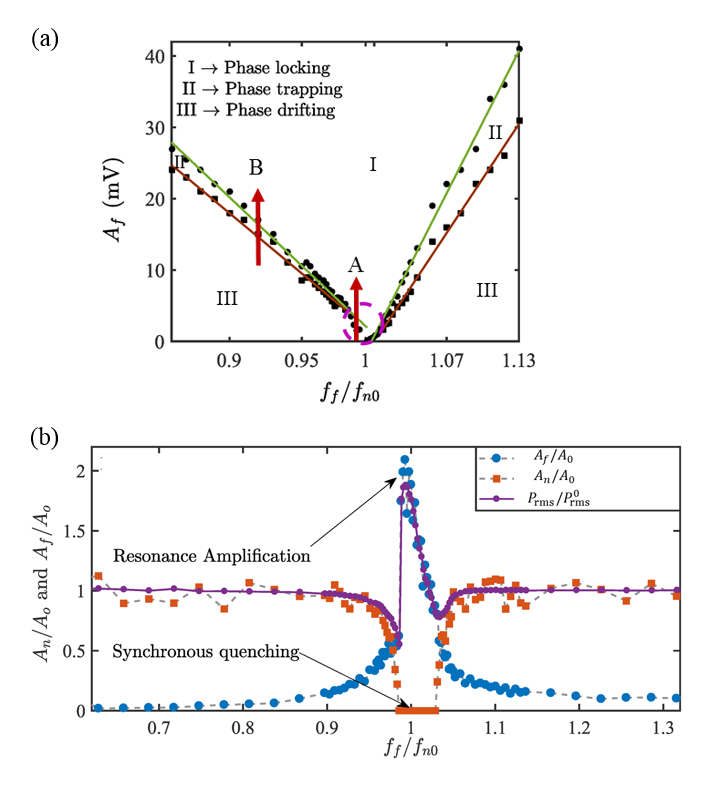}
\caption{\label{fig:20}(a) Arnold tongue highlighting the boundaries between the region of occurrence of (I) phase locking, (II) phase trapping, and (III) phase drifting in the two-parameter bifurcation plot between the amplitude of forcing ($A_f$) and the relative frequency of forcing ($f_f/f_{n0}$, where $f_f$ is the forcing frequency and $f_{n0}$ is the natural frequency of oscillation). (b) Variation of normalized amplitude response of spectral peak corresponding to self-excited oscillations ($A_n/A_0$), forcing oscillations ($A_f/A_0$), and overall acoustic pressure signal ($P_{rms}/P_{rms}^0$) in the horizontal Rijke tube. Exhibition of synchronous quenching of self-excited oscillations and resonance amplification of forcing oscillations is observed in the phase-locking region. The dotted circle in (a) represents the region where phase trapping is not observed. Reproduced with permission from Mondal \textit{et al}. \cite{mondal2019forced}.}
\end{figure}

Next, we discuss the key properties of forced synchronization of limit cycle oscillations in a Rijke tube oscillator. In Fig. \ref{fig:20}a, we show the Arnold tongue (i.e., V-shaped synchronization boundaries) observed experimentally for the forced response of limit cycle oscillations in the acoustic pressure of a horizontal Rijke tube oscillator \cite{mondal2019forced}. Inside the Arnold tongue, limit cycle oscillations in the Rijke tube are synchronized with the forcing, but outside the Arnold tongue, these oscillations are desynchronized with each other. The value of $A_f$ required for forced synchronization of the oscillator at a fixed value of $f_f$ exhibits a near linear variation with increasing frequency detuning ($|f_{n0}-f_f|$) on either side of the natural frequency ($f_{n0}$) of the oscillator. A subsequent study by Sahay \textit{et al.} \cite{sahay2020dynamics} found that an increase in the amplitude of the limit cycle oscillations in the unforced state narrows the Arnold tongue region, causing a corresponding increase in the value of forcing amplitude for forced synchronization of the oscillator. 

We notice that the occurrence of forced synchronization happens via two routes, namely the locking route and the suppression route \cite{balanov2009synchronization}, indicated by route A and route B in Fig. \ref{fig:20}a, respectively. When the difference between the forcing and the natural frequency of the oscillator is high (shown as route B in Fig. \ref{fig:20}a), the transition occurs via the route of suppression. In this route, an increase in $A_f$ causes the transition to the phase-locking state through a torus-death bifurcation, where the magnitude of natural frequency ($f_{n0}$) in the spectrum is suppressed without shifting the value of $f_{n0}$ towards the forcing frequency $f_f$ \cite{balanov2009synchronization}. On the other hand, when the difference between $f_f$ and $f_{n0}$ is low (shown as route A in Fig. \ref{fig:20}a), we observe the occurrence of the locking route to forced synchronization. The oscillator undergoes a saddle-node bifurcation to attain phase-locking, where the position of the natural frequency peak gradually shifts towards the forcing frequency with an increase in the forcing amplitude \cite{balanov2009synchronization}. 

\begin{figure}[t!]
\centering
\includegraphics[width=0.48\textwidth]{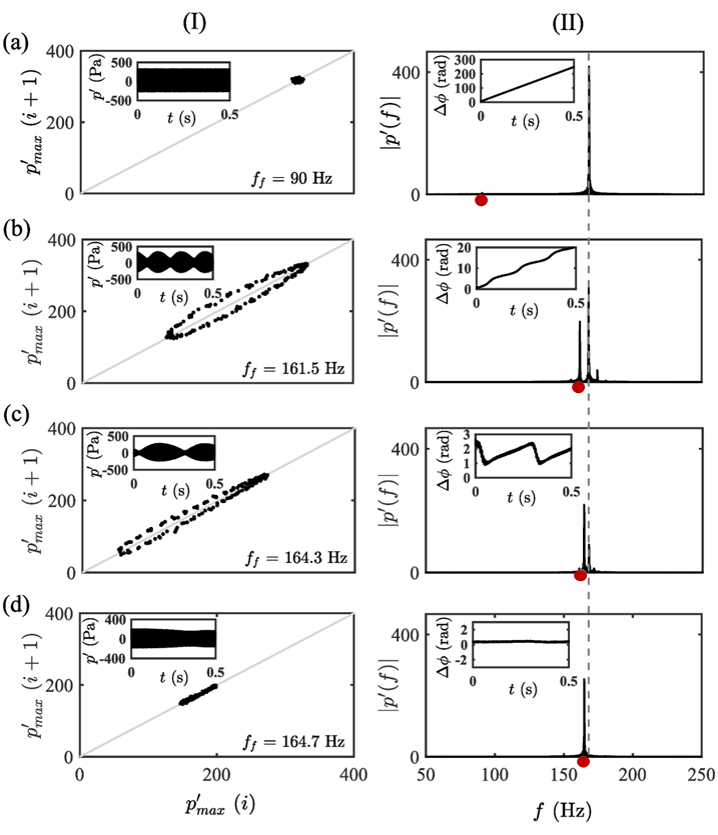}
\caption{\label{fig:21}(I) Poincaré section with an inset of the pressure time series and (II) the frequency spectrum along with an inset of the unwrapped relative phase time series, corresponding to the dynamical states of (a) phase drifting, (b) intermittent phase locking, (c) phase trapping, and (d) phase locking in a forced Rijke tube system. In (II), the red dot on the abscissa represents the forcing frequency, whereas the dashed vertical line indicates the natural frequency of limit cycle oscillations in the absence of forcing. Reproduced with permission from Mondal \textit{et al.} \cite{mondal2019forced}.}
\end{figure}

Furthermore, in the Arnold tongue (Fig. \ref{fig:20}), we notice the presence of the region of phase trapping (Fig. \ref{fig:21}c) in between the regions of phase drifting (Fig. \ref{fig:21}a) and phase locking (Fig. \ref{fig:21}d) for the suppression route, while a direct transition from phase drifting to phase locking is noticed for the locking route. Here, phase drifting is a state of desynchronization between the forced oscillator and the forcing system, where the unwrapped relative phase between the forced and forcing oscillations exhibits a continuous increase/decrease in time (inset in Fig. \ref{fig:21}a-II). The pressure signal observed during this state shows limit cycle oscillations (inset in Fig. \ref{fig:21}a-I) and the corresponding Poincaré section (first return map) shows a single small cluster of points (Fig. \ref{fig:21}a-I). During the state of phase trapping, the relative phase between the systems is bounded and oscillates about the mean phase difference (inset in Fig. \ref{fig:21}c-II). The pressure signal exhibits amplitude modulation (inset in Fig. \ref{fig:21}c-I), its amplitude spectrum shows a dominant peak at the forcing frequency (Fig. \ref{fig:21}c-II), and the Poincaré section shows a closed-loop orbit (Fig. \ref{fig:21}c-I), indicating the presence of quasiperiodic oscillations in acoustic pressure during this state. For phase locking (forced synchronization) state, the unwrapped relative phase between the forcing and forced oscillations remains constant in time (inset in Fig. \ref{fig:21}d-II), the pressure signal shows limit cycle oscillations (inset in Fig. \ref{fig:21}d-I) at the forcing frequency (Fig. \ref{fig:21}d-II), and the Poincaré section shows a single clutter of point (Fig. \ref{fig:21}d-I).

Moreover, prior to the occurrence of phase trapping, an intermediate state called intermittent phase locking is observed in the forced system. During this state, the unwrapped relative phase between the forced and the forcing systems demonstrates an alternate occurrence of epochs of phase locked and phase drifting oscillations, where the phase drifting region is associated with phase jumps covering integer multiples of $2\pi$ rad (inset in Fig. \ref{fig:21}b-II). Due to the presence of strong peaks at both the forcing and the natural frequencies (Fig. \ref{fig:21}b-II), we notice the presence of modulations (beating) in the amplitude envelope of the acoustic pressure signal (inset in Fig. \ref{fig:21}b-I), where the modulation frequency is equal to the difference between these frequency peaks. The Poincaré section of this state shows a closed-loop orbit, indicative of quasi-periodic oscillations (Fig. \ref{fig:21}b-I). 

Furthermore, we observe an interesting behavior of the simultaneous occurrence of resonance amplification of the forcing signal and the synchronous quenching of natural oscillations inside the Arnold tongue \cite{mondal2019forced}. In Fig. \ref{fig:20}b, we present the variation of normalized spectral amplitude of self-excited oscillations ($A_n/A_0$) and forced oscillations ($A_f/A_0$) along with the root-mean-square (rms) amplitude of the acoustic pressure signal ($P_{rms}/P_{rms}^0$) in a horizontal Rijke tube with increasing $f_f$ at a fixed value of $A_f$. The spectral amplitude of self-excited oscillations gradually decreases as the frequency ratio ($f_f/f_{n0}$) approaches the Arnold tongue and attains zero inside the boundaries of the Arnold tongue. This behavior is regarded as the occurrence of synchronous quenching \cite{odajima1974synchronous}. In contrast, the spectral amplitude of forced oscillations shows a gradual increase near the Arnold tongue, followed by an abrupt jump to a high amplitude as the frequency ratio enters the Arnold tongue from the left-hand-side. This behavior is attributed to the occurrence of resonant amplification \cite{odajima1974synchronous}. This is followed by a gradual decrease in the amplitude of forcing oscillations with a further increase in the frequency ratio. 

The simultaneous occurrence of synchronous quenching and resonance amplification, also referred to as synchronance (synchronization-resonance) by Mondal \textit{et al.} \cite{mondal2019forced}, leads to the dominance of the forcing signal in the final response signal of the forced acoustic pressure oscillations. When the forcing frequency is much higher than the natural frequency, both the spectral amplitude curves (i.e., $A_n/A_0$ and $A_f/A_0$) saturate and become independent of changes in the forcing frequency. The combined behavior of the two spectral amplitudes is observed in the variation of the rms value of the response pressure signal ($P_{rms}/P_{rms}^0$).

Unlike forced synchronization of limit cycle oscillations, forced synchronization of quasiperiodic oscillations happens via a complicated path and involves various variants of quasiperiodic oscillations \cite{guan2019forced,sato2020synchronization}.  The system first transitions from quasiperiodic oscillations having two dominant frequencies to another variant of quasiperiodic oscillations having three dominant frequencies (two natural frequencies and one forcing frequency). This is followed by the transition to a resonant quasiperiodicity corresponding to partial synchronization, where one of the natural frequencies undergoes synchronization, whereas the other remains desynchronized with the forcing frequency. Ultimately, the system reaches complete forced synchronization of oscillations, where both natural frequency modes synchronize with the forcing frequency. Guan \textit{et al}. \cite{guan2019forced} observed the presence of two Arnold tongues, each centered at their corresponding dominant frequencies. 

\begin{figure}
\centering
\includegraphics[width=0.45\textwidth]{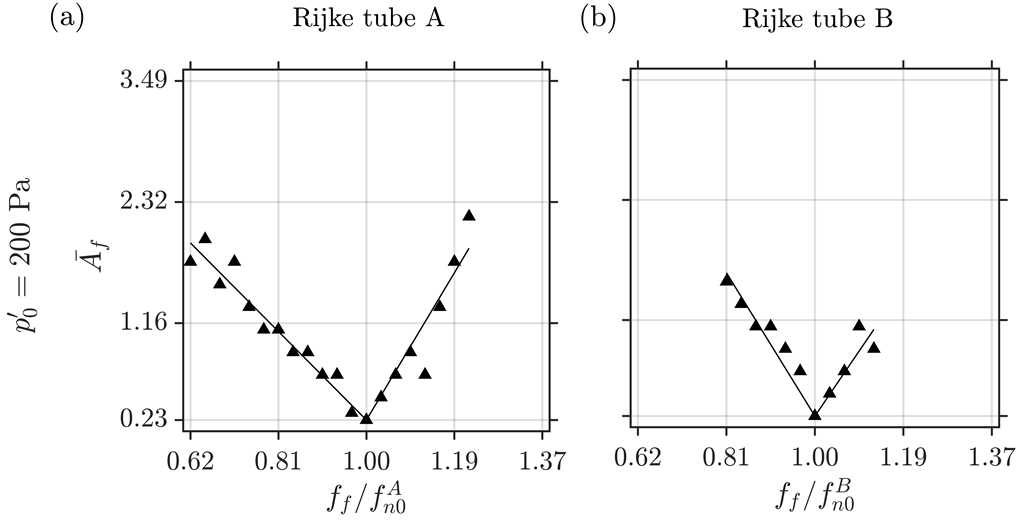}
\caption{\label{fig:22}Arnold tongue of (a) Rijke tube oscillator A and (b) Rijke tube oscillator B, where both these oscillators are mutually coupled to each other using a single coupling tube and oscillator A alone is forced using loudspeakers. Adapted with permission from Sahay \textit{et al.} \cite{sahay2020dynamics}. }
\end{figure}

Sahay \textit{et al.} \cite{sahay2020dynamics} studied the characteristics of the forced response of coupled thermoacoustic oscillators, where two horizontal Rijke tube oscillators are mutually coupled using a connecting tube and one of the two oscillators is externally forced using speakers. Figure \ref{fig:22} presents the Arnold tongue obtained from experiments through the simultaneous application of forcing and coupling in two identical Rijke tube oscillators, A and B. Here, acoustic forcing is applied only to Rijke tube A and Rijke tube B is indirectly forced through the coupling tube. The Arnold tongue of oscillator A is observed to be larger when compared to that observed for oscillator B which is not directly forced. Therefore, the region of synchronance (i.e., the combined presence of synchronous quenching and resonance amplification) is small for oscillator B when compared to oscillator A. However, in the case of non-identical Rijke tube oscillators, forced synchronization of oscillator B is rarely observed, while that of oscillator A remains nearly the same as that observed in the case of identical oscillators. 

The consequences of the interaction between coupled systems were also investigated by Zhang \textit{et al}. \cite{zhang2019devil} using the Balasubramanian-Sujith model of a horizontal Rijke tube with sinusoidal excitation. Periodic oscillations were observed when the value of $f_f$ is much lower and higher than $f_n$. In specific ranges of $f_f$, the system exhibited alternate occurrences of quasi periodicity and periodic oscillations. Furthermore, they concluded that the regime of periodic oscillations is composed of devil’s staircases. The devil’s staircase, otherwise known as the cantor function, is a monotonic continuous function mapping the set [0,1] onto itself while maintaining zero derivatives throughout the interval \cite{thomson2008elementary}.

\subsection{Synchronization between the self-excited acoustic field and the heat release rate fluctuations}

As we know, thermoacoustic instability is the result of a positive interaction between the acoustic field in the combustor and the heat release rate fluctuations in the flame. As discussed before, the Rayleigh criterion (Eq. \ref{eqtn:RC}) is a well-known condition to detect the onset of thermoacoustic instability. As per this criterion, when the mean phase difference between the acoustic pressure and the heat release rate fluctuations of the flame lies between $-\pi/2$ and $\pi/2$, the energy from the flame is periodically added to the acoustic field giving rise to thermoacoustic instabilities. Mondal \textit{et al}. \cite{mondal2017synchronous} examined the coupled behavior of these two subsystems in a laminar premixed  Rijke tube burner during the quasi-periodicity route to chaos using the framework of synchronization theory. Such an analysis may be viewed as analogous to the investigation of the coupled behavior between the human heart and respiratory or brain using synchronization theory \cite{schafer1999synchronization, mccraty2009coherent, thayer2009claude}. 

\begin{figure}[t!]
\centering
\includegraphics[width=0.47\textwidth]{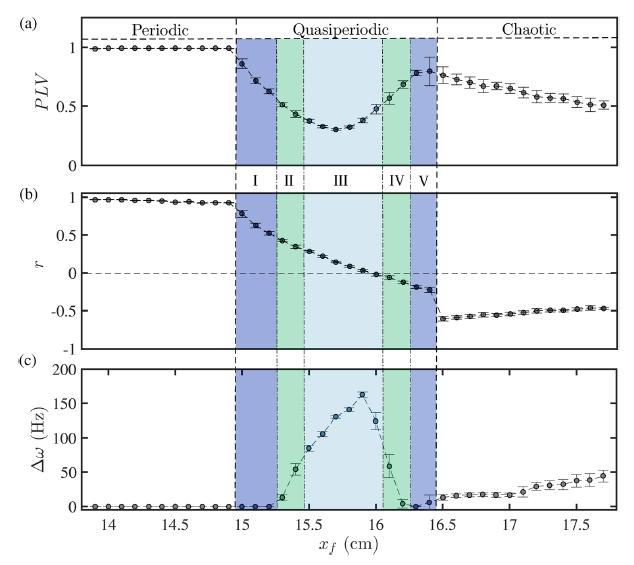}
\caption{\label{fig:23} (a) The variation of the phase locking value ($PLV$), (b) the correlation
coefficient ($r$), and (c) the relative mean frequency ($\Delta \omega$) between the acoustic pressure and heat release rate fluctuations for different regimes of quasiperiodicty route to chaos, when the location of the flame ($x_f$) inside the laminar premixed  Rijke tube burner is varied as a control parameter. Different regions of synchronization are indicated as phase locking (I and V), intermittent phase locking (II and IV), and phase drifting (III). Reproduced with permission from Mondal \textit{et al}. \cite{mondal2017synchronous}.}
\end{figure}

Mondal \textit{et al}. \cite{mondal2017synchronous} found that during the state of periodic (or limit cycle) oscillations, the acoustic pressure and heat release rate fluctuations are phase-locked (i.e., synchronized). However, in the regime of quasiperiodic oscillations, different behaviors of synchronization of these oscillators are observed which include phase-locking, phase trapping, intermittent phase locking, and phase drifting. While during chaotic oscillations, states of intermittent phase locking and phase drifting are observed. The different phase dynamics observed during these states occur primarily due to the dissimilar spectral content of two locked frequencies. They also proposed various statistical measures (see Fig. \ref{fig:23}), such as the phase locking value ($PLV$), correlation coefficient ($r$), and relative mean frequency ($\Delta \omega$) to quantitatively characterize the synchronization behavior between the acoustic pressure and the heat release rate fluctuations during self-excited states of thermoacoustic instability. These measures can help in detecting the boundaries of different states of synchronization observed during the quasiperiodicity route to chaos. 

Weng \textit{et al}. \cite{weng2020synchronization} developed a nonlinearly coupled damped oscillator model to study the coupled behavior of acoustic pressure ($p^\prime$) and heat release rate ($\dot{q^\prime}$) fluctuations in a laminar premixed Rijke tube burner. The governing equations are given below:
\begin{equation}
    \ddot{p} + \zeta_1\dot{p}(t)+\omega_{p'}^2 p(t)= C_{pq}(1-q(t-\tau_2)^2) \dot{p}(t)
\end{equation}
\begin{equation}
    \ddot{q} + \zeta_2\dot{q}(t)+\omega_{\dot q'}^2 q(t)= C_{qp}(p(t-\tau_1)^3-1)
\end{equation}
where $\tau_1= 2\pi/\omega_{p'}$,  $\tau_2 =2\pi/ \omega_{\dot q'}$, and $C_{qp}$ (or $C_{pq}$) indicates the coupling strength between $p'$ and $\dot q'$ (or $\dot q'$ and $p'$). $\omega_{p'}$, $\omega_{\dot q'}$ are the angular frequency of the acoustic and the heat release rate fluctuations, and $\zeta_1$ and $\zeta_2$ are damping terms. This model is able to capture the experimentally observed quasiperiodicity route to chaos (discussed in  Sec. \ref{sec3}.4.2) by Kabiraj \textit{et al}. \cite{kabiraj2012route} and the state of strange non-chaos identified by Premraj \textit{et al.} \cite{premraj2020strange}. Furthermore, this model also qualitatively replicates the synchronization behavior between the acoustic pressure and heat release rate fluctuations observed by Mondal \textit{et al}. \cite{mondal2017synchronous}, as shown in Fig. \ref{fig:23}.

In the preceding sections, we have discussed various behaviors and characteristics exhibited by a general nonlinear oscillator and elaborated on the existence of such dynamical behavior in Rijke tube systems. We established the vast potential of Rijke tubes in experimentally verifying complex dynamical behaviors commonly reported in literature through nonlinear oscillator models. Next, we move our focus towards various prediction and control strategies devised to warn the undesired impending thermoacoustic oscillations or to suppress them after their onset.

\section{\label{sec6}Control and prediction strategies for thermoacoustic instability}

As discussed in the previous section, thermoacoustic instabilities have been observed in the form of large amplitude self-sustained oscillations in the acoustic field of the combustor. We showed that such oscillations occur via a Hopf bifurcation in the Rijke-type thermoacoustic systems. The presence of thermoacoustic instabilities is undesirable in practical systems as they cause severe vibrations leading to heavy structural damage and loss in the performance of the engine.  Therefore, it is necessary to keep the system away from the regime of operation of thermoacoustic instability. There have been several studies dedicated to develop control strategies that can mitigate and forewarn thermoacoustic instability. Due to the simple nature and ease of handling, Rijke-type thermoacoustic systems remain the primary choice for many researchers to experiment or model novel control methodologies that can mitigate or predict thermoacoustic instabilities \cite{mondal2021mitigation,zou2021quenching}. In this section, we will summarize traditional as well as recently discovered control methodologies based on synchronization theory to suppress thermoacoustic instability in Rijke tube systems. We will also discuss recent developments in early warning technologies to forewarn critical transition to thermoacoustic instabilities in rate-dependent experiments on Rijke tube systems. 

The control strategies developed for suppression of thermoacoustic instabilities have been classified as passive and active \cite{lieuwen2005combustion,huang2009dynamics}. Passive control strategies aim at evading the occurrence of thermoacoustic instability by introducing modifications in the hardware design of the components, such as combustor geometry, fuel injection system or using acoustic dampers such as resonators or liners to remove acoustic energy from the system \cite{richards2003passive,zhao2015review,schadow1992combustion,candel2002combustion,lieuwen2005combustion,culick2006unsteady}. In contrast, active control strategies are based on interrupting the coupling between the acoustic field and the heat release rate field of the combustor through external perturbations, leading to the decay of thermoacoustic oscillations in the system \cite{mcmanus1993review,poinsot2017prediction}. Active controls are further classified into open-loop and closed-loop control (or commonly known as feedback control) depending on whether the control strategy is independent or dependent on the system response, respectively \cite{dowling2005feedback}. In the following subsections, we will discuss control strategies, such as feedback control and open-loop control, developed for the mitigation of limit cycle oscillations in a single Rijke-type thermoacoustic system. 

\subsection{Feedback control}

\begin{figure}
\centering
\includegraphics[width=0.41\textwidth]{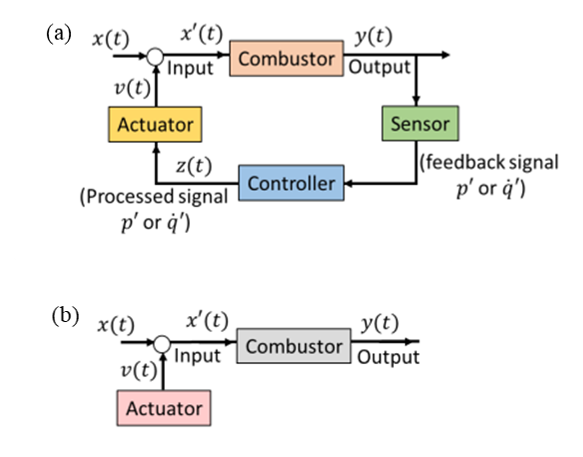}
\caption{\label{fig:24}Schematic diagrams representing the application of (a) closed-loop (or feedback) control and (b) open-loop control to mitigate limit cycle oscillations in a thermoacoustic system. }
\end{figure}

Feedback control strategy involves the usage of external actuators to perturb the inlet flow field of the Rijke tube in response to the behavior of a dynamical variable measured from the system \cite{heckl1988active,mcmanus1993review,lang1987active}. To elaborate this, we supply a portion of the information of the system to its input through controllers (Fig. \ref{fig:24}a); thus, both the input and the output of the system are dependent on each other. The controller adjusts the phase and the gain of the output signal before feeding it back into the input. Hence, this strategy involves three crucial steps \cite{dowling2005feedback,illingworth2010advances}. The acquisition of the signal of thermoacoustic instability from the system using sensors, such as a pressure transducer or a thermal sensor. This signal is then fed into a controller, where the signal is processed and supplied to an actuator. Actuators use this processed information to alter the inlet conditions of the system, thus, changing the coupling between the acoustic field and the heat release rate fluctuations, causing the mitigation of thermoacoustic instabilities. Actuation devices used in thermoacoustic systems are loudspeakers that perturb the acoustic velocity or the acoustic pressure field and fuel valves that change the heat release rate field in the Rijke tube burners \cite{li2016feedback,blonbou2000active,vaudrey2003time,rubio2015nonlinear,zalluhoglu2016delayed}. Delayed feedback has been used as a common methodology to suppress limit cycle oscillations in general oscillator systems \cite{atay1998van,reddy2000dynamics,xu2003effects} starting from the early implementation of feedback control in a laminar flame Rijke tube burner by Dines \cite{dines1984active}, Ffowcs \cite{ffowcs1984review} and Heckl \cite{heckl1988active}. Recently, such methods have been rigorously studied in the thermoacoustic literature to mitigate limit cycle oscillations \cite{lang1987active, heckl1988active, candel1992combustion, mcmanus1993review, dowling2005feedback,illingworth2010advances,dines1984active,lieuwen2005combustion}. 

\subsection{Open-loop control via asynchronous quenching}

As discussed in Sec. \ref{sec5}.2, the application of external periodic forcing to an oscillator can entrain the frequency of self-excited oscillations with forcing during the onset of forced synchronization. Further, we argued that forced synchronization can occur through the locking-route or the suppression-route (asynchronous quenching) depending on whether the frequency detuning is small or large \cite{dewan1972harmonic,balanov2009synchronization}. External periodic forcing has been commonly used in practice to suppress self-sustained oscillations in hydrodynamically unstable flows \cite{staubli1987entrainment} and wakes \cite{taira2018phase}, ionization waves \cite{ohe1974asynchronous}, oscillatory reactions \cite{fjeld1974relaxed}, transmission electrical lines \cite{shimura1967analysis}, etc., through the phenomenon of asynchronous quenching. In order to achieve such asynchronous quenching, the oscillator must be forced at a frequency far from its natural frequency \cite{dewan1972harmonic}. In the thermoacoustic literature, this method of forcing is referred to as open-loop control \cite{mcmanus1993review,candel2002combustion,evesque2003self}.

During open-loop control, external periodic forcing is used to perturb limit cycle oscillations in a thermoacoustic system at different amplitudes and frequencies. Unlike feedback control, we do not need any input from the combustor dynamics to drive the actuator in open-loop controls. The external perturbations either affect the flow field incoming to the system or affect the acoustic field developed in the system (Fig. \ref{fig:24}b). At appropriate values of the forcing parameters, external perturbations interrupt the coupling between the acoustic field and the heat release rate field, thereby quenching thermoacoustic instabilities in the system. 

\begin{figure}
\centering
\includegraphics[width=0.42\textwidth]{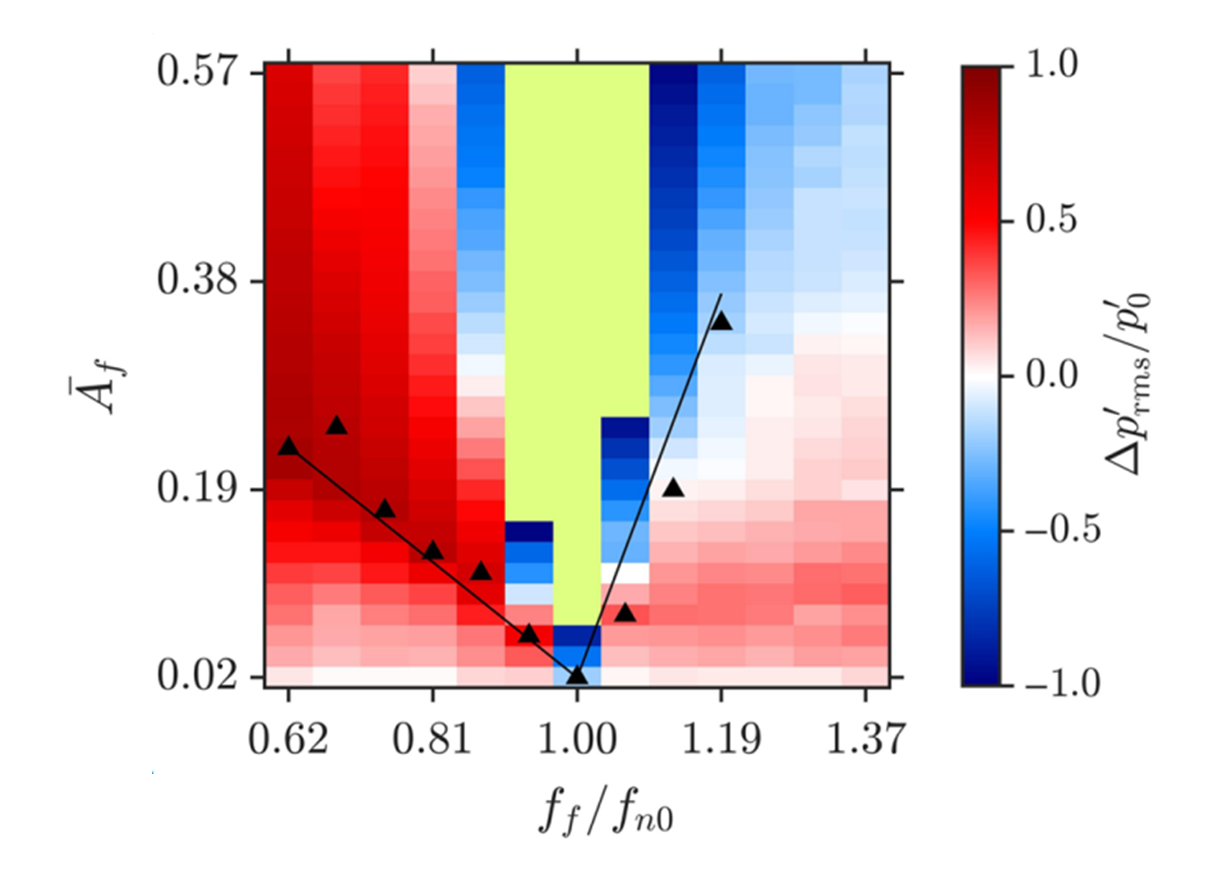}
\caption{\label{fig:25}The amplitude response (shown as the fractional change in the amplitude $\Delta p_{rms}'/p_0'$ indicated by the colormap) and synchronization properties (indicated by black lines) of limit cycle oscillations, observed experimentally in a horizontal Rijke tube oscillator. Here, $p_{rms}'$ and $p_0'$ are the root-mean-square (rms) values of acoustic pressure fluctuations in the forced and the unforced state. $\~{A}_f$ and $f_f/f_n$ are normalized forcing amplitude and forcing frequency. Adapted with permission from Sahay \textit{et al}. \cite{sahay2020dynamics}.}
\end{figure}

Recently, an approach based on forced synchronization (Sec. \ref{sec5}.2) of limit cycle oscillator has been used to explain the mitigation of thermoacoustic instabilities through open-loop controls \cite{mondal2019forced, guan2019openSymp,guan2018strange,skene2022phase,guan2019control,roy2020mechanism,sahay2020dynamics}. In Fig. \ref{fig:25}, we show the forced response of limit cycle oscillations in the acoustic pressure field of a horizontal Rijke tube for different parameters of periodic forcing generated through loudspeakers \cite{sahay2020dynamics}. In this figure, the distribution of different colors indicates the relative change in the amplitude of forced pressure fluctuations, $p_{rms}'$, against the amplitude of these oscillations in the unforced state, $p_0'$ (i.e., $\Delta p_{rms}'/p_0'$, where $\Delta p_{rms}'=p_0' - p_{rms}'$) and a V-shaped plot signifies the Arnold tongue (i.e., forced synchronization boundary). 

We notice that the introduction of forcing significantly affects the amplitude of limit cycle oscillations in the system. When forcing is applied close to the natural frequency of the oscillator, we find the occurrence of resonance amplification of synchronized limit cycle oscillations (i.e., synchronance) in the Arnold tongue, where the growth of the amplitude is greater than twice the amplitude of limit cycle oscillations in the unforced state, i.e., $\Delta p_{rms}'/p_0'<-1$ \cite{mondal2019forced,guan2019forced,sahay2020dynamics}. In contrast, when forcing is introduced at a frequency lower than the natural frequency of limit cycle oscillations (see for $f_f/f_{f0}<1$), we observe the quenching of limit cycle oscillations in the system due to asynchronous quenching \cite{minorsky1967comments}. The maximum suppression of limit cycle oscillations (i.e., $\Delta p_{rms}'/p_0' \rightarrow 1$) is observed along the Arnold tongue, i.e., at forcing parameters required for achieving forced synchronization in the system (Fig. \ref{fig:25}). Asynchronous quenching is not observed for $f_f/f_{f0}>1$ in the horizontal Rijke tube system \cite{mondal2019forced,sahay2020dynamics}; however, asynchronous quenching of limit cycle oscillations has been reported for both sides of $f_{n0}$ in a laminar premixed flame Rijke tube burner by Guan \textit{et al}. \cite{guan2019openSymp}. Furthermore, Roy \textit{et al}. \cite{roy2020mechanism} provided physical reasons behind asynchronous quenching of limit cycle oscillations by analyzing the forced response of coupled acoustic pressure and heat release rate fluctuations in the system. They found that these oscillations are locked at -90 degrees during the state of asynchronous quenching; as a result, the driving of the acoustic field by the heat release rate field is very low which in turn leads to suppression of acoustic oscillations in the system.  

\subsection{Mitigation of thermoacoustic instabilities through mutual coupling}

In Sec. \ref{sec5}.1, we saw that coupling two or more oscillators can either synchronize their oscillations or mitigate them through the amplitude death phenomenon. During amplitude death, all oscillators reach the same steady state \cite{zou2021quenching}. Similarly, coupling a system to itself via self-feedback can also quench limit cycle oscillations in a single Rijke tube oscillator \cite{srikanth2021self}. Various coupling schemes have been developed to mitigate self-excited oscillations in a system of coupled oscillators \cite{zou2021quenching, saxena2012amplitude}. Here, we will discuss the application of the amplitude death phenomenon in quenching limit cycle oscillations in a system of coupled thermoacoustic oscillators. 

Practical gas turbine engines such as can or can-annular combustors often consist of a ring array of multiple combustion units (known as “cans”), working simultaneously to provide the required thrust \cite{walsh2004gas}. Hence, these combustors tend to interact with each other and are therefore coupled through components such as the plenum chamber, the turbine stage, and cross-fire tubes \cite{farisco2017thermo}. Most of the traditional active and passive control strategies discussed before to mitigate thermoacoustic instability are expensive and are devised for an isolated combustion system. However, the application of these strategies to simultaneously mitigate thermoacoustic instabilities in a system with multiple combustors is yet to be explored in detail. 


Recent developments in the suppression of thermoacoustic instabilities in single or multiple thermoacoustic systems rely on implementing different schemes of coupling between oscillators. Towards this purpose, both experimental and theoretical studies have been performed on a single Rijke tube oscillator \cite{srikanth2021self} and a system of two coupled Rijke tube oscillators \cite{biwa2015amplitude,dange2019oscillation,sahay2020dynamics,thomas2018effecta,thomas2018effectb,sahay2020dynamics, srikanth2021dynamical}. For suppressing the limit cycle in a single system through self-delayed feedback, the acoustic field of the system is fed back into the system after a finite time delay, which is achieved experimentally by using a single coupling tube \cite{biwa2016suppression,lato2019passive,srikanth2021self}. For coupled Rijke oscillators, suppression of low amplitude limit cycle oscillations is achieved experimentally by connecting the oscillators using one \cite{biwa2015amplitude,dange2019oscillation} or two \cite{hyodo2020suppression} coupling tubes. On the other hand, suppression of high amplitude limit cycle oscillations is achieved by adding a frequency mismatch between the oscillators \cite{dange2019oscillation}. Both time delay and dissipative coupling schemes have been used to mitigate limit cycle oscillations in thermoacoustic oscillators. The coupled behavior of such oscillators is similar to that observed in Van der Pol oscillators \cite{biwa2015amplitude} and Stuart-Landau oscillators \cite{premraj2021effect}. Thus, the possibility of mitigation of limit cycle oscillations developed in single and multiple thermoacoustic systems through the mechanism of mutual coupling has emerged as a promising and cost-effective methodology.

\begin{figure}[t!]
\centering
\includegraphics[width=0.4\textwidth]{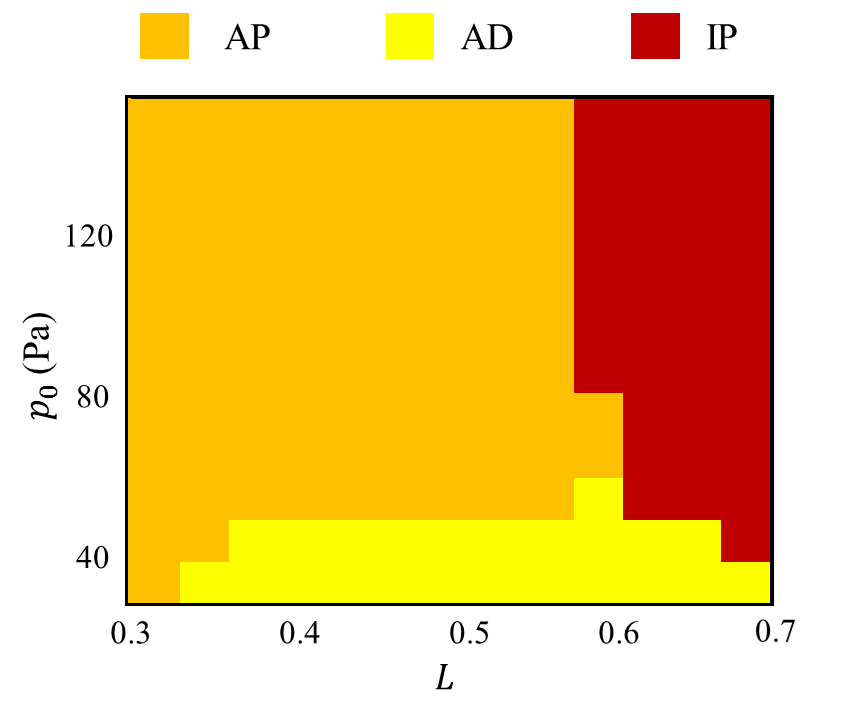}
\caption{\label{fig:26}Two-parameter bifurcation diagram between the isolated amplitude of each oscillator ($p_0$) and the length of the coupling tuhe ($L$) highlighting the dependency of the coupling and the system parameters on the occurrence of amplitude death in a system of coupled identical Rijke tube oscillators. Colors for AP, IP, and AD indicate regions of anti-phase synchronization, in-phase synchronization, and amplitude death, respectively. Reproduced with permission from Dange \textit{et al}. \cite{dange2019oscillation}.}
\end{figure}


\begin{figure}[t!]
\centering
\includegraphics[width=0.5\textwidth]{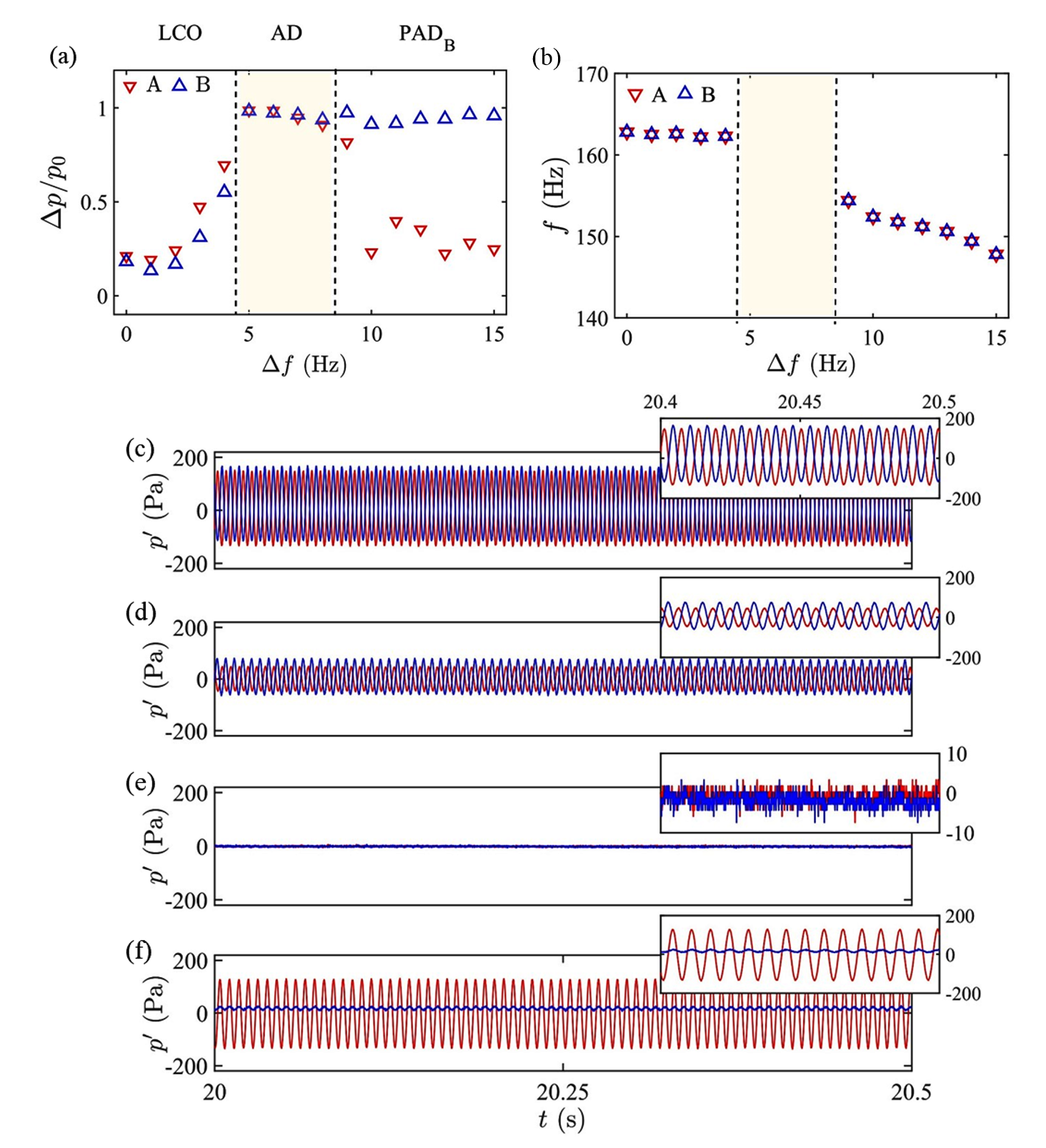}
\caption{\label{fig:27}Variation of (a) relative suppression in pressure amplitude and (b) frequency of each oscillator with frequency detuning in a pair of Rijke tube oscillators. Introduction of frequency detuning in the coupled systems engendered (c, d) enhanced suppression, (e) amplitude death and (f) partial amplitude death with an increase in the frequency detuning between the oscillators. Reproduced with permission from Dange \textit{et al}. \cite{dange2019oscillation}.}
\end{figure}

Figure \ref{fig:26} shows the dynamical behavior of a system of two identical Rijke tube oscillators coupled using a single connecting tube \cite{dange2019oscillation}. A two-parameter bifurcation diagram is shown between the amplitude of acoustic pressure fluctuations ($p_0$) in the isolated oscillator and the length of the connecting tube ($L$). We notice that low amplitude limit cycle oscillations can be easily quenched through the method of mutual coupling for a larger range of length of the coupling tube, while it is difficult to quench large amplitude limit cycle oscillations in two identical Rijke tube oscillators. The coupling of such large amplitude limit cycle oscillations results in phase-flip bifurcation on increasing the length of the coupling tube in the system. 


In order to quench high amplitude limit cycle oscillations in two Rijke tube oscillators, Dange \textit{et al.} \cite{dange2019oscillation} introduced frequency detuning in the system (see Fig. \ref{fig:27}). Frequency detuning can be introduced by varying the natural frequency of one of the oscillators, while keeping the natural frequency of the other one a constant. In a Rijke tube system, the frequency of acoustic oscillations is varied by changing the length of the duct, note that the frequency is inversely proportional to the length of the duct (i.e., $f_n \propto 1/L$). In Fig. \ref{fig:27}c,d, we notice that the mutual interaction between detuned Rijke tube oscillators having high amplitude of limit cycle oscillations facilitates a small suppression of oscillations. Increasing the detuning between the oscillators gradually increases the suppression of limit cycle oscillations (Fig. \ref{fig:27}a), and at sufficiently large detuning, the state of amplitude death is observed in the system (Fig. \ref{fig:27}e). A further increase in the detuning engenders the state of partial amplitude death (Fig. \ref{fig:27}f), where large amplitude limit cycle oscillations are restored in one Rijke tube oscillator while the other remains in a state of nearly suppressed periodic oscillations. These results are qualitatively similar to the occurrence of amplitude death and partial amplitude death in non-identical diffusively and time delay coupled weakly nonlinear oscillators \cite{atay2003total}.

Srikanth \textit{et al}. \cite{srikanth2021dynamical} theoretically studied the effect of variation in the coupling and system parameters on the occurrence of amplitude death in two time-delay coupled Rijke tube oscillators given in equations (\ref{eqtn:BSC}). Amplitude death is observed for a specific range of coupling delay ($\tau_c$), where the size of such death islands decreases with an increase in the value of coupling strength ($\mathcal{K}$). Furthermore, they found that the occurrence of amplitude death is highly dependent on the heater power ($K$), and thus the amplitude of limit cycle oscillations in the uncoupled state; an increase in the amplitude narrows the amplitude death regions and eventually suppresses them completely. They showed that the transition between amplitude death and oscillatory state (i.e., limit cycle oscillations) depends on the nature of the bifurcation of the isolated oscillator. To elaborate, this transition is explosive and hysteric for an oscillator exhibiting a subcritical Hopf bifurcation in the uncoupled state, whereas it is continuous for an oscillator undergoing a supercritical Hopf bifurcation. 

Furthermore, Srikanth \textit{et al}. \cite{srikanth2021self} extended the effectiveness of delayed acoustic coupling through a connecting tube to suppress thermoacoustic instabilities in a single Rijke tube oscillator, both experimentally and theoretically.  Thomas \textit{et al.} \cite{thomas2018effectb} investigated the effect of Gaussian white noise on the occurrence of amplitude death in the model of coupled Rijke tube oscillators and showed that the abrupt transition from the oscillatory to the steady state becomes continuous due to prebifurcation noise amplification \cite{surovyatkina2005prebifurcation}.

Having discussed various studies on the mitigation of thermoacoustic instabilities in Rijke-tube oscillators utilizing different approaches from synchronization theory, we next discuss studies that utilize Rijke tube systems to develop and demonstrate the efficacy of various early warning signals for critical transitions.


\subsection{Early warning signals (Precursors) }

As discussed in Sec. \ref{sec3}, many dynamical systems undergo abrupt transitions, also called tipping or critical transitions, as the control parameter is varied. In such systems, an early warning for the occurrence of these transitions is necessary to avoid the consequences that arise after their onset. In some cases where the variation of the control parameter is continuous, the rate at which such a parameter is varied greatly affects the performance of early warning measures for such critical transitions. In this section, we discuss various early warning measures developed in Rijke tube systems to detect the occurrence of critical transitions. Rijke tube has been utilized to evaluate the efficacy of early warning signals prior to extending the measures in other thermoacoustic systems.

\begin{figure}[t!]
\centering
\includegraphics[width=0.49\textwidth]{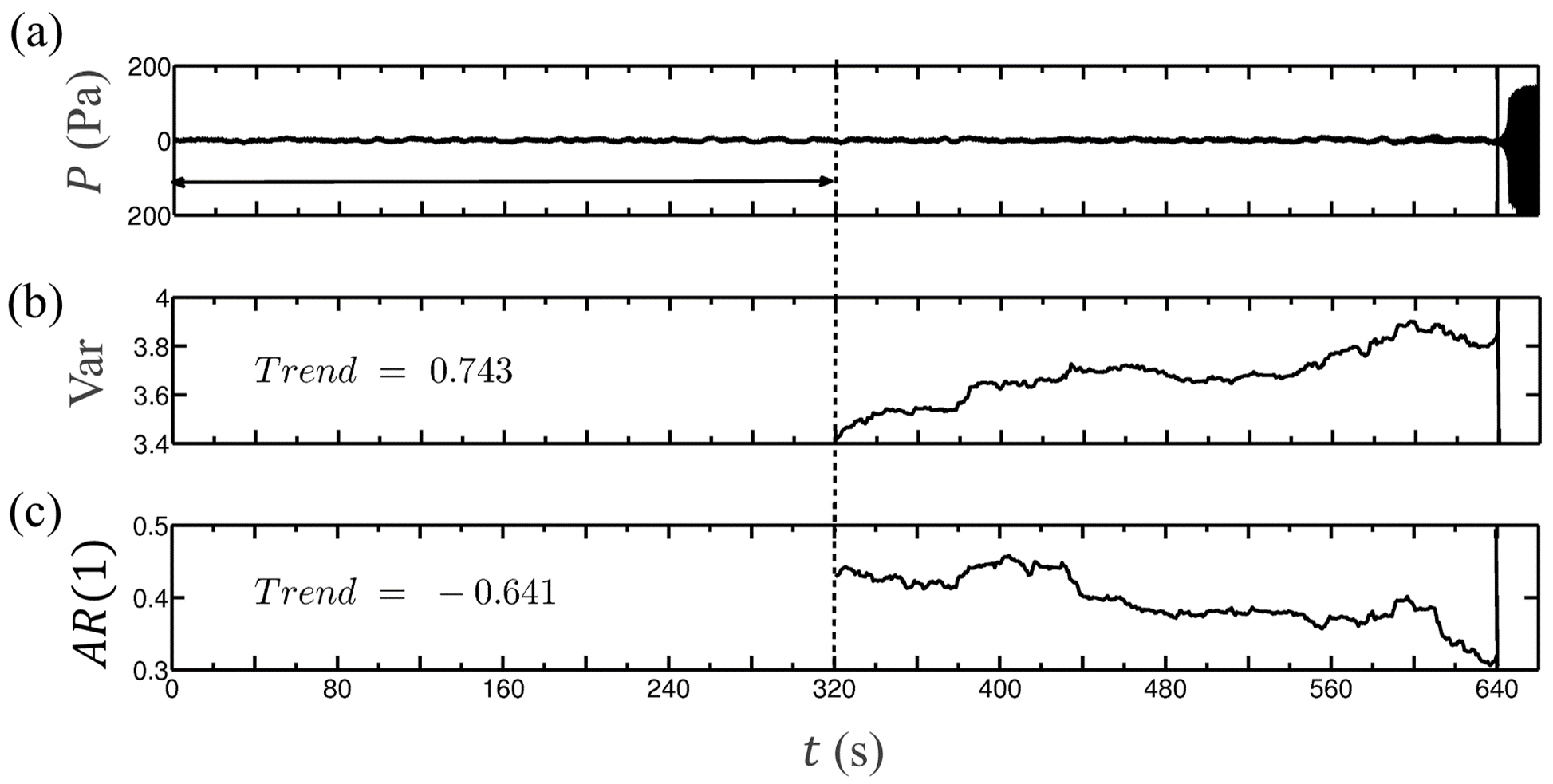}
\caption{\label{fig:28}(a) Time series of the acoustic pressure ($P$) exhibiting subcritical transition to limit cycle and the effectiveness of early warning signals such as (b) variance (Var) and (c) lag-1 autocorrelation ($AR$(1)) in identifying the onset of the impending thermoacoustic instability. The onset of such instability is marked using a solid black vertical line, whereas the dotted line represents the time step from which the measures are calculated. Adapted with permission from Gopalakrishnan \textit{et al}. \cite{gopalakrishnan2016early}.}
\end{figure}

In thermoacoustic systems, different measures have been invented to obtain early warning signals that predict the occurrence of thermoacoustic instability \cite{pavithran2021predicting}. These measures exhibit a drastic change in their values prior to the onset of such instability. Tracking the changes in the measure can help us to provide early warning for the impending instability. Thus, we can prevent the system from reaching the state of thermoacoustic instability and thereby evade its consequences. This approach goes along with the saying, ‘prevention is better than cure’. 

Gopalakrishnan \textit{et al}. \cite{gopalakrishnan2016early} applied the knowledge of a critical slowing down on approaching the tipping point in a horizontal Rijke tube to predict the occurrence of a subcritical Hopf bifurcation in the system. The phenomenon of critical slowing down is associated with the loss of stability of the system as the control parameter approaches the bifurcation point. It also indicates the slow recovery rate of the system to the external perturbations introduced close to a critical transition \cite{strogatz1994nonlinear}. Their study obtained early warning signals from variance and lag-1 autocorrelation of the acoustic pressure data from experiments as well as from the model of the Rijke tube (Fig. \ref{fig:28}). Figure \ref{fig:28}(a) shows the acoustic pressure signal obtained from the horizontal Rijke tube as the heater power is varied continuously at a fixed rate. The onset of limit cycle oscillations in the system is observed at $t=644\ \rm s$. The measures, variance and lag-1 autocorrelation, are calculated for a moving window.  Gopalakrishnan \textit{et al}. \cite{gopalakrishnan2016early} observed an increase in the variance (Fig. \ref{fig:28}b) and a decrease in the lag-1 autocorrelation (Fig. \ref{fig:28}c) well before the onset of thermoacoustic instability. Furthermore, they suggested that the variance of the signal is more robust to external noise imposed by the loudspeaker in predicting the onset of thermoacoustic instability when compared to the autocorrelation function.

\begin{figure}[t!]
\centering
\includegraphics[width=0.45\textwidth]{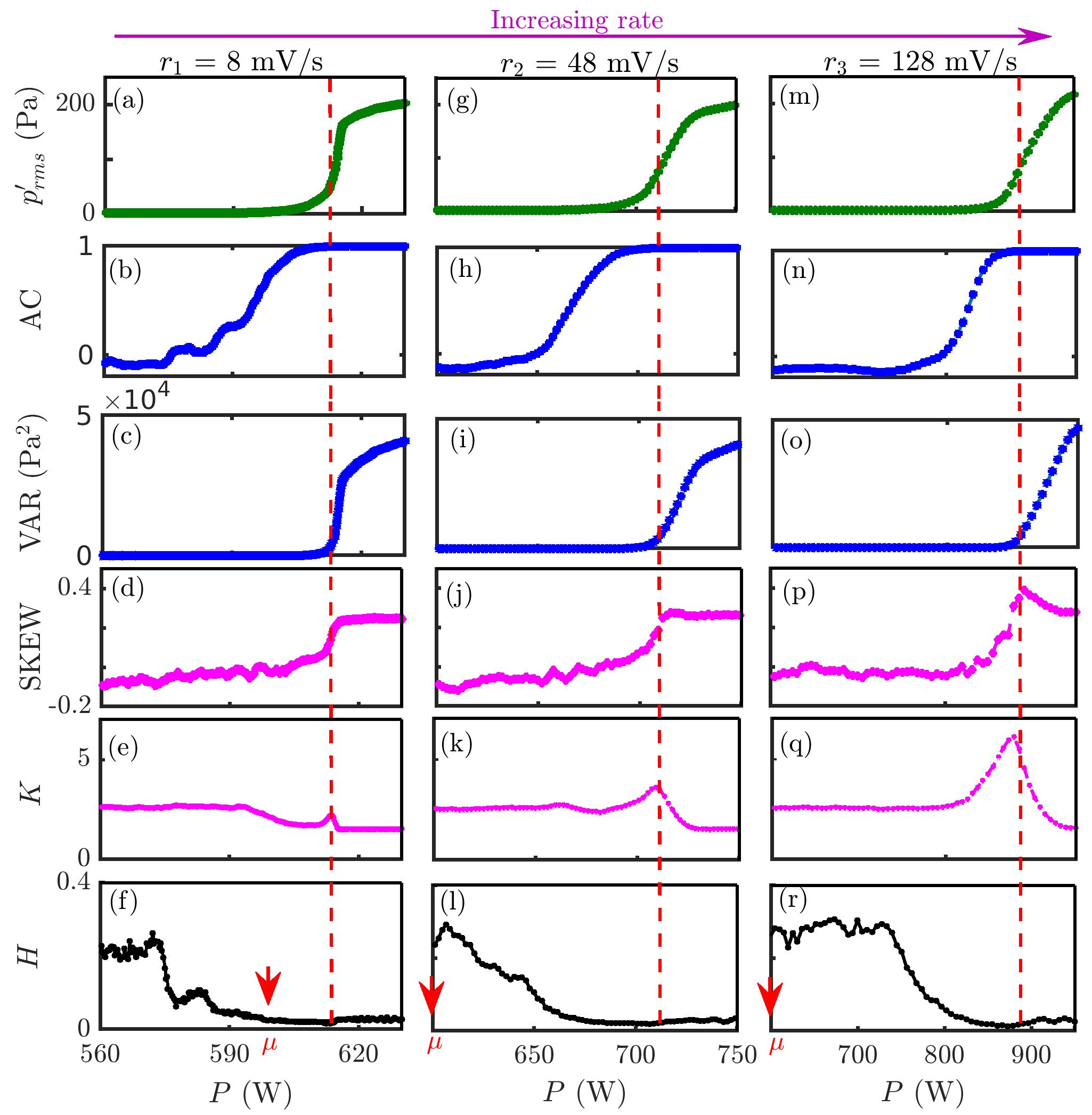}
\caption{\label{fig:29}Variation of (a,g,m) the rms amplitude of the pressure data obtained from a horizontal Rijke tube and the early warning measures such as (b,h,n) lag-1 autocorrelation, (c,i,o) variance, (d,j,p) skewness, (e,k,q) kurtosis, and (f,l,r) Hurst exponent applied to identify the impending thermoacoustic instability with respect to the control parameter, heater power. Each column corresponds to the varying rate of change of the control parameter. The dashed vertical line indicates the onset of limit cycle oscillations in the system and the red arrow points at the quasistatic Hopf point ($\mu$). Adapted with permission from Pavithran and Sujith \cite{pavithran2021effect}.}
\end{figure}

A recent study by Pavithran and Sujith \cite{pavithran2021effect} examined the impact of the rate of change of control parameter on the performance of different early warning signals of thermoacoustic instability in a horizontal Rijke tube. Various early warning measures, such as lag-1 autocorrelation ($AC$), variance ($VAR$), skewness ($SKEW$), kurtosis ($K$), and Hurst exponent ($H$), were chosen for the investigation (Fig. \ref{fig:29}). During the onset of thermoacoustic instability, the root mean square value of the acoustic pressure fluctuations indicates significant growth in the amplitude (Fig. \ref{fig:29}a,g,m). Autocorrelation (Fig. \ref{fig:29}b,h,n) and variance (Fig. \ref{fig:29}c,i,o) tend to increase much before the actual transition on approaching the onset of thermoacoustic instability due to critical slowing down \cite{scheffer2009early, dakos2012methods}. Skewness which also tends to increase on approaching the tipping point (Fig. \ref{fig:29}d,j,p) does not have any relation with critical slowing down \cite{scheffer2009early}. As the system approaches the transition to thermoacoustic instability, the skewness of the distribution changes from being negative to positive. Kurtosis ($K$) exhibits a value of 3 for a normal distribution; however, it does not show any perceivable trend during the onset of thermoacoustic instability (Fig. \ref{fig:29}e,k,q).

The Hurst exponent ($H = 2-D$, $D$ is the fractal dimension) computes the correlations in the time series \cite{hurst1951long,mandelbrot1977fractals}. When $H > 0.5$, the signal is said to be persistent, while for $H<0.5$, the signal is anti-persistent. When $H=0.5$, the signal is uncorrelated. Hence, as the system transitions from a regime of uncorrelated oscillations to a state of periodic high amplitude oscillations, the Hurst exponent decreases from 0.5 to zero, serving as a potential measure to detect the transition to thermoacoustic instability with changing fractal characteristics of the pressure signal (Fig. \ref{fig:29}f,l,r). Comparing the behavior of each of the aforementioned early warning measures at different rates of change of the control parameter, Pavithran and Sujith \cite{pavithran2021effect} noticed that measures such as lag-1 autocorrelation and Hurst exponent can predict the transition well before the tipping point for both slow and high rates; thus, providing adequate warning time for control actions (Fig.  \ref{fig:29}). 

Although these measures provide warnings about an upcoming critical transition, they cannot provide information about what bifurcation is to be expected. As the system approaches the tipping point, its dynamical behavior can be simplified into a limited number of possible “normal forms”. This, in turn, provides information about the new state (i.e., oscillatory or steady state) that may occur after the tipping point. Towards this, Bury \textit{et al}. \cite{bury2021deep} proposed a deep learning algorithm that provides early warning signals by using information about normal forms and scaling behavior of the dynamics near tipping points in a horizontal Rijke tube system. 

Lee \textit{et al.} \cite{lee2020input} proposed a framework for performing input-output system identification near a Hopf bifurcation. By using the data from the steady state behavior, they were able to predict the location and the criticality of Hopf bifurcation in a laminar Rijke tube burner and a model of the Duffing-Van der Pol oscillator perturbed with additive white noise. This novel methodology does not involve the crossing of threshold values (unlike other measures discussed before) and can be applied to various other dynamical systems that exhibit a Hopf bifurcation (or systems that can be reduced to Stuart-Landau equations). Premraj \textit{et al.} \cite{premraj2021dragon} investigated the occurrence of a catastrophic transition, such as flame blowout, in a laminar premixed  Rijke tube burner. During flame blowout, the flame ceases to exist in the combustor as the time scales of flow fluctuations become much larger than the reaction timescales in the system. After the flame blowout, the amplitude of the acoustic pressure fluctuations in the system drops to a very low value. Premraj \textit{et al.} \cite{premraj2021dragon} noticed that the occurrence of flame blowout is preceded by the existence of extreme events in the acoustic field of the system and found the presence of special kind of extreme events called the dragon-king extreme events just prior to flame blowout. Thus, the early warning to flame blowout can be provided by identifying the dragon-king extreme events in the acoustic field of the system.

Early warning measures have been investigated extensively in many systems including Rijke tubes and to date remain a topic of immense attention due to their potential applications. The readers are guided towards a recent review on critical transitions and early warning signals by Pavithran \textit{et al}. \cite{pavithran2021critical} for a more elaborate discussion.

\section{\label{sec7} Conclusions}

%
%
%
In the present review, we have introduced the Rijke tube oscillator as a novel paradigmatic oscillator to the nonlinear dynamics community. Towards this purpose, we have systematically presented the potential applications of the Rijke tube oscillator in obtaining experimental verification of various dynamical phenomena that are observed in general paradigmatic oscillators. We have shown that, depending on the operating conditions, the onset of limit cycle oscillations in a Rijke tube can happen either through subcritical or supercritical Hopf bifurcation. We have also observed the occurrence of secondary bifurcations to various dynamical phenomena such as quasiperiodic, period-$k$, chaotic, and strange nonchaotic oscillations, along with the presence of different routes to chaos in the Rijke tube system. We have further emphasized the existence of different noise-induced transitions, such as coherence resonance, stochastic bifurcation, and subcritical excitation to limit cycle oscillations in the subthreshold and bistable regimes of the system operation. We have examined mutual synchronization and forced synchronization properties of coupled and forced Rijke tube oscillators, respectively, and summarized different states of synchronization witnessed in such systems. We further discussed the application of different concepts from synchronization theory to control and mitigate the limit cycle oscillations using the phenomena of amplitude death, partial amplitude death, and asynchronous quenching in the Rijke tube systems. Finally, we have presented the use of a Rijke tube system as a platform to develop validate various early warning measures for catastrophic transitions. 

We hope that this review paves way for the dynamical systems community to utilize the Rijke tube oscillator for performing both experimental and theoretical studies in the future. Although there has been significant progress in studying the nonlinear behavior of Rijke tube oscillators over the years, there are still many interesting phenomena that are yet to be experimentally or theoretically discovered, examined and validated in Rijke tube oscillators. Next, we discuss possible areas of research that can be explored to unravel several hidden dynamical and coupled behaviors in the Rijke tube systems.

As discussed in Sec. \ref{Stochastic}, the occurrence of stochastic resonance has not been examined in Rijke type systems. Therefore, future investigations are required in order to obtain the experimental or theoretical discovery of stochastic resonance in Rijke tube oscillators. There are many theoretical studies that report the occurrence of the period-doubling route to chaos in different Rijke tube systems\cite{subramanian2010bifurcation,kashinath2014nonlinear}. However, the experimental evidence of this route to chaos is yet to be reported in a Rijke tube system. 

Furthermore, in this paper, we have restricted our discussion on the application of Rijke tube oscillators to investigate the dynamical behavior of a single and a pair of oscillators, for the variation of both systems and coupling parameters. However, we know that with an increase in the number of oscillators in a system (i.e., a network) or a change in their coupling structures (e.g., local, non-local, and global couplings or star, line, and ring topologies), we can observe many complex dynamical phenomena resulting from the coupling of oscillators \cite{manoj2021experimental, wickramasinghe2013spatially}. These phenomena include clustering, splay states, bare minimum chimera, weak chimera, amplitude death, aging, etc. in a minimal network of coupled oscillators, where the number of oscillators is less (approximately, 3 to 10). Different modeling studies in the past showed the occurrence of these phenomena in small networks of oscillators. In the future, we can construct a minimal network of Rijke tube oscillators and through appropriate coupling mechanisms, we can experimentally validate theoretically discovered phenomena. Furthermore, we have also discussed the possibility of oscillation quenching via amplitude death in a system of coupled oscillators. However, in some systems, such suppression of oscillations is undesirable. Recently, through various theoretical studies \cite{zou2021quenching}, restoring of oscillations in a system is shown by adding a processing delay factor in the coupling term of oscillators. Developing a controlled experiment on a network of Rijke tube oscillators to revoke the oscillations from the death state by varying the feedback factor in the coupling of oscillators is an interesting study worth future investigations.    

In practical systems, the control parameter of a system does not change quasi-statically; however, it varies continuously in time with different rates. Many studies have been performed investigating rate-dependent tipping (bifurcation) on a single system. Such investigations can be extended to multiple coupled oscillators. Moreover, in a system of two coupled oscillators, the effect of the rate of change of control parameters in one system on the tipping behavior of the other can be studied theoretically and validated experimentally using coupled Rijke tube systems. Furthermore, the presence of noise on the coupled behavior of Rijke tube oscillators, in terms of their transition from steady state to limit cycle oscillations or the occurrence of different dynamical states, can also be included in future investigations. 

\section*{Acknowledgements}
We are grateful to the IoE initiative (SB/2021/0845/AE/MHRD/002696), and the J. C. Bose Fellowship (No. JCB/2018/000034/SSC) from the Department of Science and Technology (DST) for the financial support.

\bibliography{References}

\begin{thebibliography}{335}%
\makeatletter
\providecommand \@ifxundefined [1]{%
 \@ifx{#1\undefined}
}%
\providecommand \@ifnum [1]{%
 \ifnum #1\expandafter \@firstoftwo
 \else \expandafter \@secondoftwo
 \fi
}%
\providecommand \@ifx [1]{%
 \ifx #1\expandafter \@firstoftwo
 \else \expandafter \@secondoftwo
 \fi
}%
\providecommand \natexlab [1]{#1}%
\providecommand \enquote  [1]{``#1''}%
\providecommand \bibnamefont  [1]{#1}%
\providecommand \bibfnamefont [1]{#1}%
\providecommand \citenamefont [1]{#1}%
\providecommand \href@noop [0]{\@secondoftwo}%
\providecommand \href [0]{\begingroup \@sanitize@url \@href}%
\providecommand \@href[1]{\@@startlink{#1}\@@href}%
\providecommand \@@href[1]{\endgroup#1\@@endlink}%
\providecommand \@sanitize@url [0]{\catcode `\\12\catcode `\$12\catcode
  `\&12\catcode `\#12\catcode `\^12\catcode `\_12\catcode `\%12\relax}%
\providecommand \@@startlink[1]{}%
\providecommand \@@endlink[0]{}%
\providecommand \url  [0]{\begingroup\@sanitize@url \@url }%
\providecommand \@url [1]{\endgroup\@href {#1}{\urlprefix }}%
\providecommand \urlprefix  [0]{URL }%
\providecommand \Eprint [0]{\href }%
\providecommand \doibase [0]{http://dx.doi.org/}%
\providecommand \selectlanguage [0]{\@gobble}%
\providecommand \bibinfo  [0]{\@secondoftwo}%
\providecommand \bibfield  [0]{\@secondoftwo}%
\providecommand \translation [1]{[#1]}%
\providecommand \BibitemOpen [0]{}%
\providecommand \bibitemStop [0]{}%
\providecommand \bibitemNoStop [0]{.\EOS\space}%
\providecommand \EOS [0]{\spacefactor3000\relax}%
\providecommand \BibitemShut  [1]{\csname bibitem#1\endcsname}%
\let\auto@bib@innerbib\@empty
\bibitem [{\citenamefont {{\AA}str{\"o}m}, \citenamefont {Klein},\ and\
  \citenamefont {Lennartsson}(2005)}]{aastrom2005bicycle}%
  \BibitemOpen
  \bibfield  {author} {\bibinfo {author} {\bibfnamefont {K.~J.}\ \bibnamefont
  {{\AA}str{\"o}m}}, \bibinfo {author} {\bibfnamefont {R.~E.}\ \bibnamefont
  {Klein}}, \ and\ \bibinfo {author} {\bibfnamefont {A.}~\bibnamefont
  {Lennartsson}},\ }\bibfield  {title} {\enquote {\bibinfo {title} {Bicycle
  dynamics and control},}\ }\href@noop {} {\bibfield  {journal} {\bibinfo
  {journal} {Control Syst. Mag.}\ }\textbf {\bibinfo {volume} {25}},\ \bibinfo
  {pages} {26--47} (\bibinfo {year} {2005})}\BibitemShut {NoStop}%
\bibitem [{\citenamefont {Robert}(2014)}]{robert2014river}%
  \BibitemOpen
  \bibfield  {author} {\bibinfo {author} {\bibfnamefont {A.}~\bibnamefont
  {Robert}},\ }\href@noop {} {\emph {\bibinfo {title} {River Processes: An
  Introduction to Fluvial Dynamics}}}\ (\bibinfo  {publisher} {Routledge},\
  \bibinfo {year} {2014})\BibitemShut {NoStop}%
\bibitem [{\citenamefont {Kundu}\ and\ \citenamefont
  {Cohen}(2002)}]{kundu2002fluid}%
  \BibitemOpen
  \bibfield  {author} {\bibinfo {author} {\bibfnamefont {P.~K.}\ \bibnamefont
  {Kundu}}\ and\ \bibinfo {author} {\bibfnamefont {I.~M.}\ \bibnamefont
  {Cohen}},\ }\bibfield  {title} {\enquote {\bibinfo {title} {Fluid
  mechanics},}\ }\href@noop {} {\  (\bibinfo {year} {2002})}\BibitemShut
  {NoStop}%
\bibitem [{\citenamefont {O’Keeffe}, \citenamefont {Hong},\ and\
  \citenamefont {Strogatz}(2017)}]{o2017oscillators}%
  \BibitemOpen
  \bibfield  {author} {\bibinfo {author} {\bibfnamefont {K.~P.}\ \bibnamefont
  {O’Keeffe}}, \bibinfo {author} {\bibfnamefont {H.}~\bibnamefont {Hong}}, \
  and\ \bibinfo {author} {\bibfnamefont {S.~H.}\ \bibnamefont {Strogatz}},\
  }\bibfield  {title} {\enquote {\bibinfo {title} {Oscillators that sync and
  swarm},}\ }\href@noop {} {\bibfield  {journal} {\bibinfo  {journal} {Nat.
  Commun.}\ }\textbf {\bibinfo {volume} {8}},\ \bibinfo {pages} {1--13}
  (\bibinfo {year} {2017})}\BibitemShut {NoStop}%
\bibitem [{\citenamefont {Nagy}\ \emph {et~al.}(2010)\citenamefont {Nagy},
  \citenamefont {{\'A}kos}, \citenamefont {Biro},\ and\ \citenamefont
  {Vicsek}}]{nagy2010hierarchical}%
  \BibitemOpen
  \bibfield  {author} {\bibinfo {author} {\bibfnamefont {M.}~\bibnamefont
  {Nagy}}, \bibinfo {author} {\bibfnamefont {Z.}~\bibnamefont {{\'A}kos}},
  \bibinfo {author} {\bibfnamefont {D.}~\bibnamefont {Biro}}, \ and\ \bibinfo
  {author} {\bibfnamefont {T.}~\bibnamefont {Vicsek}},\ }\bibfield  {title}
  {\enquote {\bibinfo {title} {Hierarchical group dynamics in pigeon flocks},}\
  }\href@noop {} {\bibfield  {journal} {\bibinfo  {journal} {Nature}\ }\textbf
  {\bibinfo {volume} {464}},\ \bibinfo {pages} {890--893} (\bibinfo {year}
  {2010})}\BibitemShut {NoStop}%
\bibitem [{\citenamefont {Hemelrijk}\ and\ \citenamefont
  {Hildenbrandt}(2012)}]{hemelrijk2012schools}%
  \BibitemOpen
  \bibfield  {author} {\bibinfo {author} {\bibfnamefont {C.~K.}\ \bibnamefont
  {Hemelrijk}}\ and\ \bibinfo {author} {\bibfnamefont {H.}~\bibnamefont
  {Hildenbrandt}},\ }\bibfield  {title} {\enquote {\bibinfo {title} {Schools of
  fish and flocks of birds: their shape and internal structure by
  self-organization},}\ }\href@noop {} {\bibfield  {journal} {\bibinfo
  {journal} {Interface Focus}\ }\textbf {\bibinfo {volume} {2}},\ \bibinfo
  {pages} {726--737} (\bibinfo {year} {2012})}\BibitemShut {NoStop}%
\bibitem [{\citenamefont {Zhisheng}\ \emph {et~al.}(2015)\citenamefont
  {Zhisheng}, \citenamefont {Guoxiong}, \citenamefont {Jianping}, \citenamefont
  {Youbin}, \citenamefont {Yimin}, \citenamefont {Weijian}, \citenamefont
  {Yanjun}, \citenamefont {Anmin}, \citenamefont {Li}, \citenamefont {Jiangyu}
  \emph {et~al.}}]{zhisheng2015global}%
  \BibitemOpen
  \bibfield  {author} {\bibinfo {author} {\bibfnamefont {A.}~\bibnamefont
  {Zhisheng}}, \bibinfo {author} {\bibfnamefont {W.}~\bibnamefont {Guoxiong}},
  \bibinfo {author} {\bibfnamefont {L.}~\bibnamefont {Jianping}}, \bibinfo
  {author} {\bibfnamefont {S.}~\bibnamefont {Youbin}}, \bibinfo {author}
  {\bibfnamefont {L.}~\bibnamefont {Yimin}}, \bibinfo {author} {\bibfnamefont
  {Z.}~\bibnamefont {Weijian}}, \bibinfo {author} {\bibfnamefont
  {C.}~\bibnamefont {Yanjun}}, \bibinfo {author} {\bibfnamefont
  {D.}~\bibnamefont {Anmin}}, \bibinfo {author} {\bibfnamefont
  {L.}~\bibnamefont {Li}}, \bibinfo {author} {\bibfnamefont {M.}~\bibnamefont
  {Jiangyu}},  \emph {et~al.},\ }\bibfield  {title} {\enquote {\bibinfo {title}
  {Global monsoon dynamics and climate change},}\ }\href@noop {} {\bibfield
  {journal} {\bibinfo  {journal} {Annu. Rev. Earth Planet. Sci.}\ }\textbf
  {\bibinfo {volume} {43}},\ \bibinfo {pages} {29--77} (\bibinfo {year}
  {2015})}\BibitemShut {NoStop}%
\bibitem [{\citenamefont {Ivanov}\ \emph {et~al.}(1999)\citenamefont {Ivanov},
  \citenamefont {Amaral}, \citenamefont {Goldberger}, \citenamefont {Havlin},
  \citenamefont {Rosenblum}, \citenamefont {Struzik},\ and\ \citenamefont
  {Stanley}}]{ivanov1999multifractality}%
  \BibitemOpen
  \bibfield  {author} {\bibinfo {author} {\bibfnamefont {P.~C.}\ \bibnamefont
  {Ivanov}}, \bibinfo {author} {\bibfnamefont {L.~A.~N.}\ \bibnamefont
  {Amaral}}, \bibinfo {author} {\bibfnamefont {A.~L.}\ \bibnamefont
  {Goldberger}}, \bibinfo {author} {\bibfnamefont {S.}~\bibnamefont {Havlin}},
  \bibinfo {author} {\bibfnamefont {M.~G.}\ \bibnamefont {Rosenblum}}, \bibinfo
  {author} {\bibfnamefont {Z.~R.}\ \bibnamefont {Struzik}}, \ and\ \bibinfo
  {author} {\bibfnamefont {H.~E.}\ \bibnamefont {Stanley}},\ }\bibfield
  {title} {\enquote {\bibinfo {title} {Multifractality in human heartbeat
  dynamics},}\ }\href@noop {} {\bibfield  {journal} {\bibinfo  {journal}
  {Nature}\ }\textbf {\bibinfo {volume} {399}},\ \bibinfo {pages} {461--465}
  (\bibinfo {year} {1999})}\BibitemShut {NoStop}%
\bibitem [{\citenamefont {Royama}(2012)}]{royama2012analytical}%
  \BibitemOpen
  \bibfield  {author} {\bibinfo {author} {\bibfnamefont {T.}~\bibnamefont
  {Royama}},\ }\href@noop {} {\emph {\bibinfo {title} {Analytical Population
  Dynamics}}},\ Vol.~\bibinfo {volume} {10}\ (\bibinfo  {publisher} {Springer
  Science \& Business Media},\ \bibinfo {year} {2012})\BibitemShut {NoStop}%
\bibitem [{\citenamefont {Turchin}(2013)}]{turchin2013complex}%
  \BibitemOpen
  \bibfield  {author} {\bibinfo {author} {\bibfnamefont {P.}~\bibnamefont
  {Turchin}},\ }\href@noop {} {\emph {\bibinfo {title} {Complex Population
  Dynamics}}}\ (\bibinfo  {publisher} {Princeton University Press},\ \bibinfo
  {year} {2013})\BibitemShut {NoStop}%
\bibitem [{\citenamefont {Lakshmanan}\ and\ \citenamefont
  {Rajaseekar}(2012)}]{lakshmanan2012nonlinear}%
  \BibitemOpen
  \bibfield  {author} {\bibinfo {author} {\bibfnamefont {M.}~\bibnamefont
  {Lakshmanan}}\ and\ \bibinfo {author} {\bibfnamefont {S.}~\bibnamefont
  {Rajaseekar}},\ }\href@noop {} {\emph {\bibinfo {title} {Nonlinear Dynamics:
  Integrability, Chaos and Patterns}}}\ (\bibinfo  {publisher} {Springer
  Science \& Business Media},\ \bibinfo {year} {2012})\BibitemShut {NoStop}%
\bibitem [{\citenamefont {Griffiths}(2005)}]{griffiths2005introduction}%
  \BibitemOpen
  \bibfield  {author} {\bibinfo {author} {\bibfnamefont {D.~J.}\ \bibnamefont
  {Griffiths}},\ }\href@noop {} {\enquote {\bibinfo {title} {Introduction to
  electrodynamics},}\ } (\bibinfo {year} {2005})\BibitemShut {NoStop}%
\bibitem [{\citenamefont {Friedland}(2012)}]{friedland2012control}%
  \BibitemOpen
  \bibfield  {author} {\bibinfo {author} {\bibfnamefont {B.}~\bibnamefont
  {Friedland}},\ }\href@noop {} {\emph {\bibinfo {title} {Control System
  Design: An Introduction to State-space Methods}}}\ (\bibinfo  {publisher}
  {Courier Corporation},\ \bibinfo {year} {2012})\BibitemShut {NoStop}%
\bibitem [{\citenamefont {Strogatz}(2004)}]{strogatz2004sync}%
  \BibitemOpen
  \bibfield  {author} {\bibinfo {author} {\bibfnamefont {S.}~\bibnamefont
  {Strogatz}},\ }\href@noop {} {\emph {\bibinfo {title} {Sync: The Emerging
  Science of Spontaneous Order}}}\ (\bibinfo  {publisher} {Penguin UK},\
  \bibinfo {year} {2004})\BibitemShut {NoStop}%
\bibitem [{\citenamefont {Kovacic}(2020)}]{kovacic2020nonlinear}%
  \BibitemOpen
  \bibfield  {author} {\bibinfo {author} {\bibfnamefont {I.}~\bibnamefont
  {Kovacic}},\ }\href@noop {} {\emph {\bibinfo {title} {Nonlinear Oscillations:
  Exact Solutions and Their Approximations}}}\ (\bibinfo  {publisher} {Springer
  Nature},\ \bibinfo {year} {2020})\BibitemShut {NoStop}%
\bibitem [{\citenamefont {Hagedorn}(1981)}]{hagedorn1981non}%
  \BibitemOpen
  \bibfield  {author} {\bibinfo {author} {\bibfnamefont {P.}~\bibnamefont
  {Hagedorn}},\ }\href@noop {} {\emph {\bibinfo {title} {Non-linear
  Oscillations}}}\ (\bibinfo  {publisher} {Oxford University Press, New York},\
  \bibinfo {year} {1981})\BibitemShut {NoStop}%
\bibitem [{\citenamefont {McIntyre}, \citenamefont {Schumacher},\ and\
  \citenamefont {Woodhouse}(1983)}]{mcintyre1983oscillations}%
  \BibitemOpen
  \bibfield  {author} {\bibinfo {author} {\bibfnamefont {M.~E.}\ \bibnamefont
  {McIntyre}}, \bibinfo {author} {\bibfnamefont {R.~T.}\ \bibnamefont
  {Schumacher}}, \ and\ \bibinfo {author} {\bibfnamefont {J.}~\bibnamefont
  {Woodhouse}},\ }\bibfield  {title} {\enquote {\bibinfo {title} {On the
  oscillations of musical instruments},}\ }\href@noop {} {\bibfield  {journal}
  {\bibinfo  {journal} {J. Acoust. Soc. Am.}\ }\textbf {\bibinfo {volume}
  {74}},\ \bibinfo {pages} {1325--1345} (\bibinfo {year} {1983})}\BibitemShut
  {NoStop}%
\bibitem [{\citenamefont {Culick}(2006)}]{culick2006unsteady}%
  \BibitemOpen
  \bibfield  {author} {\bibinfo {author} {\bibfnamefont {F.~E.~C.}\
  \bibnamefont {Culick}},\ }\href@noop {} {\enquote {\bibinfo {title} {Unsteady
  motions in combustion chambers for propulsion systems},}\ }\bibinfo {type}
  {Tech. Rep.}\ (\bibinfo  {institution} {AGARDograph, NATO/RTO-AG-AVT-039},\
  \bibinfo {year} {2006})\BibitemShut {NoStop}%
\bibitem [{\citenamefont {O{\u{g}}uzt{\"o}reli}\ and\ \citenamefont
  {Stein}(1976)}]{ouguztoreli1976effects}%
  \BibitemOpen
  \bibfield  {author} {\bibinfo {author} {\bibfnamefont {M.}~\bibnamefont
  {O{\u{g}}uzt{\"o}reli}}\ and\ \bibinfo {author} {\bibfnamefont
  {R.}~\bibnamefont {Stein}},\ }\bibfield  {title} {\enquote {\bibinfo {title}
  {The effects of multiple reflex pathways on the oscillations in
  neuro-muscular systems},}\ }\href@noop {} {\bibfield  {journal} {\bibinfo
  {journal} {J. Math. Biol.}\ }\textbf {\bibinfo {volume} {3}},\ \bibinfo
  {pages} {87--101} (\bibinfo {year} {1976})}\BibitemShut {NoStop}%
\bibitem [{\citenamefont {Cardon}\ and\ \citenamefont
  {Iberall}(1970)}]{cardon1970oscillations}%
  \BibitemOpen
  \bibfield  {author} {\bibinfo {author} {\bibfnamefont {S.}~\bibnamefont
  {Cardon}}\ and\ \bibinfo {author} {\bibfnamefont {A.}~\bibnamefont
  {Iberall}},\ }\bibfield  {title} {\enquote {\bibinfo {title} {Oscillations in
  biological systems},}\ }\href@noop {} {\bibfield  {journal} {\bibinfo
  {journal} {Biosystems}\ }\textbf {\bibinfo {volume} {3}},\ \bibinfo {pages}
  {237--249} (\bibinfo {year} {1970})}\BibitemShut {NoStop}%
\bibitem [{\citenamefont {Duncan}, \citenamefont {Duncan},\ and\ \citenamefont
  {Scott}(1997)}]{duncan1997dynamics}%
  \BibitemOpen
  \bibfield  {author} {\bibinfo {author} {\bibfnamefont {C.}~\bibnamefont
  {Duncan}}, \bibinfo {author} {\bibfnamefont {S.}~\bibnamefont {Duncan}}, \
  and\ \bibinfo {author} {\bibfnamefont {S.}~\bibnamefont {Scott}},\ }\bibfield
   {title} {\enquote {\bibinfo {title} {The dynamics of measles epidemics},}\
  }\href@noop {} {\bibfield  {journal} {\bibinfo  {journal} {Theor. Popul.
  Biol.}\ }\textbf {\bibinfo {volume} {52}},\ \bibinfo {pages} {155--163}
  (\bibinfo {year} {1997})}\BibitemShut {NoStop}%
\bibitem [{\citenamefont {Green}\ and\ \citenamefont
  {Unruh}(2006)}]{green2006failure}%
  \BibitemOpen
  \bibfield  {author} {\bibinfo {author} {\bibfnamefont {D.}~\bibnamefont
  {Green}}\ and\ \bibinfo {author} {\bibfnamefont {W.~G.}\ \bibnamefont
  {Unruh}},\ }\bibfield  {title} {\enquote {\bibinfo {title} {The failure of
  the tacoma bridge: A physical model},}\ }\href@noop {} {\bibfield  {journal}
  {\bibinfo  {journal} {Am. J. Phys.}\ }\textbf {\bibinfo {volume} {74}},\
  \bibinfo {pages} {706--716} (\bibinfo {year} {2006})}\BibitemShut {NoStop}%
\bibitem [{\citenamefont {Strogatz}\ \emph {et~al.}(2005)\citenamefont
  {Strogatz}, \citenamefont {Abrams}, \citenamefont {McRobie}, \citenamefont
  {Eckhardt},\ and\ \citenamefont {Ott}}]{strogatz2005crowd}%
  \BibitemOpen
  \bibfield  {author} {\bibinfo {author} {\bibfnamefont {S.~H.}\ \bibnamefont
  {Strogatz}}, \bibinfo {author} {\bibfnamefont {D.~M.}\ \bibnamefont
  {Abrams}}, \bibinfo {author} {\bibfnamefont {A.}~\bibnamefont {McRobie}},
  \bibinfo {author} {\bibfnamefont {B.}~\bibnamefont {Eckhardt}}, \ and\
  \bibinfo {author} {\bibfnamefont {E.}~\bibnamefont {Ott}},\ }\bibfield
  {title} {\enquote {\bibinfo {title} {Crowd synchrony on the millennium
  bridge},}\ }\href@noop {} {\bibfield  {journal} {\bibinfo  {journal}
  {Nature}\ }\textbf {\bibinfo {volume} {438}},\ \bibinfo {pages} {43--44}
  (\bibinfo {year} {2005})}\BibitemShut {NoStop}%
\bibitem [{\citenamefont {Singhose}\ \emph {et~al.}(1997)\citenamefont
  {Singhose}, \citenamefont {Singer}, \citenamefont {Derezinski~III},
  \citenamefont {Rappole~Jr},\ and\ \citenamefont
  {Pasch}}]{singhose1997method}%
  \BibitemOpen
  \bibfield  {author} {\bibinfo {author} {\bibfnamefont {W.~E.}\ \bibnamefont
  {Singhose}}, \bibinfo {author} {\bibfnamefont {N.~C.}\ \bibnamefont
  {Singer}}, \bibinfo {author} {\bibfnamefont {S.~J.}\ \bibnamefont
  {Derezinski~III}}, \bibinfo {author} {\bibfnamefont {B.~W.}\ \bibnamefont
  {Rappole~Jr}}, \ and\ \bibinfo {author} {\bibfnamefont {K.}~\bibnamefont
  {Pasch}},\ }\href@noop {} {\enquote {\bibinfo {title} {Method and apparatus
  for minimizing unwanted dynamics in a physical system},}\ } (\bibinfo {year}
  {1997}),\ \bibinfo {note} {uS Patent 5,638,267}\BibitemShut {NoStop}%
\bibitem [{\citenamefont {Ucke}\ and\ \citenamefont
  {Schlichting}(2008)}]{ucke2008oscillating}%
  \BibitemOpen
  \bibfield  {author} {\bibinfo {author} {\bibfnamefont {C.}~\bibnamefont
  {Ucke}}\ and\ \bibinfo {author} {\bibfnamefont {H.-J.}\ \bibnamefont
  {Schlichting}},\ }\bibfield  {title} {\enquote {\bibinfo {title} {Oscillating
  dolls and skyscrapers},}\ }\href@noop {} {\bibfield  {journal} {\bibinfo
  {journal} {German J. Phys Unserer Zeit}\ }\textbf {\bibinfo {volume} {39}},\
  \bibinfo {pages} {139--141} (\bibinfo {year} {2008})}\BibitemShut {NoStop}%
\bibitem [{\citenamefont {Landa}(2001)}]{landa2001regular}%
  \BibitemOpen
  \bibfield  {author} {\bibinfo {author} {\bibfnamefont {P.~S.}\ \bibnamefont
  {Landa}},\ }\href@noop {} {\emph {\bibinfo {title} {Regular and Chaotic
  Oscillations}}}\ (\bibinfo  {publisher} {Springer Science \& Business
  Media},\ \bibinfo {year} {2001})\BibitemShut {NoStop}%
\bibitem [{\citenamefont {Puu}(2013)}]{puu2013attractors}%
  \BibitemOpen
  \bibfield  {author} {\bibinfo {author} {\bibfnamefont {T.}~\bibnamefont
  {Puu}},\ }\href@noop {} {\emph {\bibinfo {title} {Attractors, Bifurcations,
  \& Chaos: Nonlinear Phenomena in Economics}}}\ (\bibinfo  {publisher}
  {Springer Science \& Business Media},\ \bibinfo {year} {2013})\BibitemShut
  {NoStop}%
\bibitem [{\citenamefont {Strogatz}(1994)}]{strogatz1994nonlinear}%
  \BibitemOpen
  \bibfield  {author} {\bibinfo {author} {\bibfnamefont {S.~H.}\ \bibnamefont
  {Strogatz}},\ }\href@noop {} {\emph {\bibinfo {title} {Nonlinear Dynamics and
  Chaos: With Applications to Physics, Biology, Chemistry, and Engineering}}}\
  (\bibinfo  {publisher} {CRC Press},\ \bibinfo {year} {1994})\BibitemShut
  {NoStop}%
\bibitem [{\citenamefont {Gleick}(1987)}]{gleick1987chaos}%
  \BibitemOpen
  \bibfield  {author} {\bibinfo {author} {\bibfnamefont {J.}~\bibnamefont
  {Gleick}},\ }\href@noop {} {\emph {\bibinfo {title} {Chaos: Making a New
  Science}}}\ (\bibinfo  {publisher} {New York, Penguin Books},\ \bibinfo
  {year} {1987})\BibitemShut {NoStop}%
\bibitem [{\citenamefont {Hilborn}(2000)}]{hilborn2000chaos}%
  \BibitemOpen
  \bibfield  {author} {\bibinfo {author} {\bibfnamefont {R.~C.}\ \bibnamefont
  {Hilborn}},\ }\href@noop {} {\emph {\bibinfo {title} {Chaos and Nonlinear
  Dynamics: An Introduction for Scientists and Engineers}}}\ (\bibinfo
  {publisher} {Oxford University Press},\ \bibinfo {year} {2000})\BibitemShut
  {NoStop}%
\bibitem [{\citenamefont {Nayfeh}\ and\ \citenamefont
  {Balachandran}(2008)}]{nayfeh2008applied}%
  \BibitemOpen
  \bibfield  {author} {\bibinfo {author} {\bibfnamefont {A.~H.}\ \bibnamefont
  {Nayfeh}}\ and\ \bibinfo {author} {\bibfnamefont {B.}~\bibnamefont
  {Balachandran}},\ }\href@noop {} {\emph {\bibinfo {title} {Applied Nonlinear
  Dynamics: Analytical, Computational, and Experimental Methods}}}\ (\bibinfo
  {publisher} {John Wiley \& Sons},\ \bibinfo {year} {2008})\BibitemShut
  {NoStop}%
\bibitem [{\citenamefont {Hale}\ and\ \citenamefont
  {Ko{\c{c}}ak}(2012)}]{hale2012dynamics}%
  \BibitemOpen
  \bibfield  {author} {\bibinfo {author} {\bibfnamefont {J.~K.}\ \bibnamefont
  {Hale}}\ and\ \bibinfo {author} {\bibfnamefont {H.}~\bibnamefont
  {Ko{\c{c}}ak}},\ }\href@noop {} {\emph {\bibinfo {title} {Dynamics and
  Bifurcations}}},\ Vol.~\bibinfo {volume} {3}\ (\bibinfo  {publisher}
  {Springer Science \& Business Media},\ \bibinfo {year} {2012})\BibitemShut
  {NoStop}%
\bibitem [{\citenamefont {Wiggins}(2013)}]{wiggins2013global}%
  \BibitemOpen
  \bibfield  {author} {\bibinfo {author} {\bibfnamefont {S.}~\bibnamefont
  {Wiggins}},\ }\href@noop {} {\emph {\bibinfo {title} {Global Bifurcations and
  Chaos: Analytical Methods}}},\ Vol.~\bibinfo {volume} {73}\ (\bibinfo
  {publisher} {Springer Science \& Business Media},\ \bibinfo {year}
  {2013})\BibitemShut {NoStop}%
\bibitem [{\citenamefont {Scheffer}(2020)}]{scheffer2020critical}%
  \BibitemOpen
  \bibfield  {author} {\bibinfo {author} {\bibfnamefont {M.}~\bibnamefont
  {Scheffer}},\ }\href@noop {} {\emph {\bibinfo {title} {Critical Transitions
  in Nature and Society}}}\ (\bibinfo  {publisher} {Princeton University
  Press},\ \bibinfo {year} {2020})\BibitemShut {NoStop}%
\bibitem [{\citenamefont {Berglund}(1998)}]{berglund1998adiabatic}%
  \BibitemOpen
  \bibfield  {author} {\bibinfo {author} {\bibfnamefont {N.}~\bibnamefont
  {Berglund}},\ }\href@noop {} {\enquote {\bibinfo {title} {Adiabatic dynamical
  systems and hysteresis},}\ }\bibinfo {type} {Tech. Rep.}\ (\bibinfo
  {institution} {EPFL},\ \bibinfo {year} {1998})\BibitemShut {NoStop}%
\bibitem [{\citenamefont {Marwan}\ \emph {et~al.}(2007)\citenamefont {Marwan},
  \citenamefont {Romano}, \citenamefont {Thiel},\ and\ \citenamefont
  {Kurths}}]{marwan2007recurrence}%
  \BibitemOpen
  \bibfield  {author} {\bibinfo {author} {\bibfnamefont {N.}~\bibnamefont
  {Marwan}}, \bibinfo {author} {\bibfnamefont {M.~C.}\ \bibnamefont {Romano}},
  \bibinfo {author} {\bibfnamefont {M.}~\bibnamefont {Thiel}}, \ and\ \bibinfo
  {author} {\bibfnamefont {J.}~\bibnamefont {Kurths}},\ }\bibfield  {title}
  {\enquote {\bibinfo {title} {Recurrence plots for the analysis of complex
  systems},}\ }\href@noop {} {\bibfield  {journal} {\bibinfo  {journal} {Phys.
  Rep.}\ }\textbf {\bibinfo {volume} {438}},\ \bibinfo {pages} {237--329}
  (\bibinfo {year} {2007})}\BibitemShut {NoStop}%
\bibitem [{\citenamefont {Parker}\ and\ \citenamefont
  {Chua}(1987)}]{parker1987chaos}%
  \BibitemOpen
  \bibfield  {author} {\bibinfo {author} {\bibfnamefont {T.~S.}\ \bibnamefont
  {Parker}}\ and\ \bibinfo {author} {\bibfnamefont {L.~O.}\ \bibnamefont
  {Chua}},\ }\bibfield  {title} {\enquote {\bibinfo {title} {Chaos: A tutorial
  for engineers},}\ }\href@noop {} {\bibfield  {journal} {\bibinfo  {journal}
  {Proc. IEEE}\ }\textbf {\bibinfo {volume} {75}},\ \bibinfo {pages}
  {982--1008} (\bibinfo {year} {1987})}\BibitemShut {NoStop}%
\bibitem [{\citenamefont {Balanov}\ \emph {et~al.}(2009)\citenamefont
  {Balanov}, \citenamefont {Janson}, \citenamefont {Postnov},\ and\
  \citenamefont {Sosnovtseva}}]{balanov2009synchronization}%
  \BibitemOpen
  \bibfield  {author} {\bibinfo {author} {\bibfnamefont {A.}~\bibnamefont
  {Balanov}}, \bibinfo {author} {\bibfnamefont {N.}~\bibnamefont {Janson}},
  \bibinfo {author} {\bibfnamefont {D.}~\bibnamefont {Postnov}}, \ and\
  \bibinfo {author} {\bibfnamefont {O.}~\bibnamefont {Sosnovtseva}},\
  }\href@noop {} {\emph {\bibinfo {title} {Synchronization: From Simple to
  Complex}}},\ Vol.~\bibinfo {volume} {17}\ (\bibinfo  {publisher} {Springer},\
  \bibinfo {year} {2009})\BibitemShut {NoStop}%
\bibitem [{\citenamefont {Pikovsky}, \citenamefont {Rosenblum},\ and\
  \citenamefont {Kurths}(2003)}]{pikovsky2003synchronization}%
  \BibitemOpen
  \bibfield  {author} {\bibinfo {author} {\bibfnamefont {A.}~\bibnamefont
  {Pikovsky}}, \bibinfo {author} {\bibfnamefont {M.}~\bibnamefont {Rosenblum}},
  \ and\ \bibinfo {author} {\bibfnamefont {J.}~\bibnamefont {Kurths}},\
  }\href@noop {} {\emph {\bibinfo {title} {Synchronization: A Universal Concept
  in Nonlinear Sciences}}},\ Vol.~\bibinfo {volume} {12}\ (\bibinfo
  {publisher} {Cambridge University Press},\ \bibinfo {year}
  {2003})\BibitemShut {NoStop}%
\bibitem [{\citenamefont {Strogatz}\ and\ \citenamefont
  {Stewart}(1993)}]{strogatz1993coupled}%
  \BibitemOpen
  \bibfield  {author} {\bibinfo {author} {\bibfnamefont {S.~H.}\ \bibnamefont
  {Strogatz}}\ and\ \bibinfo {author} {\bibfnamefont {I.}~\bibnamefont
  {Stewart}},\ }\bibfield  {title} {\enquote {\bibinfo {title} {Coupled
  oscillators and biological synchronization},}\ }\href@noop {} {\bibfield
  {journal} {\bibinfo  {journal} {Sci. Am.}\ }\textbf {\bibinfo {volume}
  {269}},\ \bibinfo {pages} {102--109} (\bibinfo {year} {1993})}\BibitemShut
  {NoStop}%
\bibitem [{\citenamefont {Awrejcewicz}(1991)}]{awrejcewicz1991bifurcation}%
  \BibitemOpen
  \bibfield  {author} {\bibinfo {author} {\bibfnamefont {J.}~\bibnamefont
  {Awrejcewicz}},\ }\href@noop {} {\emph {\bibinfo {title} {Bifurcation and
  Chaos in Coupled Oscillators}}}\ (\bibinfo  {publisher} {World Scientific},\
  \bibinfo {year} {1991})\BibitemShut {NoStop}%
\bibitem [{\citenamefont {Lakshmanan}\ and\ \citenamefont
  {Senthilkumar}(2011)}]{lakshmanan2011dynamics}%
  \BibitemOpen
  \bibfield  {author} {\bibinfo {author} {\bibfnamefont {M.}~\bibnamefont
  {Lakshmanan}}\ and\ \bibinfo {author} {\bibfnamefont {D.~V.}\ \bibnamefont
  {Senthilkumar}},\ }\href@noop {} {\emph {\bibinfo {title} {Dynamics of
  Nonlinear Time-delay Systems}}}\ (\bibinfo  {publisher} {Springer Science \&
  Business Media},\ \bibinfo {year} {2011})\BibitemShut {NoStop}%
\bibitem [{\citenamefont {Madan}(1993)}]{madan1993chua}%
  \BibitemOpen
  \bibfield  {author} {\bibinfo {author} {\bibfnamefont {R.~N.}\ \bibnamefont
  {Madan}},\ }\href@noop {} {\emph {\bibinfo {title} {Chua's Circuit: A
  Paradigm for Chaos}}},\ Vol.~\bibinfo {volume} {1}\ (\bibinfo  {publisher}
  {World Scientific},\ \bibinfo {year} {1993})\BibitemShut {NoStop}%
\bibitem [{\citenamefont {Kovacic}\ and\ \citenamefont
  {Brennan}(2011)}]{kovacic2011duffing}%
  \BibitemOpen
  \bibfield  {author} {\bibinfo {author} {\bibfnamefont {I.}~\bibnamefont
  {Kovacic}}\ and\ \bibinfo {author} {\bibfnamefont {M.~J.}\ \bibnamefont
  {Brennan}},\ }\href@noop {} {\emph {\bibinfo {title} {The Duffing Equation:
  Nonlinear Oscillators and Their Behaviour}}}\ (\bibinfo  {publisher} {John
  Wiley \& Sons},\ \bibinfo {year} {2011})\BibitemShut {NoStop}%
\bibitem [{\citenamefont {Thompson}\ and\ \citenamefont
  {Stewart}(2002)}]{thompson2002nonlinear}%
  \BibitemOpen
  \bibfield  {author} {\bibinfo {author} {\bibfnamefont {J.~M.~T.}\
  \bibnamefont {Thompson}}\ and\ \bibinfo {author} {\bibfnamefont {H.~B.}\
  \bibnamefont {Stewart}},\ }\href@noop {} {\emph {\bibinfo {title} {Nonlinear
  Dynamics and Chaos}}}\ (\bibinfo  {publisher} {John Wiley \& Sons},\ \bibinfo
  {year} {2002})\BibitemShut {NoStop}%
\bibitem [{\citenamefont {Zou}\ \emph {et~al.}(2021)\citenamefont {Zou},
  \citenamefont {Senthilkumar}, \citenamefont {Zhan},\ and\ \citenamefont
  {Kurths}}]{zou2021quenching}%
  \BibitemOpen
  \bibfield  {author} {\bibinfo {author} {\bibfnamefont {W.}~\bibnamefont
  {Zou}}, \bibinfo {author} {\bibfnamefont {D.}~\bibnamefont {Senthilkumar}},
  \bibinfo {author} {\bibfnamefont {M.}~\bibnamefont {Zhan}}, \ and\ \bibinfo
  {author} {\bibfnamefont {J.}~\bibnamefont {Kurths}},\ }\bibfield  {title}
  {\enquote {\bibinfo {title} {Quenching, aging, and reviving in coupled
  dynamical networks},}\ }\href@noop {} {\bibfield  {journal} {\bibinfo
  {journal} {Phys. Rep.}\ }\textbf {\bibinfo {volume} {931}},\ \bibinfo {pages}
  {1--72} (\bibinfo {year} {2021})}\BibitemShut {NoStop}%
\bibitem [{\citenamefont {Manoj}, \citenamefont {Pawar},\ and\ \citenamefont
  {Sujith}(2021)}]{manoj2021experimental}%
  \BibitemOpen
  \bibfield  {author} {\bibinfo {author} {\bibfnamefont {K.}~\bibnamefont
  {Manoj}}, \bibinfo {author} {\bibfnamefont {S.~A.}\ \bibnamefont {Pawar}}, \
  and\ \bibinfo {author} {\bibfnamefont {R.~I.}\ \bibnamefont {Sujith}},\
  }\bibfield  {title} {\enquote {\bibinfo {title} {Experimental investigation
  on the susceptibility of minimal networks to a change in topology and number
  of oscillators},}\ }\href@noop {} {\bibfield  {journal} {\bibinfo  {journal}
  {Phys. Rev. E}\ }\textbf {\bibinfo {volume} {103}},\ \bibinfo {pages}
  {022207} (\bibinfo {year} {2021})}\BibitemShut {NoStop}%
\bibitem [{\citenamefont {Wickramasinghe}\ and\ \citenamefont
  {Kiss}(2013{\natexlab{a}})}]{wickramasinghe2013spatially}%
  \BibitemOpen
  \bibfield  {author} {\bibinfo {author} {\bibfnamefont {M.}~\bibnamefont
  {Wickramasinghe}}\ and\ \bibinfo {author} {\bibfnamefont {I.~Z.}\
  \bibnamefont {Kiss}},\ }\bibfield  {title} {\enquote {\bibinfo {title}
  {Spatially organized dynamical states in chemical oscillator networks:
  Synchronization, dynamical differentiation, and chimera patterns},}\
  }\href@noop {} {\bibfield  {journal} {\bibinfo  {journal} {PLoS One}\
  }\textbf {\bibinfo {volume} {8}},\ \bibinfo {pages} {e80586} (\bibinfo {year}
  {2013}{\natexlab{a}})}\BibitemShut {NoStop}%
\bibitem [{\citenamefont {Wickramasinghe}\ and\ \citenamefont
  {Kiss}(2013{\natexlab{b}})}]{wickramasinghe2013synchronization}%
  \BibitemOpen
  \bibfield  {author} {\bibinfo {author} {\bibfnamefont {M.}~\bibnamefont
  {Wickramasinghe}}\ and\ \bibinfo {author} {\bibfnamefont {I.~Z.}\
  \bibnamefont {Kiss}},\ }\bibfield  {title} {\enquote {\bibinfo {title}
  {Synchronization of electrochemical oscillators with differential
  coupling},}\ }\href@noop {} {\bibfield  {journal} {\bibinfo  {journal} {Phys.
  Rev. E}\ }\textbf {\bibinfo {volume} {88}},\ \bibinfo {pages} {062911}
  (\bibinfo {year} {2013}{\natexlab{b}})}\BibitemShut {NoStop}%
\bibitem [{\citenamefont {Saxena}, \citenamefont {Prasad},\ and\ \citenamefont
  {Ramaswamy}(2012)}]{saxena2012amplitude}%
  \BibitemOpen
  \bibfield  {author} {\bibinfo {author} {\bibfnamefont {G.}~\bibnamefont
  {Saxena}}, \bibinfo {author} {\bibfnamefont {A.}~\bibnamefont {Prasad}}, \
  and\ \bibinfo {author} {\bibfnamefont {R.}~\bibnamefont {Ramaswamy}},\
  }\bibfield  {title} {\enquote {\bibinfo {title} {Amplitude death: The
  emergence of stationarity in coupled nonlinear systems},}\ }\href@noop {}
  {\bibfield  {journal} {\bibinfo  {journal} {Phys. Rep.}\ }\textbf {\bibinfo
  {volume} {521}},\ \bibinfo {pages} {205--228} (\bibinfo {year}
  {2012})}\BibitemShut {NoStop}%
\bibitem [{\citenamefont {Koseska}, \citenamefont {Volkov},\ and\ \citenamefont
  {Kurths}(2013)}]{koseska2013oscillation}%
  \BibitemOpen
  \bibfield  {author} {\bibinfo {author} {\bibfnamefont {A.}~\bibnamefont
  {Koseska}}, \bibinfo {author} {\bibfnamefont {E.}~\bibnamefont {Volkov}}, \
  and\ \bibinfo {author} {\bibfnamefont {J.}~\bibnamefont {Kurths}},\
  }\bibfield  {title} {\enquote {\bibinfo {title} {{Oscillation quenching
  mechanisms: Amplitude vs. oscillation death}},}\ }\href@noop {} {\bibfield
  {journal} {\bibinfo  {journal} {Phys. Rep.}\ }\textbf {\bibinfo {volume}
  {531}},\ \bibinfo {pages} {173--199} (\bibinfo {year} {2013})}\BibitemShut
  {NoStop}%
\bibitem [{\citenamefont {Abrams}\ and\ \citenamefont
  {Strogatz}(2004)}]{abrams2004chimera}%
  \BibitemOpen
  \bibfield  {author} {\bibinfo {author} {\bibfnamefont {D.~M.}\ \bibnamefont
  {Abrams}}\ and\ \bibinfo {author} {\bibfnamefont {S.~H.}\ \bibnamefont
  {Strogatz}},\ }\bibfield  {title} {\enquote {\bibinfo {title} {Chimera states
  for coupled oscillators},}\ }\href@noop {} {\bibfield  {journal} {\bibinfo
  {journal} {Phys. Rev. Lett.}\ }\textbf {\bibinfo {volume} {93}},\ \bibinfo
  {pages} {174102} (\bibinfo {year} {2004})}\BibitemShut {NoStop}%
\bibitem [{\citenamefont {Mondal}, \citenamefont {Unni},\ and\ \citenamefont
  {Sujith}(2017)}]{mondal2017onset}%
  \BibitemOpen
  \bibfield  {author} {\bibinfo {author} {\bibfnamefont {S.}~\bibnamefont
  {Mondal}}, \bibinfo {author} {\bibfnamefont {V.~R.}\ \bibnamefont {Unni}}, \
  and\ \bibinfo {author} {\bibfnamefont {R.~I.}\ \bibnamefont {Sujith}},\
  }\bibfield  {title} {\enquote {\bibinfo {title} {Onset of thermoacoustic
  instability in turbulent combustors: an emergence of synchronized periodicity
  through formation of chimera-like states},}\ }\href@noop {} {\bibfield
  {journal} {\bibinfo  {journal} {J. Fluid Mech.}\ }\textbf {\bibinfo {volume}
  {811}},\ \bibinfo {pages} {659--681} (\bibinfo {year} {2017})}\BibitemShut
  {NoStop}%
\bibitem [{\citenamefont {Hart}\ \emph {et~al.}(2016)\citenamefont {Hart},
  \citenamefont {Bansal}, \citenamefont {Murphy},\ and\ \citenamefont
  {Roy}}]{hart2016experimental}%
  \BibitemOpen
  \bibfield  {author} {\bibinfo {author} {\bibfnamefont {J.~D.}\ \bibnamefont
  {Hart}}, \bibinfo {author} {\bibfnamefont {K.}~\bibnamefont {Bansal}},
  \bibinfo {author} {\bibfnamefont {T.~E.}\ \bibnamefont {Murphy}}, \ and\
  \bibinfo {author} {\bibfnamefont {R.}~\bibnamefont {Roy}},\ }\bibfield
  {title} {\enquote {\bibinfo {title} {Experimental observation of chimera and
  cluster states in a minimal globally coupled network},}\ }\href@noop {}
  {\bibfield  {journal} {\bibinfo  {journal} {Chaos}\ }\textbf {\bibinfo
  {volume} {26}},\ \bibinfo {pages} {094801} (\bibinfo {year}
  {2016})}\BibitemShut {NoStop}%
\bibitem [{\citenamefont {Manoj}\ \emph {et~al.}(2019)\citenamefont {Manoj},
  \citenamefont {Pawar}, \citenamefont {Dange}, \citenamefont {Mondal},
  \citenamefont {Sujith}, \citenamefont {Surovyatkina},\ and\ \citenamefont
  {Kurths}}]{manoj2019synchronization}%
  \BibitemOpen
  \bibfield  {author} {\bibinfo {author} {\bibfnamefont {K.}~\bibnamefont
  {Manoj}}, \bibinfo {author} {\bibfnamefont {S.~A.}\ \bibnamefont {Pawar}},
  \bibinfo {author} {\bibfnamefont {S.}~\bibnamefont {Dange}}, \bibinfo
  {author} {\bibfnamefont {S.}~\bibnamefont {Mondal}}, \bibinfo {author}
  {\bibfnamefont {R.~I.}\ \bibnamefont {Sujith}}, \bibinfo {author}
  {\bibfnamefont {E.}~\bibnamefont {Surovyatkina}}, \ and\ \bibinfo {author}
  {\bibfnamefont {J.}~\bibnamefont {Kurths}},\ }\bibfield  {title} {\enquote
  {\bibinfo {title} {Synchronization route to weak chimera in four candle-flame
  oscillators},}\ }\href@noop {} {\bibfield  {journal} {\bibinfo  {journal}
  {Phys. Rev. E}\ }\textbf {\bibinfo {volume} {100}},\ \bibinfo {pages}
  {062204} (\bibinfo {year} {2019})}\BibitemShut {NoStop}%
\bibitem [{\citenamefont {Ashwin}\ and\ \citenamefont
  {Burylko}(2015)}]{ashwin2015weak}%
  \BibitemOpen
  \bibfield  {author} {\bibinfo {author} {\bibfnamefont {P.}~\bibnamefont
  {Ashwin}}\ and\ \bibinfo {author} {\bibfnamefont {O.}~\bibnamefont
  {Burylko}},\ }\bibfield  {title} {\enquote {\bibinfo {title} {Weak chimeras
  in minimal networks of coupled phase oscillators},}\ }\href@noop {}
  {\bibfield  {journal} {\bibinfo  {journal} {Chaos}\ }\textbf {\bibinfo
  {volume} {25}},\ \bibinfo {pages} {013106} (\bibinfo {year}
  {2015})}\BibitemShut {NoStop}%
\bibitem [{\citenamefont {Wojewoda}\ \emph {et~al.}(2016)\citenamefont
  {Wojewoda}, \citenamefont {Czolczynski}, \citenamefont {Maistrenko},\ and\
  \citenamefont {Kapitaniak}}]{wojewoda2016smallest}%
  \BibitemOpen
  \bibfield  {author} {\bibinfo {author} {\bibfnamefont {J.}~\bibnamefont
  {Wojewoda}}, \bibinfo {author} {\bibfnamefont {K.}~\bibnamefont
  {Czolczynski}}, \bibinfo {author} {\bibfnamefont {Y.}~\bibnamefont
  {Maistrenko}}, \ and\ \bibinfo {author} {\bibfnamefont {T.}~\bibnamefont
  {Kapitaniak}},\ }\bibfield  {title} {\enquote {\bibinfo {title} {The smallest
  chimera state for coupled pendula},}\ }\href@noop {} {\bibfield  {journal}
  {\bibinfo  {journal} {Sci. Rep.}\ }\textbf {\bibinfo {volume} {6}},\ \bibinfo
  {pages} {1--5} (\bibinfo {year} {2016})}\BibitemShut {NoStop}%
\bibitem [{\citenamefont {Pecora}\ \emph {et~al.}(2014)\citenamefont {Pecora},
  \citenamefont {Sorrentino}, \citenamefont {Hagerstrom}, \citenamefont
  {Murphy},\ and\ \citenamefont {Roy}}]{pecora2014cluster}%
  \BibitemOpen
  \bibfield  {author} {\bibinfo {author} {\bibfnamefont {L.~M.}\ \bibnamefont
  {Pecora}}, \bibinfo {author} {\bibfnamefont {F.}~\bibnamefont {Sorrentino}},
  \bibinfo {author} {\bibfnamefont {A.~M.}\ \bibnamefont {Hagerstrom}},
  \bibinfo {author} {\bibfnamefont {T.~E.}\ \bibnamefont {Murphy}}, \ and\
  \bibinfo {author} {\bibfnamefont {R.}~\bibnamefont {Roy}},\ }\bibfield
  {title} {\enquote {\bibinfo {title} {Cluster synchronization and isolated
  desynchronization in complex networks with symmetries},}\ }\href@noop {}
  {\bibfield  {journal} {\bibinfo  {journal} {Nat. Commun.}\ }\textbf {\bibinfo
  {volume} {5}},\ \bibinfo {pages} {1--8} (\bibinfo {year} {2014})}\BibitemShut
  {NoStop}%
\bibitem [{\citenamefont {Crowley}\ and\ \citenamefont
  {Field}(1986)}]{crowley1986electrically}%
  \BibitemOpen
  \bibfield  {author} {\bibinfo {author} {\bibfnamefont {M.~F.}\ \bibnamefont
  {Crowley}}\ and\ \bibinfo {author} {\bibfnamefont {R.~J.}\ \bibnamefont
  {Field}},\ }\bibfield  {title} {\enquote {\bibinfo {title} {Electrically
  coupled belousov-zhabotinskii oscillators. 1. experiments and simulations},}\
  }\href@noop {} {\bibfield  {journal} {\bibinfo  {journal} {J. Phys. Chem. A}\
  }\textbf {\bibinfo {volume} {90}},\ \bibinfo {pages} {1907--1915} (\bibinfo
  {year} {1986})}\BibitemShut {NoStop}%
\bibitem [{\citenamefont {Carroll}\ and\ \citenamefont
  {Pecora}(1995)}]{carroll1995nonlinear}%
  \BibitemOpen
  \bibfield  {author} {\bibinfo {author} {\bibfnamefont {T.~L.}\ \bibnamefont
  {Carroll}}\ and\ \bibinfo {author} {\bibfnamefont {L.~M.}\ \bibnamefont
  {Pecora}},\ }\href@noop {} {\emph {\bibinfo {title} {Nonlinear Dynamics in
  Circuits}}}\ (\bibinfo  {publisher} {World Scientific},\ \bibinfo {year}
  {1995})\BibitemShut {NoStop}%
\bibitem [{\citenamefont {W{\"u}nsche}\ \emph {et~al.}(2005)\citenamefont
  {W{\"u}nsche}, \citenamefont {Bauer}, \citenamefont {Kreissl}, \citenamefont
  {Ushakov}, \citenamefont {Korneyev}, \citenamefont {Henneberger},
  \citenamefont {Wille}, \citenamefont {Erzgr{\"a}ber}, \citenamefont {Peil},
  \citenamefont {Els{\"a}{\ss}er} \emph {et~al.}}]{wunsche2005synchronization}%
  \BibitemOpen
  \bibfield  {author} {\bibinfo {author} {\bibfnamefont {H.-J.}\ \bibnamefont
  {W{\"u}nsche}}, \bibinfo {author} {\bibfnamefont {S.}~\bibnamefont {Bauer}},
  \bibinfo {author} {\bibfnamefont {J.}~\bibnamefont {Kreissl}}, \bibinfo
  {author} {\bibfnamefont {O.}~\bibnamefont {Ushakov}}, \bibinfo {author}
  {\bibfnamefont {N.}~\bibnamefont {Korneyev}}, \bibinfo {author}
  {\bibfnamefont {F.}~\bibnamefont {Henneberger}}, \bibinfo {author}
  {\bibfnamefont {E.}~\bibnamefont {Wille}}, \bibinfo {author} {\bibfnamefont
  {H.}~\bibnamefont {Erzgr{\"a}ber}}, \bibinfo {author} {\bibfnamefont
  {M.}~\bibnamefont {Peil}}, \bibinfo {author} {\bibfnamefont {W.}~\bibnamefont
  {Els{\"a}{\ss}er}},  \emph {et~al.},\ }\bibfield  {title} {\enquote {\bibinfo
  {title} {Synchronization of delay-coupled oscillators: A study of
  semiconductor lasers},}\ }\href@noop {} {\bibfield  {journal} {\bibinfo
  {journal} {Phys. Rev. Lett.}\ }\textbf {\bibinfo {volume} {94}},\ \bibinfo
  {pages} {163901} (\bibinfo {year} {2005})}\BibitemShut {NoStop}%
\bibitem [{\citenamefont {Erzgr{\"a}ber}, \citenamefont {Wieczorek},\ and\
  \citenamefont {Krauskopf}(2009)}]{erzgraber2009locking}%
  \BibitemOpen
  \bibfield  {author} {\bibinfo {author} {\bibfnamefont {H.}~\bibnamefont
  {Erzgr{\"a}ber}}, \bibinfo {author} {\bibfnamefont {S.}~\bibnamefont
  {Wieczorek}}, \ and\ \bibinfo {author} {\bibfnamefont {B.}~\bibnamefont
  {Krauskopf}},\ }\bibfield  {title} {\enquote {\bibinfo {title} {Locking
  behavior of three coupled laser oscillators},}\ }\href@noop {} {\bibfield
  {journal} {\bibinfo  {journal} {Phys. Rev. E}\ }\textbf {\bibinfo {volume}
  {80}},\ \bibinfo {pages} {026212} (\bibinfo {year} {2009})}\BibitemShut
  {NoStop}%
\bibitem [{\citenamefont {Nkomo}, \citenamefont {Tinsley},\ and\ \citenamefont
  {Showalter}(2013)}]{nkomo2013chimera}%
  \BibitemOpen
  \bibfield  {author} {\bibinfo {author} {\bibfnamefont {S.}~\bibnamefont
  {Nkomo}}, \bibinfo {author} {\bibfnamefont {M.~R.}\ \bibnamefont {Tinsley}},
  \ and\ \bibinfo {author} {\bibfnamefont {K.}~\bibnamefont {Showalter}},\
  }\bibfield  {title} {\enquote {\bibinfo {title} {Chimera states in
  populations of nonlocally coupled chemical oscillators},}\ }\href@noop {}
  {\bibfield  {journal} {\bibinfo  {journal} {Phys. Rev. Lett.}\ }\textbf
  {\bibinfo {volume} {110}},\ \bibinfo {pages} {244102} (\bibinfo {year}
  {2013})}\BibitemShut {NoStop}%
\bibitem [{\citenamefont {Tinsley}, \citenamefont {Nkomo},\ and\ \citenamefont
  {Showalter}(2012)}]{tinsley2012chimera}%
  \BibitemOpen
  \bibfield  {author} {\bibinfo {author} {\bibfnamefont {M.~R.}\ \bibnamefont
  {Tinsley}}, \bibinfo {author} {\bibfnamefont {S.}~\bibnamefont {Nkomo}}, \
  and\ \bibinfo {author} {\bibfnamefont {K.}~\bibnamefont {Showalter}},\
  }\bibfield  {title} {\enquote {\bibinfo {title} {Chimera and phase-cluster
  states in populations of coupled chemical oscillators},}\ }\href@noop {}
  {\bibfield  {journal} {\bibinfo  {journal} {Nat. Phys.}\ }\textbf {\bibinfo
  {volume} {8}},\ \bibinfo {pages} {662--665} (\bibinfo {year}
  {2012})}\BibitemShut {NoStop}%
\bibitem [{\citenamefont {Manoj}, \citenamefont {Pawar},\ and\ \citenamefont
  {Sujith}(2018)}]{manoj2018experimental}%
  \BibitemOpen
  \bibfield  {author} {\bibinfo {author} {\bibfnamefont {K.}~\bibnamefont
  {Manoj}}, \bibinfo {author} {\bibfnamefont {S.~A.}\ \bibnamefont {Pawar}}, \
  and\ \bibinfo {author} {\bibfnamefont {R.~I.}\ \bibnamefont {Sujith}},\
  }\bibfield  {title} {\enquote {\bibinfo {title} {Experimental evidence of
  amplitude death and phase-flip bifurcation between in-phase and anti-phase
  synchronization},}\ }\href@noop {} {\bibfield  {journal} {\bibinfo  {journal}
  {Sci. Rep.}\ }\textbf {\bibinfo {volume} {8}},\ \bibinfo {pages} {1--7}
  (\bibinfo {year} {2018})}\BibitemShut {NoStop}%
\bibitem [{\citenamefont {Raun}\ \emph {et~al.}(1993)\citenamefont {Raun},
  \citenamefont {Beckstead}, \citenamefont {Finlinson},\ and\ \citenamefont
  {Brooks}}]{raun1993review}%
  \BibitemOpen
  \bibfield  {author} {\bibinfo {author} {\bibfnamefont {R.~L.}\ \bibnamefont
  {Raun}}, \bibinfo {author} {\bibfnamefont {M.~W.}\ \bibnamefont {Beckstead}},
  \bibinfo {author} {\bibfnamefont {J.~C.}\ \bibnamefont {Finlinson}}, \ and\
  \bibinfo {author} {\bibfnamefont {K.~P.}\ \bibnamefont {Brooks}},\ }\bibfield
   {title} {\enquote {\bibinfo {title} {{A review of Rijke tubes, Rijke burners
  and related devices}},}\ }\href@noop {} {\bibfield  {journal} {\bibinfo
  {journal} {Prog. Energy Combust. Sci.}\ }\textbf {\bibinfo {volume} {19}},\
  \bibinfo {pages} {313--364} (\bibinfo {year} {1993})}\BibitemShut {NoStop}%
\bibitem [{\citenamefont
  {Feldman~Jr}(1968{\natexlab{a}})}]{feldman1968reviewa}%
  \BibitemOpen
  \bibfield  {author} {\bibinfo {author} {\bibfnamefont {K.~T.}\ \bibnamefont
  {Feldman~Jr}},\ }\bibfield  {title} {\enquote {\bibinfo {title} {{Review of
  the literature on Rijke thermoacoustic phenomena}},}\ }\href@noop {}
  {\bibfield  {journal} {\bibinfo  {journal} {J. Sound Vib.}\ }\textbf
  {\bibinfo {volume} {7}},\ \bibinfo {pages} {83--89} (\bibinfo {year}
  {1968}{\natexlab{a}})}\BibitemShut {NoStop}%
\bibitem [{\citenamefont
  {Feldman~Jr}(1968{\natexlab{b}})}]{feldman1968reviewb}%
  \BibitemOpen
  \bibfield  {author} {\bibinfo {author} {\bibfnamefont {K.~T.}\ \bibnamefont
  {Feldman~Jr}},\ }\bibfield  {title} {\enquote {\bibinfo {title} {{Review of
  the literature on Sondhauss thermoacoustic phenomena}},}\ }\href@noop {}
  {\bibfield  {journal} {\bibinfo  {journal} {J. Sound Vib.}\ }\textbf
  {\bibinfo {volume} {7}},\ \bibinfo {pages} {71--82} (\bibinfo {year}
  {1968}{\natexlab{b}})}\BibitemShut {NoStop}%
\bibitem [{\citenamefont {Sarpotdar}, \citenamefont {Ananthkrishnan},\ and\
  \citenamefont {Sharma}(2003)}]{sarpotdar2003rijke}%
  \BibitemOpen
  \bibfield  {author} {\bibinfo {author} {\bibfnamefont {S.~M.}\ \bibnamefont
  {Sarpotdar}}, \bibinfo {author} {\bibfnamefont {N.}~\bibnamefont
  {Ananthkrishnan}}, \ and\ \bibinfo {author} {\bibfnamefont {S.~D.}\
  \bibnamefont {Sharma}},\ }\bibfield  {title} {\enquote {\bibinfo {title}
  {{The Rijke tube—a thermo-acoustic device}},}\ }\href@noop {} {\bibfield
  {journal} {\bibinfo  {journal} {Resonance}\ }\textbf {\bibinfo {volume}
  {8}},\ \bibinfo {pages} {59--71} (\bibinfo {year} {2003})}\BibitemShut
  {NoStop}%
\bibitem [{\citenamefont {Bisio}\ and\ \citenamefont
  {Rubatto}(1999)}]{bisio1999sondhauss}%
  \BibitemOpen
  \bibfield  {author} {\bibinfo {author} {\bibfnamefont {G.}~\bibnamefont
  {Bisio}}\ and\ \bibinfo {author} {\bibfnamefont {G.}~\bibnamefont
  {Rubatto}},\ }\bibfield  {title} {\enquote {\bibinfo {title} {{Sondhauss and
  Rijke oscillations—thermodynamic analysis, possible applications and
  analogies}},}\ }\href@noop {} {\bibfield  {journal} {\bibinfo  {journal}
  {Energy}\ }\textbf {\bibinfo {volume} {24}},\ \bibinfo {pages} {117--131}
  (\bibinfo {year} {1999})}\BibitemShut {NoStop}%
\bibitem [{\citenamefont {McManus}, \citenamefont {Poinsot},\ and\
  \citenamefont {Candel}(1993)}]{mcmanus1993review}%
  \BibitemOpen
  \bibfield  {author} {\bibinfo {author} {\bibfnamefont {K.~R.}\ \bibnamefont
  {McManus}}, \bibinfo {author} {\bibfnamefont {T.}~\bibnamefont {Poinsot}}, \
  and\ \bibinfo {author} {\bibfnamefont {S.~M.}\ \bibnamefont {Candel}},\
  }\bibfield  {title} {\enquote {\bibinfo {title} {A review of active control
  of combustion instabilities},}\ }\href@noop {} {\bibfield  {journal}
  {\bibinfo  {journal} {Prog. Energy Combust. Sci.}\ }\textbf {\bibinfo
  {volume} {19}},\ \bibinfo {pages} {1--29} (\bibinfo {year}
  {1993})}\BibitemShut {NoStop}%
\bibitem [{\citenamefont {Oran}\ and\ \citenamefont
  {Gardner}(1985)}]{oran1985chemical}%
  \BibitemOpen
  \bibfield  {author} {\bibinfo {author} {\bibfnamefont {E.~S.}\ \bibnamefont
  {Oran}}\ and\ \bibinfo {author} {\bibfnamefont {J.~H.}\ \bibnamefont
  {Gardner}},\ }\bibfield  {title} {\enquote {\bibinfo {title}
  {Chemical-acoustic interactions in combustion systems},}\ }\href@noop {}
  {\bibfield  {journal} {\bibinfo  {journal} {Prog. Energy Combust. Sci.}\
  }\textbf {\bibinfo {volume} {11}},\ \bibinfo {pages} {253--276} (\bibinfo
  {year} {1985})}\BibitemShut {NoStop}%
\bibitem [{\citenamefont {Kabiraj}, \citenamefont {Sujith},\ and\ \citenamefont
  {Wahi}(2012{\natexlab{a}})}]{kabiraj2012bifurcations}%
  \BibitemOpen
  \bibfield  {author} {\bibinfo {author} {\bibfnamefont {L.}~\bibnamefont
  {Kabiraj}}, \bibinfo {author} {\bibfnamefont {R.~I.}\ \bibnamefont {Sujith}},
  \ and\ \bibinfo {author} {\bibfnamefont {P.}~\bibnamefont {Wahi}},\
  }\bibfield  {title} {\enquote {\bibinfo {title} {Bifurcations of self-excited
  ducted laminar premixed flames},}\ }\href@noop {} {\bibfield  {journal}
  {\bibinfo  {journal} {J. Eng. Gas Turbines Power}\ }\textbf {\bibinfo
  {volume} {134}},\ \bibinfo {pages} {031502} (\bibinfo {year}
  {2012}{\natexlab{a}})}\BibitemShut {NoStop}%
\bibitem [{\citenamefont {Kashinath}, \citenamefont {Waugh},\ and\
  \citenamefont {Juniper}(2014)}]{kashinath2014nonlinear}%
  \BibitemOpen
  \bibfield  {author} {\bibinfo {author} {\bibfnamefont {K.}~\bibnamefont
  {Kashinath}}, \bibinfo {author} {\bibfnamefont {I.~C.}\ \bibnamefont
  {Waugh}}, \ and\ \bibinfo {author} {\bibfnamefont {M.~P.}\ \bibnamefont
  {Juniper}},\ }\bibfield  {title} {\enquote {\bibinfo {title} {Nonlinear
  self-excited thermoacoustic oscillations of a ducted premixed flame:
  bifurcations and routes to chaos},}\ }\href@noop {} {\bibfield  {journal}
  {\bibinfo  {journal} {J. Fluid Mech.}\ }\textbf {\bibinfo {volume} {761}},\
  \bibinfo {pages} {399--430} (\bibinfo {year} {2014})}\BibitemShut {NoStop}%
\bibitem [{\citenamefont {Vishnu}, \citenamefont {Sujith},\ and\ \citenamefont
  {Aghalayam}(2015)}]{vishnu2015role}%
  \BibitemOpen
  \bibfield  {author} {\bibinfo {author} {\bibfnamefont {R.}~\bibnamefont
  {Vishnu}}, \bibinfo {author} {\bibfnamefont {R.~I.}\ \bibnamefont {Sujith}},
  \ and\ \bibinfo {author} {\bibfnamefont {P.}~\bibnamefont {Aghalayam}},\
  }\bibfield  {title} {\enquote {\bibinfo {title} {{Role of flame dynamics on
  the bifurcation characteristics of a ducted V-flame}},}\ }\href@noop {}
  {\bibfield  {journal} {\bibinfo  {journal} {Combust. Sci. Tech.}\ }\textbf
  {\bibinfo {volume} {187}},\ \bibinfo {pages} {894--905} (\bibinfo {year}
  {2015})}\BibitemShut {NoStop}%
\bibitem [{\citenamefont {Juniper}\ and\ \citenamefont
  {Sujith}(2018)}]{juniper2018sensitivity}%
  \BibitemOpen
  \bibfield  {author} {\bibinfo {author} {\bibfnamefont {M.~P.}\ \bibnamefont
  {Juniper}}\ and\ \bibinfo {author} {\bibfnamefont {R.~I.}\ \bibnamefont
  {Sujith}},\ }\bibfield  {title} {\enquote {\bibinfo {title} {Sensitivity and
  nonlinearity of thermoacoustic oscillations},}\ }\href@noop {} {\bibfield
  {journal} {\bibinfo  {journal} {Annu. Rev. Fluid Mech.}\ }\textbf {\bibinfo
  {volume} {50}},\ \bibinfo {pages} {661--689} (\bibinfo {year}
  {2018})}\BibitemShut {NoStop}%
\bibitem [{\citenamefont {Lieuwen}(2012)}]{lieuwen2012unsteady}%
  \BibitemOpen
  \bibfield  {author} {\bibinfo {author} {\bibfnamefont {T.~C.}\ \bibnamefont
  {Lieuwen}},\ }\href@noop {} {\emph {\bibinfo {title} {Unsteady Combustor
  Physics}}}\ (\bibinfo  {publisher} {Cambridge University Press},\ \bibinfo
  {year} {2012})\BibitemShut {NoStop}%
\bibitem [{\citenamefont {Fisher}, \citenamefont {Rahman},\ and\ \citenamefont
  {Center}(2009)}]{fisher2009remembering}%
  \BibitemOpen
  \bibfield  {author} {\bibinfo {author} {\bibfnamefont {S.~C.}\ \bibnamefont
  {Fisher}}, \bibinfo {author} {\bibfnamefont {S.~A.}\ \bibnamefont {Rahman}},
  \ and\ \bibinfo {author} {\bibfnamefont {J.~C. S.~S.}\ \bibnamefont
  {Center}},\ }\bibfield  {title} {\enquote {\bibinfo {title} {Remembering
  giant},}\ }\href@noop {} {\bibfield  {journal} {\bibinfo  {journal} {NASA
  History Division, Washington, DC, USA}\ } (\bibinfo {year}
  {2009})}\BibitemShut {NoStop}%
\bibitem [{\citenamefont {Sujith}, \citenamefont {Juniper},\ and\ \citenamefont
  {Schmid}(2016)}]{sujith2016non}%
  \BibitemOpen
  \bibfield  {author} {\bibinfo {author} {\bibfnamefont {R.~I.}\ \bibnamefont
  {Sujith}}, \bibinfo {author} {\bibfnamefont {M.~P.}\ \bibnamefont {Juniper}},
  \ and\ \bibinfo {author} {\bibfnamefont {P.~J.}\ \bibnamefont {Schmid}},\
  }\bibfield  {title} {\enquote {\bibinfo {title} {Non-normality and
  nonlinearity in thermoacoustic instabilities},}\ }\href@noop {} {\bibfield
  {journal} {\bibinfo  {journal} {J. Spray Combust. Dyn.}\ }\textbf {\bibinfo
  {volume} {8}},\ \bibinfo {pages} {119--146} (\bibinfo {year}
  {2016})}\BibitemShut {NoStop}%
\bibitem [{\citenamefont {Lieuwen}\ and\ \citenamefont
  {Yang}(2005)}]{lieuwen2005combustion}%
  \BibitemOpen
  \bibfield  {author} {\bibinfo {author} {\bibfnamefont {T.~C.}\ \bibnamefont
  {Lieuwen}}\ and\ \bibinfo {author} {\bibfnamefont {V.}~\bibnamefont {Yang}},\
  }\href@noop {} {\emph {\bibinfo {title} {Combustion Instabilities in Gas
  Turbine Engines (Operational Experience, Fundamental Mechanisms and
  Modeling)}}},\ Vol.\ \bibinfo {volume} {210}\ (\bibinfo  {publisher}
  {Progress in Astronautics and Aeronautics, AIAA},\ \bibinfo {year}
  {2005})\BibitemShut {NoStop}%
\bibitem [{\citenamefont {Sujith}\ and\ \citenamefont
  {Unni}(2021)}]{sujith2020dynamical}%
  \BibitemOpen
  \bibfield  {author} {\bibinfo {author} {\bibfnamefont {R.~I.}\ \bibnamefont
  {Sujith}}\ and\ \bibinfo {author} {\bibfnamefont {V.~R.}\ \bibnamefont
  {Unni}},\ }\bibfield  {title} {\enquote {\bibinfo {title} {Dynamical systems
  and complex systems theory to study unsteady combustion},}\ }\href@noop {}
  {\bibfield  {journal} {\bibinfo  {journal} {Proc. Combust. Inst.}\ }\textbf
  {\bibinfo {volume} {38}},\ \bibinfo {pages} {3445--3462} (\bibinfo {year}
  {2021})}\BibitemShut {NoStop}%
\bibitem [{\citenamefont {Higgins}(1802)}]{higgins1802sound}%
  \BibitemOpen
  \bibfield  {author} {\bibinfo {author} {\bibfnamefont {B.}~\bibnamefont
  {Higgins}},\ }\bibfield  {title} {\enquote {\bibinfo {title} {On the sound
  produced by a current of hydrogen gas passing through a tube},}\ }\href@noop
  {} {\bibfield  {journal} {\bibinfo  {journal} {J. Nat. Phil. Chem. Arts}\
  }\textbf {\bibinfo {volume} {1}},\ \bibinfo {pages} {129--131} (\bibinfo
  {year} {1802})}\BibitemShut {NoStop}%
\bibitem [{\citenamefont {Noda}\ and\ \citenamefont
  {Ueda}(2013)}]{noda2013thermoacoustic}%
  \BibitemOpen
  \bibfield  {author} {\bibinfo {author} {\bibfnamefont {D.}~\bibnamefont
  {Noda}}\ and\ \bibinfo {author} {\bibfnamefont {Y.}~\bibnamefont {Ueda}},\
  }\bibfield  {title} {\enquote {\bibinfo {title} {A thermoacoustic oscillator
  powered by vaporized water and ethanol},}\ }\href@noop {} {\bibfield
  {journal} {\bibinfo  {journal} {Am. J. Phys.}\ }\textbf {\bibinfo {volume}
  {81}},\ \bibinfo {pages} {124--126} (\bibinfo {year} {2013})}\BibitemShut
  {NoStop}%
\bibitem [{\citenamefont {Ueda}(1974)}]{ueda1974ugetsu}%
  \BibitemOpen
  \bibfield  {author} {\bibinfo {author} {\bibfnamefont {A.}~\bibnamefont
  {Ueda}},\ }\href@noop {} {\emph {\bibinfo {title} {Ugetsu Monogatari: Tales
  of Moonlight and Rain: a Complete English Version of the Eighteenth-century
  Japanese Collection of Tales of the Supernatural}}},\ Vol.~\bibinfo {volume}
  {7}\ (\bibinfo  {publisher} {University of Washington Press},\ \bibinfo
  {year} {1974})\BibitemShut {NoStop}%
\bibitem [{\citenamefont {Lawn}\ and\ \citenamefont
  {Penelet}(2018)}]{lawn2018common}%
  \BibitemOpen
  \bibfield  {author} {\bibinfo {author} {\bibfnamefont {C.~J.}\ \bibnamefont
  {Lawn}}\ and\ \bibinfo {author} {\bibfnamefont {G.}~\bibnamefont {Penelet}},\
  }\bibfield  {title} {\enquote {\bibinfo {title} {Common features in the
  thermoacoustics of flames and engines},}\ }\href@noop {} {\bibfield
  {journal} {\bibinfo  {journal} {Int. J. Spray Combust. Dyn.}\ }\textbf
  {\bibinfo {volume} {10}},\ \bibinfo {pages} {3--37} (\bibinfo {year}
  {2018})}\BibitemShut {NoStop}%
\bibitem [{\citenamefont {Sondhauss}(1850)}]{sondhauss1850ueber}%
  \BibitemOpen
  \bibfield  {author} {\bibinfo {author} {\bibfnamefont {C.}~\bibnamefont
  {Sondhauss}},\ }\bibfield  {title} {\enquote {\bibinfo {title} {Ueber die
  schallschwingungen der luft in erhitzten glasr{\"o}hren und in gedeckten
  pfeifen von ungleicher weite},}\ }\href@noop {} {\bibfield  {journal}
  {\bibinfo  {journal} {Ann. Phys.}\ }\textbf {\bibinfo {volume} {155}},\
  \bibinfo {pages} {1--34} (\bibinfo {year} {1850})}\BibitemShut {NoStop}%
\bibitem [{\citenamefont {Rijke}(1859)}]{rijke1859lxxi}%
  \BibitemOpen
  \bibfield  {author} {\bibinfo {author} {\bibfnamefont {P.~L.}\ \bibnamefont
  {Rijke}},\ }\bibfield  {title} {\enquote {\bibinfo {title} {{LXXI. Notice of
  a new method of causing a vibration of the air contained in a tube open at
  both ends}},}\ }\href@noop {} {\bibfield  {journal} {\bibinfo  {journal}
  {Mag. J. Sci. London, Edinburgh Dublin Philos.}\ }\textbf {\bibinfo {volume}
  {17}},\ \bibinfo {pages} {419--422} (\bibinfo {year} {1859})}\BibitemShut
  {NoStop}%
\bibitem [{\citenamefont
  {Rayleigh}(1878{\natexlab{a}})}]{rayleigh1878explanation}%
  \BibitemOpen
  \bibfield  {author} {\bibinfo {author} {\bibfnamefont {J.~W.~S.}\
  \bibnamefont {Rayleigh}},\ }\bibfield  {title} {\enquote {\bibinfo {title}
  {The explanation of certain acoustical phenomena},}\ }\href@noop {}
  {\bibfield  {journal} {\bibinfo  {journal} {Nature}\ }\textbf {\bibinfo
  {volume} {18}},\ \bibinfo {pages} {319--321} (\bibinfo {year}
  {1878}{\natexlab{a}})}\BibitemShut {NoStop}%
\bibitem [{\citenamefont
  {Rayleigh}(1878{\natexlab{b}})}]{rayleigh1878instability}%
  \BibitemOpen
  \bibfield  {author} {\bibinfo {author} {\bibfnamefont {L.}~\bibnamefont
  {Rayleigh}},\ }\bibfield  {title} {\enquote {\bibinfo {title} {On the
  instability of jets},}\ }\href@noop {} {\bibfield  {journal} {\bibinfo
  {journal} {Proc. London Math. Soc.}\ }\textbf {\bibinfo {volume} {1}},\
  \bibinfo {pages} {4--13} (\bibinfo {year} {1878}{\natexlab{b}})}\BibitemShut
  {NoStop}%
\bibitem [{\citenamefont
  {Rayleigh}(1896{\natexlab{a}})}]{rayleigh1896theoretical}%
  \BibitemOpen
  \bibfield  {author} {\bibinfo {author} {\bibfnamefont {L.}~\bibnamefont
  {Rayleigh}},\ }\bibfield  {title} {\enquote {\bibinfo {title} {L. theoretical
  considerations respecting the separation of gases by diffusion and similar
  processes},}\ }\href@noop {} {\bibfield  {journal} {\bibinfo  {journal}
  {Lond. Edinb. Dublin Philos. Mag. J. Sci.}\ }\textbf {\bibinfo {volume}
  {42}},\ \bibinfo {pages} {493--498} (\bibinfo {year}
  {1896}{\natexlab{a}})}\BibitemShut {NoStop}%
\bibitem [{\citenamefont {Rayleigh}(1896{\natexlab{b}})}]{rayleigh1896theory}%
  \BibitemOpen
  \bibfield  {author} {\bibinfo {author} {\bibfnamefont {J.~W. S.~B.}\
  \bibnamefont {Rayleigh}},\ }\href@noop {} {\emph {\bibinfo {title} {The
  Theory of Sound}}},\ Vol.~\bibinfo {volume} {2}\ (\bibinfo  {publisher}
  {Macmillan},\ \bibinfo {year} {1896})\BibitemShut {NoStop}%
\bibitem [{\citenamefont {Poinsot}\ and\ \citenamefont
  {Veynante}(2005)}]{poinsot2005theoretical}%
  \BibitemOpen
  \bibfield  {author} {\bibinfo {author} {\bibfnamefont {T.}~\bibnamefont
  {Poinsot}}\ and\ \bibinfo {author} {\bibfnamefont {D.}~\bibnamefont
  {Veynante}},\ }\href@noop {} {\emph {\bibinfo {title} {Theoretical and
  Numerical Combustion}}}\ (\bibinfo  {publisher} {RT Edwards, Inc.},\ \bibinfo
  {year} {2005})\BibitemShut {NoStop}%
\bibitem [{\citenamefont {Putnam}(1971)}]{putnam1971combustion}%
  \BibitemOpen
  \bibfield  {author} {\bibinfo {author} {\bibfnamefont {A.~A.}\ \bibnamefont
  {Putnam}},\ }\href@noop {} {\emph {\bibinfo {title} {Combustion Driven
  Oscillations in Industry}}}\ (\bibinfo  {publisher} {Elsevier Publishing
  Company},\ \bibinfo {year} {1971})\BibitemShut {NoStop}%
\bibitem [{\citenamefont {Chu}(1965)}]{chu1965energy}%
  \BibitemOpen
  \bibfield  {author} {\bibinfo {author} {\bibfnamefont {B.-T.}\ \bibnamefont
  {Chu}},\ }\bibfield  {title} {\enquote {\bibinfo {title} {{On the energy
  transfer to small disturbances in fluid flow (Part I)}},}\ }\href@noop {}
  {\bibfield  {journal} {\bibinfo  {journal} {Acta Mech.}\ }\textbf {\bibinfo
  {volume} {1}},\ \bibinfo {pages} {215--234} (\bibinfo {year}
  {1965})}\BibitemShut {NoStop}%
\bibitem [{\citenamefont {Sujith}\ and\ \citenamefont
  {Pawar}(2021)}]{sujiththermoacoustic}%
  \BibitemOpen
  \bibfield  {author} {\bibinfo {author} {\bibfnamefont {R.~I.}\ \bibnamefont
  {Sujith}}\ and\ \bibinfo {author} {\bibfnamefont {S.~A.}\ \bibnamefont
  {Pawar}},\ }\href@noop {} {\emph {\bibinfo {title} {Thermoacoustic
  Instability: A Complex Systems Perspective}}}\ (\bibinfo  {publisher}
  {Springer Nature},\ \bibinfo {year} {2021})\BibitemShut {NoStop}%
\bibitem [{\citenamefont {Reynst}(1961)}]{reynst1961pulsating}%
  \BibitemOpen
  \bibfield  {author} {\bibinfo {author} {\bibfnamefont {F.~H.}\ \bibnamefont
  {Reynst}},\ }\href@noop {} {\emph {\bibinfo {title} {Pulsating Combustion:
  The Collected Works of F H Reynst}}}\ (\bibinfo  {publisher} {Pergamom
  Press},\ \bibinfo {year} {1961})\BibitemShut {NoStop}%
\bibitem [{\citenamefont {Matveev}(2003{\natexlab{a}})}]{matveev2003energy}%
  \BibitemOpen
  \bibfield  {author} {\bibinfo {author} {\bibfnamefont {K.~I.}\ \bibnamefont
  {Matveev}},\ }\bibfield  {title} {\enquote {\bibinfo {title} {{Energy
  consideration of the nonlinear effects in a Rijke tube}},}\ }\href@noop {}
  {\bibfield  {journal} {\bibinfo  {journal} {J. Fluids Struct.}\ }\textbf
  {\bibinfo {volume} {18}},\ \bibinfo {pages} {783--794} (\bibinfo {year}
  {2003}{\natexlab{a}})}\BibitemShut {NoStop}%
\bibitem [{\citenamefont {Mariappan}, \citenamefont {Sujith},\ and\
  \citenamefont {Schmid}(2015)}]{mariappan2015experimental}%
  \BibitemOpen
  \bibfield  {author} {\bibinfo {author} {\bibfnamefont {S.}~\bibnamefont
  {Mariappan}}, \bibinfo {author} {\bibfnamefont {R.~I.}\ \bibnamefont
  {Sujith}}, \ and\ \bibinfo {author} {\bibfnamefont {P.~J.}\ \bibnamefont
  {Schmid}},\ }\bibfield  {title} {\enquote {\bibinfo {title} {{Experimental
  investigation of non-normality of thermoacoustic interaction in an
  electrically heated Rijke tube}},}\ }\href@noop {} {\bibfield  {journal}
  {\bibinfo  {journal} {J. Spray Combust. Dyn.}\ }\textbf {\bibinfo {volume}
  {7}},\ \bibinfo {pages} {315--352} (\bibinfo {year} {2015})}\BibitemShut
  {NoStop}%
\bibitem [{\citenamefont {Etikyala}\ and\ \citenamefont
  {Sujith}(2017)}]{etikyala2017change}%
  \BibitemOpen
  \bibfield  {author} {\bibinfo {author} {\bibfnamefont {S.}~\bibnamefont
  {Etikyala}}\ and\ \bibinfo {author} {\bibfnamefont {R.~I.}\ \bibnamefont
  {Sujith}},\ }\bibfield  {title} {\enquote {\bibinfo {title} {Change of
  criticality in a prototypical thermoacoustic system},}\ }\href@noop {}
  {\bibfield  {journal} {\bibinfo  {journal} {Chaos}\ }\textbf {\bibinfo
  {volume} {27}},\ \bibinfo {pages} {023106} (\bibinfo {year}
  {2017})}\BibitemShut {NoStop}%
\bibitem [{\citenamefont {Munjal}(1987)}]{munjal1987acoustics}%
  \BibitemOpen
  \bibfield  {author} {\bibinfo {author} {\bibfnamefont {M.~L.}\ \bibnamefont
  {Munjal}},\ }\href@noop {} {\emph {\bibinfo {title} {Acoustics of Ducts and
  Mufflers with Application to Exhaust and Ventilation System Design}}}\
  (\bibinfo  {publisher} {John Wiley \& Sons},\ \bibinfo {year}
  {1987})\BibitemShut {NoStop}%
\bibitem [{\citenamefont {Kinsler}\ \emph {et~al.}(2000)\citenamefont
  {Kinsler}, \citenamefont {Frey}, \citenamefont {Coppens},\ and\ \citenamefont
  {Sanders}}]{kinsler2000fundamentals}%
  \BibitemOpen
  \bibfield  {author} {\bibinfo {author} {\bibfnamefont {L.~E.}\ \bibnamefont
  {Kinsler}}, \bibinfo {author} {\bibfnamefont {A.~R.}\ \bibnamefont {Frey}},
  \bibinfo {author} {\bibfnamefont {A.~B.}\ \bibnamefont {Coppens}}, \ and\
  \bibinfo {author} {\bibfnamefont {J.~V.}\ \bibnamefont {Sanders}},\
  }\href@noop {} {\emph {\bibinfo {title} {Fundamentals of Acoustics}}}\
  (\bibinfo  {publisher} {John wiley \& sons},\ \bibinfo {year}
  {2000})\BibitemShut {NoStop}%
\bibitem [{\citenamefont {Rienstra}\ and\ \citenamefont
  {Hirschberg}(2004)}]{rienstra2004introduction}%
  \BibitemOpen
  \bibfield  {author} {\bibinfo {author} {\bibfnamefont {S.~W.}\ \bibnamefont
  {Rienstra}}\ and\ \bibinfo {author} {\bibfnamefont {A.}~\bibnamefont
  {Hirschberg}},\ }\href@noop {} {\enquote {\bibinfo {title} {An introduction
  to acoustics},}\ }\bibinfo {type} {Tech. Rep.}\ (\bibinfo  {institution}
  {Eindhoven University of Technology, IWDE 92-06},\ \bibinfo {year}
  {2004})\BibitemShut {NoStop}%
\bibitem [{\citenamefont {de~Andrade}, \citenamefont {Vazquez},\ and\
  \citenamefont {Pagano}(2018)}]{de2018backstepping}%
  \BibitemOpen
  \bibfield  {author} {\bibinfo {author} {\bibfnamefont {G.~A.}\ \bibnamefont
  {de~Andrade}}, \bibinfo {author} {\bibfnamefont {R.}~\bibnamefont {Vazquez}},
  \ and\ \bibinfo {author} {\bibfnamefont {D.~J.}\ \bibnamefont {Pagano}},\
  }\bibfield  {title} {\enquote {\bibinfo {title} {Backstepping stabilization
  of a linearized ode--pde rijke tube model},}\ }\href@noop {} {\bibfield
  {journal} {\bibinfo  {journal} {Automatica}\ }\textbf {\bibinfo {volume}
  {96}},\ \bibinfo {pages} {98--109} (\bibinfo {year} {2018})}\BibitemShut
  {NoStop}%
\bibitem [{\citenamefont {de~Andrade}, \citenamefont {Vazquez},\ and\
  \citenamefont {Pagano}(2017)}]{de2017boundary}%
  \BibitemOpen
  \bibfield  {author} {\bibinfo {author} {\bibfnamefont {G.~A.}\ \bibnamefont
  {de~Andrade}}, \bibinfo {author} {\bibfnamefont {R.}~\bibnamefont {Vazquez}},
  \ and\ \bibinfo {author} {\bibfnamefont {D.~J.}\ \bibnamefont {Pagano}},\
  }\bibfield  {title} {\enquote {\bibinfo {title} {Boundary control of a rijke
  tube using irrational transfer functions with experimental validation},}\
  }\href@noop {} {\bibfield  {journal} {\bibinfo  {journal}
  {IFAC-PapersOnLine}\ }\textbf {\bibinfo {volume} {50}},\ \bibinfo {pages}
  {4528--4533} (\bibinfo {year} {2017})}\BibitemShut {NoStop}%
\bibitem [{\citenamefont {Wilhelmsen}\ and\ \citenamefont
  {Di~Meglio}(2020)}]{wilhelmsen2020observer}%
  \BibitemOpen
  \bibfield  {author} {\bibinfo {author} {\bibfnamefont {N.~C.~A.}\
  \bibnamefont {Wilhelmsen}}\ and\ \bibinfo {author} {\bibfnamefont
  {F.}~\bibnamefont {Di~Meglio}},\ }\bibfield  {title} {\enquote {\bibinfo
  {title} {An observer for the electrically heated vertical rijke tube with
  nonlinear heat release},}\ }\href@noop {} {\bibfield  {journal} {\bibinfo
  {journal} {IFAC-PapersOnLine}\ }\textbf {\bibinfo {volume} {53}},\ \bibinfo
  {pages} {4181--4188} (\bibinfo {year} {2020})}\BibitemShut {NoStop}%
\bibitem [{\citenamefont {Tandon}\ \emph {et~al.}(2020)\citenamefont {Tandon},
  \citenamefont {Pawar}, \citenamefont {Banerjee}, \citenamefont {Varghese},
  \citenamefont {Durairaj},\ and\ \citenamefont {Sujith}}]{tandon2020bursting}%
  \BibitemOpen
  \bibfield  {author} {\bibinfo {author} {\bibfnamefont {S.}~\bibnamefont
  {Tandon}}, \bibinfo {author} {\bibfnamefont {S.~A.}\ \bibnamefont {Pawar}},
  \bibinfo {author} {\bibfnamefont {S.}~\bibnamefont {Banerjee}}, \bibinfo
  {author} {\bibfnamefont {A.~J.}\ \bibnamefont {Varghese}}, \bibinfo {author}
  {\bibfnamefont {P.}~\bibnamefont {Durairaj}}, \ and\ \bibinfo {author}
  {\bibfnamefont {R.~I.}\ \bibnamefont {Sujith}},\ }\bibfield  {title}
  {\enquote {\bibinfo {title} {Bursting during intermittency route to
  thermoacoustic instability: Effects of slow--fast dynamics},}\ }\href@noop {}
  {\bibfield  {journal} {\bibinfo  {journal} {Chaos}\ }\textbf {\bibinfo
  {volume} {30}},\ \bibinfo {pages} {103112} (\bibinfo {year}
  {2020})}\BibitemShut {NoStop}%
\bibitem [{\citenamefont {Jegadeesan}\ and\ \citenamefont
  {Sujith}(2013)}]{jegadeesan2013experimental}%
  \BibitemOpen
  \bibfield  {author} {\bibinfo {author} {\bibfnamefont {V.}~\bibnamefont
  {Jegadeesan}}\ and\ \bibinfo {author} {\bibfnamefont {R.~I.}\ \bibnamefont
  {Sujith}},\ }\bibfield  {title} {\enquote {\bibinfo {title} {Experimental
  investigation of noise induced triggering in thermoacoustic systems},}\
  }\href@noop {} {\bibfield  {journal} {\bibinfo  {journal} {Proc. Combust.
  Inst.}\ }\textbf {\bibinfo {volume} {34}},\ \bibinfo {pages} {3175--3183}
  (\bibinfo {year} {2013})}\BibitemShut {NoStop}%
\bibitem [{\citenamefont {Murugesan}\ and\ \citenamefont
  {Sujith}(2018)}]{murugesan2018physical}%
  \BibitemOpen
  \bibfield  {author} {\bibinfo {author} {\bibfnamefont {M.}~\bibnamefont
  {Murugesan}}\ and\ \bibinfo {author} {\bibfnamefont {R.~I.}\ \bibnamefont
  {Sujith}},\ }\bibfield  {title} {\enquote {\bibinfo {title} {Physical
  mechanisms that cause intermittency that presages combustion instability and
  blowout in a turbulent lifted jet flame combustor},}\ }\href@noop {}
  {\bibfield  {journal} {\bibinfo  {journal} {Combust. Sci. Tech.}\ }\textbf
  {\bibinfo {volume} {190}},\ \bibinfo {pages} {312--335} (\bibinfo {year}
  {2018})}\BibitemShut {NoStop}%
\bibitem [{\citenamefont {Kabiraj}\ \emph {et~al.}(2012)\citenamefont
  {Kabiraj}, \citenamefont {Saurabh}, \citenamefont {Wahi},\ and\ \citenamefont
  {Sujith}}]{kabiraj2012route}%
  \BibitemOpen
  \bibfield  {author} {\bibinfo {author} {\bibfnamefont {L.}~\bibnamefont
  {Kabiraj}}, \bibinfo {author} {\bibfnamefont {A.}~\bibnamefont {Saurabh}},
  \bibinfo {author} {\bibfnamefont {P.}~\bibnamefont {Wahi}}, \ and\ \bibinfo
  {author} {\bibfnamefont {R.~I.}\ \bibnamefont {Sujith}},\ }\bibfield  {title}
  {\enquote {\bibinfo {title} {Route to chaos for combustion instability in
  ducted laminar premixed flames},}\ }\href@noop {} {\bibfield  {journal}
  {\bibinfo  {journal} {Chaos}\ }\textbf {\bibinfo {volume} {22}},\ \bibinfo
  {pages} {023129} (\bibinfo {year} {2012})}\BibitemShut {NoStop}%
\bibitem [{\citenamefont {Guan}, \citenamefont {Gupta},\ and\ \citenamefont
  {Li}(2020)}]{guan2020intermittency}%
  \BibitemOpen
  \bibfield  {author} {\bibinfo {author} {\bibfnamefont {Y.}~\bibnamefont
  {Guan}}, \bibinfo {author} {\bibfnamefont {V.}~\bibnamefont {Gupta}}, \ and\
  \bibinfo {author} {\bibfnamefont {L.~K.~B.}\ \bibnamefont {Li}},\ }\bibfield
  {title} {\enquote {\bibinfo {title} {Intermittency route to self-excited
  chaotic thermoacoustic oscillations},}\ }\href@noop {} {\bibfield  {journal}
  {\bibinfo  {journal} {J. Fluid Mech.}\ }\textbf {\bibinfo {volume} {894}},\
  \bibinfo {pages} {R3} (\bibinfo {year} {2020})}\BibitemShut {NoStop}%
\bibitem [{\citenamefont {Mukherjee}\ \emph {et~al.}(2015)\citenamefont
  {Mukherjee}, \citenamefont {Heckl}, \citenamefont {Bigongiari}, \citenamefont
  {Vishnu}, \citenamefont {Pawar},\ and\ \citenamefont
  {Sujith}}]{mukherjee2015nonlinear}%
  \BibitemOpen
  \bibfield  {author} {\bibinfo {author} {\bibfnamefont {N.}~\bibnamefont
  {Mukherjee}}, \bibinfo {author} {\bibfnamefont {M.}~\bibnamefont {Heckl}},
  \bibinfo {author} {\bibfnamefont {A.}~\bibnamefont {Bigongiari}}, \bibinfo
  {author} {\bibfnamefont {R.}~\bibnamefont {Vishnu}}, \bibinfo {author}
  {\bibfnamefont {S.~A.}\ \bibnamefont {Pawar}}, \ and\ \bibinfo {author}
  {\bibfnamefont {R.~I.}\ \bibnamefont {Sujith}},\ }\bibfield  {title}
  {\enquote {\bibinfo {title} {Nonlinear dynamics of a laminar v flame in a
  combustor},}\ }in\ \href@noop {} {\emph {\bibinfo {booktitle} {International
  Congress on Sound and Vibration}}},\ Vol.~\bibinfo {volume} {22}\ (\bibinfo
  {year} {2015})\BibitemShut {NoStop}%
\bibitem [{\citenamefont {Durox}\ \emph {et~al.}(2009)\citenamefont {Durox},
  \citenamefont {Schuller}, \citenamefont {Noiray},\ and\ \citenamefont
  {Candel}}]{durox2009experimental}%
  \BibitemOpen
  \bibfield  {author} {\bibinfo {author} {\bibfnamefont {D.}~\bibnamefont
  {Durox}}, \bibinfo {author} {\bibfnamefont {T.}~\bibnamefont {Schuller}},
  \bibinfo {author} {\bibfnamefont {N.}~\bibnamefont {Noiray}}, \ and\ \bibinfo
  {author} {\bibfnamefont {S.}~\bibnamefont {Candel}},\ }\bibfield  {title}
  {\enquote {\bibinfo {title} {Experimental analysis of nonlinear flame
  transfer functions for different flame geometries},}\ }\href@noop {}
  {\bibfield  {journal} {\bibinfo  {journal} {Proc. Combust. Inst.}\ }\textbf
  {\bibinfo {volume} {32}},\ \bibinfo {pages} {1391--1398} (\bibinfo {year}
  {2009})}\BibitemShut {NoStop}%
\bibitem [{\citenamefont {Kasthuri}, \citenamefont {Unni},\ and\ \citenamefont
  {Sujith}(2019)}]{kasthuri2019bursting}%
  \BibitemOpen
  \bibfield  {author} {\bibinfo {author} {\bibfnamefont {P.}~\bibnamefont
  {Kasthuri}}, \bibinfo {author} {\bibfnamefont {V.~R.}\ \bibnamefont {Unni}},
  \ and\ \bibinfo {author} {\bibfnamefont {R.~I.}\ \bibnamefont {Sujith}},\
  }\bibfield  {title} {\enquote {\bibinfo {title} {Bursting and mixed mode
  oscillations during the transition to limit cycle oscillations in a matrix
  burner},}\ }\href@noop {} {\bibfield  {journal} {\bibinfo  {journal} {Chaos}\
  }\textbf {\bibinfo {volume} {29}},\ \bibinfo {pages} {043117} (\bibinfo
  {year} {2019})}\BibitemShut {NoStop}%
\bibitem [{\citenamefont {Sardeshmukh}, \citenamefont {Bedard},\ and\
  \citenamefont {Anderson}(2017)}]{sardeshmukh2017use}%
  \BibitemOpen
  \bibfield  {author} {\bibinfo {author} {\bibfnamefont {S.}~\bibnamefont
  {Sardeshmukh}}, \bibinfo {author} {\bibfnamefont {M.}~\bibnamefont {Bedard}},
  \ and\ \bibinfo {author} {\bibfnamefont {W.}~\bibnamefont {Anderson}},\
  }\bibfield  {title} {\enquote {\bibinfo {title} {The use of oh* and ch* as
  heat release markers in combustion dynamics},}\ }\href@noop {} {\bibfield
  {journal} {\bibinfo  {journal} {Int. J. Spray Combust. Dyn.}\ }\textbf
  {\bibinfo {volume} {9}},\ \bibinfo {pages} {409--423} (\bibinfo {year}
  {2017})}\BibitemShut {NoStop}%
\bibitem [{\citenamefont {Kiefer}\ \emph {et~al.}(2008)\citenamefont {Kiefer},
  \citenamefont {Li}, \citenamefont {Zetterberg}, \citenamefont {Bai},\ and\
  \citenamefont {Ald{\'e}n}}]{kiefer2008investigation}%
  \BibitemOpen
  \bibfield  {author} {\bibinfo {author} {\bibfnamefont {J.}~\bibnamefont
  {Kiefer}}, \bibinfo {author} {\bibfnamefont {Z.}~\bibnamefont {Li}}, \bibinfo
  {author} {\bibfnamefont {J.}~\bibnamefont {Zetterberg}}, \bibinfo {author}
  {\bibfnamefont {X.-S.}\ \bibnamefont {Bai}}, \ and\ \bibinfo {author}
  {\bibfnamefont {M.}~\bibnamefont {Ald{\'e}n}},\ }\bibfield  {title} {\enquote
  {\bibinfo {title} {Investigation of local flame structures and statistics in
  partially premixed turbulent jet flames using simultaneous single-shot ch and
  oh planar laser-induced fluorescence imaging},}\ }\href@noop {} {\bibfield
  {journal} {\bibinfo  {journal} {Combust. Flame}\ }\textbf {\bibinfo {volume}
  {154}},\ \bibinfo {pages} {802--818} (\bibinfo {year} {2008})}\BibitemShut
  {NoStop}%
\bibitem [{\citenamefont {Hardalupas}\ and\ \citenamefont
  {Orain}(2004)}]{hardalupas2004local}%
  \BibitemOpen
  \bibfield  {author} {\bibinfo {author} {\bibfnamefont {Y.}~\bibnamefont
  {Hardalupas}}\ and\ \bibinfo {author} {\bibfnamefont {M.}~\bibnamefont
  {Orain}},\ }\bibfield  {title} {\enquote {\bibinfo {title} {Local
  measurements of the time-dependent heat release rate and equivalence ratio
  using chemiluminescent emission from a flame},}\ }\href@noop {} {\bibfield
  {journal} {\bibinfo  {journal} {Combust. Flame}\ }\textbf {\bibinfo {volume}
  {139}},\ \bibinfo {pages} {188--207} (\bibinfo {year} {2004})}\BibitemShut
  {NoStop}%
\bibitem [{\citenamefont {Pawar}\ \emph {et~al.}(2016)\citenamefont {Pawar},
  \citenamefont {Vishnu}, \citenamefont {Vadivukkarasan}, \citenamefont
  {Panchagnula},\ and\ \citenamefont {Sujith}}]{pawar2016intermittency}%
  \BibitemOpen
  \bibfield  {author} {\bibinfo {author} {\bibfnamefont {S.~A.}\ \bibnamefont
  {Pawar}}, \bibinfo {author} {\bibfnamefont {R.}~\bibnamefont {Vishnu}},
  \bibinfo {author} {\bibfnamefont {M.}~\bibnamefont {Vadivukkarasan}},
  \bibinfo {author} {\bibfnamefont {M.~V.}\ \bibnamefont {Panchagnula}}, \ and\
  \bibinfo {author} {\bibfnamefont {R.~I.}\ \bibnamefont {Sujith}},\ }\bibfield
   {title} {\enquote {\bibinfo {title} {Intermittency route to combustion
  instability in a laboratory spray combustor},}\ }\href@noop {} {\bibfield
  {journal} {\bibinfo  {journal} {J. Eng. Gas Turbines Power}\ }\textbf
  {\bibinfo {volume} {138}},\ \bibinfo {pages} {041505} (\bibinfo {year}
  {2016})}\BibitemShut {NoStop}%
\bibitem [{\citenamefont {Pawar}, \citenamefont {Panchagnula},\ and\
  \citenamefont {Sujith}(2019)}]{pawar2019phase}%
  \BibitemOpen
  \bibfield  {author} {\bibinfo {author} {\bibfnamefont {S.~A.}\ \bibnamefont
  {Pawar}}, \bibinfo {author} {\bibfnamefont {M.~V.}\ \bibnamefont
  {Panchagnula}}, \ and\ \bibinfo {author} {\bibfnamefont {R.~I.}\ \bibnamefont
  {Sujith}},\ }\bibfield  {title} {\enquote {\bibinfo {title} {Phase
  synchronization and collective interaction of multiple flamelets in a
  laboratory scale spray combustor},}\ }\href@noop {} {\bibfield  {journal}
  {\bibinfo  {journal} {Proc. Combust. Inst.}\ }\textbf {\bibinfo {volume}
  {37}},\ \bibinfo {pages} {5121--5128} (\bibinfo {year} {2019})}\BibitemShut
  {NoStop}%
\bibitem [{\citenamefont {Mondal}, \citenamefont {Pawar},\ and\ \citenamefont
  {Sujith}(2019)}]{mondal2019forced}%
  \BibitemOpen
  \bibfield  {author} {\bibinfo {author} {\bibfnamefont {S.}~\bibnamefont
  {Mondal}}, \bibinfo {author} {\bibfnamefont {S.~A.}\ \bibnamefont {Pawar}}, \
  and\ \bibinfo {author} {\bibfnamefont {R.~I.}\ \bibnamefont {Sujith}},\
  }\bibfield  {title} {\enquote {\bibinfo {title} {Forced synchronization and
  asynchronous quenching of periodic oscillations in a thermoacoustic
  system},}\ }\href@noop {} {\bibfield  {journal} {\bibinfo  {journal} {J.
  Fluid Mech.}\ }\textbf {\bibinfo {volume} {864}},\ \bibinfo {pages} {73--96}
  (\bibinfo {year} {2019})}\BibitemShut {NoStop}%
\bibitem [{\citenamefont {Bhattacharya}, \citenamefont {O’Connor},\ and\
  \citenamefont {Ray}(2020)}]{bhattacharya2020data}%
  \BibitemOpen
  \bibfield  {author} {\bibinfo {author} {\bibfnamefont {C.}~\bibnamefont
  {Bhattacharya}}, \bibinfo {author} {\bibfnamefont {J.}~\bibnamefont
  {O’Connor}}, \ and\ \bibinfo {author} {\bibfnamefont {A.}~\bibnamefont
  {Ray}},\ }\bibfield  {title} {\enquote {\bibinfo {title} {Data-driven
  detection and early prediction of thermoacoustic instability in a
  multi-nozzle combustor},}\ }\href@noop {} {\bibfield  {journal} {\bibinfo
  {journal} {Combust. Sci. Tech.}\ ,\ \bibinfo {pages} {1--32}} (\bibinfo
  {year} {2020})}\BibitemShut {NoStop}%
\bibitem [{\citenamefont {Sakamoto}, \citenamefont {Imamura},\ and\
  \citenamefont {Watanabe}(2007)}]{sakamoto2007improvement}%
  \BibitemOpen
  \bibfield  {author} {\bibinfo {author} {\bibfnamefont {S.-i.}\ \bibnamefont
  {Sakamoto}}, \bibinfo {author} {\bibfnamefont {Y.}~\bibnamefont {Imamura}}, \
  and\ \bibinfo {author} {\bibfnamefont {Y.}~\bibnamefont {Watanabe}},\
  }\bibfield  {title} {\enquote {\bibinfo {title} {Improvement of cooling
  effect of loop-tube-type thermoacoustic cooling system applying phase
  adjuster},}\ }\href@noop {} {\bibfield  {journal} {\bibinfo  {journal}
  {Japanese Int. J. Appl. Phys.}\ }\textbf {\bibinfo {volume} {46}},\ \bibinfo
  {pages} {4951} (\bibinfo {year} {2007})}\BibitemShut {NoStop}%
\bibitem [{\citenamefont {Yu}, \citenamefont {Jaworski},\ and\ \citenamefont
  {Abduljalil}(2010)}]{yu2010fishbone}%
  \BibitemOpen
  \bibfield  {author} {\bibinfo {author} {\bibfnamefont {Z.}~\bibnamefont
  {Yu}}, \bibinfo {author} {\bibfnamefont {A.~J.}\ \bibnamefont {Jaworski}}, \
  and\ \bibinfo {author} {\bibfnamefont {A.~S.}\ \bibnamefont {Abduljalil}},\
  }\bibfield  {title} {\enquote {\bibinfo {title} {Fishbone-like instability in
  a looped-tube thermoacoustic engine},}\ }\href@noop {} {\bibfield  {journal}
  {\bibinfo  {journal} {J. Acoust. Soc. Am.}\ }\textbf {\bibinfo {volume}
  {128}},\ \bibinfo {pages} {EL188--EL194} (\bibinfo {year}
  {2010})}\BibitemShut {NoStop}%
\bibitem [{\citenamefont {Hernandez}\ and\ \citenamefont
  {Matveev}(2010)}]{hernandez2010transition}%
  \BibitemOpen
  \bibfield  {author} {\bibinfo {author} {\bibfnamefont {R.}~\bibnamefont
  {Hernandez}}\ and\ \bibinfo {author} {\bibfnamefont {K.~I.}\ \bibnamefont
  {Matveev}},\ }\bibfield  {title} {\enquote {\bibinfo {title} {Transition to
  instability in segmented rijke tube},}\ }\href@noop {} {\bibfield  {journal}
  {\bibinfo  {journal} {Open Thermodyn. J.}\ }\textbf {\bibinfo {volume} {4}}
  (\bibinfo {year} {2010})}\BibitemShut {NoStop}%
\bibitem [{\citenamefont {Zhao}\ and\ \citenamefont
  {Morgans}(2009)}]{zhao2009tuned}%
  \BibitemOpen
  \bibfield  {author} {\bibinfo {author} {\bibfnamefont {D.}~\bibnamefont
  {Zhao}}\ and\ \bibinfo {author} {\bibfnamefont {A.~S.}\ \bibnamefont
  {Morgans}},\ }\bibfield  {title} {\enquote {\bibinfo {title} {Tuned passive
  control of combustion instabilities using multiple helmholtz resonators},}\
  }\href@noop {} {\bibfield  {journal} {\bibinfo  {journal} {J. Sound Vib.}\
  }\textbf {\bibinfo {volume} {320}},\ \bibinfo {pages} {744--757} (\bibinfo
  {year} {2009})}\BibitemShut {NoStop}%
\bibitem [{\citenamefont {Zhao}\ and\ \citenamefont
  {Chew}(2012)}]{zhao2012energy}%
  \BibitemOpen
  \bibfield  {author} {\bibinfo {author} {\bibfnamefont {D.}~\bibnamefont
  {Zhao}}\ and\ \bibinfo {author} {\bibfnamefont {Y.}~\bibnamefont {Chew}},\
  }\bibfield  {title} {\enquote {\bibinfo {title} {Energy harvesting from a
  convection-driven rijke-zhao thermoacoustic engine},}\ }\href@noop {}
  {\bibfield  {journal} {\bibinfo  {journal} {Int. J. Appl. Phys.}\ }\textbf
  {\bibinfo {volume} {112}},\ \bibinfo {pages} {114507} (\bibinfo {year}
  {2012})}\BibitemShut {NoStop}%
\bibitem [{\citenamefont {Zhao}(2013)}]{zhao2013waste}%
  \BibitemOpen
  \bibfield  {author} {\bibinfo {author} {\bibfnamefont {D.}~\bibnamefont
  {Zhao}},\ }\bibfield  {title} {\enquote {\bibinfo {title} {Waste thermal
  energy harvesting from a convection-driven rijke--zhao thermo-acoustic-piezo
  system},}\ }\href@noop {} {\bibfield  {journal} {\bibinfo  {journal} {Energy
  Convers. Manag.}\ }\textbf {\bibinfo {volume} {66}},\ \bibinfo {pages}
  {87--97} (\bibinfo {year} {2013})}\BibitemShut {NoStop}%
\bibitem [{\citenamefont {Pflaum}(1909)}]{pflaum1909versuche}%
  \BibitemOpen
  \bibfield  {author} {\bibinfo {author} {\bibfnamefont {H.}~\bibnamefont
  {Pflaum}},\ }\href@noop {} {\emph {\bibinfo {title} {Versuche Mit Einer
  Elektrischen Pfeife}}}\ (\bibinfo  {publisher} {Vieweg},\ \bibinfo {year}
  {1909})\BibitemShut {NoStop}%
\bibitem [{\citenamefont {Lehmann}(1937)}]{lehmann1937theorie}%
  \BibitemOpen
  \bibfield  {author} {\bibinfo {author} {\bibfnamefont {K.~O.}\ \bibnamefont
  {Lehmann}},\ }\bibfield  {title} {\enquote {\bibinfo {title} {{\"U}ber die
  theorie der netzt{\"o}ne (thermisch erregte schallschwingungen)},}\
  }\href@noop {} {\bibfield  {journal} {\bibinfo  {journal} {Annalen der
  Physik}\ }\textbf {\bibinfo {volume} {421}},\ \bibinfo {pages} {527--555}
  (\bibinfo {year} {1937})}\BibitemShut {NoStop}%
\bibitem [{\citenamefont {Neuringer}\ and\ \citenamefont
  {Hudson}(1952)}]{neuringer1952investigation}%
  \BibitemOpen
  \bibfield  {author} {\bibinfo {author} {\bibfnamefont {J.~L.}\ \bibnamefont
  {Neuringer}}\ and\ \bibinfo {author} {\bibfnamefont {G.~E.}\ \bibnamefont
  {Hudson}},\ }\bibfield  {title} {\enquote {\bibinfo {title} {An investigation
  of sound vibrations in a tube containing a heat source},}\ }\href@noop {}
  {\bibfield  {journal} {\bibinfo  {journal} {J. Acoust. Soc. Am.}\ }\textbf
  {\bibinfo {volume} {24}},\ \bibinfo {pages} {667--674} (\bibinfo {year}
  {1952})}\BibitemShut {NoStop}%
\bibitem [{\citenamefont {Merk}(1957{\natexlab{a}})}]{merk1957analysis}%
  \BibitemOpen
  \bibfield  {author} {\bibinfo {author} {\bibfnamefont {H.}~\bibnamefont
  {Merk}},\ }\bibfield  {title} {\enquote {\bibinfo {title} {Analysis of
  heat-driven oscillations of gas flows},}\ }\href@noop {} {\bibfield
  {journal} {\bibinfo  {journal} {Appl. Sci. Res. Sec. A}\ }\textbf {\bibinfo
  {volume} {6}},\ \bibinfo {pages} {317--336} (\bibinfo {year}
  {1957}{\natexlab{a}})}\BibitemShut {NoStop}%
\bibitem [{\citenamefont {Mugridge}(1980)}]{mugridge1980combustion}%
  \BibitemOpen
  \bibfield  {author} {\bibinfo {author} {\bibfnamefont {B.~D.}\ \bibnamefont
  {Mugridge}},\ }\bibfield  {title} {\enquote {\bibinfo {title} {Combustion
  driven oscillations},}\ }\href@noop {} {\bibfield  {journal} {\bibinfo
  {journal} {J. Sound Vib.}\ }\textbf {\bibinfo {volume} {70}},\ \bibinfo
  {pages} {437--452} (\bibinfo {year} {1980})}\BibitemShut {NoStop}%
\bibitem [{\citenamefont {Madarame}(1981)}]{madarame1981thermally}%
  \BibitemOpen
  \bibfield  {author} {\bibinfo {author} {\bibfnamefont {H.}~\bibnamefont
  {Madarame}},\ }\bibfield  {title} {\enquote {\bibinfo {title} {Thermally
  induced acoustic oscillations in a pipe: 1st report: Oscillations induced by
  plane heat source in air current},}\ }\href@noop {} {\bibfield  {journal}
  {\bibinfo  {journal} {Bull. JSME}\ }\textbf {\bibinfo {volume} {24}},\
  \bibinfo {pages} {1626--1633} (\bibinfo {year} {1981})}\BibitemShut {NoStop}%
\bibitem [{\citenamefont {Merk}(1958)}]{merk1958analysis}%
  \BibitemOpen
  \bibfield  {author} {\bibinfo {author} {\bibfnamefont {H.}~\bibnamefont
  {Merk}},\ }\bibfield  {title} {\enquote {\bibinfo {title} {Analysis of
  heat-driven oscillations of gas flows},}\ }\href@noop {} {\bibfield
  {journal} {\bibinfo  {journal} {Appl. Sci. Res. Sec. A}\ }\textbf {\bibinfo
  {volume} {7}},\ \bibinfo {pages} {175--191} (\bibinfo {year}
  {1958})}\BibitemShut {NoStop}%
\bibitem [{\citenamefont {Merk}(1957{\natexlab{b}})}]{merk1957analysisp}%
  \BibitemOpen
  \bibfield  {author} {\bibinfo {author} {\bibfnamefont {H.}~\bibnamefont
  {Merk}},\ }\bibfield  {title} {\enquote {\bibinfo {title} {An analysis of
  unstable combustion of premixed gases},}\ }in\ \href@noop {} {\emph {\bibinfo
  {booktitle} {Symposium (International) on Combustion}}},\ Vol.~\bibinfo
  {volume} {6}\ (\bibinfo {organization} {Elsevier},\ \bibinfo {year} {1957})\
  pp.\ \bibinfo {pages} {500--512}\BibitemShut {NoStop}%
\bibitem [{\citenamefont {Clarke}\ and\ \citenamefont
  {McIntosh}(1980)}]{clarke1980influence}%
  \BibitemOpen
  \bibfield  {author} {\bibinfo {author} {\bibfnamefont {J.~F.}\ \bibnamefont
  {Clarke}}\ and\ \bibinfo {author} {\bibfnamefont {A.~C.}\ \bibnamefont
  {McIntosh}},\ }\bibfield  {title} {\enquote {\bibinfo {title} {The influence
  of a flameholder on a plane flame, including its static stability},}\
  }\href@noop {} {\bibfield  {journal} {\bibinfo  {journal} {Proc. R. Soc. A:
  Math. Phys. Eng. Sci.}\ }\textbf {\bibinfo {volume} {372}},\ \bibinfo {pages}
  {367--392} (\bibinfo {year} {1980})}\BibitemShut {NoStop}%
\bibitem [{\citenamefont {McIntosh}(1985)}]{mcintosh1985cellular}%
  \BibitemOpen
  \bibfield  {author} {\bibinfo {author} {\bibfnamefont {A.}~\bibnamefont
  {McIntosh}},\ }\bibfield  {title} {\enquote {\bibinfo {title} {On the
  cellular instability of flames near porous-plug burners},}\ }\href@noop {}
  {\bibfield  {journal} {\bibinfo  {journal} {J. Fluid Mech.}\ }\textbf
  {\bibinfo {volume} {161}},\ \bibinfo {pages} {43--75} (\bibinfo {year}
  {1985})}\BibitemShut {NoStop}%
\bibitem [{\citenamefont {McIntosh}(1991)}]{mcintosh1991pressure}%
  \BibitemOpen
  \bibfield  {author} {\bibinfo {author} {\bibfnamefont {A.}~\bibnamefont
  {McIntosh}},\ }\bibfield  {title} {\enquote {\bibinfo {title} {Pressure
  disturbances of different length scales interacting with conventional
  flames},}\ }\href@noop {} {\bibfield  {journal} {\bibinfo  {journal}
  {Combust. Sci. Technol.}\ }\textbf {\bibinfo {volume} {75}},\ \bibinfo
  {pages} {287--309} (\bibinfo {year} {1991})}\BibitemShut {NoStop}%
\bibitem [{\citenamefont {Putnam}\ and\ \citenamefont
  {Dennis}(1953)}]{putnam1953organ}%
  \BibitemOpen
  \bibfield  {author} {\bibinfo {author} {\bibfnamefont {A.~A.}\ \bibnamefont
  {Putnam}}\ and\ \bibinfo {author} {\bibfnamefont {W.~R.}\ \bibnamefont
  {Dennis}},\ }\bibfield  {title} {\enquote {\bibinfo {title} {Organ-pipe
  oscillations in a flame-filled tube},}\ }in\ \href@noop {} {\emph {\bibinfo
  {booktitle} {Symposium (International) on Combustion}}},\ Vol.~\bibinfo
  {volume} {4}\ (\bibinfo {organization} {Elsevier},\ \bibinfo {year} {1953})\
  pp.\ \bibinfo {pages} {566--575}\BibitemShut {NoStop}%
\bibitem [{\citenamefont {Putnam}\ and\ \citenamefont
  {Dennis}(1954)}]{putnam1954burner}%
  \BibitemOpen
  \bibfield  {author} {\bibinfo {author} {\bibfnamefont {A.~A.}\ \bibnamefont
  {Putnam}}\ and\ \bibinfo {author} {\bibfnamefont {W.~R.}\ \bibnamefont
  {Dennis}},\ }\bibfield  {title} {\enquote {\bibinfo {title} {Burner
  oscillations of the gauze-tone type},}\ }\href@noop {} {\bibfield  {journal}
  {\bibinfo  {journal} {J. Acoust. Soc. Am.}\ }\textbf {\bibinfo {volume}
  {26}},\ \bibinfo {pages} {716--725} (\bibinfo {year} {1954})}\BibitemShut
  {NoStop}%
\bibitem [{\citenamefont {Clarke}, \citenamefont {Kassoy},\ and\ \citenamefont
  {Riley}(1984)}]{clarke1984shocks}%
  \BibitemOpen
  \bibfield  {author} {\bibinfo {author} {\bibfnamefont {J.~F.}\ \bibnamefont
  {Clarke}}, \bibinfo {author} {\bibfnamefont {D.}~\bibnamefont {Kassoy}}, \
  and\ \bibinfo {author} {\bibfnamefont {N.}~\bibnamefont {Riley}},\ }\bibfield
   {title} {\enquote {\bibinfo {title} {Shocks generated in a confined gas due
  to rapid heat addition at the boundary. i. weak shock waves},}\ }\href@noop
  {} {\bibfield  {journal} {\bibinfo  {journal} {Proc. R. Soc. A: Math. Phys.
  Eng. Sci.}\ }\textbf {\bibinfo {volume} {393}},\ \bibinfo {pages} {309--329}
  (\bibinfo {year} {1984})}\BibitemShut {NoStop}%
\bibitem [{\citenamefont {Culick}(1987)}]{culick1987note}%
  \BibitemOpen
  \bibfield  {author} {\bibinfo {author} {\bibfnamefont {F.~E.~C.}\
  \bibnamefont {Culick}},\ }\bibfield  {title} {\enquote {\bibinfo {title} {A
  note on rayleigh's criterion},}\ }\href@noop {} {\bibfield  {journal}
  {\bibinfo  {journal} {Combust. Sci. Technol.}\ }\textbf {\bibinfo {volume}
  {56}},\ \bibinfo {pages} {159--166} (\bibinfo {year} {1987})}\BibitemShut
  {NoStop}%
\bibitem [{\citenamefont {Crocco}\ and\ \citenamefont
  {Cheng}(1956)}]{crocco1956theory}%
  \BibitemOpen
  \bibfield  {author} {\bibinfo {author} {\bibfnamefont {L.}~\bibnamefont
  {Crocco}}\ and\ \bibinfo {author} {\bibfnamefont {S.-I.}\ \bibnamefont
  {Cheng}},\ }\href@noop {} {\enquote {\bibinfo {title} {Theory of combustion
  instability in liquid propellant rocket motors},}\ }\bibinfo {type} {Tech.
  Rep.}\ (\bibinfo  {institution} {Agardograph no. 8 , Butterworths Science
  Publication},\ \bibinfo {year} {1956})\BibitemShut {NoStop}%
\bibitem [{\citenamefont {Nicoli}\ and\ \citenamefont
  {Pelce}(1989)}]{nicoli1989one}%
  \BibitemOpen
  \bibfield  {author} {\bibinfo {author} {\bibfnamefont {C.}~\bibnamefont
  {Nicoli}}\ and\ \bibinfo {author} {\bibfnamefont {P.}~\bibnamefont {Pelce}},\
  }\bibfield  {title} {\enquote {\bibinfo {title} {One-dimensional model for
  the rijke tube},}\ }\href@noop {} {\bibfield  {journal} {\bibinfo  {journal}
  {J. Fluid Mech.}\ }\textbf {\bibinfo {volume} {202}},\ \bibinfo {pages}
  {83--96} (\bibinfo {year} {1989})}\BibitemShut {NoStop}%
\bibitem [{\citenamefont {King}(1914)}]{king1914xii}%
  \BibitemOpen
  \bibfield  {author} {\bibinfo {author} {\bibfnamefont {L.~V.}\ \bibnamefont
  {King}},\ }\bibfield  {title} {\enquote {\bibinfo {title} {Xii. on the
  convection of heat from small cylinders in a stream of fluid: Determination
  of the convection constants of small platinum wires with applications to
  hot-wire anemometry},}\ }\href@noop {} {\bibfield  {journal} {\bibinfo
  {journal} {Philos. Tr. R. Soc. London Ser. G}\ }\textbf {\bibinfo {volume}
  {214}},\ \bibinfo {pages} {373--432} (\bibinfo {year} {1914})}\BibitemShut
  {NoStop}%
\bibitem [{\citenamefont {Heckl}(1990)}]{heckl1990non}%
  \BibitemOpen
  \bibfield  {author} {\bibinfo {author} {\bibfnamefont {M.~A.}\ \bibnamefont
  {Heckl}},\ }\bibfield  {title} {\enquote {\bibinfo {title} {Non-linear
  acoustic effects in the rijke tube},}\ }\href@noop {} {\bibfield  {journal}
  {\bibinfo  {journal} {Acta. Acust. United Ac.}\ }\textbf {\bibinfo {volume}
  {72}},\ \bibinfo {pages} {63--71} (\bibinfo {year} {1990})}\BibitemShut
  {NoStop}%
\bibitem [{\citenamefont {Zinn}\ and\ \citenamefont
  {Lores}(1971)}]{zinn1971application}%
  \BibitemOpen
  \bibfield  {author} {\bibinfo {author} {\bibfnamefont {B.~T.}\ \bibnamefont
  {Zinn}}\ and\ \bibinfo {author} {\bibfnamefont {M.~E.}\ \bibnamefont
  {Lores}},\ }\bibfield  {title} {\enquote {\bibinfo {title} {{Application of
  the Galerkin method in the solution of non-linear axial combustion
  instability problems in liquid rockets}},}\ }\href@noop {} {\bibfield
  {journal} {\bibinfo  {journal} {Combust. Sci. Tech.}\ }\textbf {\bibinfo
  {volume} {4}},\ \bibinfo {pages} {269--278} (\bibinfo {year}
  {1971})}\BibitemShut {NoStop}%
\bibitem [{\citenamefont {Zinn}\ and\ \citenamefont
  {Powell}(1970)}]{zinn1970application}%
  \BibitemOpen
  \bibfield  {author} {\bibinfo {author} {\bibfnamefont {B.}~\bibnamefont
  {Zinn}}\ and\ \bibinfo {author} {\bibfnamefont {E.}~\bibnamefont {Powell}},\
  }\bibfield  {title} {\enquote {\bibinfo {title} {Application of the galerkin
  method in the solution of combustion-instability problems},}\ }in\ \href@noop
  {} {\emph {\bibinfo {booktitle} {Propulsion Re-Entry Physics}}}\ (\bibinfo
  {publisher} {Elsevier},\ \bibinfo {year} {1970})\ pp.\ \bibinfo {pages}
  {59--73}\BibitemShut {NoStop}%
\bibitem [{\citenamefont {Zinn}\ and\ \citenamefont
  {Powell}(1971)}]{zinn1971nonlinear}%
  \BibitemOpen
  \bibfield  {author} {\bibinfo {author} {\bibfnamefont {B.~T.}\ \bibnamefont
  {Zinn}}\ and\ \bibinfo {author} {\bibfnamefont {E.~A.}\ \bibnamefont
  {Powell}},\ }\bibfield  {title} {\enquote {\bibinfo {title} {Nonlinear
  combustion instability in liquid-propellant rocket engines},}\ }in\
  \href@noop {} {\emph {\bibinfo {booktitle} {Symposium (International) on
  Combustion}}},\ Vol.~\bibinfo {volume} {13}\ (\bibinfo {organization}
  {Elsevier},\ \bibinfo {year} {1971})\ pp.\ \bibinfo {pages}
  {491--503}\BibitemShut {NoStop}%
\bibitem [{\citenamefont {Balasubramanian}\ and\ \citenamefont
  {Sujith}(2008{\natexlab{a}})}]{balasubramanian2008thermoacoustic}%
  \BibitemOpen
  \bibfield  {author} {\bibinfo {author} {\bibfnamefont {K.}~\bibnamefont
  {Balasubramanian}}\ and\ \bibinfo {author} {\bibfnamefont {R.~I.}\
  \bibnamefont {Sujith}},\ }\bibfield  {title} {\enquote {\bibinfo {title}
  {{Thermoacoustic instability in a Rijke tube: Non-normality and
  nonlinearity}},}\ }\href@noop {} {\bibfield  {journal} {\bibinfo  {journal}
  {Phys. Fluids}\ }\textbf {\bibinfo {volume} {20}},\ \bibinfo {pages} {044103}
  (\bibinfo {year} {2008}{\natexlab{a}})}\BibitemShut {NoStop}%
\bibitem [{\citenamefont {Subramanian}\ \emph {et~al.}(2010)\citenamefont
  {Subramanian}, \citenamefont {Mariappan}, \citenamefont {Sujith},\ and\
  \citenamefont {Wahi}}]{subramanian2010bifurcation}%
  \BibitemOpen
  \bibfield  {author} {\bibinfo {author} {\bibfnamefont {P.}~\bibnamefont
  {Subramanian}}, \bibinfo {author} {\bibfnamefont {S.}~\bibnamefont
  {Mariappan}}, \bibinfo {author} {\bibfnamefont {R.~I.}\ \bibnamefont
  {Sujith}}, \ and\ \bibinfo {author} {\bibfnamefont {P.}~\bibnamefont
  {Wahi}},\ }\bibfield  {title} {\enquote {\bibinfo {title} {{Bifurcation
  analysis of thermoacoustic instability in a horizontal Rijke tube}},}\
  }\href@noop {} {\bibfield  {journal} {\bibinfo  {journal} {J. Spray Combust.
  Dyn.}\ }\textbf {\bibinfo {volume} {2}},\ \bibinfo {pages} {325--355}
  (\bibinfo {year} {2010})}\BibitemShut {NoStop}%
\bibitem [{\citenamefont {Subramanian}, \citenamefont {Sujith},\ and\
  \citenamefont {Wahi}(2013)}]{subramanian2013subcritical}%
  \BibitemOpen
  \bibfield  {author} {\bibinfo {author} {\bibfnamefont {P.}~\bibnamefont
  {Subramanian}}, \bibinfo {author} {\bibfnamefont {R.~I.}\ \bibnamefont
  {Sujith}}, \ and\ \bibinfo {author} {\bibfnamefont {P.}~\bibnamefont
  {Wahi}},\ }\bibfield  {title} {\enquote {\bibinfo {title} {Subcritical
  bifurcation and bistability in thermoacoustic systems},}\ }\href@noop {}
  {\bibfield  {journal} {\bibinfo  {journal} {J. Fluid Mech.}\ }\textbf
  {\bibinfo {volume} {715}},\ \bibinfo {pages} {210--238} (\bibinfo {year}
  {2013})}\BibitemShut {NoStop}%
\bibitem [{\citenamefont {García-Morales}\ and\ \citenamefont
  {Krischer}(2012)}]{garcia2012complex}%
  \BibitemOpen
  \bibfield  {author} {\bibinfo {author} {\bibfnamefont {V.}~\bibnamefont
  {García-Morales}}\ and\ \bibinfo {author} {\bibfnamefont {K.}~\bibnamefont
  {Krischer}},\ }\bibfield  {title} {\enquote {\bibinfo {title} {The complex
  ginzburg–landau equation: an introduction},}\ }\href {\doibase
  10.1080/00107514.2011.642554} {\bibfield  {journal} {\bibinfo  {journal}
  {Contemp. Phys.}\ }\textbf {\bibinfo {volume} {53}},\ \bibinfo {pages}
  {79--95} (\bibinfo {year} {2012})}\BibitemShut {NoStop}%
\bibitem [{\citenamefont {Orchini}, \citenamefont {Rigas},\ and\ \citenamefont
  {Juniper}(2016)}]{orchini2016weakly}%
  \BibitemOpen
  \bibfield  {author} {\bibinfo {author} {\bibfnamefont {A.}~\bibnamefont
  {Orchini}}, \bibinfo {author} {\bibfnamefont {G.}~\bibnamefont {Rigas}}, \
  and\ \bibinfo {author} {\bibfnamefont {M.~P.}\ \bibnamefont {Juniper}},\
  }\bibfield  {title} {\enquote {\bibinfo {title} {Weakly nonlinear analysis of
  thermoacoustic bifurcations in the rijke tube},}\ }\href@noop {} {\bibfield
  {journal} {\bibinfo  {journal} {J. Fluid Mech.}\ }\textbf {\bibinfo {volume}
  {805}},\ \bibinfo {pages} {523--550} (\bibinfo {year} {2016})}\BibitemShut
  {NoStop}%
\bibitem [{\citenamefont {Magri}\ and\ \citenamefont
  {Juniper}(2013)}]{magri2013sensitivity}%
  \BibitemOpen
  \bibfield  {author} {\bibinfo {author} {\bibfnamefont {L.}~\bibnamefont
  {Magri}}\ and\ \bibinfo {author} {\bibfnamefont {M.~P.}\ \bibnamefont
  {Juniper}},\ }\bibfield  {title} {\enquote {\bibinfo {title} {Sensitivity
  analysis of a time-delayed thermo-acoustic system via an adjoint-based
  approach},}\ }\href@noop {} {\bibfield  {journal} {\bibinfo  {journal} {J.
  Fluid Mech.}\ }\textbf {\bibinfo {volume} {719}},\ \bibinfo {pages}
  {183--202} (\bibinfo {year} {2013})}\BibitemShut {NoStop}%
\bibitem [{\citenamefont {Magri}\ and\ \citenamefont
  {Juniper}(2014)}]{magri2014adjoint}%
  \BibitemOpen
  \bibfield  {author} {\bibinfo {author} {\bibfnamefont {L.}~\bibnamefont
  {Magri}}\ and\ \bibinfo {author} {\bibfnamefont {M.~P.}\ \bibnamefont
  {Juniper}},\ }\bibfield  {title} {\enquote {\bibinfo {title} {Adjoint-based
  linear analysis in reduced-order thermo-acoustic models},}\ }\href@noop {}
  {\bibfield  {journal} {\bibinfo  {journal} {Int. J. Spray Combust. Dyn.}\
  }\textbf {\bibinfo {volume} {6}},\ \bibinfo {pages} {225--246} (\bibinfo
  {year} {2014})}\BibitemShut {NoStop}%
\bibitem [{\citenamefont {Guckenheimer}\ and\ \citenamefont
  {Holmes}(2013)}]{guckenheimer2013nonlinear}%
  \BibitemOpen
  \bibfield  {author} {\bibinfo {author} {\bibfnamefont {J.}~\bibnamefont
  {Guckenheimer}}\ and\ \bibinfo {author} {\bibfnamefont {P.}~\bibnamefont
  {Holmes}},\ }\href@noop {} {\emph {\bibinfo {title} {Nonlinear Oscillations,
  Dynamical Systems, and Bifurcations of Vector Fields}}},\ Vol.~\bibinfo
  {volume} {42}\ (\bibinfo  {publisher} {Springer Science \& Business Media},\
  \bibinfo {year} {2013})\BibitemShut {NoStop}%
\bibitem [{\citenamefont {Marsden}\ and\ \citenamefont
  {McCracken}(2012)}]{marsden2012hopf}%
  \BibitemOpen
  \bibfield  {author} {\bibinfo {author} {\bibfnamefont {J.~E.}\ \bibnamefont
  {Marsden}}\ and\ \bibinfo {author} {\bibfnamefont {M.}~\bibnamefont
  {McCracken}},\ }\href@noop {} {\emph {\bibinfo {title} {The Hopf Bifurcation
  and its Applications}}},\ Vol.~\bibinfo {volume} {19}\ (\bibinfo  {publisher}
  {Springer Science \& Business Media},\ \bibinfo {year} {2012})\BibitemShut
  {NoStop}%
\bibitem [{\citenamefont
  {Matveev}(2003{\natexlab{b}})}]{matveev2003thermoacoustic}%
  \BibitemOpen
  \bibfield  {author} {\bibinfo {author} {\bibfnamefont {K.~I.}\ \bibnamefont
  {Matveev}},\ }\emph {\bibinfo {title} {{Thermoacoustic instabilities in the
  Rijke tube: experiments and modeling}}},\ \href@noop {} {Ph.D. thesis},\
  \bibinfo  {school} {California Institute of Technology, USA} (\bibinfo {year}
  {2003}{\natexlab{b}})\BibitemShut {NoStop}%
\bibitem [{\citenamefont {Mariappan}(2012)}]{mariappan2012theoretical}%
  \BibitemOpen
  \bibfield  {author} {\bibinfo {author} {\bibfnamefont {S.}~\bibnamefont
  {Mariappan}},\ }\emph {\bibinfo {title} {Theoretical and experimental
  investigation of the non-normal nature of thermoacoustic interactions}},\
  \href@noop {} {Ph.D. thesis},\ \bibinfo  {school} {PhD thesis, Indian
  Institute of Technology, Madras} (\bibinfo {year} {2012})\BibitemShut
  {NoStop}%
\bibitem [{\citenamefont {Juniper}(2011)}]{juniper2011triggering}%
  \BibitemOpen
  \bibfield  {author} {\bibinfo {author} {\bibfnamefont {M.~P.}\ \bibnamefont
  {Juniper}},\ }\bibfield  {title} {\enquote {\bibinfo {title} {{Triggering in
  the horizontal Rijke tube: non-normality, transient growth and bypass
  transition}},}\ }\href@noop {} {\bibfield  {journal} {\bibinfo  {journal} {J.
  Fluid Mech.}\ }\textbf {\bibinfo {volume} {667}},\ \bibinfo {pages}
  {272--308} (\bibinfo {year} {2011})}\BibitemShut {NoStop}%
\bibitem [{\citenamefont {Gopalakrishnan}\ and\ \citenamefont
  {Sujith}(2014)}]{gopalakrishnan2014influence}%
  \BibitemOpen
  \bibfield  {author} {\bibinfo {author} {\bibfnamefont {E.~A.}\ \bibnamefont
  {Gopalakrishnan}}\ and\ \bibinfo {author} {\bibfnamefont {R.~I.}\
  \bibnamefont {Sujith}},\ }\bibfield  {title} {\enquote {\bibinfo {title}
  {{Influence of system parameters on the hysteresis characteristics of a
  horizontal Rijke tube}},}\ }\href@noop {} {\bibfield  {journal} {\bibinfo
  {journal} {J. Spray Combust. Dyn.}\ }\textbf {\bibinfo {volume} {6}},\
  \bibinfo {pages} {293--316} (\bibinfo {year} {2014})}\BibitemShut {NoStop}%
\bibitem [{\citenamefont {Ashwin}\ \emph {et~al.}(2012)\citenamefont {Ashwin},
  \citenamefont {Wieczorek}, \citenamefont {Vitolo},\ and\ \citenamefont
  {Cox}}]{ashwin2012tipping}%
  \BibitemOpen
  \bibfield  {author} {\bibinfo {author} {\bibfnamefont {P.}~\bibnamefont
  {Ashwin}}, \bibinfo {author} {\bibfnamefont {S.}~\bibnamefont {Wieczorek}},
  \bibinfo {author} {\bibfnamefont {R.}~\bibnamefont {Vitolo}}, \ and\ \bibinfo
  {author} {\bibfnamefont {P.}~\bibnamefont {Cox}},\ }\bibfield  {title}
  {\enquote {\bibinfo {title} {Tipping points in open systems: bifurcation,
  noise-induced and rate-dependent examples in the climate system},}\
  }\href@noop {} {\bibfield  {journal} {\bibinfo  {journal} {Philos. Tr. R.
  Soc. A}\ }\textbf {\bibinfo {volume} {370}},\ \bibinfo {pages} {1166--1184}
  (\bibinfo {year} {2012})}\BibitemShut {NoStop}%
\bibitem [{\citenamefont {Scheffer}\ \emph {et~al.}(2009)\citenamefont
  {Scheffer}, \citenamefont {Bascompte}, \citenamefont {Brock}, \citenamefont
  {Brovkin}, \citenamefont {Carpenter}, \citenamefont {Dakos}, \citenamefont
  {Held}, \citenamefont {Van~Nes}, \citenamefont {Rietkerk},\ and\
  \citenamefont {Sugihara}}]{scheffer2009early}%
  \BibitemOpen
  \bibfield  {author} {\bibinfo {author} {\bibfnamefont {M.}~\bibnamefont
  {Scheffer}}, \bibinfo {author} {\bibfnamefont {J.}~\bibnamefont {Bascompte}},
  \bibinfo {author} {\bibfnamefont {W.~A.}\ \bibnamefont {Brock}}, \bibinfo
  {author} {\bibfnamefont {V.}~\bibnamefont {Brovkin}}, \bibinfo {author}
  {\bibfnamefont {S.~R.}\ \bibnamefont {Carpenter}}, \bibinfo {author}
  {\bibfnamefont {V.}~\bibnamefont {Dakos}}, \bibinfo {author} {\bibfnamefont
  {H.}~\bibnamefont {Held}}, \bibinfo {author} {\bibfnamefont {E.~H.}\
  \bibnamefont {Van~Nes}}, \bibinfo {author} {\bibfnamefont {M.}~\bibnamefont
  {Rietkerk}}, \ and\ \bibinfo {author} {\bibfnamefont {G.}~\bibnamefont
  {Sugihara}},\ }\bibfield  {title} {\enquote {\bibinfo {title} {Early-warning
  signals for critical transitions},}\ }\href@noop {} {\bibfield  {journal}
  {\bibinfo  {journal} {Nature}\ }\textbf {\bibinfo {volume} {461}},\ \bibinfo
  {pages} {53--59} (\bibinfo {year} {2009})}\BibitemShut {NoStop}%
\bibitem [{\citenamefont {Pavithran}\ and\ \citenamefont
  {Sujith}(2021)}]{pavithran2021effect}%
  \BibitemOpen
  \bibfield  {author} {\bibinfo {author} {\bibfnamefont {I.}~\bibnamefont
  {Pavithran}}\ and\ \bibinfo {author} {\bibfnamefont {R.~I.}\ \bibnamefont
  {Sujith}},\ }\bibfield  {title} {\enquote {\bibinfo {title} {Effect of rate
  of change of parameter on early warning signals for critical transitions},}\
  }\href@noop {} {\bibfield  {journal} {\bibinfo  {journal} {Chaos}\ }\textbf
  {\bibinfo {volume} {31}},\ \bibinfo {pages} {013116} (\bibinfo {year}
  {2021})}\BibitemShut {NoStop}%
\bibitem [{\citenamefont {Thompson}\ and\ \citenamefont
  {Sieber}(2011{\natexlab{a}})}]{thompson2011predicting}%
  \BibitemOpen
  \bibfield  {author} {\bibinfo {author} {\bibfnamefont {J.~M.~T.}\
  \bibnamefont {Thompson}}\ and\ \bibinfo {author} {\bibfnamefont
  {J.}~\bibnamefont {Sieber}},\ }\bibfield  {title} {\enquote {\bibinfo {title}
  {Predicting climate tipping as a noisy bifurcation: a review},}\ }\href@noop
  {} {\bibfield  {journal} {\bibinfo  {journal} {Int. J. Bifurc. Chaos}\
  }\textbf {\bibinfo {volume} {21}},\ \bibinfo {pages} {399--423} (\bibinfo
  {year} {2011}{\natexlab{a}})}\BibitemShut {NoStop}%
\bibitem [{\citenamefont {Thompson}\ and\ \citenamefont
  {Sieber}(2011{\natexlab{b}})}]{thompson2011climate}%
  \BibitemOpen
  \bibfield  {author} {\bibinfo {author} {\bibfnamefont {J.~M.~T.}\
  \bibnamefont {Thompson}}\ and\ \bibinfo {author} {\bibfnamefont
  {J.}~\bibnamefont {Sieber}},\ }\bibfield  {title} {\enquote {\bibinfo {title}
  {Climate tipping as a noisy bifurcation: a predictive technique},}\
  }\href@noop {} {\bibfield  {journal} {\bibinfo  {journal} {IMA J. Appl.
  Math.}\ }\textbf {\bibinfo {volume} {76}},\ \bibinfo {pages} {27--46}
  (\bibinfo {year} {2011}{\natexlab{b}})}\BibitemShut {NoStop}%
\bibitem [{\citenamefont {Tony}\ \emph {et~al.}(2017)\citenamefont {Tony},
  \citenamefont {Subarna}, \citenamefont {Syamkumar}, \citenamefont {Sudha},
  \citenamefont {Akshay}, \citenamefont {Gopalakrishnan}, \citenamefont
  {Surovyatkina},\ and\ \citenamefont {Sujith}}]{tony2017experimental}%
  \BibitemOpen
  \bibfield  {author} {\bibinfo {author} {\bibfnamefont {J.}~\bibnamefont
  {Tony}}, \bibinfo {author} {\bibfnamefont {S.}~\bibnamefont {Subarna}},
  \bibinfo {author} {\bibfnamefont {K.~S.}\ \bibnamefont {Syamkumar}}, \bibinfo
  {author} {\bibfnamefont {G.}~\bibnamefont {Sudha}}, \bibinfo {author}
  {\bibfnamefont {S.}~\bibnamefont {Akshay}}, \bibinfo {author} {\bibfnamefont
  {E.~A.}\ \bibnamefont {Gopalakrishnan}}, \bibinfo {author} {\bibfnamefont
  {E.}~\bibnamefont {Surovyatkina}}, \ and\ \bibinfo {author} {\bibfnamefont
  {R.~I.}\ \bibnamefont {Sujith}},\ }\bibfield  {title} {\enquote {\bibinfo
  {title} {Experimental investigation on preconditioned rate induced tipping in
  a thermoacoustic system},}\ }\href@noop {} {\bibfield  {journal} {\bibinfo
  {journal} {Sci. Rep.}\ }\textbf {\bibinfo {volume} {7}},\ \bibinfo {pages}
  {1--7} (\bibinfo {year} {2017})}\BibitemShut {NoStop}%
\bibitem [{\citenamefont {Unni}\ \emph {et~al.}(2019)\citenamefont {Unni},
  \citenamefont {Gopalakrishnan}, \citenamefont {Syamkumar}, \citenamefont
  {Sujith}, \citenamefont {Surovyatkina},\ and\ \citenamefont
  {Kurths}}]{unni2019interplay}%
  \BibitemOpen
  \bibfield  {author} {\bibinfo {author} {\bibfnamefont {V.~R.}\ \bibnamefont
  {Unni}}, \bibinfo {author} {\bibfnamefont {E.~A.}\ \bibnamefont
  {Gopalakrishnan}}, \bibinfo {author} {\bibfnamefont {K.~S.}\ \bibnamefont
  {Syamkumar}}, \bibinfo {author} {\bibfnamefont {R.~I.}\ \bibnamefont
  {Sujith}}, \bibinfo {author} {\bibfnamefont {E.}~\bibnamefont
  {Surovyatkina}}, \ and\ \bibinfo {author} {\bibfnamefont {J.}~\bibnamefont
  {Kurths}},\ }\bibfield  {title} {\enquote {\bibinfo {title} {Interplay
  between random fluctuations and rate dependent phenomena at slow passage to
  limit-cycle oscillations in a bistable thermoacoustic system},}\ }\href@noop
  {} {\bibfield  {journal} {\bibinfo  {journal} {Chaos}\ }\textbf {\bibinfo
  {volume} {29}},\ \bibinfo {pages} {031102} (\bibinfo {year}
  {2019})}\BibitemShut {NoStop}%
\bibitem [{\citenamefont {Zhang}\ \emph {et~al.}(2020)\citenamefont {Zhang},
  \citenamefont {Xu}, \citenamefont {Liu},\ and\ \citenamefont
  {Kurths}}]{zhang2020rate}%
  \BibitemOpen
  \bibfield  {author} {\bibinfo {author} {\bibfnamefont {X.}~\bibnamefont
  {Zhang}}, \bibinfo {author} {\bibfnamefont {Y.}~\bibnamefont {Xu}}, \bibinfo
  {author} {\bibfnamefont {Q.}~\bibnamefont {Liu}}, \ and\ \bibinfo {author}
  {\bibfnamefont {J.}~\bibnamefont {Kurths}},\ }\bibfield  {title} {\enquote
  {\bibinfo {title} {Rate-dependent tipping-delay phenomenon in a
  thermoacoustic system with colored noise},}\ }\href@noop {} {\bibfield
  {journal} {\bibinfo  {journal} {Sci. China Technol. Sci.}\ }\textbf {\bibinfo
  {volume} {63}},\ \bibinfo {pages} {2315--2327} (\bibinfo {year}
  {2020})}\BibitemShut {NoStop}%
\bibitem [{\citenamefont {Bury}\ \emph {et~al.}(2021)\citenamefont {Bury},
  \citenamefont {Sujith}, \citenamefont {Pavithran}, \citenamefont {Scheffer},
  \citenamefont {Lenton}, \citenamefont {Anand},\ and\ \citenamefont
  {Bauch}}]{bury2021deep}%
  \BibitemOpen
  \bibfield  {author} {\bibinfo {author} {\bibfnamefont {T.~M.}\ \bibnamefont
  {Bury}}, \bibinfo {author} {\bibfnamefont {R.~I.}\ \bibnamefont {Sujith}},
  \bibinfo {author} {\bibfnamefont {I.}~\bibnamefont {Pavithran}}, \bibinfo
  {author} {\bibfnamefont {M.}~\bibnamefont {Scheffer}}, \bibinfo {author}
  {\bibfnamefont {T.~M.}\ \bibnamefont {Lenton}}, \bibinfo {author}
  {\bibfnamefont {M.}~\bibnamefont {Anand}}, \ and\ \bibinfo {author}
  {\bibfnamefont {C.~T.}\ \bibnamefont {Bauch}},\ }\bibfield  {title} {\enquote
  {\bibinfo {title} {Deep learning for early warning signals of tipping
  points},}\ }\href@noop {} {\bibfield  {journal} {\bibinfo  {journal} {Proc.
  Natl. Acad. Sci.}\ }\textbf {\bibinfo {volume} {118}} (\bibinfo {year}
  {2021})}\BibitemShut {NoStop}%
\bibitem [{\citenamefont {Weng}\ \emph {et~al.}(2016)\citenamefont {Weng},
  \citenamefont {Li}, \citenamefont {Zhong},\ and\ \citenamefont
  {Zhu}}]{weng2016investigation}%
  \BibitemOpen
  \bibfield  {author} {\bibinfo {author} {\bibfnamefont {F.}~\bibnamefont
  {Weng}}, \bibinfo {author} {\bibfnamefont {S.}~\bibnamefont {Li}}, \bibinfo
  {author} {\bibfnamefont {D.}~\bibnamefont {Zhong}}, \ and\ \bibinfo {author}
  {\bibfnamefont {M.}~\bibnamefont {Zhu}},\ }\bibfield  {title} {\enquote
  {\bibinfo {title} {Investigation of self-sustained beating oscillations in a
  rijke burner},}\ }\href@noop {} {\bibfield  {journal} {\bibinfo  {journal}
  {Combust. Flame}\ }\textbf {\bibinfo {volume} {166}},\ \bibinfo {pages}
  {181--191} (\bibinfo {year} {2016})}\BibitemShut {NoStop}%
\bibitem [{\citenamefont {Omelchenko}, \citenamefont {Rosenblum},\ and\
  \citenamefont {Pikovsky}(2010)}]{omelchenko2010synchronization}%
  \BibitemOpen
  \bibfield  {author} {\bibinfo {author} {\bibfnamefont {I.}~\bibnamefont
  {Omelchenko}}, \bibinfo {author} {\bibfnamefont {M.}~\bibnamefont
  {Rosenblum}}, \ and\ \bibinfo {author} {\bibfnamefont {A.}~\bibnamefont
  {Pikovsky}},\ }\bibfield  {title} {\enquote {\bibinfo {title}
  {Synchronization of slow-fast systems},}\ }\href@noop {} {\bibfield
  {journal} {\bibinfo  {journal} {Eur. Phys. J. Spec. Top.}\ }\textbf {\bibinfo
  {volume} {191}},\ \bibinfo {pages} {3--14} (\bibinfo {year}
  {2010})}\BibitemShut {NoStop}%
\bibitem [{\citenamefont {Bertram}\ and\ \citenamefont
  {Rubin}(2017)}]{bertram2017multi}%
  \BibitemOpen
  \bibfield  {author} {\bibinfo {author} {\bibfnamefont {R.}~\bibnamefont
  {Bertram}}\ and\ \bibinfo {author} {\bibfnamefont {J.~E.}\ \bibnamefont
  {Rubin}},\ }\bibfield  {title} {\enquote {\bibinfo {title} {Multi-timescale
  systems and fast-slow analysis},}\ }\href@noop {} {\bibfield  {journal}
  {\bibinfo  {journal} {Math. Biosci.}\ }\textbf {\bibinfo {volume} {287}},\
  \bibinfo {pages} {105--121} (\bibinfo {year} {2017})}\BibitemShut {NoStop}%
\bibitem [{\citenamefont {Kasthuri}\ \emph {et~al.}(2020)\citenamefont
  {Kasthuri}, \citenamefont {Pavithran}, \citenamefont {Krishnan},
  \citenamefont {Pawar}, \citenamefont {Sujith}, \citenamefont {Gejji},
  \citenamefont {Anderson}, \citenamefont {Marwan},\ and\ \citenamefont
  {Kurths}}]{kasthuri2020recurrence}%
  \BibitemOpen
  \bibfield  {author} {\bibinfo {author} {\bibfnamefont {P.}~\bibnamefont
  {Kasthuri}}, \bibinfo {author} {\bibfnamefont {I.}~\bibnamefont {Pavithran}},
  \bibinfo {author} {\bibfnamefont {A.}~\bibnamefont {Krishnan}}, \bibinfo
  {author} {\bibfnamefont {S.~A.}\ \bibnamefont {Pawar}}, \bibinfo {author}
  {\bibfnamefont {R.~I.}\ \bibnamefont {Sujith}}, \bibinfo {author}
  {\bibfnamefont {R.}~\bibnamefont {Gejji}}, \bibinfo {author} {\bibfnamefont
  {W.}~\bibnamefont {Anderson}}, \bibinfo {author} {\bibfnamefont
  {N.}~\bibnamefont {Marwan}}, \ and\ \bibinfo {author} {\bibfnamefont
  {J.}~\bibnamefont {Kurths}},\ }\bibfield  {title} {\enquote {\bibinfo {title}
  {Recurrence analysis of slow--fast systems},}\ }\href@noop {} {\bibfield
  {journal} {\bibinfo  {journal} {Chaos}\ }\textbf {\bibinfo {volume} {30}},\
  \bibinfo {pages} {063152} (\bibinfo {year} {2020})}\BibitemShut {NoStop}%
\bibitem [{Note1()}]{Note1}%
  \BibitemOpen
  \bibinfo {note} {Kabiraj and Sujith \cite {kabiraj2012nonlinearJFM} were the
  first to report the term intermittency in the context of thermoacoustics in a
  ducted laminar premixed flame Rijke tube burner. They observed the occurrence
  of intermittency prior to flame blowout in the system and not prior to the
  onset of limit cycle oscillations.}\BibitemShut {Stop}%
\bibitem [{\citenamefont {Weng}, \citenamefont {Zhu},\ and\ \citenamefont
  {Jing}(2014)}]{weng2014beat}%
  \BibitemOpen
  \bibfield  {author} {\bibinfo {author} {\bibfnamefont {F.}~\bibnamefont
  {Weng}}, \bibinfo {author} {\bibfnamefont {M.}~\bibnamefont {Zhu}}, \ and\
  \bibinfo {author} {\bibfnamefont {L.}~\bibnamefont {Jing}},\ }\bibfield
  {title} {\enquote {\bibinfo {title} {Beat: a nonlinear thermoacoustic
  instability in rijke burners},}\ }\href@noop {} {\bibfield  {journal}
  {\bibinfo  {journal} {Int. J. Spray Combust. Dyn.}\ }\textbf {\bibinfo
  {volume} {6}},\ \bibinfo {pages} {247--266} (\bibinfo {year}
  {2014})}\BibitemShut {NoStop}%
\bibitem [{\citenamefont {Lei}\ and\ \citenamefont
  {Turan}(2009)}]{lei2009nonlinear}%
  \BibitemOpen
  \bibfield  {author} {\bibinfo {author} {\bibfnamefont {S.}~\bibnamefont
  {Lei}}\ and\ \bibinfo {author} {\bibfnamefont {A.}~\bibnamefont {Turan}},\
  }\bibfield  {title} {\enquote {\bibinfo {title} {Nonlinear/chaotic behaviour
  in thermo-acoustic instability},}\ }\href@noop {} {\bibfield  {journal}
  {\bibinfo  {journal} {Combust. Theor. Model.}\ }\textbf {\bibinfo {volume}
  {13}},\ \bibinfo {pages} {541--557} (\bibinfo {year} {2009})}\BibitemShut
  {NoStop}%
\bibitem [{\citenamefont {Premraj}\ \emph {et~al.}(2020)\citenamefont
  {Premraj}, \citenamefont {Pawar}, \citenamefont {Kabiraj},\ and\
  \citenamefont {Sujith}}]{premraj2020strange}%
  \BibitemOpen
  \bibfield  {author} {\bibinfo {author} {\bibfnamefont {D.}~\bibnamefont
  {Premraj}}, \bibinfo {author} {\bibfnamefont {S.~A.}\ \bibnamefont {Pawar}},
  \bibinfo {author} {\bibfnamefont {L.}~\bibnamefont {Kabiraj}}, \ and\
  \bibinfo {author} {\bibfnamefont {R.~I.}\ \bibnamefont {Sujith}},\ }\bibfield
   {title} {\enquote {\bibinfo {title} {Strange nonchaos in self-excited
  singing flames},}\ }\href@noop {} {\bibfield  {journal} {\bibinfo  {journal}
  {Europhys. Lett.}\ }\textbf {\bibinfo {volume} {128}},\ \bibinfo {pages}
  {54005} (\bibinfo {year} {2020})}\BibitemShut {NoStop}%
\bibitem [{\citenamefont {Kabiraj}\ and\ \citenamefont
  {Sujith}(2012)}]{kabiraj2012nonlinearJFM}%
  \BibitemOpen
  \bibfield  {author} {\bibinfo {author} {\bibfnamefont {L.}~\bibnamefont
  {Kabiraj}}\ and\ \bibinfo {author} {\bibfnamefont {R.~I.}\ \bibnamefont
  {Sujith}},\ }\bibfield  {title} {\enquote {\bibinfo {title} {Nonlinear
  self-excited thermoacoustic oscillations: intermittency and flame blowout},}\
  }\href@noop {} {\bibfield  {journal} {\bibinfo  {journal} {J. Fluid Mech.}\
  }\textbf {\bibinfo {volume} {713}},\ \bibinfo {pages} {376--397} (\bibinfo
  {year} {2012})}\BibitemShut {NoStop}%
\bibitem [{\citenamefont {Premraj}\ \emph
  {et~al.}(2021{\natexlab{a}})\citenamefont {Premraj}, \citenamefont {Suresh},
  \citenamefont {Pawar}, \citenamefont {Kabiraj}, \citenamefont {Prasad},\ and\
  \citenamefont {Sujith}}]{premraj2021dragon}%
  \BibitemOpen
  \bibfield  {author} {\bibinfo {author} {\bibfnamefont {D.}~\bibnamefont
  {Premraj}}, \bibinfo {author} {\bibfnamefont {K.}~\bibnamefont {Suresh}},
  \bibinfo {author} {\bibfnamefont {S.~A.}\ \bibnamefont {Pawar}}, \bibinfo
  {author} {\bibfnamefont {L.}~\bibnamefont {Kabiraj}}, \bibinfo {author}
  {\bibfnamefont {A.}~\bibnamefont {Prasad}}, \ and\ \bibinfo {author}
  {\bibfnamefont {R.~I.}\ \bibnamefont {Sujith}},\ }\bibfield  {title}
  {\enquote {\bibinfo {title} {Dragon-king extreme events as precursors for
  catastrophic transition},}\ }\href@noop {} {\bibfield  {journal} {\bibinfo
  {journal} {Europhys. Lett.}\ }\textbf {\bibinfo {volume} {134}},\ \bibinfo
  {pages} {34006} (\bibinfo {year} {2021}{\natexlab{a}})}\BibitemShut {NoStop}%
\bibitem [{\citenamefont {Schuster}\ and\ \citenamefont
  {Just}(2006)}]{schuster2006deterministic}%
  \BibitemOpen
  \bibfield  {author} {\bibinfo {author} {\bibfnamefont {H.~G.}\ \bibnamefont
  {Schuster}}\ and\ \bibinfo {author} {\bibfnamefont {W.}~\bibnamefont
  {Just}},\ }\href@noop {} {\emph {\bibinfo {title} {Deterministic Chaos: An
  Introduction}}}\ (\bibinfo  {publisher} {John Wiley \& Sons},\ \bibinfo
  {year} {2006})\BibitemShut {NoStop}%
\bibitem [{\citenamefont {Kabiraj}, \citenamefont {Sujith},\ and\ \citenamefont
  {Wahi}(2012{\natexlab{b}})}]{kabiraj2012investigating}%
  \BibitemOpen
  \bibfield  {author} {\bibinfo {author} {\bibfnamefont {L.}~\bibnamefont
  {Kabiraj}}, \bibinfo {author} {\bibfnamefont {R.~I.}\ \bibnamefont {Sujith}},
  \ and\ \bibinfo {author} {\bibfnamefont {P.}~\bibnamefont {Wahi}},\
  }\bibfield  {title} {\enquote {\bibinfo {title} {Investigating the dynamics
  of combustion-driven oscillations leading to lean blowout},}\ }\href@noop {}
  {\bibfield  {journal} {\bibinfo  {journal} {Fluid Dyn. Res.}\ }\textbf
  {\bibinfo {volume} {44}},\ \bibinfo {pages} {031408} (\bibinfo {year}
  {2012}{\natexlab{b}})}\BibitemShut {NoStop}%
\bibitem [{\citenamefont {Guan}\ \emph
  {et~al.}(2019{\natexlab{a}})\citenamefont {Guan}, \citenamefont {Li},
  \citenamefont {Ahn},\ and\ \citenamefont {Kim}}]{guan2019chaos}%
  \BibitemOpen
  \bibfield  {author} {\bibinfo {author} {\bibfnamefont {Y.}~\bibnamefont
  {Guan}}, \bibinfo {author} {\bibfnamefont {L.~K.~B.}\ \bibnamefont {Li}},
  \bibinfo {author} {\bibfnamefont {B.}~\bibnamefont {Ahn}}, \ and\ \bibinfo
  {author} {\bibfnamefont {K.~T.}\ \bibnamefont {Kim}},\ }\bibfield  {title}
  {\enquote {\bibinfo {title} {Chaos, synchronization, and desynchronization in
  a liquid-fueled diffusion-flame combustor with an intrinsic hydrodynamic
  mode},}\ }\href@noop {} {\bibfield  {journal} {\bibinfo  {journal} {Chaos}\
  }\textbf {\bibinfo {volume} {29}},\ \bibinfo {pages} {053124} (\bibinfo
  {year} {2019}{\natexlab{a}})}\BibitemShut {NoStop}%
\bibitem [{\citenamefont {Premraj}\ \emph
  {et~al.}(2021{\natexlab{b}})\citenamefont {Premraj}, \citenamefont {Manoj},
  \citenamefont {Pawar},\ and\ \citenamefont {Sujith}}]{premraj2021effect}%
  \BibitemOpen
  \bibfield  {author} {\bibinfo {author} {\bibfnamefont {D.}~\bibnamefont
  {Premraj}}, \bibinfo {author} {\bibfnamefont {K.}~\bibnamefont {Manoj}},
  \bibinfo {author} {\bibfnamefont {S.~A.}\ \bibnamefont {Pawar}}, \ and\
  \bibinfo {author} {\bibfnamefont {R.~I.}\ \bibnamefont {Sujith}},\ }\bibfield
   {title} {\enquote {\bibinfo {title} {Effect of amplitude and frequency of
  limit cycle oscillators on their coupled and forced dynamics},}\ }\href@noop
  {} {\bibfield  {journal} {\bibinfo  {journal} {Nonlinear Dyn.}\ }\textbf
  {\bibinfo {volume} {103}},\ \bibinfo {pages} {1439--1452} (\bibinfo {year}
  {2021}{\natexlab{b}})}\BibitemShut {NoStop}%
\bibitem [{\citenamefont {Kabiraj}(2012)}]{kabiraj2012intermittencyThesis}%
  \BibitemOpen
  \bibfield  {author} {\bibinfo {author} {\bibfnamefont {L.}~\bibnamefont
  {Kabiraj}},\ }\emph {\bibinfo {title} {Intermittency and route to chaos in
  thermoacoustic oscillations}},\ \href@noop {} {Ph.D. thesis},\ \bibinfo
  {school} {Indian Institute of Technology Madras, India} (\bibinfo {year}
  {2012})\BibitemShut {NoStop}%
\bibitem [{\citenamefont {Pikovsky}\ and\ \citenamefont
  {Feudel}(1995)}]{pikovsky1995characterizing}%
  \BibitemOpen
  \bibfield  {author} {\bibinfo {author} {\bibfnamefont {A.~S.}\ \bibnamefont
  {Pikovsky}}\ and\ \bibinfo {author} {\bibfnamefont {U.}~\bibnamefont
  {Feudel}},\ }\bibfield  {title} {\enquote {\bibinfo {title} {Characterizing
  strange nonchaotic attractors},}\ }\href@noop {} {\bibfield  {journal}
  {\bibinfo  {journal} {Chaos}\ }\textbf {\bibinfo {volume} {5}},\ \bibinfo
  {pages} {253--260} (\bibinfo {year} {1995})}\BibitemShut {NoStop}%
\bibitem [{\citenamefont {Heagy}\ and\ \citenamefont
  {Hammel}(1994)}]{heagy1994birth}%
  \BibitemOpen
  \bibfield  {author} {\bibinfo {author} {\bibfnamefont {J.}~\bibnamefont
  {Heagy}}\ and\ \bibinfo {author} {\bibfnamefont {S.}~\bibnamefont {Hammel}},\
  }\bibfield  {title} {\enquote {\bibinfo {title} {The birth of strange
  nonchaotic attractors},}\ }\href@noop {} {\bibfield  {journal} {\bibinfo
  {journal} {Phys. D: Nonlinear Phenom.}\ }\textbf {\bibinfo {volume} {70}},\
  \bibinfo {pages} {140--153} (\bibinfo {year} {1994})}\BibitemShut {NoStop}%
\bibitem [{\citenamefont {Ditto}\ \emph {et~al.}(1990)\citenamefont {Ditto},
  \citenamefont {Spano}, \citenamefont {Savage}, \citenamefont {Rauseo},
  \citenamefont {Heagy},\ and\ \citenamefont {Ott}}]{ditto1990experimental}%
  \BibitemOpen
  \bibfield  {author} {\bibinfo {author} {\bibfnamefont {W.}~\bibnamefont
  {Ditto}}, \bibinfo {author} {\bibfnamefont {M.}~\bibnamefont {Spano}},
  \bibinfo {author} {\bibfnamefont {H.}~\bibnamefont {Savage}}, \bibinfo
  {author} {\bibfnamefont {S.}~\bibnamefont {Rauseo}}, \bibinfo {author}
  {\bibfnamefont {J.}~\bibnamefont {Heagy}}, \ and\ \bibinfo {author}
  {\bibfnamefont {E.}~\bibnamefont {Ott}},\ }\bibfield  {title} {\enquote
  {\bibinfo {title} {Experimental observation of a strange nonchaotic
  attractor},}\ }\href@noop {} {\bibfield  {journal} {\bibinfo  {journal}
  {Phys. Rev. Lett.}\ }\textbf {\bibinfo {volume} {65}},\ \bibinfo {pages}
  {533} (\bibinfo {year} {1990})}\BibitemShut {NoStop}%
\bibitem [{\citenamefont {Lindner}\ \emph {et~al.}(2015)\citenamefont
  {Lindner}, \citenamefont {Kohar}, \citenamefont {Kia}, \citenamefont
  {Hippke}, \citenamefont {Learned},\ and\ \citenamefont
  {Ditto}}]{lindner2015strange}%
  \BibitemOpen
  \bibfield  {author} {\bibinfo {author} {\bibfnamefont {J.~F.}\ \bibnamefont
  {Lindner}}, \bibinfo {author} {\bibfnamefont {V.}~\bibnamefont {Kohar}},
  \bibinfo {author} {\bibfnamefont {B.}~\bibnamefont {Kia}}, \bibinfo {author}
  {\bibfnamefont {M.}~\bibnamefont {Hippke}}, \bibinfo {author} {\bibfnamefont
  {J.~G.}\ \bibnamefont {Learned}}, \ and\ \bibinfo {author} {\bibfnamefont
  {W.~L.}\ \bibnamefont {Ditto}},\ }\bibfield  {title} {\enquote {\bibinfo
  {title} {Strange nonchaotic stars},}\ }\href@noop {} {\bibfield  {journal}
  {\bibinfo  {journal} {Phys. Rev. Lett.}\ }\textbf {\bibinfo {volume} {114}},\
  \bibinfo {pages} {054101} (\bibinfo {year} {2015})}\BibitemShut {NoStop}%
\bibitem [{\citenamefont {Guan}, \citenamefont {Murugesan},\ and\ \citenamefont
  {Li}(2018)}]{guan2018strange}%
  \BibitemOpen
  \bibfield  {author} {\bibinfo {author} {\bibfnamefont {Y.}~\bibnamefont
  {Guan}}, \bibinfo {author} {\bibfnamefont {M.}~\bibnamefont {Murugesan}}, \
  and\ \bibinfo {author} {\bibfnamefont {L.~K.~B.}\ \bibnamefont {Li}},\
  }\bibfield  {title} {\enquote {\bibinfo {title} {Strange nonchaotic and
  chaotic attractors in a self-excited thermoacoustic oscillator subjected to
  external periodic forcing},}\ }\href@noop {} {\bibfield  {journal} {\bibinfo
  {journal} {Chaos}\ }\textbf {\bibinfo {volume} {28}},\ \bibinfo {pages}
  {093109} (\bibinfo {year} {2018})}\BibitemShut {NoStop}%
\bibitem [{\citenamefont {Weng}\ \emph {et~al.}(2020)\citenamefont {Weng},
  \citenamefont {Unni}, \citenamefont {Sujith},\ and\ \citenamefont
  {Saha}}]{weng2020synchronization}%
  \BibitemOpen
  \bibfield  {author} {\bibinfo {author} {\bibfnamefont {Y.}~\bibnamefont
  {Weng}}, \bibinfo {author} {\bibfnamefont {V.~R.}\ \bibnamefont {Unni}},
  \bibinfo {author} {\bibfnamefont {R.~I.}\ \bibnamefont {Sujith}}, \ and\
  \bibinfo {author} {\bibfnamefont {A.}~\bibnamefont {Saha}},\ }\bibfield
  {title} {\enquote {\bibinfo {title} {Synchronization framework for modeling
  transition to thermoacoustic instability in laminar combustors},}\
  }\href@noop {} {\bibfield  {journal} {\bibinfo  {journal} {Nonlinear Dyn.}\
  }\textbf {\bibinfo {volume} {100}},\ \bibinfo {pages} {3295–--3306}
  (\bibinfo {year} {2020})}\BibitemShut {NoStop}%
\bibitem [{\citenamefont {Sastry}\ and\ \citenamefont
  {Hijab}(1981)}]{sastry1981bifurcation}%
  \BibitemOpen
  \bibfield  {author} {\bibinfo {author} {\bibfnamefont {S.}~\bibnamefont
  {Sastry}}\ and\ \bibinfo {author} {\bibfnamefont {O.}~\bibnamefont {Hijab}},\
  }\bibfield  {title} {\enquote {\bibinfo {title} {Bifurcation in the presence
  of small noise},}\ }\href@noop {} {\bibfield  {journal} {\bibinfo  {journal}
  {Syst. Control. Lett.}\ }\textbf {\bibinfo {volume} {1}},\ \bibinfo {pages}
  {159--167} (\bibinfo {year} {1981})}\BibitemShut {NoStop}%
\bibitem [{\citenamefont {Sleeman}(1988)}]{sleeman1988period}%
  \BibitemOpen
  \bibfield  {author} {\bibinfo {author} {\bibfnamefont {B.}~\bibnamefont
  {Sleeman}},\ }\bibfield  {title} {\enquote {\bibinfo {title} {Period-doubling
  bifurcations leading to chaos in discrete models of biology},}\ }\href@noop
  {} {\bibfield  {journal} {\bibinfo  {journal} {Math. Med. Biol.}\ }\textbf
  {\bibinfo {volume} {5}},\ \bibinfo {pages} {21--31} (\bibinfo {year}
  {1988})}\BibitemShut {NoStop}%
\bibitem [{\citenamefont {Geest}\ \emph {et~al.}(1992)\citenamefont {Geest},
  \citenamefont {Steinmetz}, \citenamefont {Larter},\ and\ \citenamefont
  {Olsen}}]{geest1992period}%
  \BibitemOpen
  \bibfield  {author} {\bibinfo {author} {\bibfnamefont {T.}~\bibnamefont
  {Geest}}, \bibinfo {author} {\bibfnamefont {C.~G.}\ \bibnamefont
  {Steinmetz}}, \bibinfo {author} {\bibfnamefont {R.}~\bibnamefont {Larter}}, \
  and\ \bibinfo {author} {\bibfnamefont {L.~F.}\ \bibnamefont {Olsen}},\
  }\bibfield  {title} {\enquote {\bibinfo {title} {Period-doubling bifurcations
  and chaos in an enzyme reaction},}\ }\href@noop {} {\bibfield  {journal}
  {\bibinfo  {journal} {J. Phys. Chem. A}\ }\textbf {\bibinfo {volume} {96}},\
  \bibinfo {pages} {5678--5680} (\bibinfo {year} {1992})}\BibitemShut {NoStop}%
\bibitem [{\citenamefont {Cheung}\ and\ \citenamefont
  {Wong}(1987)}]{cheung1987chaotic}%
  \BibitemOpen
  \bibfield  {author} {\bibinfo {author} {\bibfnamefont {P.}~\bibnamefont
  {Cheung}}\ and\ \bibinfo {author} {\bibfnamefont {A.}~\bibnamefont {Wong}},\
  }\bibfield  {title} {\enquote {\bibinfo {title} {Chaotic behavior and period
  doubling in plasmas},}\ }\href@noop {} {\bibfield  {journal} {\bibinfo
  {journal} {Phys. Rev. Lett.}\ }\textbf {\bibinfo {volume} {59}},\ \bibinfo
  {pages} {551} (\bibinfo {year} {1987})}\BibitemShut {NoStop}%
\bibitem [{\citenamefont {Ye}, \citenamefont {Li},\ and\ \citenamefont
  {McInerney}(1993)}]{ye1993period}%
  \BibitemOpen
  \bibfield  {author} {\bibinfo {author} {\bibfnamefont {J.}~\bibnamefont
  {Ye}}, \bibinfo {author} {\bibfnamefont {H.}~\bibnamefont {Li}}, \ and\
  \bibinfo {author} {\bibfnamefont {J.~G.}\ \bibnamefont {McInerney}},\
  }\bibfield  {title} {\enquote {\bibinfo {title} {Period-doubling route to
  chaos in a semiconductor laser with weak optical feedback},}\ }\href@noop {}
  {\bibfield  {journal} {\bibinfo  {journal} {Phys. Rev. A}\ }\textbf {\bibinfo
  {volume} {47}},\ \bibinfo {pages} {2249} (\bibinfo {year}
  {1993})}\BibitemShut {NoStop}%
\bibitem [{\citenamefont {Stone}(1993)}]{stone1993period}%
  \BibitemOpen
  \bibfield  {author} {\bibinfo {author} {\bibfnamefont {L.}~\bibnamefont
  {Stone}},\ }\bibfield  {title} {\enquote {\bibinfo {title} {Period-doubling
  reversals and chaos in simple ecological models},}\ }\href@noop {} {\bibfield
   {journal} {\bibinfo  {journal} {Nature}\ }\textbf {\bibinfo {volume}
  {365}},\ \bibinfo {pages} {617--620} (\bibinfo {year} {1993})}\BibitemShut
  {NoStop}%
\bibitem [{\citenamefont {Simpson}\ \emph {et~al.}(1994)\citenamefont
  {Simpson}, \citenamefont {Liu}, \citenamefont {Gavrielides}, \citenamefont
  {Kovanis},\ and\ \citenamefont {Alsing}}]{simpson1994period}%
  \BibitemOpen
  \bibfield  {author} {\bibinfo {author} {\bibfnamefont {T.}~\bibnamefont
  {Simpson}}, \bibinfo {author} {\bibfnamefont {J.}~\bibnamefont {Liu}},
  \bibinfo {author} {\bibfnamefont {A.}~\bibnamefont {Gavrielides}}, \bibinfo
  {author} {\bibfnamefont {V.}~\bibnamefont {Kovanis}}, \ and\ \bibinfo
  {author} {\bibfnamefont {P.}~\bibnamefont {Alsing}},\ }\bibfield  {title}
  {\enquote {\bibinfo {title} {Period-doubling route to chaos in a
  semiconductor laser subject to optical injection},}\ }\href@noop {}
  {\bibfield  {journal} {\bibinfo  {journal} {Appl. Phys. Lett.}\ }\textbf
  {\bibinfo {volume} {64}},\ \bibinfo {pages} {3539--3541} (\bibinfo {year}
  {1994})}\BibitemShut {NoStop}%
\bibitem [{\citenamefont {Feigenbaum}(1979)}]{feigenbaum1979universal}%
  \BibitemOpen
  \bibfield  {author} {\bibinfo {author} {\bibfnamefont {M.~J.}\ \bibnamefont
  {Feigenbaum}},\ }\bibfield  {title} {\enquote {\bibinfo {title} {The
  universal metric properties of nonlinear transformations},}\ }\href@noop {}
  {\bibfield  {journal} {\bibinfo  {journal} {J. Stat. Phys.}\ }\textbf
  {\bibinfo {volume} {21}},\ \bibinfo {pages} {669--706} (\bibinfo {year}
  {1979})}\BibitemShut {NoStop}%
\bibitem [{\citenamefont {Mondal}, \citenamefont {Pawar},\ and\ \citenamefont
  {Sujith}(2017)}]{mondal2017synchronous}%
  \BibitemOpen
  \bibfield  {author} {\bibinfo {author} {\bibfnamefont {S.}~\bibnamefont
  {Mondal}}, \bibinfo {author} {\bibfnamefont {S.~A.}\ \bibnamefont {Pawar}}, \
  and\ \bibinfo {author} {\bibfnamefont {R.~I.}\ \bibnamefont {Sujith}},\
  }\bibfield  {title} {\enquote {\bibinfo {title} {Synchronous behaviour of two
  interacting oscillatory systems undergoing quasiperiodic route to chaos},}\
  }\href@noop {} {\bibfield  {journal} {\bibinfo  {journal} {Chaos}\ }\textbf
  {\bibinfo {volume} {27}},\ \bibinfo {pages} {103119} (\bibinfo {year}
  {2017})}\BibitemShut {NoStop}%
\bibitem [{\citenamefont {Ruelle}\ and\ \citenamefont
  {Takens}(1971)}]{ruelle1971nature}%
  \BibitemOpen
  \bibfield  {author} {\bibinfo {author} {\bibfnamefont {D.}~\bibnamefont
  {Ruelle}}\ and\ \bibinfo {author} {\bibfnamefont {F.}~\bibnamefont
  {Takens}},\ }\bibfield  {title} {\enquote {\bibinfo {title} {On the nature of
  turbulence},}\ }\href@noop {} {\bibfield  {journal} {\bibinfo  {journal} {Les
  Rencontres Physiciens-Math{\'e}maticiens de Strasbourg-RCP25}\ }\textbf
  {\bibinfo {volume} {12}},\ \bibinfo {pages} {1--44} (\bibinfo {year}
  {1971})}\BibitemShut {NoStop}%
\bibitem [{\citenamefont {Newhouse}, \citenamefont {Ruelle},\ and\
  \citenamefont {Takens}(1978)}]{newhouse1978occurrence}%
  \BibitemOpen
  \bibfield  {author} {\bibinfo {author} {\bibfnamefont {S.}~\bibnamefont
  {Newhouse}}, \bibinfo {author} {\bibfnamefont {D.}~\bibnamefont {Ruelle}}, \
  and\ \bibinfo {author} {\bibfnamefont {F.}~\bibnamefont {Takens}},\
  }\bibfield  {title} {\enquote {\bibinfo {title} {{Occurrence of strange
  AxiomA attractors near quasi periodic flows on $T^m$, $m\geqq3$}},}\
  }\href@noop {} {\bibfield  {journal} {\bibinfo  {journal} {Commun. Math.
  Phys.}\ }\textbf {\bibinfo {volume} {64}},\ \bibinfo {pages} {35--40}
  (\bibinfo {year} {1978})}\BibitemShut {NoStop}%
\bibitem [{\citenamefont {Pomeau}\ and\ \citenamefont
  {Manneville}(1980)}]{pomeau1980intermittent}%
  \BibitemOpen
  \bibfield  {author} {\bibinfo {author} {\bibfnamefont {Y.}~\bibnamefont
  {Pomeau}}\ and\ \bibinfo {author} {\bibfnamefont {P.}~\bibnamefont
  {Manneville}},\ }\bibfield  {title} {\enquote {\bibinfo {title} {Intermittent
  transition to turbulence in dissipative dynamical systems},}\ }\href@noop {}
  {\bibfield  {journal} {\bibinfo  {journal} {Commun. Math. Phys.}\ }\textbf
  {\bibinfo {volume} {74}},\ \bibinfo {pages} {189--197} (\bibinfo {year}
  {1980})}\BibitemShut {NoStop}%
\bibitem [{\citenamefont {Horsthemke}\ and\ \citenamefont
  {Lefever}(1984)}]{horsthemke1984noise}%
  \BibitemOpen
  \bibfield  {author} {\bibinfo {author} {\bibfnamefont {W.}~\bibnamefont
  {Horsthemke}}\ and\ \bibinfo {author} {\bibfnamefont {R.}~\bibnamefont
  {Lefever}},\ }\bibfield  {title} {\enquote {\bibinfo {title} {Noise-induced
  transitions in physics, chemistry, and biology},}\ }in\ \href@noop {} {\emph
  {\bibinfo {booktitle} {Noise-Induced Transitions: Theory and Applications in
  Physics, Chemistry, and Biology}}}\ (\bibinfo  {publisher} {Springer},\
  \bibinfo {year} {1984})\ pp.\ \bibinfo {pages} {164--200}\BibitemShut
  {NoStop}%
\bibitem [{\citenamefont {Moss}\ and\ \citenamefont
  {McClintock}(1989)}]{moss1989noise}%
  \BibitemOpen
  \bibfield  {author} {\bibinfo {author} {\bibfnamefont {F.}~\bibnamefont
  {Moss}}\ and\ \bibinfo {author} {\bibfnamefont {P.~V.~E.}\ \bibnamefont
  {McClintock}},\ }\href@noop {} {\emph {\bibinfo {title} {Noise in Nonlinear
  Dynamical Systems}}},\ Vol.~\bibinfo {volume} {2}\ (\bibinfo  {publisher}
  {Cambridge University Press},\ \bibinfo {year} {1989})\BibitemShut {NoStop}%
\bibitem [{\citenamefont {Van~den Broeck}, \citenamefont {Parrondo},\ and\
  \citenamefont {Toral}(1994)}]{van1994noise}%
  \BibitemOpen
  \bibfield  {author} {\bibinfo {author} {\bibfnamefont {C.}~\bibnamefont
  {Van~den Broeck}}, \bibinfo {author} {\bibfnamefont {J.~M.~R.}\ \bibnamefont
  {Parrondo}}, \ and\ \bibinfo {author} {\bibfnamefont {R.}~\bibnamefont
  {Toral}},\ }\bibfield  {title} {\enquote {\bibinfo {title} {Noise-induced
  nonequilibrium phase transition},}\ }\href@noop {} {\bibfield  {journal}
  {\bibinfo  {journal} {Phys. Rev. Lett.}\ }\textbf {\bibinfo {volume} {73}},\
  \bibinfo {pages} {3395} (\bibinfo {year} {1994})}\BibitemShut {NoStop}%
\bibitem [{\citenamefont {Gammaitoni}\ \emph {et~al.}(1998)\citenamefont
  {Gammaitoni}, \citenamefont {H{\"a}nggi}, \citenamefont {Jung},\ and\
  \citenamefont {Marchesoni}}]{gammaitoni1998stochastic}%
  \BibitemOpen
  \bibfield  {author} {\bibinfo {author} {\bibfnamefont {L.}~\bibnamefont
  {Gammaitoni}}, \bibinfo {author} {\bibfnamefont {P.}~\bibnamefont
  {H{\"a}nggi}}, \bibinfo {author} {\bibfnamefont {P.}~\bibnamefont {Jung}}, \
  and\ \bibinfo {author} {\bibfnamefont {F.}~\bibnamefont {Marchesoni}},\
  }\bibfield  {title} {\enquote {\bibinfo {title} {Stochastic resonance},}\
  }\href@noop {} {\bibfield  {journal} {\bibinfo  {journal} {Rev. Mod. Phys.}\
  }\textbf {\bibinfo {volume} {70}},\ \bibinfo {pages} {223} (\bibinfo {year}
  {1998})}\BibitemShut {NoStop}%
\bibitem [{\citenamefont {Garc{\'\i}a-Ojalvo}\ and\ \citenamefont
  {Sancho}(2012)}]{garcia2012noise}%
  \BibitemOpen
  \bibfield  {author} {\bibinfo {author} {\bibfnamefont {J.}~\bibnamefont
  {Garc{\'\i}a-Ojalvo}}\ and\ \bibinfo {author} {\bibfnamefont
  {J.}~\bibnamefont {Sancho}},\ }\href@noop {} {\emph {\bibinfo {title} {Noise
  in Spatially Extended Systems}}}\ (\bibinfo  {publisher} {Springer Science \&
  Business Media},\ \bibinfo {year} {2012})\BibitemShut {NoStop}%
\bibitem [{\citenamefont {Arnold}(1995)}]{arnold1995random}%
  \BibitemOpen
  \bibfield  {author} {\bibinfo {author} {\bibfnamefont {L.}~\bibnamefont
  {Arnold}},\ }\bibfield  {title} {\enquote {\bibinfo {title} {Random dynamical
  systems},}\ }in\ \href@noop {} {\emph {\bibinfo {booktitle} {Dynamical
  Systems}}}\ (\bibinfo  {publisher} {Springer},\ \bibinfo {year} {1995})\ pp.\
  \bibinfo {pages} {1--43}\BibitemShut {NoStop}%
\bibitem [{\citenamefont {Kabiraj}, \citenamefont {Vishnoi},\ and\
  \citenamefont {Saurabh}(2020)}]{kabiraj2020review}%
  \BibitemOpen
  \bibfield  {author} {\bibinfo {author} {\bibfnamefont {L.}~\bibnamefont
  {Kabiraj}}, \bibinfo {author} {\bibfnamefont {N.}~\bibnamefont {Vishnoi}}, \
  and\ \bibinfo {author} {\bibfnamefont {A.}~\bibnamefont {Saurabh}},\
  }\bibfield  {title} {\enquote {\bibinfo {title} {A review on noise-induced
  dynamics of thermoacoustic systems},}\ }in\ \href@noop {} {\emph {\bibinfo
  {booktitle} {Dynamics and Control of Energy Systems}}}\ (\bibinfo
  {publisher} {Springer},\ \bibinfo {year} {2020})\ pp.\ \bibinfo {pages}
  {265--281}\BibitemShut {NoStop}%
\bibitem [{\citenamefont {Lindner}\ and\ \citenamefont
  {Schimansky-Geier}(2000)}]{lindner2000coherence}%
  \BibitemOpen
  \bibfield  {author} {\bibinfo {author} {\bibfnamefont {B.}~\bibnamefont
  {Lindner}}\ and\ \bibinfo {author} {\bibfnamefont {L.}~\bibnamefont
  {Schimansky-Geier}},\ }\bibfield  {title} {\enquote {\bibinfo {title}
  {Coherence and stochastic resonance in a two-state system},}\ }\href@noop {}
  {\bibfield  {journal} {\bibinfo  {journal} {Phys. Rev. E}\ }\textbf {\bibinfo
  {volume} {61}},\ \bibinfo {pages} {6103} (\bibinfo {year}
  {2000})}\BibitemShut {NoStop}%
\bibitem [{\citenamefont {Gang}\ \emph {et~al.}(1993)\citenamefont {Gang},
  \citenamefont {Ditzinger}, \citenamefont {Ning},\ and\ \citenamefont
  {Haken}}]{gang1993stochastic}%
  \BibitemOpen
  \bibfield  {author} {\bibinfo {author} {\bibfnamefont {H.}~\bibnamefont
  {Gang}}, \bibinfo {author} {\bibfnamefont {T.}~\bibnamefont {Ditzinger}},
  \bibinfo {author} {\bibfnamefont {C.-Z.}\ \bibnamefont {Ning}}, \ and\
  \bibinfo {author} {\bibfnamefont {H.}~\bibnamefont {Haken}},\ }\bibfield
  {title} {\enquote {\bibinfo {title} {Stochastic resonance without external
  periodic force},}\ }\href@noop {} {\bibfield  {journal} {\bibinfo  {journal}
  {Phys. Rev. Lett.}\ }\textbf {\bibinfo {volume} {71}},\ \bibinfo {pages}
  {807} (\bibinfo {year} {1993})}\BibitemShut {NoStop}%
\bibitem [{\citenamefont {Waugh}\ and\ \citenamefont
  {Juniper}(2011)}]{waugh2011triggeringa}%
  \BibitemOpen
  \bibfield  {author} {\bibinfo {author} {\bibfnamefont {I.~C.}\ \bibnamefont
  {Waugh}}\ and\ \bibinfo {author} {\bibfnamefont {M.~P.}\ \bibnamefont
  {Juniper}},\ }\bibfield  {title} {\enquote {\bibinfo {title} {Triggering in a
  thermoacoustic system with stochastic noise},}\ }\href@noop {} {\bibfield
  {journal} {\bibinfo  {journal} {J. Spray Combust. Dyn.}\ }\textbf {\bibinfo
  {volume} {3}},\ \bibinfo {pages} {225--241} (\bibinfo {year}
  {2011})}\BibitemShut {NoStop}%
\bibitem [{\citenamefont {Waugh}, \citenamefont {Geu{\ss}},\ and\ \citenamefont
  {Juniper}(2011)}]{waugh2011triggeringb}%
  \BibitemOpen
  \bibfield  {author} {\bibinfo {author} {\bibfnamefont {I.}~\bibnamefont
  {Waugh}}, \bibinfo {author} {\bibfnamefont {M.}~\bibnamefont {Geu{\ss}}}, \
  and\ \bibinfo {author} {\bibfnamefont {M.}~\bibnamefont {Juniper}},\
  }\bibfield  {title} {\enquote {\bibinfo {title} {Triggering, bypass
  transition and the effect of noise on a linearly stable thermoacoustic
  system},}\ }\href@noop {} {\bibfield  {journal} {\bibinfo  {journal} {Proc.
  Combust. Inst.}\ }\textbf {\bibinfo {volume} {33}},\ \bibinfo {pages}
  {2945--2952} (\bibinfo {year} {2011})}\BibitemShut {NoStop}%
\bibitem [{\citenamefont {Kabiraj}\ \emph {et~al.}(2015)\citenamefont
  {Kabiraj}, \citenamefont {Steinert}, \citenamefont {Saurabh},\ and\
  \citenamefont {Paschereit}}]{kabiraj2015coherence}%
  \BibitemOpen
  \bibfield  {author} {\bibinfo {author} {\bibfnamefont {L.}~\bibnamefont
  {Kabiraj}}, \bibinfo {author} {\bibfnamefont {R.}~\bibnamefont {Steinert}},
  \bibinfo {author} {\bibfnamefont {A.}~\bibnamefont {Saurabh}}, \ and\
  \bibinfo {author} {\bibfnamefont {C.~O.}\ \bibnamefont {Paschereit}},\
  }\bibfield  {title} {\enquote {\bibinfo {title} {Coherence resonance in a
  thermoacoustic system},}\ }\href@noop {} {\bibfield  {journal} {\bibinfo
  {journal} {Phys. Rev. E}\ }\textbf {\bibinfo {volume} {92}},\ \bibinfo
  {pages} {042909} (\bibinfo {year} {2015})}\BibitemShut {NoStop}%
\bibitem [{\citenamefont {Saurabh}\ \emph {et~al.}(2017)\citenamefont
  {Saurabh}, \citenamefont {Kabiraj}, \citenamefont {Steinert},\ and\
  \citenamefont {Oliver~Paschereit}}]{saurabh2017noise}%
  \BibitemOpen
  \bibfield  {author} {\bibinfo {author} {\bibfnamefont {A.}~\bibnamefont
  {Saurabh}}, \bibinfo {author} {\bibfnamefont {L.}~\bibnamefont {Kabiraj}},
  \bibinfo {author} {\bibfnamefont {R.}~\bibnamefont {Steinert}}, \ and\
  \bibinfo {author} {\bibfnamefont {C.}~\bibnamefont {Oliver~Paschereit}},\
  }\bibfield  {title} {\enquote {\bibinfo {title} {Noise-induced dynamics in
  the subthreshold region in thermoacoustic systems},}\ }\href@noop {}
  {\bibfield  {journal} {\bibinfo  {journal} {J. Eng. Gas Turbines Power}\
  }\textbf {\bibinfo {volume} {139}} (\bibinfo {year} {2017})}\BibitemShut
  {NoStop}%
\bibitem [{\citenamefont {Gupta}\ \emph {et~al.}(2017)\citenamefont {Gupta},
  \citenamefont {Saurabh}, \citenamefont {Paschereit},\ and\ \citenamefont
  {Kabiraj}}]{gupta2017numerical}%
  \BibitemOpen
  \bibfield  {author} {\bibinfo {author} {\bibfnamefont {V.}~\bibnamefont
  {Gupta}}, \bibinfo {author} {\bibfnamefont {A.}~\bibnamefont {Saurabh}},
  \bibinfo {author} {\bibfnamefont {C.~O.}\ \bibnamefont {Paschereit}}, \ and\
  \bibinfo {author} {\bibfnamefont {L.}~\bibnamefont {Kabiraj}},\ }\bibfield
  {title} {\enquote {\bibinfo {title} {Numerical results on noise-induced
  dynamics in the subthreshold regime for thermoacoustic systems},}\
  }\href@noop {} {\bibfield  {journal} {\bibinfo  {journal} {J. Sound Vib.}\
  }\textbf {\bibinfo {volume} {390}},\ \bibinfo {pages} {55--66} (\bibinfo
  {year} {2017})}\BibitemShut {NoStop}%
\bibitem [{\citenamefont {Lee}\ \emph {et~al.}(2020)\citenamefont {Lee},
  \citenamefont {Guan}, \citenamefont {Gupta},\ and\ \citenamefont
  {Li}}]{lee2020input}%
  \BibitemOpen
  \bibfield  {author} {\bibinfo {author} {\bibfnamefont {M.}~\bibnamefont
  {Lee}}, \bibinfo {author} {\bibfnamefont {Y.}~\bibnamefont {Guan}}, \bibinfo
  {author} {\bibfnamefont {V.}~\bibnamefont {Gupta}}, \ and\ \bibinfo {author}
  {\bibfnamefont {L.~K.~B.}\ \bibnamefont {Li}},\ }\bibfield  {title} {\enquote
  {\bibinfo {title} {{Input-output system identification of a thermoacoustic
  oscillator near a Hopf bifurcation using only fixed-point data}},}\
  }\href@noop {} {\bibfield  {journal} {\bibinfo  {journal} {Phys. Rev. E}\
  }\textbf {\bibinfo {volume} {101}},\ \bibinfo {pages} {013102} (\bibinfo
  {year} {2020})}\BibitemShut {NoStop}%
\bibitem [{\citenamefont {Gopalakrishnan}\ \emph
  {et~al.}(2016{\natexlab{a}})\citenamefont {Gopalakrishnan}, \citenamefont
  {Tony}, \citenamefont {Sreelekha},\ and\ \citenamefont
  {Sujith}}]{gopalakrishnan2016stochastic}%
  \BibitemOpen
  \bibfield  {author} {\bibinfo {author} {\bibfnamefont {E.~A.}\ \bibnamefont
  {Gopalakrishnan}}, \bibinfo {author} {\bibfnamefont {J.}~\bibnamefont
  {Tony}}, \bibinfo {author} {\bibfnamefont {E.}~\bibnamefont {Sreelekha}}, \
  and\ \bibinfo {author} {\bibfnamefont {R.~I.}\ \bibnamefont {Sujith}},\
  }\bibfield  {title} {\enquote {\bibinfo {title} {Stochastic bifurcations in a
  prototypical thermoacoustic system},}\ }\href@noop {} {\bibfield  {journal}
  {\bibinfo  {journal} {Phys. Rev. E}\ }\textbf {\bibinfo {volume} {94}},\
  \bibinfo {pages} {022203} (\bibinfo {year} {2016}{\natexlab{a}})}\BibitemShut
  {NoStop}%
\bibitem [{\citenamefont {Li}, \citenamefont {Zhao},\ and\ \citenamefont
  {Shi}(2019)}]{li2019coherence}%
  \BibitemOpen
  \bibfield  {author} {\bibinfo {author} {\bibfnamefont {X.}~\bibnamefont
  {Li}}, \bibinfo {author} {\bibfnamefont {D.}~\bibnamefont {Zhao}}, \ and\
  \bibinfo {author} {\bibfnamefont {B.}~\bibnamefont {Shi}},\ }\bibfield
  {title} {\enquote {\bibinfo {title} {Coherence resonance and stochastic
  bifurcation behaviors of simplified standing-wave thermoacoustic systems},}\
  }\href@noop {} {\bibfield  {journal} {\bibinfo  {journal} {J. Acoust. Soc.
  Am.}\ }\textbf {\bibinfo {volume} {145}},\ \bibinfo {pages} {692--702}
  (\bibinfo {year} {2019})}\BibitemShut {NoStop}%
\bibitem [{\citenamefont {Zakharova}\ \emph {et~al.}(2013)\citenamefont
  {Zakharova}, \citenamefont {Feoktistov}, \citenamefont {Vadivasova},\ and\
  \citenamefont {Sch{\"o}ll}}]{zakharova2013coherence}%
  \BibitemOpen
  \bibfield  {author} {\bibinfo {author} {\bibfnamefont {A.}~\bibnamefont
  {Zakharova}}, \bibinfo {author} {\bibfnamefont {A.}~\bibnamefont
  {Feoktistov}}, \bibinfo {author} {\bibfnamefont {T.}~\bibnamefont
  {Vadivasova}}, \ and\ \bibinfo {author} {\bibfnamefont {E.}~\bibnamefont
  {Sch{\"o}ll}},\ }\bibfield  {title} {\enquote {\bibinfo {title} {Coherence
  resonance and stochastic synchronization in a nonlinear circuit near a
  subcritical hopf bifurcation},}\ }\href@noop {} {\bibfield  {journal}
  {\bibinfo  {journal} {Eur. Phys. J. Spec. Top.}\ }\textbf {\bibinfo {volume}
  {222}},\ \bibinfo {pages} {2481--2495} (\bibinfo {year} {2013})}\BibitemShut
  {NoStop}%
\bibitem [{\citenamefont {Pikovsky}\ and\ \citenamefont
  {Kurths}(1997)}]{pikovsky1997coherence}%
  \BibitemOpen
  \bibfield  {author} {\bibinfo {author} {\bibfnamefont {A.~S.}\ \bibnamefont
  {Pikovsky}}\ and\ \bibinfo {author} {\bibfnamefont {J.}~\bibnamefont
  {Kurths}},\ }\bibfield  {title} {\enquote {\bibinfo {title} {Coherence
  resonance in a noise--driven excitable system},}\ }\href@noop {} {\bibfield
  {journal} {\bibinfo  {journal} {Phys. Rev. Lett.}\ }\textbf {\bibinfo
  {volume} {78}},\ \bibinfo {pages} {775--778} (\bibinfo {year}
  {1997})}\BibitemShut {NoStop}%
\bibitem [{\citenamefont {Ushakov}\ \emph {et~al.}(2005)\citenamefont
  {Ushakov}, \citenamefont {W{\"u}nsche}, \citenamefont {Henneberger},
  \citenamefont {Khovanov}, \citenamefont {Schimansky-Geier},\ and\
  \citenamefont {Zaks}}]{ushakov2005coherence}%
  \BibitemOpen
  \bibfield  {author} {\bibinfo {author} {\bibfnamefont {O.}~\bibnamefont
  {Ushakov}}, \bibinfo {author} {\bibfnamefont {H.-J.}\ \bibnamefont
  {W{\"u}nsche}}, \bibinfo {author} {\bibfnamefont {F.}~\bibnamefont
  {Henneberger}}, \bibinfo {author} {\bibfnamefont {I.}~\bibnamefont
  {Khovanov}}, \bibinfo {author} {\bibfnamefont {L.}~\bibnamefont
  {Schimansky-Geier}}, \ and\ \bibinfo {author} {\bibfnamefont
  {M.}~\bibnamefont {Zaks}},\ }\bibfield  {title} {\enquote {\bibinfo {title}
  {Coherence resonance near a hopf bifurcation},}\ }\href@noop {} {\bibfield
  {journal} {\bibinfo  {journal} {Phys. Rev. Lett.}\ }\textbf {\bibinfo
  {volume} {95}},\ \bibinfo {pages} {123903} (\bibinfo {year}
  {2005})}\BibitemShut {NoStop}%
\bibitem [{\citenamefont {Zakharova}\ \emph {et~al.}(2010)\citenamefont
  {Zakharova}, \citenamefont {Vadivasova}, \citenamefont {Anishchenko},
  \citenamefont {Koseska},\ and\ \citenamefont
  {Kurths}}]{zakharova2010stochastic}%
  \BibitemOpen
  \bibfield  {author} {\bibinfo {author} {\bibfnamefont {A.}~\bibnamefont
  {Zakharova}}, \bibinfo {author} {\bibfnamefont {T.}~\bibnamefont
  {Vadivasova}}, \bibinfo {author} {\bibfnamefont {V.}~\bibnamefont
  {Anishchenko}}, \bibinfo {author} {\bibfnamefont {A.}~\bibnamefont
  {Koseska}}, \ and\ \bibinfo {author} {\bibfnamefont {J.}~\bibnamefont
  {Kurths}},\ }\bibfield  {title} {\enquote {\bibinfo {title} {Stochastic
  bifurcations and coherencelike resonance in a self-sustained bistable noisy
  oscillator},}\ }\href@noop {} {\bibfield  {journal} {\bibinfo  {journal}
  {Phys. Rev. E}\ }\textbf {\bibinfo {volume} {81}},\ \bibinfo {pages} {011106}
  (\bibinfo {year} {2010})}\BibitemShut {NoStop}%
\bibitem [{\citenamefont {Wellens}, \citenamefont {Shatokhin},\ and\
  \citenamefont {Buchleitner}(2003)}]{wellens2003stochastic}%
  \BibitemOpen
  \bibfield  {author} {\bibinfo {author} {\bibfnamefont {T.}~\bibnamefont
  {Wellens}}, \bibinfo {author} {\bibfnamefont {V.}~\bibnamefont {Shatokhin}},
  \ and\ \bibinfo {author} {\bibfnamefont {A.}~\bibnamefont {Buchleitner}},\
  }\bibfield  {title} {\enquote {\bibinfo {title} {Stochastic resonance},}\
  }\href@noop {} {\bibfield  {journal} {\bibinfo  {journal} {Rep. Prog. Phys.}\
  }\textbf {\bibinfo {volume} {67}},\ \bibinfo {pages} {45} (\bibinfo {year}
  {2003})}\BibitemShut {NoStop}%
\bibitem [{\citenamefont {Moss}, \citenamefont {Ward},\ and\ \citenamefont
  {Sannita}(2004)}]{moss2004stochastic}%
  \BibitemOpen
  \bibfield  {author} {\bibinfo {author} {\bibfnamefont {F.}~\bibnamefont
  {Moss}}, \bibinfo {author} {\bibfnamefont {L.~M.}\ \bibnamefont {Ward}}, \
  and\ \bibinfo {author} {\bibfnamefont {W.~G.}\ \bibnamefont {Sannita}},\
  }\bibfield  {title} {\enquote {\bibinfo {title} {Stochastic resonance and
  sensory information processing: a tutorial and review of application},}\
  }\href@noop {} {\bibfield  {journal} {\bibinfo  {journal} {Clin.
  Neurophysiol.}\ }\textbf {\bibinfo {volume} {115}},\ \bibinfo {pages}
  {267--281} (\bibinfo {year} {2004})}\BibitemShut {NoStop}%
\bibitem [{\citenamefont {Benzi}\ \emph {et~al.}(1982)\citenamefont {Benzi},
  \citenamefont {Parisi}, \citenamefont {Sutera},\ and\ \citenamefont
  {Vulpiani}}]{benzi1982stochastic}%
  \BibitemOpen
  \bibfield  {author} {\bibinfo {author} {\bibfnamefont {R.}~\bibnamefont
  {Benzi}}, \bibinfo {author} {\bibfnamefont {G.}~\bibnamefont {Parisi}},
  \bibinfo {author} {\bibfnamefont {A.}~\bibnamefont {Sutera}}, \ and\ \bibinfo
  {author} {\bibfnamefont {A.}~\bibnamefont {Vulpiani}},\ }\bibfield  {title}
  {\enquote {\bibinfo {title} {Stochastic resonance in climatic change},}\
  }\href@noop {} {\bibfield  {journal} {\bibinfo  {journal} {Tellus}\ }\textbf
  {\bibinfo {volume} {34}},\ \bibinfo {pages} {10--16} (\bibinfo {year}
  {1982})}\BibitemShut {NoStop}%
\bibitem [{\citenamefont {H{\"a}nggi}(2002)}]{hanggi2002stochastic}%
  \BibitemOpen
  \bibfield  {author} {\bibinfo {author} {\bibfnamefont {P.}~\bibnamefont
  {H{\"a}nggi}},\ }\bibfield  {title} {\enquote {\bibinfo {title} {Stochastic
  resonance in biology},}\ }\href@noop {} {\bibfield  {journal} {\bibinfo
  {journal} {ChemPhysChem}\ }\textbf {\bibinfo {volume} {3}},\ \bibinfo {pages}
  {285--290} (\bibinfo {year} {2002})}\BibitemShut {NoStop}%
\bibitem [{\citenamefont {Juel}, \citenamefont {Darbyshire},\ and\
  \citenamefont {Mullin}(1997)}]{juel1997effect}%
  \BibitemOpen
  \bibfield  {author} {\bibinfo {author} {\bibfnamefont {A.}~\bibnamefont
  {Juel}}, \bibinfo {author} {\bibfnamefont {A.~G.}\ \bibnamefont
  {Darbyshire}}, \ and\ \bibinfo {author} {\bibfnamefont {T.}~\bibnamefont
  {Mullin}},\ }\bibfield  {title} {\enquote {\bibinfo {title} {The effect of
  noise on pitchfork and hopf bifurcations},}\ }\href@noop {} {\bibfield
  {journal} {\bibinfo  {journal} {Proc. R. Soc. A: Math. Phys. Eng. Sci.}\
  }\textbf {\bibinfo {volume} {453}},\ \bibinfo {pages} {2627--2647} (\bibinfo
  {year} {1997})}\BibitemShut {NoStop}%
\bibitem [{\citenamefont {Namachchivaya}(1990)}]{namachchivaya1990stochastic}%
  \BibitemOpen
  \bibfield  {author} {\bibinfo {author} {\bibfnamefont {N.~S.}\ \bibnamefont
  {Namachchivaya}},\ }\bibfield  {title} {\enquote {\bibinfo {title}
  {Stochastic bifurcation},}\ }\href@noop {} {\bibfield  {journal} {\bibinfo
  {journal} {Appl. Math. Comput.}\ }\textbf {\bibinfo {volume} {38}},\ \bibinfo
  {pages} {101--159} (\bibinfo {year} {1990})}\BibitemShut {NoStop}%
\bibitem [{\citenamefont {Crauel}\ and\ \citenamefont
  {Gundlach}(1999)}]{crauel1999stochastic}%
  \BibitemOpen
  \bibfield  {author} {\bibinfo {author} {\bibfnamefont {H.}~\bibnamefont
  {Crauel}}\ and\ \bibinfo {author} {\bibfnamefont {M.}~\bibnamefont
  {Gundlach}},\ }\href@noop {} {\emph {\bibinfo {title} {Stochastic
  Dynamics}}}\ (\bibinfo  {publisher} {Springer Science \& Business Media},\
  \bibinfo {year} {1999})\BibitemShut {NoStop}%
\bibitem [{\citenamefont {Song}\ \emph {et~al.}(2010)\citenamefont {Song},
  \citenamefont {Phenix}, \citenamefont {Abedi}, \citenamefont {Scott},
  \citenamefont {Ingalls}, \citenamefont {K{\ae}rn},\ and\ \citenamefont
  {Perkins}}]{song2010estimating}%
  \BibitemOpen
  \bibfield  {author} {\bibinfo {author} {\bibfnamefont {C.}~\bibnamefont
  {Song}}, \bibinfo {author} {\bibfnamefont {H.}~\bibnamefont {Phenix}},
  \bibinfo {author} {\bibfnamefont {V.}~\bibnamefont {Abedi}}, \bibinfo
  {author} {\bibfnamefont {M.}~\bibnamefont {Scott}}, \bibinfo {author}
  {\bibfnamefont {B.~P.}\ \bibnamefont {Ingalls}}, \bibinfo {author}
  {\bibfnamefont {M.}~\bibnamefont {K{\ae}rn}}, \ and\ \bibinfo {author}
  {\bibfnamefont {T.~J.}\ \bibnamefont {Perkins}},\ }\bibfield  {title}
  {\enquote {\bibinfo {title} {Estimating the stochastic bifurcation structure
  of cellular networks},}\ }\href@noop {} {\bibfield  {journal} {\bibinfo
  {journal} {PLoS Comput. Biol.}\ }\textbf {\bibinfo {volume} {6}},\ \bibinfo
  {pages} {e1000699} (\bibinfo {year} {2010})}\BibitemShut {NoStop}%
\bibitem [{\citenamefont {Billings}\ \emph {et~al.}(2004)\citenamefont
  {Billings}, \citenamefont {Schwartz}, \citenamefont {Morgan}, \citenamefont
  {Bollt}, \citenamefont {Meucci},\ and\ \citenamefont
  {Allaria}}]{billings2004stochastic}%
  \BibitemOpen
  \bibfield  {author} {\bibinfo {author} {\bibfnamefont {L.}~\bibnamefont
  {Billings}}, \bibinfo {author} {\bibfnamefont {I.~B.}\ \bibnamefont
  {Schwartz}}, \bibinfo {author} {\bibfnamefont {D.~S.}\ \bibnamefont
  {Morgan}}, \bibinfo {author} {\bibfnamefont {E.~M.}\ \bibnamefont {Bollt}},
  \bibinfo {author} {\bibfnamefont {R.}~\bibnamefont {Meucci}}, \ and\ \bibinfo
  {author} {\bibfnamefont {E.}~\bibnamefont {Allaria}},\ }\bibfield  {title}
  {\enquote {\bibinfo {title} {Stochastic bifurcation in a driven laser system:
  Experiment and theory},}\ }\href@noop {} {\bibfield  {journal} {\bibinfo
  {journal} {Phys. Rev. E}\ }\textbf {\bibinfo {volume} {70}},\ \bibinfo
  {pages} {026220} (\bibinfo {year} {2004})}\BibitemShut {NoStop}%
\bibitem [{\citenamefont {Jin}, \citenamefont {Sun},\ and\ \citenamefont
  {Xu}(2022)}]{jin2022stochastic}%
  \BibitemOpen
  \bibfield  {author} {\bibinfo {author} {\bibfnamefont {C.}~\bibnamefont
  {Jin}}, \bibinfo {author} {\bibfnamefont {Z.}~\bibnamefont {Sun}}, \ and\
  \bibinfo {author} {\bibfnamefont {W.}~\bibnamefont {Xu}},\ }\bibfield
  {title} {\enquote {\bibinfo {title} {Stochastic bifurcations and its
  regulation in a rijke tube model},}\ }\href@noop {} {\bibfield  {journal}
  {\bibinfo  {journal} {Chaos, Solitons \& Fractals}\ }\textbf {\bibinfo
  {volume} {154}},\ \bibinfo {pages} {111650} (\bibinfo {year}
  {2022})}\BibitemShut {NoStop}%
\bibitem [{\citenamefont {Clavin}, \citenamefont {Kim},\ and\ \citenamefont
  {Williams}(1994)}]{clavin1994turbulence}%
  \BibitemOpen
  \bibfield  {author} {\bibinfo {author} {\bibfnamefont {P.}~\bibnamefont
  {Clavin}}, \bibinfo {author} {\bibfnamefont {J.~S.}\ \bibnamefont {Kim}}, \
  and\ \bibinfo {author} {\bibfnamefont {F.~A.}\ \bibnamefont {Williams}},\
  }\bibfield  {title} {\enquote {\bibinfo {title} {Turbulence-induced noise
  effects on high-frequency combustion instabilities},}\ }\href@noop {}
  {\bibfield  {journal} {\bibinfo  {journal} {Combust. Sci. Tech.}\ }\textbf
  {\bibinfo {volume} {96}},\ \bibinfo {pages} {61--84} (\bibinfo {year}
  {1994})}\BibitemShut {NoStop}%
\bibitem [{\citenamefont {Noiray}\ and\ \citenamefont
  {Schuermans}(2013)}]{noiray2013deterministic}%
  \BibitemOpen
  \bibfield  {author} {\bibinfo {author} {\bibfnamefont {N.}~\bibnamefont
  {Noiray}}\ and\ \bibinfo {author} {\bibfnamefont {B.}~\bibnamefont
  {Schuermans}},\ }\bibfield  {title} {\enquote {\bibinfo {title}
  {{Deterministic quantities characterizing noise driven Hopf bifurcations in
  gas turbine combustors}},}\ }\href@noop {} {\bibfield  {journal} {\bibinfo
  {journal} {Int. J. Nonlin. Mech.}\ }\textbf {\bibinfo {volume} {50}},\
  \bibinfo {pages} {152--163} (\bibinfo {year} {2013})}\BibitemShut {NoStop}%
\bibitem [{\citenamefont {Gopalakrishnan}\ and\ \citenamefont
  {Sujith}(2015)}]{gopalakrishnan2015effect}%
  \BibitemOpen
  \bibfield  {author} {\bibinfo {author} {\bibfnamefont {E.~A.}\ \bibnamefont
  {Gopalakrishnan}}\ and\ \bibinfo {author} {\bibfnamefont {R.~I.}\
  \bibnamefont {Sujith}},\ }\bibfield  {title} {\enquote {\bibinfo {title}
  {Effect of external noise on the hysteresis characteristics of a
  thermoacoustic system},}\ }\href@noop {} {\bibfield  {journal} {\bibinfo
  {journal} {J. Fluid Mech.}\ }\textbf {\bibinfo {volume} {776}},\ \bibinfo
  {pages} {334--353} (\bibinfo {year} {2015})}\BibitemShut {NoStop}%
\bibitem [{\citenamefont {Li}, \citenamefont {Zhao},\ and\ \citenamefont
  {Yang}(2017)}]{li2017experimental}%
  \BibitemOpen
  \bibfield  {author} {\bibinfo {author} {\bibfnamefont {X.}~\bibnamefont
  {Li}}, \bibinfo {author} {\bibfnamefont {D.}~\bibnamefont {Zhao}}, \ and\
  \bibinfo {author} {\bibfnamefont {X.}~\bibnamefont {Yang}},\ }\bibfield
  {title} {\enquote {\bibinfo {title} {Experimental and theoretical bifurcation
  study of a nonlinear standing-wave thermoacoustic system},}\ }\href@noop {}
  {\bibfield  {journal} {\bibinfo  {journal} {Energy}\ }\textbf {\bibinfo
  {volume} {135}},\ \bibinfo {pages} {553--562} (\bibinfo {year}
  {2017})}\BibitemShut {NoStop}%
\bibitem [{\citenamefont {Li}, \citenamefont {Zhao},\ and\ \citenamefont
  {Li}(2018)}]{li2018effects}%
  \BibitemOpen
  \bibfield  {author} {\bibinfo {author} {\bibfnamefont {X.}~\bibnamefont
  {Li}}, \bibinfo {author} {\bibfnamefont {D.}~\bibnamefont {Zhao}}, \ and\
  \bibinfo {author} {\bibfnamefont {X.}~\bibnamefont {Li}},\ }\bibfield
  {title} {\enquote {\bibinfo {title} {Effects of background noises on
  nonlinear dynamics of a modelled thermoacoustic combustor},}\ }\href@noop {}
  {\bibfield  {journal} {\bibinfo  {journal} {J. Acoust. Soc. Am.}\ }\textbf
  {\bibinfo {volume} {143}},\ \bibinfo {pages} {60--70} (\bibinfo {year}
  {2018})}\BibitemShut {NoStop}%
\bibitem [{\citenamefont {Ananthkrishnan}, \citenamefont {Deo},\ and\
  \citenamefont {Culick}(2005)}]{ananthkrishnan2005reduced}%
  \BibitemOpen
  \bibfield  {author} {\bibinfo {author} {\bibfnamefont {N.}~\bibnamefont
  {Ananthkrishnan}}, \bibinfo {author} {\bibfnamefont {S.}~\bibnamefont {Deo}},
  \ and\ \bibinfo {author} {\bibfnamefont {F.~E.~C.}\ \bibnamefont {Culick}},\
  }\bibfield  {title} {\enquote {\bibinfo {title} {Reduced-order modeling and
  dynamics of nonlinear acoustic waves in a combustion chamber},}\ }\href@noop
  {} {\bibfield  {journal} {\bibinfo  {journal} {Combust. Sci. Tech.}\ }\textbf
  {\bibinfo {volume} {177}},\ \bibinfo {pages} {221--248} (\bibinfo {year}
  {2005})}\BibitemShut {NoStop}%
\bibitem [{\citenamefont {Chapman}(2002)}]{chapman2002subcritical}%
  \BibitemOpen
  \bibfield  {author} {\bibinfo {author} {\bibfnamefont {S.~J.}\ \bibnamefont
  {Chapman}},\ }\bibfield  {title} {\enquote {\bibinfo {title} {Subcritical
  transition in channel flows},}\ }\href@noop {} {\bibfield  {journal}
  {\bibinfo  {journal} {J. Fluid Mech.}\ }\textbf {\bibinfo {volume} {451}},\
  \bibinfo {pages} {35--97} (\bibinfo {year} {2002})}\BibitemShut {NoStop}%
\bibitem [{\citenamefont {Baggett}\ and\ \citenamefont
  {Trefethen}(1997)}]{baggett1997low}%
  \BibitemOpen
  \bibfield  {author} {\bibinfo {author} {\bibfnamefont {J.~S.}\ \bibnamefont
  {Baggett}}\ and\ \bibinfo {author} {\bibfnamefont {L.~N.}\ \bibnamefont
  {Trefethen}},\ }\bibfield  {title} {\enquote {\bibinfo {title}
  {Low-dimensional models of subcritical transition to turbulence},}\
  }\href@noop {} {\bibfield  {journal} {\bibinfo  {journal} {Phys. Fluids}\
  }\textbf {\bibinfo {volume} {9}},\ \bibinfo {pages} {1043--1053} (\bibinfo
  {year} {1997})}\BibitemShut {NoStop}%
\bibitem [{\citenamefont {Ryan}\ and\ \citenamefont
  {Johnson}(1959)}]{ryan1959transistion}%
  \BibitemOpen
  \bibfield  {author} {\bibinfo {author} {\bibfnamefont {N.}~\bibnamefont
  {Ryan}}\ and\ \bibinfo {author} {\bibfnamefont {M.}~\bibnamefont {Johnson}},\
  }\bibfield  {title} {\enquote {\bibinfo {title} {Transistion from laminar to
  turbulent flow in pipes},}\ }\href@noop {} {\bibfield  {journal} {\bibinfo
  {journal} {AIChE Journal}\ }\textbf {\bibinfo {volume} {5}},\ \bibinfo
  {pages} {433--435} (\bibinfo {year} {1959})}\BibitemShut {NoStop}%
\bibitem [{\citenamefont {Rott}(1990)}]{rott1990note}%
  \BibitemOpen
  \bibfield  {author} {\bibinfo {author} {\bibfnamefont {N.}~\bibnamefont
  {Rott}},\ }\bibfield  {title} {\enquote {\bibinfo {title} {Note on the
  history of the reynolds number},}\ }\href@noop {} {\bibfield  {journal}
  {\bibinfo  {journal} {Annu. Rev. Fluid Mech.}\ }\textbf {\bibinfo {volume}
  {22}},\ \bibinfo {pages} {1--12} (\bibinfo {year} {1990})}\BibitemShut
  {NoStop}%
\bibitem [{\citenamefont {Schmid}(2007)}]{schmid2007nonmodal}%
  \BibitemOpen
  \bibfield  {author} {\bibinfo {author} {\bibfnamefont {P.~J.}\ \bibnamefont
  {Schmid}},\ }\bibfield  {title} {\enquote {\bibinfo {title} {Nonmodal
  stability theory},}\ }\href@noop {} {\bibfield  {journal} {\bibinfo
  {journal} {Annu. Rev. Fluid Mech.}\ }\textbf {\bibinfo {volume} {39}},\
  \bibinfo {pages} {129--162} (\bibinfo {year} {2007})}\BibitemShut {NoStop}%
\bibitem [{\citenamefont {Kedia}, \citenamefont {Nagaraja},\ and\ \citenamefont
  {Sujith}(2008)}]{kedia2008impact}%
  \BibitemOpen
  \bibfield  {author} {\bibinfo {author} {\bibfnamefont {K.~S.}\ \bibnamefont
  {Kedia}}, \bibinfo {author} {\bibfnamefont {S.~B.}\ \bibnamefont {Nagaraja}},
  \ and\ \bibinfo {author} {\bibfnamefont {R.~I.}\ \bibnamefont {Sujith}},\
  }\bibfield  {title} {\enquote {\bibinfo {title} {Impact of linear coupling on
  thermoacoustic instabilities},}\ }\href@noop {} {\bibfield  {journal}
  {\bibinfo  {journal} {Combust. Sci. Technol.}\ }\textbf {\bibinfo {volume}
  {180}},\ \bibinfo {pages} {1588--1612} (\bibinfo {year} {2008})}\BibitemShut
  {NoStop}%
\bibitem [{\citenamefont {Zhao}(2012)}]{zhao2012transient}%
  \BibitemOpen
  \bibfield  {author} {\bibinfo {author} {\bibfnamefont {D.}~\bibnamefont
  {Zhao}},\ }\bibfield  {title} {\enquote {\bibinfo {title} {{Transient growth
  of flow disturbances in triggering a Rijke tube combustion instability}},}\
  }\href@noop {} {\bibfield  {journal} {\bibinfo  {journal} {Combust. Flame}\
  }\textbf {\bibinfo {volume} {159}},\ \bibinfo {pages} {2126--2137} (\bibinfo
  {year} {2012})}\BibitemShut {NoStop}%
\bibitem [{\citenamefont {Selimefendigil}, \citenamefont {Sujith},\ and\
  \citenamefont {Polifke}(2011)}]{selimefendigil2011identification}%
  \BibitemOpen
  \bibfield  {author} {\bibinfo {author} {\bibfnamefont {F.}~\bibnamefont
  {Selimefendigil}}, \bibinfo {author} {\bibfnamefont {R.~I.}\ \bibnamefont
  {Sujith}}, \ and\ \bibinfo {author} {\bibfnamefont {W.}~\bibnamefont
  {Polifke}},\ }\bibfield  {title} {\enquote {\bibinfo {title} {Identification
  of heat transfer dynamics for non-modal analysis of thermoacoustic
  stability},}\ }\href@noop {} {\bibfield  {journal} {\bibinfo  {journal}
  {Appl. Math. Comput.}\ }\textbf {\bibinfo {volume} {217}},\ \bibinfo {pages}
  {5134--5150} (\bibinfo {year} {2011})}\BibitemShut {NoStop}%
\bibitem [{\citenamefont {Balasubramanian}\ and\ \citenamefont
  {Sujith}(2008{\natexlab{b}})}]{balasubramanian2008non}%
  \BibitemOpen
  \bibfield  {author} {\bibinfo {author} {\bibfnamefont {K.}~\bibnamefont
  {Balasubramanian}}\ and\ \bibinfo {author} {\bibfnamefont {R.~I.}\
  \bibnamefont {Sujith}},\ }\bibfield  {title} {\enquote {\bibinfo {title}
  {Non-normality and nonlinearity in combustion--acoustic interaction in
  diffusion flames},}\ }\href@noop {} {\bibfield  {journal} {\bibinfo
  {journal} {J. Fluid Mech.}\ }\textbf {\bibinfo {volume} {594}},\ \bibinfo
  {pages} {29--57} (\bibinfo {year} {2008}{\natexlab{b}})}\BibitemShut
  {NoStop}%
\bibitem [{\citenamefont {Zhang}\ \emph {et~al.}(2015)\citenamefont {Zhang},
  \citenamefont {Zhao}, \citenamefont {Li}, \citenamefont {Ji}, \citenamefont
  {Li},\ and\ \citenamefont {Li}}]{zhang2015transient}%
  \BibitemOpen
  \bibfield  {author} {\bibinfo {author} {\bibfnamefont {Z.}~\bibnamefont
  {Zhang}}, \bibinfo {author} {\bibfnamefont {D.}~\bibnamefont {Zhao}},
  \bibinfo {author} {\bibfnamefont {S.}~\bibnamefont {Li}}, \bibinfo {author}
  {\bibfnamefont {C.}~\bibnamefont {Ji}}, \bibinfo {author} {\bibfnamefont
  {X.}~\bibnamefont {Li}}, \ and\ \bibinfo {author} {\bibfnamefont
  {J.}~\bibnamefont {Li}},\ }\bibfield  {title} {\enquote {\bibinfo {title}
  {Transient energy growth of acoustic disturbances in triggering
  self-sustained thermoacoustic oscillations},}\ }\href@noop {} {\bibfield
  {journal} {\bibinfo  {journal} {Energy}\ }\textbf {\bibinfo {volume} {82}},\
  \bibinfo {pages} {370--381} (\bibinfo {year} {2015})}\BibitemShut {NoStop}%
\bibitem [{\citenamefont {Mangesius}\ and\ \citenamefont
  {Polifke}(2011)}]{mangesius2011discrete}%
  \BibitemOpen
  \bibfield  {author} {\bibinfo {author} {\bibfnamefont {H.}~\bibnamefont
  {Mangesius}}\ and\ \bibinfo {author} {\bibfnamefont {W.}~\bibnamefont
  {Polifke}},\ }\bibfield  {title} {\enquote {\bibinfo {title} {A
  discrete-time, state-space approach for the investigation of non-normal
  effects in thermoacoustic systems},}\ }\href@noop {} {\bibfield  {journal}
  {\bibinfo  {journal} {Int. J. Spray Combust. Dyn.}\ }\textbf {\bibinfo
  {volume} {3}},\ \bibinfo {pages} {331--350} (\bibinfo {year}
  {2011})}\BibitemShut {NoStop}%
\bibitem [{\citenamefont {Wieczorek}\ \emph {et~al.}(2011)\citenamefont
  {Wieczorek}, \citenamefont {Sensiau}, \citenamefont {Polifke},\ and\
  \citenamefont {Nicoud}}]{wieczorek2011assessing}%
  \BibitemOpen
  \bibfield  {author} {\bibinfo {author} {\bibfnamefont {K.}~\bibnamefont
  {Wieczorek}}, \bibinfo {author} {\bibfnamefont {C.}~\bibnamefont {Sensiau}},
  \bibinfo {author} {\bibfnamefont {W.}~\bibnamefont {Polifke}}, \ and\
  \bibinfo {author} {\bibfnamefont {F.}~\bibnamefont {Nicoud}},\ }\bibfield
  {title} {\enquote {\bibinfo {title} {Assessing non-normal effects in
  thermoacoustic systems with mean flow},}\ }\href@noop {} {\bibfield
  {journal} {\bibinfo  {journal} {Phys. Fluids}\ }\textbf {\bibinfo {volume}
  {23}},\ \bibinfo {pages} {107103} (\bibinfo {year} {2011})}\BibitemShut
  {NoStop}%
\bibitem [{\citenamefont {Li}\ and\ \citenamefont {Zhao}(2015)}]{li2015mean}%
  \BibitemOpen
  \bibfield  {author} {\bibinfo {author} {\bibfnamefont {X.}~\bibnamefont
  {Li}}\ and\ \bibinfo {author} {\bibfnamefont {D.}~\bibnamefont {Zhao}},\
  }\bibfield  {title} {\enquote {\bibinfo {title} {Mean temperature effect on a
  thermoacoustic system stability and non-normality},}\ }\href@noop {}
  {\bibfield  {journal} {\bibinfo  {journal} {J. Low Freq. Noise Vib. Act.
  Control}\ }\textbf {\bibinfo {volume} {34}},\ \bibinfo {pages} {185--200}
  (\bibinfo {year} {2015})}\BibitemShut {NoStop}%
\bibitem [{\citenamefont {Li}, \citenamefont {Zhao},\ and\ \citenamefont
  {Yang}(2016)}]{li2016effect}%
  \BibitemOpen
  \bibfield  {author} {\bibinfo {author} {\bibfnamefont {L.}~\bibnamefont
  {Li}}, \bibinfo {author} {\bibfnamefont {D.}~\bibnamefont {Zhao}}, \ and\
  \bibinfo {author} {\bibfnamefont {X.}~\bibnamefont {Yang}},\ }\bibfield
  {title} {\enquote {\bibinfo {title} {Effect of entropy waves on transient
  energy growth of flow disturbances in triggering thermoacoustic
  instability},}\ }\href@noop {} {\bibfield  {journal} {\bibinfo  {journal}
  {Int. J. Heat Mass Transf.}\ }\textbf {\bibinfo {volume} {99}},\ \bibinfo
  {pages} {219--233} (\bibinfo {year} {2016})}\BibitemShut {NoStop}%
\bibitem [{\citenamefont {Kulkarni}, \citenamefont {Balasubramanian},\ and\
  \citenamefont {Sujith}(2011)}]{kulkarni2011non}%
  \BibitemOpen
  \bibfield  {author} {\bibinfo {author} {\bibfnamefont {R.}~\bibnamefont
  {Kulkarni}}, \bibinfo {author} {\bibfnamefont {K.}~\bibnamefont
  {Balasubramanian}}, \ and\ \bibinfo {author} {\bibfnamefont {R.~I.}\
  \bibnamefont {Sujith}},\ }\bibfield  {title} {\enquote {\bibinfo {title}
  {Non-normality and its consequences in active control of thermoacoustic
  instabilities},}\ }\href@noop {} {\bibfield  {journal} {\bibinfo  {journal}
  {J. Fluid Mech.}\ }\textbf {\bibinfo {volume} {670}},\ \bibinfo {pages}
  {130--149} (\bibinfo {year} {2011})}\BibitemShut {NoStop}%
\bibitem [{\citenamefont {Zhang}\ and\ \citenamefont
  {Guan}(2015)}]{zhang2015feedback}%
  \BibitemOpen
  \bibfield  {author} {\bibinfo {author} {\bibfnamefont {Z.}~\bibnamefont
  {Zhang}}\ and\ \bibinfo {author} {\bibfnamefont {D.}~\bibnamefont {Guan}},\
  }\bibfield  {title} {\enquote {\bibinfo {title} {Feedback control of
  rijke-type thermoacoustic oscillations transient growth},}\ }\href@noop {}
  {\bibfield  {journal} {\bibinfo  {journal} {J. Low Freq. Noise Vib. Act.
  Control}\ }\textbf {\bibinfo {volume} {34}},\ \bibinfo {pages} {219--232}
  (\bibinfo {year} {2015})}\BibitemShut {NoStop}%
\bibitem [{\citenamefont {Kabiraj}\ and\ \citenamefont
  {Sujith}(2011)}]{kabiraj2011investigation}%
  \BibitemOpen
  \bibfield  {author} {\bibinfo {author} {\bibfnamefont {L.}~\bibnamefont
  {Kabiraj}}\ and\ \bibinfo {author} {\bibfnamefont {R.~I.}\ \bibnamefont
  {Sujith}},\ }\bibfield  {title} {\enquote {\bibinfo {title} {Investigation of
  subcritical instability in ducted premixed flames},}\ }in\ \href@noop {}
  {\emph {\bibinfo {booktitle} {Turbo Expo: Power for Land, Sea, and Air}}},\
  Vol.\ \bibinfo {volume} {54624}\ (\bibinfo {year} {2011})\ pp.\ \bibinfo
  {pages} {969--977}\BibitemShut {NoStop}%
\bibitem [{\citenamefont {Lieuwen}(2002)}]{lieuwen2002experimental}%
  \BibitemOpen
  \bibfield  {author} {\bibinfo {author} {\bibfnamefont {T.~C.}\ \bibnamefont
  {Lieuwen}},\ }\bibfield  {title} {\enquote {\bibinfo {title} {Experimental
  investigation of limit-cycle oscillations in an unstable gas turbine
  combustor},}\ }\href@noop {} {\bibfield  {journal} {\bibinfo  {journal} {J.
  Propuls. Power}\ }\textbf {\bibinfo {volume} {18}},\ \bibinfo {pages}
  {61--67} (\bibinfo {year} {2002})}\BibitemShut {NoStop}%
\bibitem [{\citenamefont {Lieuwen}\ and\ \citenamefont
  {Banaszuk}(2005)}]{lieuwen2005background}%
  \BibitemOpen
  \bibfield  {author} {\bibinfo {author} {\bibfnamefont {T.}~\bibnamefont
  {Lieuwen}}\ and\ \bibinfo {author} {\bibfnamefont {A.}~\bibnamefont
  {Banaszuk}},\ }\bibfield  {title} {\enquote {\bibinfo {title} {Background
  noise effects on combustor stability},}\ }\href@noop {} {\bibfield  {journal}
  {\bibinfo  {journal} {J. Propuls. Power}\ }\textbf {\bibinfo {volume} {21}},\
  \bibinfo {pages} {25--31} (\bibinfo {year} {2005})}\BibitemShut {NoStop}%
\bibitem [{\citenamefont {Pikovsky}\ and\ \citenamefont
  {Maistrenko}(2012)}]{pikovsky2012synchronization}%
  \BibitemOpen
  \bibfield  {author} {\bibinfo {author} {\bibfnamefont {A.}~\bibnamefont
  {Pikovsky}}\ and\ \bibinfo {author} {\bibfnamefont {Y.}~\bibnamefont
  {Maistrenko}},\ }\href@noop {} {\emph {\bibinfo {title} {Synchronization:
  Theory and Application}}},\ Vol.\ \bibinfo {volume} {109}\ (\bibinfo
  {publisher} {Springer Science \& Business Media},\ \bibinfo {year}
  {2012})\BibitemShut {NoStop}%
\bibitem [{\citenamefont {Boccaletti}\ \emph {et~al.}(2018)\citenamefont
  {Boccaletti}, \citenamefont {Pisarchik}, \citenamefont {Del~Genio},\ and\
  \citenamefont {Amann}}]{boccaletti2018synchronization}%
  \BibitemOpen
  \bibfield  {author} {\bibinfo {author} {\bibfnamefont {S.}~\bibnamefont
  {Boccaletti}}, \bibinfo {author} {\bibfnamefont {A.~N.}\ \bibnamefont
  {Pisarchik}}, \bibinfo {author} {\bibfnamefont {C.~I.}\ \bibnamefont
  {Del~Genio}}, \ and\ \bibinfo {author} {\bibfnamefont {A.}~\bibnamefont
  {Amann}},\ }\href@noop {} {\emph {\bibinfo {title} {Synchronization: From
  Coupled Systems to Complex Networks}}}\ (\bibinfo  {publisher} {Cambridge
  University Press},\ \bibinfo {year} {2018})\BibitemShut {NoStop}%
\bibitem [{\citenamefont {Srikanth}\ \emph
  {et~al.}(2021{\natexlab{a}})\citenamefont {Srikanth}, \citenamefont {Pawar},
  \citenamefont {Manoj},\ and\ \citenamefont {Sujith}}]{srikanth2021dynamical}%
  \BibitemOpen
  \bibfield  {author} {\bibinfo {author} {\bibfnamefont {S.}~\bibnamefont
  {Srikanth}}, \bibinfo {author} {\bibfnamefont {S.~A.}\ \bibnamefont {Pawar}},
  \bibinfo {author} {\bibfnamefont {K.}~\bibnamefont {Manoj}}, \ and\ \bibinfo
  {author} {\bibfnamefont {R.~I.}\ \bibnamefont {Sujith}},\ }\bibfield  {title}
  {\enquote {\bibinfo {title} {Dynamical states and bifurcations in coupled
  thermoacoustic oscillators},}\ }\href@noop {} {\bibfield  {journal} {\bibinfo
   {journal} {arXiv:2109.09600}\ } (\bibinfo {year}
  {2021}{\natexlab{a}})}\BibitemShut {NoStop}%
\bibitem [{\citenamefont {Dewan}(1972)}]{dewan1972harmonic}%
  \BibitemOpen
  \bibfield  {author} {\bibinfo {author} {\bibfnamefont {E.}~\bibnamefont
  {Dewan}},\ }\bibfield  {title} {\enquote {\bibinfo {title} {Harmonic
  entrainment of van der pol oscillations: Phase locking and asynchronous
  quenching},}\ }\href@noop {} {\bibfield  {journal} {\bibinfo  {journal} {IEEE
  Trans. Autom. Control}\ }\textbf {\bibinfo {volume} {17}},\ \bibinfo {pages}
  {655--663} (\bibinfo {year} {1972})}\BibitemShut {NoStop}%
\bibitem [{\citenamefont {Keen}\ and\ \citenamefont
  {Fletcher}(1970)}]{keen1970suppression}%
  \BibitemOpen
  \bibfield  {author} {\bibinfo {author} {\bibfnamefont {B.~E.}\ \bibnamefont
  {Keen}}\ and\ \bibinfo {author} {\bibfnamefont {W.~H.~W.}\ \bibnamefont
  {Fletcher}},\ }\bibfield  {title} {\enquote {\bibinfo {title} {Suppression of
  a plasma instability by the method of "asynchronous quenching"},}\
  }\href@noop {} {\bibfield  {journal} {\bibinfo  {journal} {Phys. Rev. Lett.}\
  }\textbf {\bibinfo {volume} {24}},\ \bibinfo {pages} {130--134} (\bibinfo
  {year} {1970})}\BibitemShut {NoStop}%
\bibitem [{\citenamefont {Quiroga}\ \emph {et~al.}(2002)\citenamefont
  {Quiroga}, \citenamefont {Kraskov}, \citenamefont {Kreuz},\ and\
  \citenamefont {Grassberger}}]{quiroga2002performance}%
  \BibitemOpen
  \bibfield  {author} {\bibinfo {author} {\bibfnamefont {R.~Q.}\ \bibnamefont
  {Quiroga}}, \bibinfo {author} {\bibfnamefont {A.}~\bibnamefont {Kraskov}},
  \bibinfo {author} {\bibfnamefont {T.}~\bibnamefont {Kreuz}}, \ and\ \bibinfo
  {author} {\bibfnamefont {P.}~\bibnamefont {Grassberger}},\ }\bibfield
  {title} {\enquote {\bibinfo {title} {Performance of different synchronization
  measures in real data: a case study on electroencephalographic signals},}\
  }\href@noop {} {\bibfield  {journal} {\bibinfo  {journal} {Phys. Rev. E}\
  }\textbf {\bibinfo {volume} {65}},\ \bibinfo {pages} {041903} (\bibinfo
  {year} {2002})}\BibitemShut {NoStop}%
\bibitem [{\citenamefont {Granovetter}(1978)}]{granovetter1978threshold}%
  \BibitemOpen
  \bibfield  {author} {\bibinfo {author} {\bibfnamefont {M.}~\bibnamefont
  {Granovetter}},\ }\bibfield  {title} {\enquote {\bibinfo {title} {Threshold
  models of collective behavior},}\ }\href@noop {} {\bibfield  {journal}
  {\bibinfo  {journal} {Am. J. Sociol.}\ }\textbf {\bibinfo {volume} {83}},\
  \bibinfo {pages} {1420--1443} (\bibinfo {year} {1978})}\BibitemShut {NoStop}%
\bibitem [{\citenamefont {Mosekilde}, \citenamefont {Maistrenko},\ and\
  \citenamefont {Postnov}(2002)}]{mosekilde2002chaotic}%
  \BibitemOpen
  \bibfield  {author} {\bibinfo {author} {\bibfnamefont {E.}~\bibnamefont
  {Mosekilde}}, \bibinfo {author} {\bibfnamefont {Y.}~\bibnamefont
  {Maistrenko}}, \ and\ \bibinfo {author} {\bibfnamefont {D.}~\bibnamefont
  {Postnov}},\ }\href@noop {} {\emph {\bibinfo {title} {Chaotic
  Synchronization: Applications to Living Systems}}},\ Vol.~\bibinfo {volume}
  {42}\ (\bibinfo  {publisher} {World Scientific},\ \bibinfo {year}
  {2002})\BibitemShut {NoStop}%
\bibitem [{\citenamefont {Blasius}, \citenamefont {Huppert},\ and\
  \citenamefont {Stone}(1999)}]{blasius1999complex}%
  \BibitemOpen
  \bibfield  {author} {\bibinfo {author} {\bibfnamefont {B.}~\bibnamefont
  {Blasius}}, \bibinfo {author} {\bibfnamefont {A.}~\bibnamefont {Huppert}}, \
  and\ \bibinfo {author} {\bibfnamefont {L.}~\bibnamefont {Stone}},\ }\bibfield
   {title} {\enquote {\bibinfo {title} {Complex dynamics and phase
  synchronization in spatially extended ecological systems},}\ }\href@noop {}
  {\bibfield  {journal} {\bibinfo  {journal} {Nature}\ }\textbf {\bibinfo
  {volume} {399}},\ \bibinfo {pages} {354--–359} (\bibinfo {year}
  {1999})}\BibitemShut {NoStop}%
\bibitem [{\citenamefont {Srikanth}\ \emph
  {et~al.}(2021{\natexlab{b}})\citenamefont {Srikanth}, \citenamefont {Sahay},
  \citenamefont {Pawar}, \citenamefont {Manoj},\ and\ \citenamefont
  {Sujith}}]{srikanth2021self}%
  \BibitemOpen
  \bibfield  {author} {\bibinfo {author} {\bibfnamefont {S.}~\bibnamefont
  {Srikanth}}, \bibinfo {author} {\bibfnamefont {A.}~\bibnamefont {Sahay}},
  \bibinfo {author} {\bibfnamefont {S.~A.}\ \bibnamefont {Pawar}}, \bibinfo
  {author} {\bibfnamefont {K.}~\bibnamefont {Manoj}}, \ and\ \bibinfo {author}
  {\bibfnamefont {R.~I.}\ \bibnamefont {Sujith}},\ }\bibfield  {title}
  {\enquote {\bibinfo {title} {Self-coupling: An effective method to mitigate
  thermoacoustic instability},}\ }\href@noop {} {\bibfield  {journal} {\bibinfo
   {journal} {arXiv:2112.14152}\ } (\bibinfo {year}
  {2021}{\natexlab{b}})}\BibitemShut {NoStop}%
\bibitem [{\citenamefont {Thomas}\ \emph
  {et~al.}(2018{\natexlab{a}})\citenamefont {Thomas}, \citenamefont {Mondal},
  \citenamefont {Pawar},\ and\ \citenamefont {Sujith}}]{thomas2018effecta}%
  \BibitemOpen
  \bibfield  {author} {\bibinfo {author} {\bibfnamefont {N.}~\bibnamefont
  {Thomas}}, \bibinfo {author} {\bibfnamefont {S.}~\bibnamefont {Mondal}},
  \bibinfo {author} {\bibfnamefont {S.~A.}\ \bibnamefont {Pawar}}, \ and\
  \bibinfo {author} {\bibfnamefont {R.~I.}\ \bibnamefont {Sujith}},\ }\bibfield
   {title} {\enquote {\bibinfo {title} {Effect of time-delay and dissipative
  coupling on amplitude death in coupled thermoacoustic oscillators},}\
  }\href@noop {} {\bibfield  {journal} {\bibinfo  {journal} {Chaos}\ }\textbf
  {\bibinfo {volume} {28}},\ \bibinfo {pages} {033119} (\bibinfo {year}
  {2018}{\natexlab{a}})}\BibitemShut {NoStop}%
\bibitem [{\citenamefont {Thomas}\ \emph
  {et~al.}(2018{\natexlab{b}})\citenamefont {Thomas}, \citenamefont {Mondal},
  \citenamefont {Pawar},\ and\ \citenamefont {Sujith}}]{thomas2018effectb}%
  \BibitemOpen
  \bibfield  {author} {\bibinfo {author} {\bibfnamefont {N.}~\bibnamefont
  {Thomas}}, \bibinfo {author} {\bibfnamefont {S.}~\bibnamefont {Mondal}},
  \bibinfo {author} {\bibfnamefont {S.~A.}\ \bibnamefont {Pawar}}, \ and\
  \bibinfo {author} {\bibfnamefont {R.~I.}\ \bibnamefont {Sujith}},\ }\bibfield
   {title} {\enquote {\bibinfo {title} {Effect of noise amplification during
  the transition to amplitude death in coupled thermoacoustic oscillators},}\
  }\href@noop {} {\bibfield  {journal} {\bibinfo  {journal} {Chaos}\ }\textbf
  {\bibinfo {volume} {28}},\ \bibinfo {pages} {093116} (\bibinfo {year}
  {2018}{\natexlab{b}})}\BibitemShut {NoStop}%
\bibitem [{\citenamefont {Dange}\ \emph {et~al.}(2019)\citenamefont {Dange},
  \citenamefont {Manoj}, \citenamefont {Banerjee}, \citenamefont {Pawar},
  \citenamefont {Mondal},\ and\ \citenamefont {Sujith}}]{dange2019oscillation}%
  \BibitemOpen
  \bibfield  {author} {\bibinfo {author} {\bibfnamefont {S.}~\bibnamefont
  {Dange}}, \bibinfo {author} {\bibfnamefont {K.}~\bibnamefont {Manoj}},
  \bibinfo {author} {\bibfnamefont {S.}~\bibnamefont {Banerjee}}, \bibinfo
  {author} {\bibfnamefont {S.~A.}\ \bibnamefont {Pawar}}, \bibinfo {author}
  {\bibfnamefont {S.}~\bibnamefont {Mondal}}, \ and\ \bibinfo {author}
  {\bibfnamefont {R.~I.}\ \bibnamefont {Sujith}},\ }\bibfield  {title}
  {\enquote {\bibinfo {title} {Oscillation quenching and phase-flip bifurcation
  in coupled thermoacoustic systems},}\ }\href@noop {} {\bibfield  {journal}
  {\bibinfo  {journal} {Chaos}\ }\textbf {\bibinfo {volume} {29}},\ \bibinfo
  {pages} {093135} (\bibinfo {year} {2019})}\BibitemShut {NoStop}%
\bibitem [{\citenamefont {Hyodo}, \citenamefont {Iwasaki},\ and\ \citenamefont
  {Biwa}(2020)}]{hyodo2020suppression}%
  \BibitemOpen
  \bibfield  {author} {\bibinfo {author} {\bibfnamefont {H.}~\bibnamefont
  {Hyodo}}, \bibinfo {author} {\bibfnamefont {M.}~\bibnamefont {Iwasaki}}, \
  and\ \bibinfo {author} {\bibfnamefont {T.}~\bibnamefont {Biwa}},\ }\bibfield
  {title} {\enquote {\bibinfo {title} {{Suppression of Rijke tube oscillations
  by delay coupling}},}\ }\href@noop {} {\bibfield  {journal} {\bibinfo
  {journal} {J. Appl. Phys.}\ }\textbf {\bibinfo {volume} {128}},\ \bibinfo
  {pages} {094902} (\bibinfo {year} {2020})}\BibitemShut {NoStop}%
\bibitem [{\citenamefont {Sahay}\ \emph {et~al.}(2021)\citenamefont {Sahay},
  \citenamefont {Roy}, \citenamefont {Pawar},\ and\ \citenamefont
  {Sujith}}]{sahay2020dynamics}%
  \BibitemOpen
  \bibfield  {author} {\bibinfo {author} {\bibfnamefont {A.}~\bibnamefont
  {Sahay}}, \bibinfo {author} {\bibfnamefont {A.}~\bibnamefont {Roy}}, \bibinfo
  {author} {\bibfnamefont {S.~A.}\ \bibnamefont {Pawar}}, \ and\ \bibinfo
  {author} {\bibfnamefont {R.~I.}\ \bibnamefont {Sujith}},\ }\bibfield  {title}
  {\enquote {\bibinfo {title} {Dynamics of coupled thermoacoustic oscillators
  under asymmetric forcing},}\ }\href@noop {} {\bibfield  {journal} {\bibinfo
  {journal} {Phys. Rev. Appl.}\ }\textbf {\bibinfo {volume} {15}},\ \bibinfo
  {pages} {044011} (\bibinfo {year} {2021})}\BibitemShut {NoStop}%
\bibitem [{\citenamefont {Jegal}\ \emph {et~al.}(2019)\citenamefont {Jegal},
  \citenamefont {Moon}, \citenamefont {Gu}, \citenamefont {Li},\ and\
  \citenamefont {Kim}}]{jegal2019mutual}%
  \BibitemOpen
  \bibfield  {author} {\bibinfo {author} {\bibfnamefont {H.}~\bibnamefont
  {Jegal}}, \bibinfo {author} {\bibfnamefont {K.}~\bibnamefont {Moon}},
  \bibinfo {author} {\bibfnamefont {J.}~\bibnamefont {Gu}}, \bibinfo {author}
  {\bibfnamefont {L.~K.~B.}\ \bibnamefont {Li}}, \ and\ \bibinfo {author}
  {\bibfnamefont {K.~T.}\ \bibnamefont {Kim}},\ }\bibfield  {title} {\enquote
  {\bibinfo {title} {Mutual synchronization of two lean-premixed gas turbine
  combustors: Phase locking and amplitude death},}\ }\href@noop {} {\bibfield
  {journal} {\bibinfo  {journal} {Combust. Flame}\ }\textbf {\bibinfo {volume}
  {206}},\ \bibinfo {pages} {424--437} (\bibinfo {year} {2019})}\BibitemShut
  {NoStop}%
\bibitem [{\citenamefont {Ghirardo}\ \emph {et~al.}(2019)\citenamefont
  {Ghirardo}, \citenamefont {Di~Giovine}, \citenamefont {Moeck},\ and\
  \citenamefont {Bothien}}]{ghirardo2019thermoacoustics}%
  \BibitemOpen
  \bibfield  {author} {\bibinfo {author} {\bibfnamefont {G.}~\bibnamefont
  {Ghirardo}}, \bibinfo {author} {\bibfnamefont {C.}~\bibnamefont
  {Di~Giovine}}, \bibinfo {author} {\bibfnamefont {J.~P.}\ \bibnamefont
  {Moeck}}, \ and\ \bibinfo {author} {\bibfnamefont {M.~R.}\ \bibnamefont
  {Bothien}},\ }\bibfield  {title} {\enquote {\bibinfo {title} {Thermoacoustics
  of can-annular combustors},}\ }\href@noop {} {\bibfield  {journal} {\bibinfo
  {journal} {J. Eng. Gas Turbine Power}\ }\textbf {\bibinfo {volume} {141}},\
  \bibinfo {pages} {011007} (\bibinfo {year} {2019})}\BibitemShut {NoStop}%
\bibitem [{\citenamefont {Farisco}, \citenamefont {Panek},\ and\ \citenamefont
  {Kok}(2017)}]{farisco2017thermo}%
  \BibitemOpen
  \bibfield  {author} {\bibinfo {author} {\bibfnamefont {F.}~\bibnamefont
  {Farisco}}, \bibinfo {author} {\bibfnamefont {L.}~\bibnamefont {Panek}}, \
  and\ \bibinfo {author} {\bibfnamefont {J.~B.~W.}\ \bibnamefont {Kok}},\
  }\bibfield  {title} {\enquote {\bibinfo {title} {Thermo-acoustic cross-talk
  between cans in a can-annular combustor},}\ }\href@noop {} {\bibfield
  {journal} {\bibinfo  {journal} {Int. J. Spray Combust. Dyn.}\ }\textbf
  {\bibinfo {volume} {9}},\ \bibinfo {pages} {452--469} (\bibinfo {year}
  {2017})}\BibitemShut {NoStop}%
\bibitem [{\citenamefont {Moon}\ \emph {et~al.}(2020)\citenamefont {Moon},
  \citenamefont {Guan}, \citenamefont {Li},\ and\ \citenamefont
  {Kim}}]{moon2020mutual}%
  \BibitemOpen
  \bibfield  {author} {\bibinfo {author} {\bibfnamefont {K.}~\bibnamefont
  {Moon}}, \bibinfo {author} {\bibfnamefont {Y.}~\bibnamefont {Guan}}, \bibinfo
  {author} {\bibfnamefont {L.~K.~B.}\ \bibnamefont {Li}}, \ and\ \bibinfo
  {author} {\bibfnamefont {K.~T.}\ \bibnamefont {Kim}},\ }\bibfield  {title}
  {\enquote {\bibinfo {title} {Mutual synchronization of two flame-driven
  thermoacoustic oscillators: Dissipative and time-delayed coupling effects},}\
  }\href@noop {} {\bibfield  {journal} {\bibinfo  {journal} {Chaos}\ }\textbf
  {\bibinfo {volume} {30}},\ \bibinfo {pages} {023110} (\bibinfo {year}
  {2020})}\BibitemShut {NoStop}%
\bibitem [{\citenamefont {Moon}\ \emph {et~al.}(2019)\citenamefont {Moon},
  \citenamefont {Jegal}, \citenamefont {Gu},\ and\ \citenamefont
  {Kim}}]{moon2019combustion}%
  \BibitemOpen
  \bibfield  {author} {\bibinfo {author} {\bibfnamefont {K.}~\bibnamefont
  {Moon}}, \bibinfo {author} {\bibfnamefont {H.}~\bibnamefont {Jegal}},
  \bibinfo {author} {\bibfnamefont {J.}~\bibnamefont {Gu}}, \ and\ \bibinfo
  {author} {\bibfnamefont {K.~T.}\ \bibnamefont {Kim}},\ }\bibfield  {title}
  {\enquote {\bibinfo {title} {Combustion-acoustic interactions through
  cross-talk area between adjacent model gas turbine combustors},}\ }\href@noop
  {} {\bibfield  {journal} {\bibinfo  {journal} {Combust. Flame}\ }\textbf
  {\bibinfo {volume} {202}},\ \bibinfo {pages} {405--416} (\bibinfo {year}
  {2019})}\BibitemShut {NoStop}%
\bibitem [{\citenamefont {Biwa}, \citenamefont {Tozuka},\ and\ \citenamefont
  {Yazaki}(2015)}]{biwa2015amplitude}%
  \BibitemOpen
  \bibfield  {author} {\bibinfo {author} {\bibfnamefont {T.}~\bibnamefont
  {Biwa}}, \bibinfo {author} {\bibfnamefont {S.}~\bibnamefont {Tozuka}}, \ and\
  \bibinfo {author} {\bibfnamefont {T.}~\bibnamefont {Yazaki}},\ }\bibfield
  {title} {\enquote {\bibinfo {title} {Amplitude death in coupled
  thermoacoustic oscillators},}\ }\href@noop {} {\bibfield  {journal} {\bibinfo
   {journal} {Phys. Rev. Appl.}\ }\textbf {\bibinfo {volume} {3}},\ \bibinfo
  {pages} {034006} (\bibinfo {year} {2015})}\BibitemShut {NoStop}%
\bibitem [{\citenamefont {Prasad}\ \emph {et~al.}(2006)\citenamefont {Prasad},
  \citenamefont {Kurths}, \citenamefont {Dana},\ and\ \citenamefont
  {Ramaswamy}}]{prasad2006phase}%
  \BibitemOpen
  \bibfield  {author} {\bibinfo {author} {\bibfnamefont {A.}~\bibnamefont
  {Prasad}}, \bibinfo {author} {\bibfnamefont {J.}~\bibnamefont {Kurths}},
  \bibinfo {author} {\bibfnamefont {S.~K.}\ \bibnamefont {Dana}}, \ and\
  \bibinfo {author} {\bibfnamefont {R.}~\bibnamefont {Ramaswamy}},\ }\bibfield
  {title} {\enquote {\bibinfo {title} {Phase-flip bifurcation induced by time
  delay},}\ }\href@noop {} {\bibfield  {journal} {\bibinfo  {journal} {Phys.
  Rev. E}\ }\textbf {\bibinfo {volume} {74}},\ \bibinfo {pages} {035204}
  (\bibinfo {year} {2006})}\BibitemShut {NoStop}%
\bibitem [{\citenamefont {Prasad}\ \emph {et~al.}(2008)\citenamefont {Prasad},
  \citenamefont {Dana}, \citenamefont {Karnatak}, \citenamefont {Kurths},
  \citenamefont {Blasius},\ and\ \citenamefont
  {Ramaswamy}}]{prasad2008universal}%
  \BibitemOpen
  \bibfield  {author} {\bibinfo {author} {\bibfnamefont {A.}~\bibnamefont
  {Prasad}}, \bibinfo {author} {\bibfnamefont {S.~K.}\ \bibnamefont {Dana}},
  \bibinfo {author} {\bibfnamefont {R.}~\bibnamefont {Karnatak}}, \bibinfo
  {author} {\bibfnamefont {J.}~\bibnamefont {Kurths}}, \bibinfo {author}
  {\bibfnamefont {B.}~\bibnamefont {Blasius}}, \ and\ \bibinfo {author}
  {\bibfnamefont {R.}~\bibnamefont {Ramaswamy}},\ }\bibfield  {title} {\enquote
  {\bibinfo {title} {Universal occurrence of the phase-flip bifurcation in
  time-delay coupled systems},}\ }\href@noop {} {\bibfield  {journal} {\bibinfo
   {journal} {Chaos}\ }\textbf {\bibinfo {volume} {18}},\ \bibinfo {pages}
  {023111} (\bibinfo {year} {2008})}\BibitemShut {NoStop}%
\bibitem [{\citenamefont {Odajima}, \citenamefont {Nishida},\ and\
  \citenamefont {Hatta}(1974)}]{odajima1974synchronous}%
  \BibitemOpen
  \bibfield  {author} {\bibinfo {author} {\bibfnamefont {K.}~\bibnamefont
  {Odajima}}, \bibinfo {author} {\bibfnamefont {Y.}~\bibnamefont {Nishida}}, \
  and\ \bibinfo {author} {\bibfnamefont {Y.}~\bibnamefont {Hatta}},\ }\bibfield
   {title} {\enquote {\bibinfo {title} {Synchronous quenching of drift-wave
  instability},}\ }\href@noop {} {\bibfield  {journal} {\bibinfo  {journal}
  {Phys. Fluids}\ }\textbf {\bibinfo {volume} {17}},\ \bibinfo {pages}
  {1631--1633} (\bibinfo {year} {1974})}\BibitemShut {NoStop}%
\bibitem [{\citenamefont {Kashinath}, \citenamefont {Li},\ and\ \citenamefont
  {Juniper}(2018)}]{kashinath2018forced}%
  \BibitemOpen
  \bibfield  {author} {\bibinfo {author} {\bibfnamefont {K.}~\bibnamefont
  {Kashinath}}, \bibinfo {author} {\bibfnamefont {L.~K.~B.}\ \bibnamefont
  {Li}}, \ and\ \bibinfo {author} {\bibfnamefont {M.~P.}\ \bibnamefont
  {Juniper}},\ }\bibfield  {title} {\enquote {\bibinfo {title} {Forced
  synchronization of periodic and aperiodic thermoacoustic oscillations:
  lock-in, bifurcations and open-loop control},}\ }\href@noop {} {\bibfield
  {journal} {\bibinfo  {journal} {J. Fluid Mech.}\ }\textbf {\bibinfo {volume}
  {838}},\ \bibinfo {pages} {690--714} (\bibinfo {year} {2018})}\BibitemShut
  {NoStop}%
\bibitem [{\citenamefont {Guan}\ \emph
  {et~al.}(2019{\natexlab{b}})\citenamefont {Guan}, \citenamefont {He},
  \citenamefont {Murugesan}, \citenamefont {Li}, \citenamefont {Liu},\ and\
  \citenamefont {Li}}]{guan2019control}%
  \BibitemOpen
  \bibfield  {author} {\bibinfo {author} {\bibfnamefont {Y.}~\bibnamefont
  {Guan}}, \bibinfo {author} {\bibfnamefont {W.}~\bibnamefont {He}}, \bibinfo
  {author} {\bibfnamefont {M.}~\bibnamefont {Murugesan}}, \bibinfo {author}
  {\bibfnamefont {Q.}~\bibnamefont {Li}}, \bibinfo {author} {\bibfnamefont
  {P.}~\bibnamefont {Liu}}, \ and\ \bibinfo {author} {\bibfnamefont {L.~K.~B.}\
  \bibnamefont {Li}},\ }\bibfield  {title} {\enquote {\bibinfo {title} {Control
  of self-excited thermoacoustic oscillations using transient forcing,
  hysteresis and mode switching},}\ }\href@noop {} {\bibfield  {journal}
  {\bibinfo  {journal} {Combust. Flame}\ }\textbf {\bibinfo {volume} {202}},\
  \bibinfo {pages} {262--275} (\bibinfo {year}
  {2019}{\natexlab{b}})}\BibitemShut {NoStop}%
\bibitem [{\citenamefont {Guan}\ \emph
  {et~al.}(2019{\natexlab{c}})\citenamefont {Guan}, \citenamefont {Gupta},
  \citenamefont {Wan},\ and\ \citenamefont {Li}}]{guan2019forced}%
  \BibitemOpen
  \bibfield  {author} {\bibinfo {author} {\bibfnamefont {Y.}~\bibnamefont
  {Guan}}, \bibinfo {author} {\bibfnamefont {V.}~\bibnamefont {Gupta}},
  \bibinfo {author} {\bibfnamefont {M.}~\bibnamefont {Wan}}, \ and\ \bibinfo
  {author} {\bibfnamefont {L.~K.~B.}\ \bibnamefont {Li}},\ }\bibfield  {title}
  {\enquote {\bibinfo {title} {Forced synchronization of quasiperiodic
  oscillations in a thermoacoustic system},}\ }\href@noop {} {\bibfield
  {journal} {\bibinfo  {journal} {J. Fluid Mech.}\ }\textbf {\bibinfo {volume}
  {879}},\ \bibinfo {pages} {390--421} (\bibinfo {year}
  {2019}{\natexlab{c}})}\BibitemShut {NoStop}%
\bibitem [{\citenamefont {Roy}\ \emph {et~al.}(2020)\citenamefont {Roy},
  \citenamefont {Mondal}, \citenamefont {Pawar},\ and\ \citenamefont
  {Sujith}}]{roy2020mechanism}%
  \BibitemOpen
  \bibfield  {author} {\bibinfo {author} {\bibfnamefont {A.}~\bibnamefont
  {Roy}}, \bibinfo {author} {\bibfnamefont {S.}~\bibnamefont {Mondal}},
  \bibinfo {author} {\bibfnamefont {S.~A.}\ \bibnamefont {Pawar}}, \ and\
  \bibinfo {author} {\bibfnamefont {R.~I.}\ \bibnamefont {Sujith}},\ }\bibfield
   {title} {\enquote {\bibinfo {title} {On the mechanism of open-loop control
  of thermoacoustic instability in a laminar premixed combustor},}\ }\href@noop
  {} {\bibfield  {journal} {\bibinfo  {journal} {J. Fluid Mech.}\ }\textbf
  {\bibinfo {volume} {884}},\ \bibinfo {pages} {A2} (\bibinfo {year}
  {2020})}\BibitemShut {NoStop}%
\bibitem [{\citenamefont {Sato}\ \emph {et~al.}(2020)\citenamefont {Sato},
  \citenamefont {Hyodo}, \citenamefont {Biwa},\ and\ \citenamefont
  {Delage}}]{sato2020synchronization}%
  \BibitemOpen
  \bibfield  {author} {\bibinfo {author} {\bibfnamefont {M.}~\bibnamefont
  {Sato}}, \bibinfo {author} {\bibfnamefont {H.}~\bibnamefont {Hyodo}},
  \bibinfo {author} {\bibfnamefont {T.}~\bibnamefont {Biwa}}, \ and\ \bibinfo
  {author} {\bibfnamefont {R.}~\bibnamefont {Delage}},\ }\bibfield  {title}
  {\enquote {\bibinfo {title} {Synchronization of thermoacoustic quasiperiodic
  oscillation by periodic external force},}\ }\href@noop {} {\bibfield
  {journal} {\bibinfo  {journal} {Chaos}\ }\textbf {\bibinfo {volume} {30}},\
  \bibinfo {pages} {063130} (\bibinfo {year} {2020})}\BibitemShut {NoStop}%
\bibitem [{\citenamefont {Zhang}\ \emph {et~al.}(2019)\citenamefont {Zhang},
  \citenamefont {Zhang}, \citenamefont {Fan},\ and\ \citenamefont
  {He}}]{zhang2019devil}%
  \BibitemOpen
  \bibfield  {author} {\bibinfo {author} {\bibfnamefont {Z.}~\bibnamefont
  {Zhang}}, \bibinfo {author} {\bibfnamefont {J.}~\bibnamefont {Zhang}},
  \bibinfo {author} {\bibfnamefont {P.}~\bibnamefont {Fan}}, \ and\ \bibinfo
  {author} {\bibfnamefont {Y.}~\bibnamefont {He}},\ }\bibfield  {title}
  {\enquote {\bibinfo {title} {Devil’s staircases in a thermoacoustic system
  with sinusoidal excitations},}\ }\href@noop {} {\bibfield  {journal}
  {\bibinfo  {journal} {Eur. Phys. J. Spec. Top.}\ }\textbf {\bibinfo {volume}
  {228}},\ \bibinfo {pages} {1891--1901} (\bibinfo {year} {2019})}\BibitemShut
  {NoStop}%
\bibitem [{\citenamefont {Thomson}, \citenamefont {Bruckner},\ and\
  \citenamefont {Bruckner}(2008)}]{thomson2008elementary}%
  \BibitemOpen
  \bibfield  {author} {\bibinfo {author} {\bibfnamefont {B.~S.}\ \bibnamefont
  {Thomson}}, \bibinfo {author} {\bibfnamefont {J.~B.}\ \bibnamefont
  {Bruckner}}, \ and\ \bibinfo {author} {\bibfnamefont {A.~M.}\ \bibnamefont
  {Bruckner}},\ }\href@noop {} {\emph {\bibinfo {title} {Elementary Real
  Analysis}}},\ Vol.~\bibinfo {volume} {1}\ (\bibinfo  {publisher}
  {ClassicalRealAnalysis. com},\ \bibinfo {year} {2008})\BibitemShut {NoStop}%
\bibitem [{\citenamefont {Sch{\"a}fer}\ \emph {et~al.}(1999)\citenamefont
  {Sch{\"a}fer}, \citenamefont {Rosenblum}, \citenamefont {Abel},\ and\
  \citenamefont {Kurths}}]{schafer1999synchronization}%
  \BibitemOpen
  \bibfield  {author} {\bibinfo {author} {\bibfnamefont {C.}~\bibnamefont
  {Sch{\"a}fer}}, \bibinfo {author} {\bibfnamefont {M.~G.}\ \bibnamefont
  {Rosenblum}}, \bibinfo {author} {\bibfnamefont {H.-H.}\ \bibnamefont {Abel}},
  \ and\ \bibinfo {author} {\bibfnamefont {J.}~\bibnamefont {Kurths}},\
  }\bibfield  {title} {\enquote {\bibinfo {title} {Synchronization in the human
  cardiorespiratory system},}\ }\href@noop {} {\bibfield  {journal} {\bibinfo
  {journal} {Phys. Rev. E}\ }\textbf {\bibinfo {volume} {60}},\ \bibinfo
  {pages} {857} (\bibinfo {year} {1999})}\BibitemShut {NoStop}%
\bibitem [{\citenamefont {McCraty}\ \emph {et~al.}(2009)\citenamefont
  {McCraty}, \citenamefont {Atkinson}, \citenamefont {Tomasino},\ and\
  \citenamefont {Bradley}}]{mccraty2009coherent}%
  \BibitemOpen
  \bibfield  {author} {\bibinfo {author} {\bibfnamefont {R.}~\bibnamefont
  {McCraty}}, \bibinfo {author} {\bibfnamefont {M.}~\bibnamefont {Atkinson}},
  \bibinfo {author} {\bibfnamefont {D.}~\bibnamefont {Tomasino}}, \ and\
  \bibinfo {author} {\bibfnamefont {R.~T.}\ \bibnamefont {Bradley}},\
  }\bibfield  {title} {\enquote {\bibinfo {title} {The coherent heart
  heart-brain interactions, psychophysiological coherence, and the emergence of
  system-wide order},}\ }\href@noop {} {\bibfield  {journal} {\bibinfo
  {journal} {Integral Review}\ }\textbf {\bibinfo {volume} {5}} (\bibinfo
  {year} {2009})}\BibitemShut {NoStop}%
\bibitem [{\citenamefont {Thayer}\ and\ \citenamefont
  {Lane}(2009)}]{thayer2009claude}%
  \BibitemOpen
  \bibfield  {author} {\bibinfo {author} {\bibfnamefont {J.~F.}\ \bibnamefont
  {Thayer}}\ and\ \bibinfo {author} {\bibfnamefont {R.~D.}\ \bibnamefont
  {Lane}},\ }\bibfield  {title} {\enquote {\bibinfo {title} {Claude bernard and
  the heart--brain connection: Further elaboration of a model of neurovisceral
  integration},}\ }\href@noop {} {\bibfield  {journal} {\bibinfo  {journal}
  {Neurosci. Biobehav. Rev.}\ }\textbf {\bibinfo {volume} {33}},\ \bibinfo
  {pages} {81--88} (\bibinfo {year} {2009})}\BibitemShut {NoStop}%
\bibitem [{\citenamefont {Mondal}\ and\ \citenamefont
  {Thomas}(2021)}]{mondal2021mitigation}%
  \BibitemOpen
  \bibfield  {author} {\bibinfo {author} {\bibfnamefont {S.}~\bibnamefont
  {Mondal}}\ and\ \bibinfo {author} {\bibfnamefont {N.}~\bibnamefont
  {Thomas}},\ }\bibfield  {title} {\enquote {\bibinfo {title} {Mitigation of
  thermoacoustic instability through amplitude death: Model and experiments},}\
  }in\ \href@noop {} {\emph {\bibinfo {booktitle} {Sustainable Development for
  Energy, Power, and Propulsion}}}\ (\bibinfo  {publisher} {Springer},\
  \bibinfo {year} {2021})\ pp.\ \bibinfo {pages} {287--322}\BibitemShut
  {NoStop}%
\bibitem [{\citenamefont {Huang}\ and\ \citenamefont
  {Yang}(2009)}]{huang2009dynamics}%
  \BibitemOpen
  \bibfield  {author} {\bibinfo {author} {\bibfnamefont {Y.}~\bibnamefont
  {Huang}}\ and\ \bibinfo {author} {\bibfnamefont {V.}~\bibnamefont {Yang}},\
  }\bibfield  {title} {\enquote {\bibinfo {title} {Dynamics and stability of
  lean-premixed swirl-stabilized combustion},}\ }\href@noop {} {\bibfield
  {journal} {\bibinfo  {journal} {Prog. Energy Combust. Sci.}\ }\textbf
  {\bibinfo {volume} {35}},\ \bibinfo {pages} {293--364} (\bibinfo {year}
  {2009})}\BibitemShut {NoStop}%
\bibitem [{\citenamefont {Richards}, \citenamefont {Straub},\ and\
  \citenamefont {Robey}(2003)}]{richards2003passive}%
  \BibitemOpen
  \bibfield  {author} {\bibinfo {author} {\bibfnamefont {G.~A.}\ \bibnamefont
  {Richards}}, \bibinfo {author} {\bibfnamefont {D.~L.}\ \bibnamefont
  {Straub}}, \ and\ \bibinfo {author} {\bibfnamefont {E.~H.}\ \bibnamefont
  {Robey}},\ }\bibfield  {title} {\enquote {\bibinfo {title} {Passive control
  of combustion dynamics in stationary gas turbines},}\ }\href@noop {}
  {\bibfield  {journal} {\bibinfo  {journal} {J. Propuls. Power}\ }\textbf
  {\bibinfo {volume} {19}},\ \bibinfo {pages} {795--810} (\bibinfo {year}
  {2003})}\BibitemShut {NoStop}%
\bibitem [{\citenamefont {Zhao}\ and\ \citenamefont
  {Li}(2015)}]{zhao2015review}%
  \BibitemOpen
  \bibfield  {author} {\bibinfo {author} {\bibfnamefont {D.}~\bibnamefont
  {Zhao}}\ and\ \bibinfo {author} {\bibfnamefont {X.~Y.}\ \bibnamefont {Li}},\
  }\bibfield  {title} {\enquote {\bibinfo {title} {A review of acoustic dampers
  applied to combustion chambers in aerospace industry},}\ }\href@noop {}
  {\bibfield  {journal} {\bibinfo  {journal} {Prog. Aerosp. Sci.}\ }\textbf
  {\bibinfo {volume} {74}},\ \bibinfo {pages} {114--130} (\bibinfo {year}
  {2015})}\BibitemShut {NoStop}%
\bibitem [{\citenamefont {Schadow}\ and\ \citenamefont
  {Gutmark}(1992)}]{schadow1992combustion}%
  \BibitemOpen
  \bibfield  {author} {\bibinfo {author} {\bibfnamefont {K.~C.}\ \bibnamefont
  {Schadow}}\ and\ \bibinfo {author} {\bibfnamefont {E.}~\bibnamefont
  {Gutmark}},\ }\bibfield  {title} {\enquote {\bibinfo {title} {Combustion
  instability related to vortex shedding in dump combustors and their passive
  control},}\ }\href@noop {} {\bibfield  {journal} {\bibinfo  {journal} {Prog.
  Energy Combust. Sci.}\ }\textbf {\bibinfo {volume} {18}},\ \bibinfo {pages}
  {117--132} (\bibinfo {year} {1992})}\BibitemShut {NoStop}%
\bibitem [{\citenamefont {Candel}(2002)}]{candel2002combustion}%
  \BibitemOpen
  \bibfield  {author} {\bibinfo {author} {\bibfnamefont {S.}~\bibnamefont
  {Candel}},\ }\bibfield  {title} {\enquote {\bibinfo {title} {Combustion
  dynamics and control: {P}rogress and challenges},}\ }\href@noop {} {\bibfield
   {journal} {\bibinfo  {journal} {Proc. Combust. Inst.}\ }\textbf {\bibinfo
  {volume} {29}},\ \bibinfo {pages} {1--28} (\bibinfo {year}
  {2002})}\BibitemShut {NoStop}%
\bibitem [{\citenamefont {Poinsot}(2017)}]{poinsot2017prediction}%
  \BibitemOpen
  \bibfield  {author} {\bibinfo {author} {\bibfnamefont {T.}~\bibnamefont
  {Poinsot}},\ }\bibfield  {title} {\enquote {\bibinfo {title} {Prediction and
  control of combustion instabilities in real engines},}\ }\href@noop {}
  {\bibfield  {journal} {\bibinfo  {journal} {Proc. Combust. Inst.}\ }\textbf
  {\bibinfo {volume} {36}},\ \bibinfo {pages} {1--28} (\bibinfo {year}
  {2017})}\BibitemShut {NoStop}%
\bibitem [{\citenamefont {Dowling}\ and\ \citenamefont
  {Morgans}(2005)}]{dowling2005feedback}%
  \BibitemOpen
  \bibfield  {author} {\bibinfo {author} {\bibfnamefont {A.~P.}\ \bibnamefont
  {Dowling}}\ and\ \bibinfo {author} {\bibfnamefont {A.~S.}\ \bibnamefont
  {Morgans}},\ }\bibfield  {title} {\enquote {\bibinfo {title} {Feedback
  control of combustion oscillations},}\ }\href@noop {} {\bibfield  {journal}
  {\bibinfo  {journal} {Annu. Rev. Fluid Mech.}\ }\textbf {\bibinfo {volume}
  {37}},\ \bibinfo {pages} {151--182} (\bibinfo {year} {2005})}\BibitemShut
  {NoStop}%
\bibitem [{\citenamefont {Heckl}(1988)}]{heckl1988active}%
  \BibitemOpen
  \bibfield  {author} {\bibinfo {author} {\bibfnamefont {M.~A.}\ \bibnamefont
  {Heckl}},\ }\bibfield  {title} {\enquote {\bibinfo {title} {{Active control
  of the noise from a Rijke tube}},}\ }\href@noop {} {\bibfield  {journal}
  {\bibinfo  {journal} {J. Sound Vib.}\ }\textbf {\bibinfo {volume} {124}},\
  \bibinfo {pages} {117--133} (\bibinfo {year} {1988})}\BibitemShut {NoStop}%
\bibitem [{\citenamefont {Lang}, \citenamefont {Poinsot},\ and\ \citenamefont
  {Candel}(1987)}]{lang1987active}%
  \BibitemOpen
  \bibfield  {author} {\bibinfo {author} {\bibfnamefont {W.}~\bibnamefont
  {Lang}}, \bibinfo {author} {\bibfnamefont {T.}~\bibnamefont {Poinsot}}, \
  and\ \bibinfo {author} {\bibfnamefont {S.}~\bibnamefont {Candel}},\
  }\bibfield  {title} {\enquote {\bibinfo {title} {Active control of combustion
  instability},}\ }\href@noop {} {\bibfield  {journal} {\bibinfo  {journal}
  {Combust. Flame}\ }\textbf {\bibinfo {volume} {70}},\ \bibinfo {pages}
  {281--289} (\bibinfo {year} {1987})}\BibitemShut {NoStop}%
\bibitem [{\citenamefont {Illingworth}\ and\ \citenamefont
  {Morgans}(2010)}]{illingworth2010advances}%
  \BibitemOpen
  \bibfield  {author} {\bibinfo {author} {\bibfnamefont {S.~J.}\ \bibnamefont
  {Illingworth}}\ and\ \bibinfo {author} {\bibfnamefont {A.~S.}\ \bibnamefont
  {Morgans}},\ }\bibfield  {title} {\enquote {\bibinfo {title} {Advances in
  feedback control of the rijke tube thermoacoustic instability.}}\ }\href@noop
  {} {\bibfield  {journal} {\bibinfo  {journal} {Int. J. Flow Control.}\
  }\textbf {\bibinfo {volume} {2}} (\bibinfo {year} {2010})}\BibitemShut
  {NoStop}%
\bibitem [{\citenamefont {Li}\ and\ \citenamefont
  {Zhao}(2016)}]{li2016feedback}%
  \BibitemOpen
  \bibfield  {author} {\bibinfo {author} {\bibfnamefont {X.}~\bibnamefont
  {Li}}\ and\ \bibinfo {author} {\bibfnamefont {D.}~\bibnamefont {Zhao}},\
  }\bibfield  {title} {\enquote {\bibinfo {title} {Feedback control of
  self-sustained nonlinear combustion oscillations},}\ }\href@noop {}
  {\bibfield  {journal} {\bibinfo  {journal} {J. Eng. Gas Turbine Power}\
  }\textbf {\bibinfo {volume} {138}} (\bibinfo {year} {2016})}\BibitemShut
  {NoStop}%
\bibitem [{\citenamefont {Blonbou}\ \emph {et~al.}(2000)\citenamefont
  {Blonbou}, \citenamefont {Laverdant}, \citenamefont {Zaleski},\ and\
  \citenamefont {Kuentzmann}}]{blonbou2000active}%
  \BibitemOpen
  \bibfield  {author} {\bibinfo {author} {\bibfnamefont {R.}~\bibnamefont
  {Blonbou}}, \bibinfo {author} {\bibfnamefont {A.}~\bibnamefont {Laverdant}},
  \bibinfo {author} {\bibfnamefont {S.}~\bibnamefont {Zaleski}}, \ and\
  \bibinfo {author} {\bibfnamefont {P.}~\bibnamefont {Kuentzmann}},\ }\bibfield
   {title} {\enquote {\bibinfo {title} {Active control of combustion
  instabilities on a rijke tube using neural networks},}\ }\href@noop {}
  {\bibfield  {journal} {\bibinfo  {journal} {Proc. Combust. Inst.}\ }\textbf
  {\bibinfo {volume} {28}},\ \bibinfo {pages} {747--755} (\bibinfo {year}
  {2000})}\BibitemShut {NoStop}%
\bibitem [{\citenamefont {Vaudrey}, \citenamefont {Baumann},\ and\
  \citenamefont {Saunders}(2003)}]{vaudrey2003time}%
  \BibitemOpen
  \bibfield  {author} {\bibinfo {author} {\bibfnamefont {M.~A.}\ \bibnamefont
  {Vaudrey}}, \bibinfo {author} {\bibfnamefont {W.~T.}\ \bibnamefont
  {Baumann}}, \ and\ \bibinfo {author} {\bibfnamefont {W.~R.}\ \bibnamefont
  {Saunders}},\ }\bibfield  {title} {\enquote {\bibinfo {title} {Time-averaged
  gradient control of thermoacoustic instabilities},}\ }\href@noop {}
  {\bibfield  {journal} {\bibinfo  {journal} {J. Propuls. Power}\ }\textbf
  {\bibinfo {volume} {19}},\ \bibinfo {pages} {830--836} (\bibinfo {year}
  {2003})}\BibitemShut {NoStop}%
\bibitem [{\citenamefont {Rubio-Hervas}, \citenamefont {Zhao},\ and\
  \citenamefont {Reyhanoglu}(2015)}]{rubio2015nonlinear}%
  \BibitemOpen
  \bibfield  {author} {\bibinfo {author} {\bibfnamefont {J.}~\bibnamefont
  {Rubio-Hervas}}, \bibinfo {author} {\bibfnamefont {D.}~\bibnamefont {Zhao}},
  \ and\ \bibinfo {author} {\bibfnamefont {M.}~\bibnamefont {Reyhanoglu}},\
  }\bibfield  {title} {\enquote {\bibinfo {title} {Nonlinear feedback control
  of self-sustained thermoacoustic oscillations},}\ }\href@noop {} {\bibfield
  {journal} {\bibinfo  {journal} {Aerosp. Sci. Technol.}\ }\textbf {\bibinfo
  {volume} {41}},\ \bibinfo {pages} {209--215} (\bibinfo {year}
  {2015})}\BibitemShut {NoStop}%
\bibitem [{\citenamefont {Zalluhoglu}, \citenamefont {Kammer},\ and\
  \citenamefont {Olgac}(2016)}]{zalluhoglu2016delayed}%
  \BibitemOpen
  \bibfield  {author} {\bibinfo {author} {\bibfnamefont {U.}~\bibnamefont
  {Zalluhoglu}}, \bibinfo {author} {\bibfnamefont {A.~S.}\ \bibnamefont
  {Kammer}}, \ and\ \bibinfo {author} {\bibfnamefont {N.}~\bibnamefont
  {Olgac}},\ }\bibfield  {title} {\enquote {\bibinfo {title} {Delayed feedback
  control laws for rijke tube thermoacoustic instability, synthesis, and
  experimental validation},}\ }\href@noop {} {\bibfield  {journal} {\bibinfo
  {journal} {IEEE Trans. Control Syst. Technol .}\ }\textbf {\bibinfo {volume}
  {24}},\ \bibinfo {pages} {1861--1868} (\bibinfo {year} {2016})}\BibitemShut
  {NoStop}%
\bibitem [{\citenamefont {Atay}(1998)}]{atay1998van}%
  \BibitemOpen
  \bibfield  {author} {\bibinfo {author} {\bibfnamefont {F.~M.}\ \bibnamefont
  {Atay}},\ }\bibfield  {title} {\enquote {\bibinfo {title} {Van der pol's
  oscillator under delayed feedback},}\ }\href@noop {} {\bibfield  {journal}
  {\bibinfo  {journal} {J. Sound Vib.}\ }\textbf {\bibinfo {volume} {218}},\
  \bibinfo {pages} {333--339} (\bibinfo {year} {1998})}\BibitemShut {NoStop}%
\bibitem [{\citenamefont {Reddy}, \citenamefont {Sen},\ and\ \citenamefont
  {Johnston}(2000)}]{reddy2000dynamics}%
  \BibitemOpen
  \bibfield  {author} {\bibinfo {author} {\bibfnamefont {D.~R.}\ \bibnamefont
  {Reddy}}, \bibinfo {author} {\bibfnamefont {A.}~\bibnamefont {Sen}}, \ and\
  \bibinfo {author} {\bibfnamefont {G.~L.}\ \bibnamefont {Johnston}},\
  }\bibfield  {title} {\enquote {\bibinfo {title} {Dynamics of a limit cycle
  oscillator under time delayed linear and nonlinear feedbacks},}\ }\href@noop
  {} {\bibfield  {journal} {\bibinfo  {journal} {Phys. D: Nonlinear Phenom.}\
  }\textbf {\bibinfo {volume} {144}},\ \bibinfo {pages} {335--357} (\bibinfo
  {year} {2000})}\BibitemShut {NoStop}%
\bibitem [{\citenamefont {Xu}\ and\ \citenamefont
  {Chung}(2003)}]{xu2003effects}%
  \BibitemOpen
  \bibfield  {author} {\bibinfo {author} {\bibfnamefont {J.}~\bibnamefont
  {Xu}}\ and\ \bibinfo {author} {\bibfnamefont {K.}~\bibnamefont {Chung}},\
  }\bibfield  {title} {\enquote {\bibinfo {title} {Effects of time delayed
  position feedback on a van der pol--duffing oscillator},}\ }\href@noop {}
  {\bibfield  {journal} {\bibinfo  {journal} {Phys. D: Nonlinear Phenom.}\
  }\textbf {\bibinfo {volume} {180}},\ \bibinfo {pages} {17--39} (\bibinfo
  {year} {2003})}\BibitemShut {NoStop}%
\bibitem [{\citenamefont {Dines}(1984)}]{dines1984active}%
  \BibitemOpen
  \bibfield  {author} {\bibinfo {author} {\bibfnamefont {P.~J.}\ \bibnamefont
  {Dines}},\ }\emph {\bibinfo {title} {Active control of flame noise}},\
  \href@noop {} {Ph.D. thesis},\ \bibinfo  {school} {University of Cambridge,
  UK} (\bibinfo {year} {1984})\BibitemShut {NoStop}%
\bibitem [{\citenamefont {Ffowcs~Williams}(1984)}]{ffowcs1984review}%
  \BibitemOpen
  \bibfield  {author} {\bibinfo {author} {\bibfnamefont {J.}~\bibnamefont
  {Ffowcs~Williams}},\ }\bibfield  {title} {\enquote {\bibinfo {title} {Review
  lecture-anti-sound},}\ }\href@noop {} {\bibfield  {journal} {\bibinfo
  {journal} {Proc. R. Soc. A: Math. Phys. Eng. Sci.}\ }\textbf {\bibinfo
  {volume} {395}},\ \bibinfo {pages} {63--88} (\bibinfo {year}
  {1984})}\BibitemShut {NoStop}%
\bibitem [{\citenamefont {Candel}(1992)}]{candel1992combustion}%
  \BibitemOpen
  \bibfield  {author} {\bibinfo {author} {\bibfnamefont {S.~M.}\ \bibnamefont
  {Candel}},\ }\bibfield  {title} {\enquote {\bibinfo {title} {Combustion
  instabilities coupled by pressure waves and their active control},}\ }in\
  \href@noop {} {\emph {\bibinfo {booktitle} {Symposium (International) on
  Combustion}}},\ Vol.~\bibinfo {volume} {24}\ (\bibinfo {organization}
  {Elsevier},\ \bibinfo {year} {1992})\ pp.\ \bibinfo {pages}
  {1277--1296}\BibitemShut {NoStop}%
\bibitem [{\citenamefont {Staubli}(1987)}]{staubli1987entrainment}%
  \BibitemOpen
  \bibfield  {author} {\bibinfo {author} {\bibfnamefont {T.}~\bibnamefont
  {Staubli}},\ }\bibfield  {title} {\enquote {\bibinfo {title} {Entrainment of
  self-sustained flow oscillations: phase locking or asynchronous quenching?}}\
  }\href@noop {} {\bibfield  {journal} {\bibinfo  {journal} {J. Appl. Mech.}\
  }\textbf {\bibinfo {volume} {54}},\ \bibinfo {pages} {706--712} (\bibinfo
  {year} {1987})}\BibitemShut {NoStop}%
\bibitem [{\citenamefont {Taira}\ and\ \citenamefont
  {Nakao}(2018)}]{taira2018phase}%
  \BibitemOpen
  \bibfield  {author} {\bibinfo {author} {\bibfnamefont {K.}~\bibnamefont
  {Taira}}\ and\ \bibinfo {author} {\bibfnamefont {H.}~\bibnamefont {Nakao}},\
  }\bibfield  {title} {\enquote {\bibinfo {title} {Phase-response analysis of
  synchronization for periodic flows},}\ }\href@noop {} {\bibfield  {journal}
  {\bibinfo  {journal} {J. Fluid Mech.}\ }\textbf {\bibinfo {volume} {846}}
  (\bibinfo {year} {2018})}\BibitemShut {NoStop}%
\bibitem [{\citenamefont {Ohe}\ and\ \citenamefont
  {Takeda}(1974)}]{ohe1974asynchronous}%
  \BibitemOpen
  \bibfield  {author} {\bibinfo {author} {\bibfnamefont {K.}~\bibnamefont
  {Ohe}}\ and\ \bibinfo {author} {\bibfnamefont {S.}~\bibnamefont {Takeda}},\
  }\bibfield  {title} {\enquote {\bibinfo {title} {Asynchronous quenching and
  resonance excitation of ionization waves in positive columns},}\ }\href@noop
  {} {\bibfield  {journal} {\bibinfo  {journal} {Contrib. Plasma Phys.}\
  }\textbf {\bibinfo {volume} {14}},\ \bibinfo {pages} {55--65} (\bibinfo
  {year} {1974})}\BibitemShut {NoStop}%
\bibitem [{\citenamefont {Fjeld}(1974)}]{fjeld1974relaxed}%
  \BibitemOpen
  \bibfield  {author} {\bibinfo {author} {\bibfnamefont {M.}~\bibnamefont
  {Fjeld}},\ }\bibfield  {title} {\enquote {\bibinfo {title} {Relaxed controls
  in asynchronous quenching and dynamical optimization},}\ }\href@noop {}
  {\bibfield  {journal} {\bibinfo  {journal} {Chem. Eng. Sci.}\ }\textbf
  {\bibinfo {volume} {29}},\ \bibinfo {pages} {921--933} (\bibinfo {year}
  {1974})}\BibitemShut {NoStop}%
\bibitem [{\citenamefont {Shimura}(1967)}]{shimura1967analysis}%
  \BibitemOpen
  \bibfield  {author} {\bibinfo {author} {\bibfnamefont {M.}~\bibnamefont
  {Shimura}},\ }\bibfield  {title} {\enquote {\bibinfo {title} {Analysis of
  some nonlinear phenomena in a transmission line},}\ }\href@noop {} {\bibfield
   {journal} {\bibinfo  {journal} {IEEE Trans. Circuits Syst. I Regul. Pap.}\
  }\textbf {\bibinfo {volume} {14}},\ \bibinfo {pages} {60--68} (\bibinfo
  {year} {1967})}\BibitemShut {NoStop}%
\bibitem [{\citenamefont {Evesque}, \citenamefont {Dowling},\ and\
  \citenamefont {Annaswamy}(2003)}]{evesque2003self}%
  \BibitemOpen
  \bibfield  {author} {\bibinfo {author} {\bibfnamefont {S.}~\bibnamefont
  {Evesque}}, \bibinfo {author} {\bibfnamefont {A.~P.}\ \bibnamefont
  {Dowling}}, \ and\ \bibinfo {author} {\bibfnamefont {A.~M.}\ \bibnamefont
  {Annaswamy}},\ }\bibfield  {title} {\enquote {\bibinfo {title} {Self-tuning
  regulators for combustion oscillations},}\ }\href@noop {} {\bibfield
  {journal} {\bibinfo  {journal} {Proc. R. Soc. A: Math. Phys. Eng. Sci.}\
  }\textbf {\bibinfo {volume} {459}},\ \bibinfo {pages} {1709--1749} (\bibinfo
  {year} {2003})}\BibitemShut {NoStop}%
\bibitem [{\citenamefont {Guan}\ \emph
  {et~al.}(2019{\natexlab{d}})\citenamefont {Guan}, \citenamefont {Gupta},
  \citenamefont {Kashinath},\ and\ \citenamefont {Li}}]{guan2019openSymp}%
  \BibitemOpen
  \bibfield  {author} {\bibinfo {author} {\bibfnamefont {Y.}~\bibnamefont
  {Guan}}, \bibinfo {author} {\bibfnamefont {V.}~\bibnamefont {Gupta}},
  \bibinfo {author} {\bibfnamefont {K.}~\bibnamefont {Kashinath}}, \ and\
  \bibinfo {author} {\bibfnamefont {L.~K.~B.}\ \bibnamefont {Li}},\ }\bibfield
  {title} {\enquote {\bibinfo {title} {Open-loop control of periodic
  thermoacoustic oscillations: experiments and low-order modelling in a
  synchronization framework},}\ }\href@noop {} {\bibfield  {journal} {\bibinfo
  {journal} {Proc. Combust. Inst.}\ }\textbf {\bibinfo {volume} {37}},\
  \bibinfo {pages} {5315--5323} (\bibinfo {year}
  {2019}{\natexlab{d}})}\BibitemShut {NoStop}%
\bibitem [{\citenamefont {Skene}\ and\ \citenamefont
  {Taira}(2022)}]{skene2022phase}%
  \BibitemOpen
  \bibfield  {author} {\bibinfo {author} {\bibfnamefont {C.~S.}\ \bibnamefont
  {Skene}}\ and\ \bibinfo {author} {\bibfnamefont {K.}~\bibnamefont {Taira}},\
  }\bibfield  {title} {\enquote {\bibinfo {title} {Phase-reduction analysis of
  periodic thermoacoustic oscillations in a rijke tube},}\ }\href@noop {}
  {\bibfield  {journal} {\bibinfo  {journal} {J. Fluid Mech.}\ }\textbf
  {\bibinfo {volume} {933}} (\bibinfo {year} {2022})}\BibitemShut {NoStop}%
\bibitem [{\citenamefont {Minorsky}(1967)}]{minorsky1967comments}%
  \BibitemOpen
  \bibfield  {author} {\bibinfo {author} {\bibfnamefont {N.}~\bibnamefont
  {Minorsky}},\ }\bibfield  {title} {\enquote {\bibinfo {title} {{Comments" On
  asynchronous quenching"}},}\ }\href@noop {} {\bibfield  {journal} {\bibinfo
  {journal} {IEEE Trans. Autom. Control}\ }\textbf {\bibinfo {volume} {12}},\
  \bibinfo {pages} {225--227} (\bibinfo {year} {1967})}\BibitemShut {NoStop}%
\bibitem [{\citenamefont {Walsh}\ and\ \citenamefont
  {Fletcher}(2004)}]{walsh2004gas}%
  \BibitemOpen
  \bibfield  {author} {\bibinfo {author} {\bibfnamefont {P.~P.}\ \bibnamefont
  {Walsh}}\ and\ \bibinfo {author} {\bibfnamefont {P.}~\bibnamefont
  {Fletcher}},\ }\href@noop {} {\emph {\bibinfo {title} {Gas Turbine
  Performance}}}\ (\bibinfo  {publisher} {John Wiley \& Sons},\ \bibinfo {year}
  {2004})\BibitemShut {NoStop}%
\bibitem [{\citenamefont {Biwa}\ \emph {et~al.}(2016)\citenamefont {Biwa},
  \citenamefont {Sawada}, \citenamefont {Hyodo},\ and\ \citenamefont
  {Kato}}]{biwa2016suppression}%
  \BibitemOpen
  \bibfield  {author} {\bibinfo {author} {\bibfnamefont {T.}~\bibnamefont
  {Biwa}}, \bibinfo {author} {\bibfnamefont {Y.}~\bibnamefont {Sawada}},
  \bibinfo {author} {\bibfnamefont {H.}~\bibnamefont {Hyodo}}, \ and\ \bibinfo
  {author} {\bibfnamefont {S.}~\bibnamefont {Kato}},\ }\bibfield  {title}
  {\enquote {\bibinfo {title} {Suppression of spontaneous gas oscillations by
  acoustic self-feedback},}\ }\href@noop {} {\bibfield  {journal} {\bibinfo
  {journal} {Phys. Rev. Appl.}\ }\textbf {\bibinfo {volume} {6}},\ \bibinfo
  {pages} {044020} (\bibinfo {year} {2016})}\BibitemShut {NoStop}%
\bibitem [{\citenamefont {Lato}, \citenamefont {Mohany},\ and\ \citenamefont
  {Hassan}(2019)}]{lato2019passive}%
  \BibitemOpen
  \bibfield  {author} {\bibinfo {author} {\bibfnamefont {T.}~\bibnamefont
  {Lato}}, \bibinfo {author} {\bibfnamefont {A.}~\bibnamefont {Mohany}}, \ and\
  \bibinfo {author} {\bibfnamefont {M.}~\bibnamefont {Hassan}},\ }\bibfield
  {title} {\enquote {\bibinfo {title} {A passive damping device for suppressing
  acoustic pressure pulsations: The infinity tube},}\ }\href@noop {} {\bibfield
   {journal} {\bibinfo  {journal} {J. Acoust. Soc. Am.}\ }\textbf {\bibinfo
  {volume} {146}},\ \bibinfo {pages} {4534--4544} (\bibinfo {year}
  {2019})}\BibitemShut {NoStop}%
\bibitem [{\citenamefont {Atay}(2003)}]{atay2003total}%
  \BibitemOpen
  \bibfield  {author} {\bibinfo {author} {\bibfnamefont {F.~M.}\ \bibnamefont
  {Atay}},\ }\bibfield  {title} {\enquote {\bibinfo {title} {Total and partial
  amplitude death in networks of diffusively coupled oscillators},}\
  }\href@noop {} {\bibfield  {journal} {\bibinfo  {journal} {Physica D}\
  }\textbf {\bibinfo {volume} {183}},\ \bibinfo {pages} {1--18} (\bibinfo
  {year} {2003})}\BibitemShut {NoStop}%
\bibitem [{\citenamefont
  {Surovyatkina}(2005)}]{surovyatkina2005prebifurcation}%
  \BibitemOpen
  \bibfield  {author} {\bibinfo {author} {\bibfnamefont {E.}~\bibnamefont
  {Surovyatkina}},\ }\bibfield  {title} {\enquote {\bibinfo {title}
  {Prebifurcation noise amplification and noise-dependent hysteresis as
  indicators of bifurcations in nonlinear geophysical systems},}\ }\href@noop
  {} {\bibfield  {journal} {\bibinfo  {journal} {Nonlinear Process. Geophys.}\
  }\textbf {\bibinfo {volume} {12}},\ \bibinfo {pages} {25--29} (\bibinfo
  {year} {2005})}\BibitemShut {NoStop}%
\bibitem [{\citenamefont {Gopalakrishnan}\ \emph
  {et~al.}(2016{\natexlab{b}})\citenamefont {Gopalakrishnan}, \citenamefont
  {Sharma}, \citenamefont {John}, \citenamefont {Dutta},\ and\ \citenamefont
  {Sujith}}]{gopalakrishnan2016early}%
  \BibitemOpen
  \bibfield  {author} {\bibinfo {author} {\bibfnamefont {E.~A.}\ \bibnamefont
  {Gopalakrishnan}}, \bibinfo {author} {\bibfnamefont {Y.}~\bibnamefont
  {Sharma}}, \bibinfo {author} {\bibfnamefont {T.}~\bibnamefont {John}},
  \bibinfo {author} {\bibfnamefont {P.~S.}\ \bibnamefont {Dutta}}, \ and\
  \bibinfo {author} {\bibfnamefont {R.~I.}\ \bibnamefont {Sujith}},\ }\bibfield
   {title} {\enquote {\bibinfo {title} {Early warning signals for critical
  transitions in a thermoacoustic system},}\ }\href@noop {} {\bibfield
  {journal} {\bibinfo  {journal} {Sci. Rep.}\ }\textbf {\bibinfo {volume}
  {6}},\ \bibinfo {pages} {35310} (\bibinfo {year}
  {2016}{\natexlab{b}})}\BibitemShut {NoStop}%
\bibitem [{\citenamefont {Pavithran}\ \emph {et~al.}(2021)\citenamefont
  {Pavithran}, \citenamefont {Unni}, \citenamefont {Varghese}, \citenamefont
  {Sujith}, \citenamefont {Saha}, \citenamefont {Marwan},\ and\ \citenamefont
  {Kurths}}]{pavithran2021predicting}%
  \BibitemOpen
  \bibfield  {author} {\bibinfo {author} {\bibfnamefont {I.}~\bibnamefont
  {Pavithran}}, \bibinfo {author} {\bibfnamefont {V.~R.}\ \bibnamefont {Unni}},
  \bibinfo {author} {\bibfnamefont {A.~J.}\ \bibnamefont {Varghese}}, \bibinfo
  {author} {\bibfnamefont {R.~I.}\ \bibnamefont {Sujith}}, \bibinfo {author}
  {\bibfnamefont {A.}~\bibnamefont {Saha}}, \bibinfo {author} {\bibfnamefont
  {N.}~\bibnamefont {Marwan}}, \ and\ \bibinfo {author} {\bibfnamefont
  {J.}~\bibnamefont {Kurths}},\ }\bibfield  {title} {\enquote {\bibinfo {title}
  {Predicting the amplitude of thermoacoustic instability using universal
  scaling behaviour},}\ }in\ \href@noop {} {\emph {\bibinfo {booktitle}
  {Proceedings of ASME Turbo Expo 2021: Turbomachinery Technical Conference and
  Exposition}}},\ \bibinfo {series and number} {\bibinfo {number}
  {GT2021-60074}}\ (\bibinfo {year} {2021})\BibitemShut {NoStop}%
\bibitem [{\citenamefont {Dakos}\ \emph {et~al.}(2012)\citenamefont {Dakos},
  \citenamefont {Carpenter}, \citenamefont {Brock}, \citenamefont {Ellison},
  \citenamefont {Guttal}, \citenamefont {Ives}, \citenamefont {Kefi},
  \citenamefont {Livina}, \citenamefont {Seekell}, \citenamefont {van Nes}
  \emph {et~al.}}]{dakos2012methods}%
  \BibitemOpen
  \bibfield  {author} {\bibinfo {author} {\bibfnamefont {V.}~\bibnamefont
  {Dakos}}, \bibinfo {author} {\bibfnamefont {S.~R.}\ \bibnamefont
  {Carpenter}}, \bibinfo {author} {\bibfnamefont {W.~A.}\ \bibnamefont
  {Brock}}, \bibinfo {author} {\bibfnamefont {A.~M.}\ \bibnamefont {Ellison}},
  \bibinfo {author} {\bibfnamefont {V.}~\bibnamefont {Guttal}}, \bibinfo
  {author} {\bibfnamefont {A.~R.}\ \bibnamefont {Ives}}, \bibinfo {author}
  {\bibfnamefont {S.}~\bibnamefont {Kefi}}, \bibinfo {author} {\bibfnamefont
  {V.}~\bibnamefont {Livina}}, \bibinfo {author} {\bibfnamefont {D.~A.}\
  \bibnamefont {Seekell}}, \bibinfo {author} {\bibfnamefont {E.~H.}\
  \bibnamefont {van Nes}},  \emph {et~al.},\ }\bibfield  {title} {\enquote
  {\bibinfo {title} {Methods for detecting early warnings of critical
  transitions in time series illustrated using simulated ecological data},}\
  }\href@noop {} {\bibfield  {journal} {\bibinfo  {journal} {PLoS One}\
  }\textbf {\bibinfo {volume} {7}},\ \bibinfo {pages} {e41010} (\bibinfo {year}
  {2012})}\BibitemShut {NoStop}%
\bibitem [{\citenamefont {Hurst}(1951)}]{hurst1951long}%
  \BibitemOpen
  \bibfield  {author} {\bibinfo {author} {\bibfnamefont {H.~E.}\ \bibnamefont
  {Hurst}},\ }\bibfield  {title} {\enquote {\bibinfo {title} {Long-term storage
  capacity of reservoirs},}\ }\href@noop {} {\bibfield  {journal} {\bibinfo
  {journal} {Trans. Amer. Soc. Civil Eng.}\ }\textbf {\bibinfo {volume}
  {116}},\ \bibinfo {pages} {770--799} (\bibinfo {year} {1951})}\BibitemShut
  {NoStop}%
\bibitem [{\citenamefont {Mandelbrot}(1977)}]{mandelbrot1977fractals}%
  \BibitemOpen
  \bibfield  {author} {\bibinfo {author} {\bibfnamefont {B.~B.}\ \bibnamefont
  {Mandelbrot}},\ }\href@noop {} {\emph {\bibinfo {title} {Fractals - Form,
  Chance and Dimension}}}\ (\bibinfo  {publisher} {Freeman, San Francisco},\
  \bibinfo {year} {1977})\BibitemShut {NoStop}%
\bibitem [{\citenamefont {Pavithran}, \citenamefont {Unni},\ and\ \citenamefont
  {Sujith}(2021)}]{pavithran2021critical}%
  \BibitemOpen
  \bibfield  {author} {\bibinfo {author} {\bibfnamefont {I.}~\bibnamefont
  {Pavithran}}, \bibinfo {author} {\bibfnamefont {V.~R.}\ \bibnamefont {Unni}},
  \ and\ \bibinfo {author} {\bibfnamefont {R.~I.}\ \bibnamefont {Sujith}},\
  }\bibfield  {title} {\enquote {\bibinfo {title} {Critical transitions and
  their early warning signals in thermoacoustic systems},}\ }\href@noop {}
  {\bibfield  {journal} {\bibinfo  {journal} {Eur. Phys. J. Spec. Top.}\
  }\textbf {\bibinfo {volume} {230}},\ \bibinfo {pages} {3411--3432} (\bibinfo
  {year} {2021})}\BibitemShut {NoStop}%
\end{thebibliography}%

\end{document}